\documentclass[]{article}
\usepackage{amssymb}
\usepackage{amsmath}
\usepackage{mathtools}
\usepackage{subfig}
\usepackage{graphicx}
\usepackage[margin=1.0in]{geometry}
\usepackage{placeins}
\usepackage{color}
\usepackage{soul}
\usepackage{hyperref}
\usepackage[utf8]{inputenc}
\usepackage{slashed}
\usepackage{multirow}
\usepackage{mathrsfs}
\usepackage{fixltx2e}
\usepackage{setspace}
\usepackage{enumerate}
\usepackage{bm}
\usepackage{authblk}

\usepackage{CJKutf8}

\newcommand{\Knaught}{\smash{\overline{K}}\vphantom{K}^{0}}
\newcommand{\bigO}{\mathcal{O}}
\newcommand{\SMOM}{\mathrm{SMOM}}
\newcommand{\SMOMg}{\mathrm{SMOM}_{\gamma^{\mu}}}
\DeclareMathOperator{\tr}{tr}

\begin{document}

\title{\textbf{The Low Energy Constants of $\bm{SU(2)}$ Partially Quenched Chiral Perturbation Theory from $\bm{N_{f} = 2+1}$ Domain Wall QCD}}

\date{}

\begin{CJK*}{UTF8}{}
\CJKfamily{min}

\author[1]{P.A.~Boyle}
\author[2]{N.H.~Christ}
\author[3]{N.~Garron}
\author[4]{C.~Jung}
\author[5]{A.~J\"{u}ttner}
\author[6]{C.~Kelly}
\author[4]{R.D.~Mawhinney}
\author[4]{G.~McGlynn}
\author[4]{D.J.~Murphy}
\author[7,8,6]{S.~Ohta (太田滋生)}
\author[5]{A.~Portelli\footnote{Present address: SUPA, School of Physics, The University of Edinburgh, Edinburgh EH9 3JZ, UK}}
\author[5]{C.T.~Sachrajda}
\affil[1]{\normalsize \textit{SUPA, School of Physics, The University of Edinburgh, Edinburgh EH9 3JZ, UK}}
\affil[2]{\normalsize \textit{Department of Physics, Columbia University, New York, NY 10027, USA}}
\affil[3]{\normalsize \textit{Centre for Mathematical Sciences, Plymouth University,
Plymouth, PL4 8AA, UK}}
\affil[4]{\normalsize \textit{Department of Physics, Brookhaven National Laboratory, Upton, NY 11973, USA}}
\affil[5]{\normalsize \textit{School of Physics and Astronomy, University of Southampton, Southampton SO17 1BJ, UK}}
\affil[6]{\normalsize \textit{RIKEN-BNL Research Center, Brookhaven National Laboratory, Upton, NY 11973, USA}}
\affil[7]{\normalsize \textit{Theory Center, KEK, Tsukuba, Ibaraki 305-0801, Japan}}
\affil[8]{\normalsize \textit{Department of Particle and Nuclear Physics, SOKENDAI, Hayama, Kanagawa 240-0193, Japan}}

\begin{flushright}
{\large Edinburgh 2015/26, KEK-TH-1861, RBRC-1151}
\end{flushright}

{\let \newpage \relax \maketitle}
\end{CJK*}


\begin{abstract}
	
We have performed fits of the pseudoscalar masses and decay constants, from a variety of RBC-UKQCD domain wall fermion ensembles, to $SU(2)$ partially quenched chiral perturbation theory at next-to leading order (NLO) and next-to-next-to leading order (NNLO). We report values for 9 NLO and 8 linearly independent combinations of NNLO partially quenched low energy constants, which we compare to other lattice and phenomenological determinations. We discuss the size of successive terms in the chiral expansion and use our large set of low energy constants to make predictions for mass splittings due to QCD isospin breaking effects and the S-wave $\pi \pi$ scattering lengths. We conclude that, for the range of pseudoscalar masses explored in this work, $115~\mathrm{MeV} \lesssim m_{\rm PS} \lesssim 430~\mathrm{MeV}$, the NNLO $SU(2)$ expansion is quite robust and can fit lattice data with percent-scale accuracy.

\end{abstract}


\section{Introduction}

Effective field theories (EFT) formalize the intuitive idea that to understand physics at a particular energy scale $E$, the full details of physics at much higher energy scales $\Lambda \gg E$ are not needed. After identifying the relevant degrees of freedom associated with scale $E$, one can write down a low-energy approximation, which differs from the full theory up to corrections which are powers in $E/\Lambda$. If the separation of scales is large, the approximation is arbitrarily good, and the precise form of the $E/\Lambda$ corrections need not be specified. In practice, high energy degrees of freedom do not need to be integrated out of the theory explicitly: it suffices to write down the most general low-energy effective Lagrangian containing all terms consistent with the symmetries of the full theory~\cite{Weinberg1979327}. An early, successful example is the Fermi theory of $\beta$ decay, which can be regarded as a low-energy approximation to the standard model obtained by integrating out the $W$ boson~\cite{9780511524370}. Effective field theories are widely employed in modern physics, and the standard model itself is an EFT likely modified by some yet-unknown new physics at sufficiently high energies. Renormalization plays an important role in defining effective field theories, both in understanding how heavy particle masses can enter at low scales via the Appelquist and Carazzone decoupling theorem~\cite{PhysRevD.11.2856}, and in handling higher loop calculations in the low energy effective Lagrangian. Correctly matching EFTs across particle mass thresholds is a crucial detail of precision calculations in the standard model~\cite{Georgi:1994qn}. \\

In this paper we discuss the physics of light pseudoscalar mesons, which played an important role in the development of the theory of the strong interactions --- Quantum Chromodynamics (QCD) --- and in the development of effective field theory techniques in general. The EFT of the light pseudoscalar mesons --- Chiral Perturbation Theory (ChPT) --- is both a prototypical example of an EFT and a theory whose corrections in powers of $E/\Lambda$ can be determined, since lattice techniques enable direct QCD calculations. These correction terms contain ``low energy constants'' (LECs) which must be determined by matching to QCD. In this paper we fit lattice QCD data for the light pseudoscalar mesons to the corresponding ChPT formulas to determine the LECs and to gain information about the accuracy of ChPT as an approximation to QCD at low energies. While the primary focus is the physics of QCD, it is also of general interest to explore a system where the reliability of calculating in an EFT truncated to some order --- we consider next-to leading order (NLO) and next-to-next-to leading order (NNLO) --- can be directly tested against calculations in the full theory. \\

QCD is highly nonlinear in the low energy regime, and lattice QCD provides the only known technique for calculating hadronic properties from first principles\footnote{A perturbative expansion in powers of the strong coupling constant, $g_{s}$, is only useful at very high energies.}. The QCD vacuum dynamically breaks the $SU(N_{f})_{L} \times SU(N_{f})_{R}$ chiral symmetry of QCD with $N_f$ massless quarks, at least when $N_{f} \leq 6$, giving rise to $N_f^2 -1$ pseudo-Goldstone bosons, which are the pseudoscalar mesons. The scale of this meson physics is lighter than the scales of other phenomenon in QCD provided the quark masses are not too large, suggesting an effective field theory description (ChPT). For quarks of nonzero mass, one is naturally led to consider an effective field theory expansion in powers of the masses and momenta. One obtains $SU(2)$ ChPT~\cite{Gasser:1983yg} or $SU(3)$ ChPT~\cite{Gasser:1984gg} depending on whether or not the strange quark is included. The $SU(2)$ theory allows for explicit calculations of pion physics, while the $SU(3)$ theory describes the pseudoscalar meson octet $(\pi, K, \eta)$. The matching of ChPT to QCD is encapsulated in the a priori unknown LECs, which parametrize the contributions from the various operators appearing in the ChPT Lagrangian.  \\

Historically, ChPT has been an important tool for lattice QCD practitioners, as the limitations of available computational resources required the use of unphysically heavy quarks to make calculations practical. Until recently, a typical lattice calculation was performed at several, heavy values of the input quark masses, and then extrapolated with ChPT to the quark masses found in nature to make physical predictions. This is the approach taken in all but the most recent of the RBC-UKQCD collaboration's domain wall QCD simulations~\cite{Allton:2008pn, Aoki:2010dy, Arthur:2012opa}. The reliability of ChPT as an approximation to QCD at the heavy, simulated points was largely left as an open question by these studies. \\

Recent advances in algorithms and computers have enabled computations directly at physical quark masses, minimizing the need for sophisticated chiral extrapolations. In the RBC-UKQCD collaboration's recent analysis of two physical mass M\"{o}bius domain wall fermion ensembles~\cite{Blum:2014tka} $SU(2)$ ChPT was only used to correct for small mistunings in the simulation parameters, resulting in modest $\bigO(1\%)$ corrections to the simulated pseudoscalar masses and decay constants. While ChPT-based extrapolations may no longer be necessary in lattice QCD, the availability of lattice data ranging from physical to much heavier than physical quark mass allows for a complementary study of the applicability of ChPT as a low energy approximation to QCD. In this paper we seek to:
\begin{enumerate}
	\item Determine as many of the low energy constants of $SU(2)$ ChPT as possible from our data, and
	\item Systematically study the behavior and range of applicability of the $SU(2)$ ChPT expansion up to next-to-next-to leading order (NNLO).
\end{enumerate}
Exploratory fits of an earlier RBC-UKQCD domain wall QCD data set to NNLO $SU(2)$ ChPT were first performed in Ref.~\cite{Mawhinney:2009jy}, but suffered from numerical instabilities. More recently, the BMW collaboration has studied the pion mass and decay constant in $SU(2)$ ChPT up to NNLO using staggered~\cite{Borsanyi:2012zv} and Wilson~\cite{Durr:2013goa} fermions. Fits of the pion mass, decay constant, and vector form factor computed using $\mathcal{O}(a)$-improved Wilson quarks to NNLO $SU(2)$ ChPT were performed by Brandt, J\"{u}ttner, and Wittig~\cite{Brandt:2013dua}. Our domain wall fermion analyses complement these studies, providing an additional fermion discretization with excellent chiral symmetry properties. In addition, we perform our fits using the more general formalism of partially quenched chiral perturbation theory (PQChPT), from which we can also readily extract the low energy constants of ordinary ChPT. An analogous study of fits of RBC/UKQCD domain wall fermion data to $SU(3)$ partially quenched ChPT at NLO and NNLO will be the topic of a subsequent paper \cite{su3_paper}. \\

We briefly discuss some of the issues that arise in fitting our data to ChPT, which we will elaborate on in later sections. First, given that perturbative expansions of four-dimensional field theories generally produce asymptotic series rather than convergent series, the ChPT expansion is expected not to be convergent, with new counterterms arising at each loop order due to the non-renormalizability of the theory. One can hope that the series has the correct hierarchy to give accurate results when truncated to the first few orders --- {\em i.e.} that each subsequent term is of smaller magnitude than the one that precedes it --- for the range of quark masses probed in a typical lattice simulation, but this is not guaranteed. Second, if a large data set with quark masses less than some bound is fit to a given order of ChPT, statistical tests of the goodness of fit will become arbitrarily poor as the statistical resolution of the data is improved. This occurs because truncations of the ChPT expansion are only an approximation to QCD --- eventually the data will be more accurate than the ChPT expansion can describe at a given order unless additional, higher order terms are added. This means that statistical goodness of fit criterion may initially show a reasonable fit to a small data set --- when the statistical errors exceed the systematic errors from truncating the expansion --- and then produce arbitrarily poor fits as more measurements are added and the statistical errors become smaller than the truncation errors. Finally, our fit procedure only gives us a self-consistent view of the properties of the expansion: we have data corresponding to a particular range of quark masses, which we fit to ChPT, and then ask whether the resulting expansion is sensible. While we have some freedom to vary the range of quark masses included in our fits, lattice QCD can, in principle, provide arbitrarily accurate data at arbitrarily small quark masses. For the time being we remain far from that situation.


\section{Partially Quenched Chiral Perturbation Theory at Next-to-Next-to-Leading Order}

The basic degrees of freedom in QCD are the quark fields, $q_{f}$, which transform in the fundamental representation of (color) $SU(3)$ and carry a flavor index $f$, and the gluon fields, $A^{a}_{\mu}$, which transform in the adjoint representation of (color) $SU(3)$ and mediate the strong nuclear force. In the limit of vanishing quark masses, the QCD Lagrangian with $N_{f}$ flavors of quarks
\begin{equation}
\label{eqn:qcd_lagrangian}
\mathcal{L}_{QCD} = -\frac{1}{4} G_{\mu \nu}^{a} G_{a}^{\mu \nu} + \sum\limits_{f} \overline{q}_{f} i \slashed{D} q_{f}
\end{equation}
has an exact $SU(N_{f})_{L} \times SU(N_{f})_{R}$ symmetry\footnote{Naively, the classical Lagrangian~\eqref{eqn:qcd_lagrangian} has an even larger $U(N_{f})_{L} \times U(N_{f})_{R}$ symmetry, but the $U(1)_{A}$ component is broken by the chiral anomaly and fails to be a symmetry of the quantum theory.}. This symmetry is spontaneously broken down to a single $SU(N_{f})_{V}$ subgroup by the QCD vacuum, giving rise to $N_{f}^{2} - 1$ Goldstone bosons: these are the pions $(\pi^{+}, \pi^{0}, \pi^{-})$ for $N_{f} = 2$, and
the pseudoscalar octet\footnote{We use the notation $\eta_{8}$ to emphasize that this is the pseudo-Goldstone boson associated with the eighth generator of $SU(3)$, not the physical $\eta$ meson detected in particle experiments. In reality flavor $SU(3)$ is not an exact symmetry of nature, and the states $\eta_{1} = (u \overline{u} + d \overline{d} + s \overline{s})/\sqrt{3}$ and $\eta_{8} = (u \overline{u} + d \overline{d} - 2 s \overline{s})/\sqrt{6}$ mix to form the physical $\eta$ and $\eta'$.} $(\pi^{+}, \pi^{0}, \pi^{-}, K^{+}, K^{0}, \Knaught, K^{-}, \eta_{8})$ for $N_{f} = 3$. The full $SU(N_{f})_{L} \times SU(N_{f})_{R}$ symmetry of the massless Lagrangian is also explicitly broken by the nonzero masses of the quarks in nature, generating masses for the (pseudo-)Goldstone bosons. \\

ChPT is the low-energy effective theory whose degrees of freedom are precisely the Goldstone bosons of QCD. The Goldstone fields can be parametrized in the exponential representation
\begin{equation}
U(x) = \exp \left( \frac{i}{f} \phi(x) \right), \hspace{0.4cm} \phi(x) \in su(N_{f})
\end{equation}
with
\begin{equation}
\phi(x) = \left( \begin{array}{cc} \frac{1}{\sqrt{2}} \pi^{0} & \pi^{+} \\ \pi^{-} & -\frac{1}{\sqrt{2}} \pi^{0} \end{array} \right)
\end{equation}
for the $SU(2)$ theory, and
\begin{equation}
\phi(x) = \left( \begin{array}{ccc} \frac{1}{\sqrt{2}} \pi^{0} + \frac{1}{\sqrt{6}} \eta_{8} & \pi^{+} & K^{+} \\ \pi^{-} & -\frac{1}{\sqrt{2}} \pi^{0} + \frac{1}{\sqrt{6}} \eta_{8} & K^{0} \\ K^{-} & \Knaught & - \frac{2}{\sqrt{6}} \eta_{8} \end{array} \right)
\end{equation}
for the $SU(3)$ theory. A \textit{chiral order} is assigned to each term by counting the number of derivatives of $U$ which enter: $\partial^{n} U \sim p^{n}$, where $p$ corresponds to external momenta carried by the Goldstone bosons. One can then systematically construct the ChPT Lagrangian order-by-order in this power counting scheme
\begin{equation}
\mathcal{L}_{\rm ChPT} = \underbrace{\mathcal{L}_{\rm ChPT}^{(2)}}_{\rm LO} + \underbrace{\mathcal{L}_{\rm ChPT}^{(4)}}_{\rm NLO} + \underbrace{\mathcal{L}_{\rm ChPT}^{(6)}}_{\rm NNLO}  + \cdots
\end{equation}
by writing down all possible terms of $\bigO(p^{n})$,
\begin{equation}
\mathcal{L}_{\rm ChPT}^{(n)} = \sum_{i} \alpha_{i} \mathscr{O}_{i}^{(n)},
\end{equation} 
where $\alpha_{i} \in \mathbb{R}$ are the low energy constants, and $\mathscr{O}_{i}^{(n)} \sim p^{n}$ is constructed from $U$ and its derivatives, and is invariant under the $SU(N_{f})_{L} \times SU(N_{f})_{R}$ symmetry. Gasser and Leutwyler further showed that by coupling the quark mass matrix, vector and axial currents, and scalar and pseudoscalar densities to the ChPT Lagrangian as external sources one can elegantly reproduce the Ward identities of QCD by taking appropriate functional derivatives~\cite{Gasser:1983yg, Gasser:1984gg}. While this construction produces the most general effective Lagrangian consistent with the underlying symmetries of QCD, the numerical values of the low energy constants (LECs) are a priori unknown, and must be determined phenomenologically or by fits to lattice simulations. \\

The first detailed, next-to-leading order ChPT calculations were performed by Gasser and Leutwyler in Ref.~\cite{Gasser:1983yg} for the $SU(2)$ case, and Ref.~\cite{Gasser:1984gg} for the $SU(3)$ case. They compute a number of two-point and four-point correlation functions which allow them to determine the pseudoscalar masses and decay constants, scattering lengths, and other low-energy observables of interest. These calculations were then extended to NNLO in~\cite{Fearing:1994ga}, where the $\bigO(p^{6})$ Lagrangian was first explicitly constructed, and in Ref.~\cite{Bijnens:1997vq} ($SU(2)$) and Ref.~\cite{Amoros:1999dp} ($SU(3)$). We will make use of two further generalizations of chiral perturbation theory: finite volume ChPT and partially quenched ChPT. \\

In finite volume (FV) ChPT the spatial $\mathbb{R}^{3}$ of Minkowski spacetime is replaced with a cubic box of volume $L^{3}$. This discretizes the allowed momentum states, requiring continuous integrals over momenta to be replaced with sums. Corrections to infinite volume ChPT results can be computed as functions of $L$, and must vanish in the $L \rightarrow \infty$ limit. Since, in a typical lattice QCD simulation, the pion correlation length is comparable to $L$, finite volume effects are often one of the dominant systematic errors when trying to make physical predictions, and FV ChPT is important to remove or bound these errors. In our fits we parametrize the chiral ans\"{a}tze for the pseudoscalar masses and decay constants as
\begin{equation}
\begin{split}
m_{xy}^{2} &= (m_{xy}^{2})^{\infty} + \Delta_{m_{x y}^{2}}^{L} \\
f_{xy} &= (f_{xy})^{\infty} + \Delta_{f_{x y}}^{L}
\end{split}
\end{equation}
where $(X)^{\infty}$ denotes the infinite volume result, and $\Delta_{X}^{L} \equiv (X)^{L} - (X)^{\infty}$ is the finite volume correction for a box of size $L$. Explicit formulae for $\Delta_{X}^{L}$ are known to NNLO~\cite{Colangelo:2005gd,Colangelo:2006mp,Bijnens:2014dea}, but we will only make use of the NLO results summarized in the appendices of Ref.~\cite{Allton:2008pn} for our fits. \\

\textit{Partial quenching} is a technique used in lattice simulations to lower the simulated pion mass without substantially increasing computational cost. On the lattice one is free to independently vary the sea and valence quark masses: the former enter the fermion determinant used to generate gauge field configurations, and the latter appear in fermion propagators when computing correlation functions. In practice $m_{\rm val} < m_{\rm sea}$ is often used since reducing the sea quark masses is more expensive than reducing the valence quark masses. One can regard partially quenched QCD as a theory in its own right, which reduces to ordinary QCD in the unitary limit $m_{\rm val} = m_{\rm sea}$. \\

In the framework of ChPT partial quenching is included analytically by generalizing to a supersymmetric theory with $N_{\rm sea}$ and $N_{\rm val}$ sea and valence quarks, respectively. The theory also contains $N_{\rm val}$ unphysical bosonic ghost quarks which exactly cancel the contributions from the fermionic valence quarks to closed fermion loops. The $SU(N_{f})_{L} \times SU(N_{f})_{R}$ symmetry of ordinary massless QCD is promoted to a graded $SU(N_{\rm val} + N_{\rm sea} | N_{\rm val})_{L} \times SU(N_{\rm val} + N_{\rm sea} | N_{\rm val})_{R}$ symmetry, and the most general effective Lagrangian consistent with this symmetry is constructed order-by-order, in analogy to ordinary ChPT. The original construction of the PQChPT Lagrangian is discussed in Ref.~\cite{PhysRevD.49.486}, and in Ref.~\cite{Sharpe:1997by} NLO expressions for the pion mass and decay constant are calculated. For our NLO PQChPT fits we use the explicit $SU(2)$ formulae collected in Ref.~\cite{Allton:2008pn}. Bijnens, Danielsson, and L\"{a}hde further generalized the PQChPT expressions for the partially quenched pseudoscalar masses and decay constants to NNLO: these calculations are presented in Ref.~\cite{Bijnens:2005pa} for the $SU(2)$ case and Ref.~\cite{Bijnens:2004hk, Bijnens:2005ae, Bijnens:2006jv} for the $SU(3)$ case. We make use of Fortran codes provided by Bijnens to compute these expressions in our NNLO fits. By explicitly taking the unitary limit $m_{\rm val} = m_{\rm sea}$ in the PQChPT Lagrangian and matching to the ChPT Lagrangian one can write down explicit relations between the PQChPT and ChPT LECs. We collect these results in Appendix~\ref{appendix:pqchpt_to_chpt}. \\ 

In Table~\ref{tab:lec_counting} we summarize the counting of LECs up to NNLO in $SU(2)$ and $SU(3)$ ChPT and PQChPT, and introduce our notation.

\begin{table}[!ht]
\centering
\begin{tabular}{c||c|c|c|c}
\hline
\hline
\rule{0cm}{0.4cm} & ChPT & ChPT & PQChPT & PQChPT \\
$N_{f}$ & 2 & 3 & 2 & 3 \\
\hline
\rule{0cm}{0.4cm} LO & $B,f$ & $B_{0},f_{0}$ & $B,f$ & $B_{0},f_{0}$ \\
\hline
\rule{0cm}{0.4cm} \multirow{2}{*}{NLO} & $l_{i}$ & $L_{i}$ & $\hat{L}_{i}^{(2)}$ & $\hat{L}_{i}^{(3)}$ \\
\rule{0cm}{0.4cm} & 7 & 10 & 11 & 11 \\
\hline
\rule{0cm}{0.4cm} \multirow{2}{*}{NNLO} & $c_{i}$ & $C_{i}$ & $\hat{K}_{i}^{(2)}$ & $\hat{K}_{i}^{(3)}$ \\
\rule{0cm}{0.4cm} & 53 & 90 & 112 & 112 \\
\hline
\hline
\end{tabular}
\caption{Counting of the LECs in ChPT and PQChPT up to NNLO, from~\cite{Bijnens:2006zp}. The notations $\{l_{i}, c_{i}\}$ for the $SU(2)$ ChPT LECs and $\{L_{i}, C_{i}\}$ for the $SU(3)$ ChPT LECs are conventional in the literature. Similarly, we use the notation $\{ \hat{L}_{i}^{(N_{f})}, \hat{K}_{i}^{(N_{f})} \}$ to distinguish the more general partially quenched LECs.}
\label{tab:lec_counting}
\end{table}
\FloatBarrier


\section{Lattice Setup}

In this analysis we make use of a number of RBC/UKQCD domain wall fermion ensembles with a wide range of unitary pion masses, $117 \,\, \mathrm{MeV} \le m_{\pi} \le 432 \,\, \mathrm{MeV}$, physical volumes, $(2.005(11) \,\, \mathrm{fm})^{3} \le L^{3} \le (6.43(26) \,\, \mathrm{fm})^{3}$, and inverse lattice spacings, $0.98(4) \,\, \mathrm{GeV} \le a^{-1} \le 3.14(2) \,\, \mathrm{GeV}$. In all cases we work in the isospin symmetric limit of QCD, with two, degenerate dynamical light quark flavors of bare mass $m_{l}$, and a single dynamical heavy flavor of bare mass $m_{h}$ ($N_{f} = 2 + 1$). Many of these ensembles have been analyzed in earlier publications which describe the ensemble generation, fits to extract the spectrum, and earlier chiral extrapolations based on NLO chiral perturbation theory~\cite{Allton:2008pn, Aoki:2010dy, Arthur:2012opa, Blum:2014tka}. We also include two new M\"{o}bius domain wall fermion ensembles; details of the ensemble generation and fits to extract the spectrum are discussed in Appendix~\ref{appendix:new_ensembles}. \\

In Table~\ref{tab:ensembles_input} we list the 12 ensembles included in this analysis and summarize the actions and input parameters. In all cases we use the Iwasaki gauge action (I)~\cite{Iwasaki:1984cj}, and on some ensembles supplement this with the dislocation suppressing determinant ratio (I+DSDR)~\cite{Vranas:2006zk, Renfrew:2009wu}. The DSDR term suppresses dislocations (``tears'') in the gauge field, representing tunneling between different topological sectors, that give rise to enhanced chiral symmetry breaking in domain wall fermion calculations, and occur more frequently at strong coupling. We simulate QCD with $N_{f} = 2+1$ quark flavors using the domain wall fermion formalism, with either the Shamir (DWF)~\cite{Kaplan:1992bt, Shamir:1993zy} or M\"{o}bius (MDWF)~\cite{Brower:2004xi, Brower:2005qw, Brower:2012vk} kernel. The details of how the low-energy QCD spectrum has been extracted from fits to various Green's functions can be found in Ref.~\cite{Allton:2008pn} for the 24I ensembles, Ref.~\cite{Aoki:2010dy} for the 32I ensembles, Ref.~\cite{Arthur:2012opa} for the 32ID ensembles, Ref.~\cite{Blum:2014tka} for the 48I, 64I, and 32I-fine ensembles, and in Appendix~\ref{appendix:new_ensembles} for the 32ID-M1 and 32ID-M2 ensembles. In addition, detailed discussions of the M\"{o}bius kernel and the properties of MDWF simulations of QCD can be found in Ref.~\cite{Blum:2014tka}. \\

\begin{table}[h]
\centering
\begin{tabular}{c||c|cccc|cc}
\hline
\hline
\rule{0cm}{0.4cm} Ensemble & Action & $\beta$ & $L^{3} \times T \times L_{s}$ & $a m_{l}$ & $a m_{h}$ & $m_{\pi} L$ & $m_{\pi}$ (MeV) \\
\hline
\rule{0cm}{0.4cm} \multirow{2}{*}{24I} & DWF+I & 2.13 & $24^{3} \times 64 \times 16$ & 0.005 & 0.04 & 4.568(13) & 339.6(1.2) \\
\rule{0cm}{0.4cm} & DWF+I & 2.13 & $24^{3} \times 64 \times 16$ & 0.01 & 0.04 & 5.814(12) & 432.2(1.4) \\
\hline
\rule{0cm}{0.4cm} \multirow{3}{*}{32I} & DWF+I & 2.25 & $32^{3} \times 64 \times 16$ & 0.004 & 0.03 & 4.062(11) & 302.0(1.1) \\
\rule{0cm}{0.4cm} & DWF+I & 2.25 & $32^{3} \times 64 \times 16$ & 0.006 & 0.03 & 4.8377(82) & 359.7(1.2) \\
\rule{0cm}{0.4cm} & DWF+I & 2.25 & $32^{3} \times 64 \times 16$ & 0.008 & 0.03 & 5.526(12) & 410.8(1.5) \\
\hline
\rule{0cm}{0.4cm} \multirow{2}{*}{32ID} & DWF+I+DSDR & 1.75 & $32^{3} \times 64 \times 32$ & 0.001 & 0.046 & 3.9992(69) & 172.7(9) \\
\rule{0cm}{0.4cm} & DWF+I+DSDR & 1.75 & $32^{3} \times 64 \times 32$ & 0.0042 & 0.046 & 5.7918(79) & 250.1(1.2) \\
\hline
\rule{0cm}{0.4cm} 32I-fine & DWF+I & 2.37 & $32^{3} \times 64 \times 12$ & 0.0047 & 0.0186 & 3.773(42) & 370.1(4.4) \\
\hline
\rule{0cm}{0.4cm} 48I & MDWF+I & 2.13 & $48^{3} \times 96 \times 24$ & 0.00078 & 0.0362 & 3.8633(63) & 139.1(4) \\
\hline
\rule{0cm}{0.4cm} 64I & MDWF+I & 2.25 & $64^{3} \times 128 \times 12$ & 0.000678 & 0.02661 & 3.7778(84) & 139.0(5) \\
\hline
\rule{0cm}{0.4cm} 32ID-M1 & MDWF+I+DSDR & 1.633 & $32^{3} \times 64 \times 24$ & 0.00022 & 0.0596 & 3.780(15) & 117.3(4.4) \\
\hline
\rule{0cm}{0.4cm} 32ID-M2 & MDWF+I+DSDR & 1.943 & $32^{3} \times 64 \times 12$ & 0.00478 & 0.03297 & 6.236(21) & 401.0(2.3) \\
\hline
\hline
\end{tabular}
\caption{Summary of ensembles included in this analysis and input parameters. Here $\beta$ is the gauge coupling, $L^{3} \times T \times L_{s}$ is the lattice volume decomposed into the length of the spatial ($L$), temporal ($T$), and fifth ($L_{s}$) dimensions, and $a m_{l}$ and $a m_{h}$ are the bare, input light and heavy quark masses. The value of $m_{\pi}$ quoted is the unitary pion mass in physical units, where we have used the lattice spacings listed in table~\ref{tab:ensembles_gf}.}
\label{tab:ensembles_input}
\end{table}

In Appendix~\ref{appendix:lattice_fits} we list fit values at the simulated quark masses in lattice units for the pseudoscalar masses and decay constants, $\Omega$ baryon mass, residual mass, and Wilson flow scales on each ensemble. On the older 24I, 32I, and 32ID ensembles these measurements were performed for a number of different partially quenched valence quark mass combinations which are listed explicitly in the appendix. In addition, reweighting in the dynamical heavy quark mass was used to determine the $m_{h}$ dependence and allow for a small, linear interpolation from the simulated $m_{h}$ to the physical value. On the newer ensembles --- 32I-fine, 48I, 64I, 32ID-M1, and 32ID-M2 --- we perform a single set of unitary measurements of the same observables, and do not reweight in $m_{h}$.


\section{The Global Fit Procedure}
\label{sec:canonical_gf}

In Ref.~\cite{Aoki:2010dy, Arthur:2012opa, Blum:2014tka} we have developed a ``global fit" procedure for performing a combined chiral fit and continuum extrapolation of lattice data, the details of which we will summarize here. The global fit also allows us to convert predictions from our simulations, which are performed in dimensionless lattice units, into physical units by determining the lattice spacing $a$ on each ensemble. While we have historically focused on using this construction to make physical predictions from our simulations, viewing chiral perturbation theory as a tool to parametrize the quark mass dependence of low-energy QCD observables, here we will adopt a slightly different view and regard the fit to ChPT itself as our primary interest. \\

Our canonical global fit, which we have most recently used in Ref.~\cite{Blum:2014tka}, includes the pion and kaon masses\footnote{Note: we work in the isospin symmetric limit of QCD, where $m_{u} = m_{d} \equiv m_{l}$, and neglect electromagnetic corrections. In this limit the charged and neutral pions are degenerate, as are the charged and neutral kaons, so we can speak unambiguously of ``the pion'' and ``the kaon''.} $m_{\pi}$ and $m_{K}$, the pion and kaon decay constants $f_{\pi}$ and $f_{K}$, the omega baryon mass $m_{\Omega}$, and the Wilson flow scales \cite{Luscher:2010iy} $t_{0}^{1/2}$ and $w_{0}$. Partially quenched next-to-leading order $SU(2)$ chiral perturbation theory with finite volume corrections is used to perform the chiral fit to the valence quark ($m_{x}$, $m_{y}$) and light dynamical quark ($m_{l}$) mass dependence of $m_{\pi}$ and $f_{\pi}$. The input dynamical heavy quark mass is carefully tuned during the ensemble generation to closely correspond to the physical strange quark mass, however, any slight mistuning introduces small errors in our simulated values of $m_{\pi}$ and $f_{\pi}$, which are not described by $SU(2)$ PQChPT. We account for this by reweighting (see Section II.D of Ref.~\cite{Aoki:2010dy}) in the heavy quark determinant to generate a series of values of each observable for several $m_{h}$ near the simulated mass, and then supplement the chiral $SU(2)$ ansatz with a term linear in $m_{h}$, allowing us to interpolate the reweighted data to the physical strange quark mass. NLO $SU(2)$ heavy meson PQChPT with finite volume corrections \cite{Allton:2008pn, Roessl:1999iu} is used for $m_{K}$ and $f_{K}$. The chiral fits to $m_{\Omega}$ and the Wilson flow scales are performed using a simple analytic ansatz which is linear in the quark masses. Discretization effects are included by adding a term linear in $a^{2}$ to each fit form, allowing us to ultimately take the continuum limit $a \rightarrow 0$. The raw simulation data is in dimensionless lattice units which are different for each ensemble, reflecting the different (physical) lattice spacings. We account for this by performing the chiral fits in the bare, dimensionless lattice units of a single reference ensemble, which we choose to be our $32^{3} \times 64$ Iwasaki (32I) lattice (Table~\ref{tab:ensembles_input}). We introduce additional fit parameters
\begin{equation}
\label{eqn:def_latt_ratios}
R_{a}^{e} \equiv \frac{a^{r}}{a^{e}}, \hspace{0.4cm} Z^{e}_{l} \equiv \frac{1}{R_{a}^{e}} \frac{\left( a \tilde{m}_{l} \right)^{r}}{\left( a \tilde{m}_{l} \right)^{e}}, \hspace{0.4cm} Z^{e}_{h} \equiv \frac{1}{R_{a}^{e}} \frac{\left( a \tilde{m}_{h} \right)^{r}}{\left( a \tilde{m}_{h} \right)^{e}}
\end{equation}
to convert between bare lattice units on the reference ensemble $r$ and other ensembles $e$, where $a$ is the lattice spacing and $\tilde{m}_{q} = m_{q} + m_{\rm res}$ is the total quark mass\footnote{In the domain wall fermion formalism a finite fifth dimension introduces a small chiral symmetry breaking, leading to an additive renormalization of the input quark masses by $m_{\rm res}$ (the residual mass). In Appendix~\ref{appendix:new_ensembles} we briefly discuss how $m_{\rm res}$ is extracted.}. \\

The chiral ans\"{a}tze discussed above reflect a simultaneous expansion in the quark masses, lattice volume ($L$), and lattice spacing ($a$), about the infinite volume, continuum, chiral limit. Our power-counting scheme counts the dominant discretization term --- which is proportional to $a^{2}$ for domain wall fermions --- as the same order as the NLO continuum PQChPT corrections. While we include continuum PQChPT terms up to $\mathcal{O}(p^{6})$ in our NNLO fits, cross terms proportional to $X^{\rm NLO} \times \Delta_{X}^{\rm NLO}$ and $X^{\rm NLO} \times a^{2}$ are neglected since they are higher-order in our power-counting, and are empirically observed to be small. The full chiral ansatz for $X \in  \{ m_{\pi}^{2}, f_{\pi} \}$, for example, including the finite volume and $a^{2}$ terms, has the generic form
\begin{equation}
X(\tilde{m}_{q}, L, a^{2}) \simeq X_{0} 
\left( \vphantom{1 + X^{\rm NLO}(\tilde{m}_{q}) + X^{\rm NNLO}(\tilde{m}_{q}) + \Delta_{X}^{\rm NLO}(\tilde{m}_{q}, L) + c_{X} a^{2} } \right.
1 + \underbrace{ X^{\rm NLO}(\tilde{m}_{q}) + X^{\rm NNLO}(\tilde{m}_{q}) }_{\rm NNLO \,\, Continuum \,\, PQChPT} + \underbrace{\Delta_{X}^{\rm NLO}(\tilde{m}_{q}, L)}_{\rm NLO \,\, FV \,\, corrections} + \underbrace{c_{X} a^{2}}_{\rm Lattice \,\, spacing}
\left. \vphantom{1 + X^{\rm NLO}(\tilde{m}_{q}) + X^{\rm NNLO}(\tilde{m}_{q}) + \Delta_{X}^{\rm NLO}(\tilde{m}_{q}, L) + c_{X} a^{2} } \right)
\end{equation}
where $X_{0}$ is the leading order value of $X$ in the continuum and infinite-volume limits, and ``$\simeq$'' denotes equality up to truncation of higher order terms. Since the Iwasaki and I+DSDR actions have, in general, different discretization errors for a given value of the lattice spacing, we fit independent $a^{2}$ coefficients for each observable $X$, denoted $c_{X}^{I}$ and $c_{X}^{ID}$, respectively. The NLO $SU(2)$ ans\"{a}tze are written in complete detail in Appendix H of Ref.~\cite{Blum:2014tka}; the generalization to NNLO is straightforward. Appendix B of the same reference also discusses how to write a given chiral ansatz in our dimensionless formalism. \\

The procedure for performing a global fit is as follows:
\begin{enumerate}
\item The valence quark mass dependence of $m_{\rm res}$ is fit to a linear ansatz on each ensemble. We then extrapolate $m_{\rm res}$ to the chiral limit $m_{q} \rightarrow 0$, and use this value in the remainder of the analysis.
\item A simultaneous chiral/continuum fit of $m_{\pi}^{2}$, $m_{K}^{2}$, $f_{\pi}$, $f_{K}$, $m_{\Omega}$, $t_{0}^{1/2}$ and $w_{0}$ is performed on all ensembles using the ans\"{a}tze described in the preceding paragraph. The quark mass dependence is parametrized in terms of $\tilde{m}_{q} = m_{q} + m_{\rm res}$. This step also determines the ratios of lattice scales $R_{a}^{e}$ and $Z_{\{l,h\}}^{e}$ and the dependence on $a^{2}$.
\item Three of the quantities from 2 are defined to have no $a^{2}$ corrections and establish our continuum scaling trajectory by matching onto their known, physical values\footnote{For reference, our values for the ``physical'', isospin symmetric masses and decay constants, excluding QED effects, are: $m_{\pi}^{\rm phys} = 135.0 \, \mathrm{MeV}$ (PDG $\pi^{0}$ mass), $m_{K}^{\rm phys} = 495.7 \, \mathrm{MeV}$ (average of the PDG $K^{0}$ and $K^{\pm}$ masses), $m_{\Omega}^{\rm phys} = 1672.45 \, \mathrm{MeV}$ (PDG $\Omega^{-}$ mass), $f_{\pi}^{\rm phys} = 130.4 \, \mathrm{MeV}$ (PDG $\pi^{-}$ decay constant), and $f_{K}^{\rm phys} = 156.1 \, \mathrm{MeV}$ (PDG $K^{-}$ decay constant) \cite{1674-1137-38-9-090001}.}. In the analysis of \cite{Blum:2014tka} we have used $m_{\pi}$, $m_{K}$, and $m_{\Omega}$, and implemented this condition by numerically inverting the chiral fit to determine input bare valence quark masses $m_{l}^{\rm phys}$ and $m_{h}^{\rm phys}$ such that the ratios $m_{\pi}/m_{\Omega}$ and $m_{K}/m_{\Omega}$ take their physical values.
\item From 3 we obtain $m_{\Omega}$ at $m_{l}^{\rm phys}$ and $m_{h}^{\rm phys}$ on the reference ensemble; we then use the ratio $m_{\Omega}^{r} / m_{\Omega}^{\rm phys}$ to determine the lattice spacing $a^{r}$ in physical units. Together with the ratios of lattice scales from 2 we can determine the lattice spacings on the other ensembles, as well as extrapolate observables to the physical quark mass, continuum limit in physical units. 
\end{enumerate}
The fits described in steps 1 and 2 are performed using uncorrelated nonlinear $\chi^{2}$ minimization with the Levenberg-Marquardt algorithm \cite{levenberg44, doi:10.1137/0111030}. Due to the large number of data points in our fits we have a very nearly singular correlation matrix that we cannot reliably invert, as would be required to perform fits with a fully correlated $\chi^{2}$; we show an example of one of our correlation matrices in Appendix \ref{appendix:chisq_normalized}. As a result, the $\chi^{2}/\mathrm{dof}$ that we present cannot be interpreted as the goodness-of-fit, and instead we will present histograms showing the distribution of the data around our fit. These histograms provide a simple summary of the fit quality, and, in particular, highlight any data that is far from the fit function. The numerical inversion in step 3 is performed by minimizing
\begin{equation}
\chi^{2} = \left[ \left( \frac{m_{\pi}}{m_{\Omega}} \right)(\tilde{m}_{l},\tilde{m}_{h}) - \left( \frac{m_{\pi}}{m_{\Omega}} \right)^{\rm PDG}  \right]^{2} + \left[ \left( \frac{m_{K}}{m_{\Omega}} \right)(\tilde{m}_{l},\tilde{m}_{h}) - \left( \frac{m_{K}}{m_{\Omega}} \right)^{\rm PDG}  \right]^{2} ,
\end{equation}
where PDG denotes the experimental value from \cite{1674-1137-38-9-090001}. Statistical errors on the fit parameters are computed using the superjackknife resampling technique \cite{Bratt:2010jn}. The choices of which quantities are used to determine the physical quark masses in step 3 and the lattice spacing in step 4 are arbitrary, and all results should agree in the continuum limit regardless of this choice. \\

The matching to our chosen scaling trajectory results in values of the physical quark masses, $m_l^{\rm phys}$ and $m_h^{\rm phys}$, as well as corresponding values of the leading-order chiral parameter $B$, that are normalized in the native units of our 32I ensemble. In order to be useful to others, these quantities must be renormalized into a more convenient scheme such as $\overline{\rm MS}$. As described in Refs.~\cite{Aoki:2010dy,Arthur:2012opa,Blum:2014tka} we achieve this by first renormalizing in variants of the non-perturbative Rome-Southampton regularization-invariant momentum scheme with symmetric kinematics (RI/SMOM)~\cite{Martinelli:1994ty,Aoki:2007xm,Arthur:2010ht,Sturm:2009kb,Arthur:2011cn}. The matching factors between these schemes and $\overline{\rm MS}$ can be computed using standard continuum perturbation theory with dimensional regularization applied at a high energy scale, typically $\mu \sim 3$ GeV, at which perturbation theory is known to be reliable. We use the RI/SMOM intermediate scheme for our central values. The only significant systematic error on the result is due to the truncation of the perturbative series to two-loop order in the computation of the RI/SMOM$\to\overline{\rm MS}$ matching factors. In order to estimate the size of this effect we compare the resulting $\overline{\rm MS}$ values to those computed using the RI/SMOM$_{\gamma^\mu}$ intermediate scheme, taking the full difference as a conservative estimate\footnote{For more detail regarding the SMOM and $\mathrm{SMOM}_{\gamma^{\mu}}$ schemes we refer the reader to Refs. \cite{Sturm:2009kb,Aoki:2010dy}.}. \\

Renormalized quark masses are obtained by taking the product
\begin{equation}
m_f^{\overline{\rm MS}} = Z_m^{\overline{\rm MS},\ {\rm 32I}}m_f^{\rm phys} + {\cal O}(a^2)\,,
\end{equation}
where $f\in \{l,h\}$ and $Z_m^{\overline{\rm MS},\ {\rm 32I}}$ is the quark mass renormalization coefficient computed on the 32I ensemble. This determination of $m_{f}^{\overline{\rm MS}}$ contains ${\rm O}(a^2)$ errors because the renormalization factors have only been computed at a single lattice spacing. Using the quantities $Z_l$ and $Z_h$ defined in Eqn.~\eqref{eqn:def_latt_ratios}, we can also compute the renormalized physical quark mass using renormalization factors calculated on the 24I ensemble as follows:
\begin{equation}
m_f^{\overline{\rm MS}} = \frac{ Z_m^{\overline{\rm MS},\ {\rm 24I}} }{ Z_f^{\rm 24I} } m_f^{\rm phys}+ {\cal O}(a^2)\,,
\end{equation}
Combining these two equations, we can compute a value for the quark mass that is free from ${\cal O}(a^2)$ errors:
\begin{equation}
m_f^{\overline{\rm MS}} = Z^{\overline{\rm MS}}_{mf} m_f^{\rm phys} + {\cal O}(a^4)\,,
\end{equation}
where 
\begin{equation}
Z_{mf}^{\overline{\rm MS}} = \lim_{a\to 0}\left\{ Z_m^{\overline{\rm MS}}(a)/Z_f(a) \right\}\,,
\end{equation}
and the $a \rightarrow 0$ limit is taken by performing a linear extrapolation using the two available lattice spacings. Similarly, the renormalized value of $B$ can be obtained as
\begin{equation}
B^{\overline{\rm MS}} = B^{\rm fit}/Z_{ml}^{\overline{\rm MS}}\,.
\end{equation}
Note that the fact that domain wall fermions are non-perturbatively ${\cal O}(a)$ improved and have good chiral symmetry eliminates dependence on odd-powers of the lattice spacing. \\

For this analysis we use the values of $Z_{ml}$ and $Z_{mh}$ computed in Ref.~\cite{Blum:2014tka}, and for more details we refer the reader to Section V.C and Appendix F of that work. Note that the calculation of these quantities necessarily involves the computed values of the lattice spacing, which differ between the various fits we perform. For the analyses presented in this document we do not recompute $Z_{mf}$ for each fit; however our lattice spacings are all in excellent agreement with those in the aforementioned work, hence we choose to neglect the small systematic error associated with this mismatch. \\

While the fits discussed in this work are in many ways an extension of the analysis presented in Ref.~\cite{Blum:2014tka}, there are a few important differences we would like to emphasize. First, in Ref.~\cite{Blum:2014tka} chiral perturbation theory was used only to make modest, $\bigO(1\%)$ corrections to the spectrum computed on the physical quark mass $48^{3} \times 96$ (48I) and $64^{3} \times 128$ (64I) lattices. This was achieved using an overweighting procedure, in which the contributions
\begin{equation}
\label{eqn:weighted_chisq}
\chi_{e}^{2} = \alpha_{e} \sum_{i} \left( \frac{y_{e}^{i} - f_{e}^{i}}{\sigma_{e}^{i}} \right)^{2}
\end{equation} 
to $\chi^{2} = \sum_{e} \chi_{e}^{2}$ from each ensemble were multiplied by tunable, independent parameters $\alpha_{e}$. By choosing $\alpha_{\rm 48I}, \alpha_{\rm 64I} \gg 1$ and $\alpha_{e} = 1$ otherwise, the fit was effectively forced to pass through the 48I and 64I data, using information from the other ensembles only to make a small correction to the physical point. In this work we are interested more generally in the applicability of chiral perturbation theory to describe the quark mass dependence of the QCD spectrum, and thus we do not employ overweighting. Second, in Ref.~\cite{Blum:2014tka} the Wilson flow scales $t_{0}^{1/2}$ and $w_{0}$ were introduced into the global fit procedure, which we do not include in any of the fits presented in Section~\ref{sec:su2_fits}. While the inclusion of the Wilson flow scales leads to a marked improvement in the determination of the lattice spacings, they do not constrain the ChPT LECs, and are unnecessary for our computationally demanding NNLO fits. \\

Since the 32ID-M1 and 32ID-M2 lattices have not appeared in our earlier global fit analyses, we have updated our canonical global fit from Ref.~\cite{Blum:2014tka} to include these ensembles and determine their properties. We note that even though the 32ID-M2 ensemble has a relatively heavy unitary pion mass ($m_{\pi} = 401.0(2.3) \, \mathrm{MeV}$) that lies outside the 370 MeV cut used in this fit, the overweighting procedure results in a fit that is insensitive to heavy ensembles, and we can safely assume that this discrepancy will not lead to any significant systematics. This provides an explicit check that our fits in this work, including the new ensembles, are consistent with our earlier work, and we indeed see that the lattice spacings and other parameters are consistent with Ref.~\cite{Blum:2014tka}. This fit also establishes a baseline relative to the global fit performed in Ref.~\cite{Blum:2014tka}, by which we can judge the consistency of the new fits discussed in Section~\ref{sec:su2_fits}. The values we obtain for the physical box sizes, lattice spacings, and residual mass in the chiral limit are summarized in Table~\ref{tab:ensembles_gf}. \\

\begin{table}[h]
\centering
\begin{tabular}{c||ccccc}
\hline
\hline
\rule{0cm}{0.4cm} Ensemble & $L$ (fm) & $a^{-1}$ (GeV) & $a m_{l}^{\rm phys}$ & $a m_{h}^{\rm phys}$ & $a m_{\text{res}}$ \\
\hline
\rule{0cm}{0.4cm} 24I & 2.6496(73) & 1.7844(49) & -0.001770(79) & 0.03225(18) & 0.003038(78) \\
\hline
\rule{0cm}{0.4cm} 32I & 2.6466(93) & 2.3820(84) & 0.000261(13) & 0.02480(18) & 0.000662(11) \\
\hline
\rule{0cm}{0.4cm} 32ID & 4.573(22) & 1.3784(68) & -0.000106(16) & 0.04625(48) & 0.0018478(73) \\
\hline
\rule{0cm}{0.4cm} 32I-fine & 2.005(11) & 3.144(17) & 0.000057(16) & 0.01846(32) & 0.0006300(59) \\
\hline
\rule{0cm}{0.4cm} 48I & 5.468(12) & 1.7293(36) & 0.0006982(80) & 0.03580(16) & 0.0006102(40) \\
\hline
\rule{0cm}{0.4cm} 64I & 5.349(16) & 2.3572(69) & 0.0006213(77) & 0.02542(17) & 0.0003116(23) \\
\hline
\rule{0cm}{0.4cm} 32ID-M1 & 6.43(26) & 0.981(39) & 0.00107(26) & 0.0850(68) & 0.002170(16) \\
\hline
\rule{0cm}{0.4cm} 32ID-M2 & 3.067(16) & 2.055(11) & -0.003429(16) & 0.02358(33) & 0.0044660(46) \\
\hline
\hline 
\end{tabular}
\caption{Physical box sizes, inverse lattice spacings, bare, unrenormalized quark masses, and residual mass in the chiral limit for the ensembles included in this work. These numbers are obtained by repeating the global fit analysis published in Ref.~\cite{Blum:2014tka}, including the new 32ID-M1 and 32ID-M2 ensembles.}
\label{tab:ensembles_gf}
\end{table}


\section{Fits to $SU(2)$ PQChPT}
\label{sec:su2_fits}

In this section we discuss global fits based on $SU(2)$ partially quenched chiral perturbation theory. These fits include:
\begin{enumerate}
\item The pion mass and decay constant, fit to NLO or NNLO PQChPT, with NLO finite volume corrections in both cases.
\item The kaon mass and decay constant, fit to NLO heavy-meson PQChPT with NLO finite volume corrections.
\item The $\Omega$ baryon mass, fit to a linear, analytic ansatz.
\end{enumerate}
$m_{\pi}$, $m_{K}$, and $m_{\Omega}$  are used as the three inputs to determine the physical quark masses and lattice spacings; this leaves $f_{\pi}$ and $f_{K}$ as predictions. We consider two different cuts on the heaviest unitary pion mass included in the fit: 370 MeV and 450 MeV. Any ensemble with a unitary pion mass greater than the cut is excluded from the fit completely. Likewise, all partially quenched ``pion'' measurements with $m_{x y} > m_{\pi}^{\rm cut}$ are excluded even if the unitary pion mass is within the cut. The data we use for the fits with a 370 MeV cut is the same as the data used in the fits with a 370 MeV cut in Ref.~\cite{Blum:2014tka}, with the addition of the new 32ID-M1 ensemble. We do not include any additional kaon or $\Omega$ baryon data when we raise the mass cut, since these quantities are described by NLO (kaon) or linear ($\Omega$) ansatz\"{a}e in all of the fits that we have performed --- the heavier 450 MeV cut is intended to test the full partially quenched NNLO expressions for $m_{\pi}$ and $f_{\pi}$ by using all of our available data. \\

In Sections~\ref{sec:su2_fit_parameters}-\ref{sec:su2_chiral_fit} we present the fit results, including our values for the partially quenched NLO and NNLO LECs. In Section~\ref{sec:su2_convergence} we examine the range of applicability of NNLO $SU(2)$ ChPT and the relative sizes of the terms in the chiral expansion. Finally, in Section~\ref{sec:su2_predictions} we compute the unquenched $SU(2)$ ChPT LECs from these results, and also discuss other predictions we can make from $SU(2)$ ChPT. All fits discussed in this section were performed by minimizing the uncorrelated $\chi^{2}$; in Appendix~\ref{appendix:chisq_normalized} we repeat the fits using a weighted $\chi^{2}$ to explore systematic effects associated with correlations in the data. These weighted fits are also defined by Eqn.~\eqref{eqn:weighted_chisq}, but rather than choosing $\alpha_{e} \gg 1$ to overweight the physical point ensembles as we did in Ref.~\cite{Blum:2014tka}, here we underweight the 24I, 32I, and 32ID ensembles by a factor $\alpha_{e} = 1/N_{e}$, where $N_{e}$ is the number of nondegenerate (partially quenched) pseudoscalar mass measurements on ensemble $e$. This has the effect of capturing some of the most important correlations --- those between partially quenched measurements with different combinations of valence quarks on a given ensemble, and between reweightings in $m_{h}$ of the same observable --- as we argue in Appendix~\ref{appendix:chisq_normalized}, while avoiding the numerical instabilities that plague fully correlated fits.

\subsection{Fit Parameters}
\label{sec:su2_fit_parameters}

Tables~\ref{tab:su2_ainv_mq_chisq} -~\ref{tab:more_su2_parameters} summarize the fit parameters, including a statistical error computed with the superjackknife resampling technique~\cite{Blum:2014tka}. These include the $\chi^{2}/$dof, physical quark masses, and inverse lattice spacings in physical units (Table~\ref{tab:su2_ainv_mq_chisq}), the ratios of quark masses and lattice spacings between the reference 32I ensemble and the other ensembles (Table~\ref{tab:su2_z_r}), the PQChPT LECs (Table~\ref{tab:su2_pqchpt_lecs}), and additional fit parameters describing the continuum and chiral scaling of the kaon and $\Omega$ baryon data (Table~\ref{tab:more_su2_parameters}). We generally observe excellent consistency comparing ensemble properties across the fits we have performed --- the physical quark masses and lattice spacings from Table~\ref{tab:su2_ainv_mq_chisq}, and the ratios of lattice scales from Table~\ref{tab:su2_z_r}, for example --- with the notable exception of the NLO fit with a 450 MeV cut, for which we observe systematic shifts outside our statistical errors. This is not surprising, however, since we do not expect NLO ChPT to accurately describe the lattice data up to such a heavy scale, and indeed we see a large increase in the $\chi^{2}/$dof for this particular fit. \\

While NLO fits constrain the four LECS $\{ \hat{L}_{4}^{(2)}, \hat{L}_{5}^{(2)}, \hat{L}_{6}^{(2)}, \hat{L}_{8}^{(2)} \}$, NNLO fits constrain nine NLO LECs --- $\{ \hat{L}_{i}^{(2)} \}_{i=0}^{8}$ --- as well as eight linear combinations of twelve NNLO LECs, which are listed explicitly in Table~\ref{tab:su2_pqchpt_lecs}. We have set $\hat{K}_{22}^{(2)} = \hat{K}_{27}^{(2)} = \hat{K}_{39}^{(2)} = \hat{K}_{40}^{(2)} = 0$ when we perform the fits for simplicity, so that each linear combination reduces to a single, independent LEC. We also impose the constraint\footnote{We have experimented with fits where $\hat{L}_{11}^{(2)}$ is left as a free parameter, but we find that $\hat{L}_{11}^{(2)} \ne -l_{4}/4$ well outside of statistics.} $\hat{L}_{11}^{(2)} = -l_{4}/4$, which is required for the PQChPT Lagrangian to reduce to the unquenched ChPT Lagrangian in the unitary limit~\cite{Bijnens:1999hw}. We perform independent fits at the two chiral scales $\Lambda_{\chi} = 770 \, \mathrm{MeV}$ and $\Lambda_{\chi} = 1 \, \mathrm{GeV}$, and report the PQChPT LECs at both scales. Since $\hat{L}_{7}^{(2)}$ and $\hat{L}_{8}^{(2)}$ are scale-independent, comparing the results for the fit with $\Lambda_{\chi} = 770 \, \mathrm{MeV}$ and the fit with $\Lambda_{\chi} = 1 \, \mathrm{GeV}$ provides a further consistency check. \\

We note that the 32ID-M1 ensemble has previously appeared in Ref.~\cite{Bhattacharya:2014ara}, where a simple estimate of the lattice spacing --- $a = m_{\Omega} / m_{\Omega}^{\rm PDG}$, with $m_{\Omega}$ at the simulated heavy quark mass --- was used to convert the spectrum from lattice units to physical units. We find a 10\% discrepancy between this lattice spacing and the lattice spacings obtained from our global fits and reported in Table~\ref{tab:su2_ainv_mq_chisq}. This arises from the 33\% difference between the input bare heavy quark mass $a m_{h} = 0.0596$ and the physical bare heavy quark masses determined from the global fits (also reported in Table~\ref{tab:su2_ainv_mq_chisq}): there is an $\mathcal{O}(10\%)$ shift in the ratio $m_{\Omega} / m_{\Omega}^{\rm PDG}$ when $m_{\Omega}$ is adjusted from the simulated point to the physical point. \\

We note that $Z_{l} = Z_{h} = R_{a} = 1$ by definition on the 32I ensemble. We have constrained $Z_{l}^{64 I} = Z_{h}^{64 I} = 1$ since the M\"{o}bius parameters and gauge coupling on the 64I ensemble have been chosen such that the 64I action is identical to the 32I action up to small chiral symmetry breaking effects. As we argue in Ref.~\cite{Blum:2014tka}, these chiral symmetry breaking effects lead to a small shift in the lattice spacings, so we do not constrain $R_{a}^{64 I} = 1$. Likewise, we constrain $Z_{l}^{24I} = Z_{l}^{48I}$ and $Z_{h}^{24I} = Z_{h}^{48I}$ for the same reason, but do not set $R_{a}^{24 I} = R_{a}^{48 I}$. \\

The observation that $Z_{l}, Z_{h} \sim 0.7$ for the 32ID-M1 ensemble in Table~\ref{tab:su2_z_r} suggests that this lattice is at sufficiently strong coupling that the five-dimensional domain wall fermion fields are no longer tightly bound to the domain walls, and instead leak into the fifth ($s$) dimension. As a result, somewhat larger input masses are required to achieve the same effective mass for the physical four-dimensional quark fields defined on the domain walls. We choose to include this ensemble in our fits since we do not observe any significant systematics if it is removed, and it is our only ensemble with lighter-than-physical pions, which probes the regime where chiral curvature is most pronounced.

\vspace*{\fill}
\begin{table}[h]
\centering
\resizebox{\columnwidth}{!}{
\begin{tabular}{cc||cc|cc}
\hline
\hline
 \rule{0cm}{0.4cm} & & NLO ($370 \, \mathrm{MeV}$ cut) & NLO ($450 \, \mathrm{MeV}$ cut) & NNLO ($370 \, \mathrm{MeV}$ cut) & NNLO ($450 \, \mathrm{MeV}$ cut) \\
\hline
 \rule{0cm}{0.4cm} & $\chi^2/$dof & 0.36(10)  & 1.14(27)  & 0.21(9)  & 0.29(10)  \\
\hline
 \rule{0cm}{0.4cm}\multirow{3}{*}{24I} & $a m_{l}^{\rm phys}$ & -0.001774(82)  & -0.001764(77)  & -0.001772(81)  & -0.001767(80) \\
 & $a m_{h}^{\rm phys}$ & 0.03209(40)  & 0.03239(32)  & 0.03210(38)  & 0.03219(35)  \\
 & $a^{-1}$ & 1.784(14) GeV & 1.781(12) GeV & 1.784(13) GeV & 1.782(13) GeV \\
\hline
 \rule{0cm}{0.4cm}\multirow{3}{*}{32I} & $a m_{l}^{\rm phys}$ & 0.000272(15)  & 0.000244(18)  & 0.000282(14)  & 0.000282(14)  \\
 & $a m_{h}^{\rm phys}$ & 0.02512(29)  & 0.02424(43)  & 0.02537(27)  & 0.02550(27)  \\
 & $a^{-1}$ & 2.360(17) GeV & 2.405(22) GeV & 2.349(16) GeV & 2.344(16) GeV \\
\hline
 \rule{0cm}{0.4cm}\multirow{3}{*}{32ID} & $a m_{l}^{\rm phys}$ & -0.000098(20)  & -0.000105(21)  & -0.000098(20)  & -0.000097(18)  \\
 & $a m_{h}^{\rm phys}$ & 0.04652(58)  & 0.04633(61)  & 0.04637(53)  & 0.04624(50)  \\
 & $a^{-1}$ & 1.374(8) GeV & 1.377(9) GeV & 1.376(8) GeV & 1.377(7) GeV \\
\hline
 \rule{0cm}{0.4cm}\multirow{3}{*}{32I-fine}& $a m_{l}^{\rm phys}$ & 0.000091(32)  & 0.000059(32)  & 0.000098(32)  & 0.000095(32)  \\
 & $a m_{h}^{\rm phys}$ & 0.01936(67)  & 0.01784(66)  & 0.01977(68)  & 0.01993(70)  \\
 & $a^{-1}$ & 3.079(44) GeV & 3.176(48) GeV & 3.059(44) GeV & 3.051(43) GeV \\
\hline
 \rule{0cm}{0.4cm}\multirow{3}{*}{48I} & $a m_{l}^{\rm phys}$ & 0.000685(14)  & 0.000706(12)  & 0.000688(13)  & 0.000695(13)  \\
 & $a m_{h}^{\rm phys}$ & 0.03547(33)  & 0.03595(24)  & 0.03550(31)  & 0.03562(27)  \\
 & $a^{-1}$ & 1.737(8) GeV & 1.726(6) GeV & 1.736(7) GeV & 1.733(6) GeV \\
\hline
 \rule{0cm}{0.4cm}\multirow{3}{*}{64I} & $a m_{l}^{\rm phys}$ & 0.000625(10)  & 0.000604(15)  & 0.0006352(92)  & 0.000635(10)  \\
 & $a m_{h}^{\rm phys}$ & 0.02556(23)  & 0.02486(40)  & 0.02579(21)  & 0.02590(21)  \\
 & $a^{-1}$ & 2.352(9) GeV & 2.379(17) GeV & 2.343(8) GeV & 2.339(8) GeV \\
\hline
 \rule{0cm}{0.4cm}\multirow{3}{*}{32ID-M1} & $a m_{l}^{\rm phys}$ & 0.00094(12)  & 0.00110(12)  & 0.00087(11)  & 0.00086(11)  \\
 & $a m_{h}^{\rm phys}$ & 0.0823(35)  & 0.0860(32)  & 0.0800(30)  & 0.0797(30)  \\
 & $a^{-1}$ & 1.002(20) GeV & 0.978(17) GeV & 1.015(17) GeV & 1.017(18) GeV \\
\hline
 \rule{0cm}{0.4cm}\multirow{3}{*}{32ID-M2} & $a m_{l}^{\rm phys}$ & ---  & -0.003404(35)  & ---  & -0.003367(37)  \\
 & $a m_{h}^{\rm phys}$ & ---  & 0.02486(97)  & ---  & 0.0255(11)  \\
 & $a^{-1}$ & --- & 2.025(34) GeV & --- & 1.990(35) GeV \\
\hline
\hline
\end{tabular}
}
\caption{The (uncorrelated) $\chi^{2}$/dof, unrenormalized physical quark masses in bare lattice units (without $m_{\mathrm{res}}$ included), and the values of the inverse lattice spacing $a^{-1}$ in physical units, obtained from fits to $SU(2)$ PQChPT with the stated pion mass cuts.}
\label{tab:su2_ainv_mq_chisq}
\end{table}
\vspace*{\fill}

\begin{table}[h]
\centering
\resizebox{\columnwidth}{!}{
\begin{tabular}{cc||cc|cc}
\hline
\hline
\rule{0cm}{0.4cm} & & NLO ($370 \, \mathrm{MeV}$ cut) & NLO ($450 \, \mathrm{MeV}$ cut) & NNLO ($370 \, \mathrm{MeV}$ cut) & NNLO ($450 \, \mathrm{MeV}$ cut) \\
\hline
 \rule{0cm}{0.4cm}\multirow{3}{*}{24I} & $Z_{l}$ & 0.980(11) & 0.959(11) & 0.9842(97) & 0.979(10) \\
 & $Z_{h}$ & 0.9711(82) & 0.950(10) & 0.9756(78) & 0.9770(73) \\
 & $R_{a}$ & 0.7561(61) & 0.7402(73) & 0.7596(58) & 0.7604(56) \\
 \hline
 \rule{0cm}{0.4cm}\multirow{3}{*}{32I} & $Z_{l}$ & $\equiv 1.0$ & $\equiv 1.0$ & $\equiv 1.0$ & $\equiv 1.0$ \\
 & $Z_{h}$ & $\equiv 1.0$ & $\equiv 1.0$ & $\equiv 1.0$ & $\equiv 1.0$ \\
 & $R_{a}$ & $\equiv 1.0$ & $\equiv 1.0$ & $\equiv 1.0$ & $\equiv 1.0$ \\
 \hline
 \rule{0cm}{0.4cm}\multirow{3}{*}{32ID} & $Z_{l}$ & 0.9162(79) & 0.908(11) & 0.9212(76) & 0.9186(84) \\
 & $Z_{h}$ & 0.9157(66) & 0.9028(99) & 0.9218(61) & 0.9258(59) \\
 & $R_{a}$ & 0.5822(45) & 0.5725(64) & 0.5858(41) & 0.5877(40) \\
 \hline
 \rule{0cm}{0.4cm}\multirow{3}{*}{32I-fine} & $Z_{l}$ & 0.994(30) & 0.995(31) & 0.995(30) & 1.001(30) \\
 & $Z_{h}$ & 0.989(21) & 1.021(20) & 0.980(21) & 0.978(21) \\
 & $R_{a}$ & 1.305(16) & 1.320(16) & 1.302(16) & 1.302(16) \\
 \hline
 \rule{0cm}{0.4cm}\multirow{3}{*}{48I} & $Z_{l}$ & 0.980(11) & 0.959(11) & 0.9842(97) & 0.979(10) \\
 & $Z_{h}$ & 0.9711(82) & 0.950(10) & 0.9756(78) & 0.9770(73) \\
 & $R_{a}$ & 0.7360(69) & 0.7174(76) & 0.7391(65) & 0.7393(62) \\
 \hline
 \rule{0cm}{0.4cm}\multirow{3}{*}{64I} & $Z_{l}$ & $\equiv 1.0$ & $\equiv 1.0$ & $\equiv 1.0$ & $\equiv 1.0$ \\
 & $Z_{h}$ & $\equiv 1.0$ & $\equiv 1.0$ & $\equiv 1.0$ & $\equiv 1.0$ \\
 & $R_{a}$ & 0.9968(57) & 0.9892(52) & 0.9973(57) & 0.9981(57) \\
 \hline
 \rule{0cm}{0.4cm}\multirow{3}{*}{32ID-M1} & $Z_{l}$ & 0.708(15) & 0.682(15) & 0.720(14) & 0.719(14) \\
 & $Z_{h}$ & 0.719(15) & 0.694(15) & 0.733(13) & 0.737(13) \\
 & $R_{a}$ & 0.4246(83) & 0.4067(77) & 0.4321(74) & 0.4338(74) \\
 \hline
 \rule{0cm}{0.4cm}\multirow{3}{*}{32ID-M2} & $Z_{l}$ & --- & 1.013(13) & --- & 1.013(16) \\
 & $Z_{h}$ & --- & 1.009(14) & --- & 1.028(18) \\
 & $R_{a}$ & --- & 0.8419(97) & --- & 0.849(12) \\
\hline
\hline
\end{tabular}
}
\caption{Ratios of lattice spacings ($R_{a}$) and light and heavy quark masses ($Z_{l}$, $Z_{h}$) between each ensemble and the reference 32I ensemble.}
\label{tab:su2_z_r}
\end{table}

\begin{table}[h]
\centering
\resizebox{\columnwidth}{!}{
\begin{tabular}{cc||cc|cc}
\hline
\hline
\rule{0cm}{0.4cm} LEC & $\Lambda_{\chi}$ & NLO ($370 \, \mathrm{MeV}$ cut) & NLO ($450 \, \mathrm{MeV}$ cut) & NNLO ($370 \, \mathrm{MeV}$ cut) & NNLO ($450 \, \mathrm{MeV}$ cut) \\
\hline
 \rule{0cm}{0.4cm}$B$ & \multirow{2}{*}{---} & 4.229(35) GeV & 4.270(41) GeV & 4.189(43) GeV & 4.203(44) GeV \\
 $f$ & & 0.1213(15) GeV & 0.1236(20) GeV & 0.1207(17) GeV & 0.1215(16) GeV \\
\hline
 \rule{0cm}{0.4cm}$10^{3} \hat{L}_{0}^{(2)}$ & \multirow{10}{*}{1 GeV} & --- & --- & -3.8(2.5) & 1.0(1.1) \\
 $10^{3} \hat{L}_{1}^{(2)}$ & & --- & --- & 0.52(71) & -0.62(52) \\
 $10^{3} \hat{L}_{2}^{(2)}$ & & --- & --- & -4.1(1.7) & 0.06(74) \\
 $10^{3} \hat{L}_{3}^{(2)}$ & & --- & --- & 1.1(1.4) & -1.56(87) \\
 $10^{3} \hat{L}_{4}^{(2)}$ & & -0.211(79) & -0.038(51) & -0.31(25) & -0.56(22) \\
 $10^{3} \hat{L}_{5}^{(2)}$ & & 0.438(72) & 0.501(43) & 0.37(34) & 0.60(28) \\
 $10^{3} \hat{L}_{6}^{(2)}$ & & -0.175(48) & -0.054(31) & -0.19(13) & -0.38(10) \\
 $10^{3} \hat{L}_{7}^{(2)}$ & & --- & --- & -1.30(48) & -0.75(27) \\
 $10^{3} \hat{L}_{8}^{(2)}$ & & 0.594(36) & 0.581(22) & 0.52(16) & 0.69(13) \\
\hline
\rule{0cm}{0.4cm}$10^{3} \hat{L}_{0}^{(2)}$ & \multirow{10}{*}{770 MeV} & --- & --- & -3.7(2.8) & 1.1(1.1) \\
$10^{3} \hat{L}_{1}^{(2)}$ & & --- & --- & 0.63(90) & -0.52(53) \\
$10^{3} \hat{L}_{2}^{(2)}$ & & --- & --- & -3.9(2.0) & 0.27(78) \\
$10^{3} \hat{L}_{3}^{(2)}$ & & --- & --- & 1.3(1.3) & -1.42(85) \\
$10^{3} \hat{L}_{4}^{(2)}$ & & -0.004(79) & 0.169(51) & -0.10(27) & -0.35(22) \\
$10^{3} \hat{L}_{5}^{(2)}$ & & 0.852(72) & 0.915(43) & 0.78(35) & 1.02(28) \\
$10^{3} \hat{L}_{6}^{(2)}$ & & -0.019(48) & 0.101(31) & -0.04(14) & -0.23(10) \\
$10^{3} \hat{L}_{7}^{(2)}$ & & --- & --- & -1.30(52) & -0.75(26) \\
$10^{3} \hat{L}_{8}^{(2)}$ & & 0.594(36) & 0.581(22) & 0.52(17) & 0.69(13) \\
\hline
 \rule{0cm}{0.4cm}$10^{6} \left( \hat{K}_{17}^{(2)}-\hat{K}_{39}^{(2)} \right)$ & \multirow{10}{*}{1 GeV} & --- & --- & -7.3(2.0) & -7.6(1.1) \\
 $10^{6} \left( \hat{K}_{18}^{(2)}+6\hat{K}_{27}^{(2)}-\hat{K}_{40}^{(2)} \right)$ & & --- & --- & 14.5(7.9) & 19.2(4.7) \\
 $10^{6} \hat{K}_{19}^{(2)}$ & & --- & --- & 11(16) & -0.9(4.2) \\
 $10^{6} \hat{K}_{20}^{(2)}$ & & --- & --- & -12(10) & -3.2(2.8) \\
 $10^{6} \left( \hat{K}_{21}^{(2)}+2K_{22}^{(2)} \right)$ & & --- & --- & -7.6(6.9) & 4.9(4.1) \\
 $10^{6} \hat{K}_{23}^{(2)}$ & & --- & --- & -12.4(3.5) & -2.8(1.4) \\
 $10^{6} \hat{K}_{25}^{(2)}$ & & --- & --- & 6.7(4.5) & 1.3(1.7) \\
 $10^{6} \left( \hat{K}_{26}^{(2)}+6\hat{K}_{27}^{(2)} \right)$ & & --- & --- & 3.2(6.8) & 11.2(3.6) \\
\hline
 \rule{0cm}{0.4cm}$10^{6} \left( \hat{K}_{17}^{(2)}-\hat{K}_{39}^{(2)} \right)$ & \multirow{10}{*}{770 MeV} & --- & --- & -6.2(1.5) & -5.33(77) \\
 $10^{6} \left( \hat{K}_{18}^{(2)}+6\hat{K}_{27}^{(2)}-\hat{K}_{40}^{(2)} \right)$ & & --- & --- & 8.3(6.6) & 14.5(3.9) \\
 $10^{6} \hat{K}_{19}^{(2)}$ & & --- & --- & 3(12) & -3.9(2.3) \\
 $10^{6} \hat{K}_{20}^{(2)}$ & & --- & --- & -5.0(7.5) & 0.0(1.8) \\
 $10^{6} \left( \hat{K}_{21}^{(2)}+2K_{22}^{(2)} \right)$ & & --- & --- & -6.8(7.0) & 6.2(3.2) \\
 $10^{6} \hat{K}_{23}^{(2)}$ & & --- & --- & -7.2(3.2) & -0.2(1.2) \\
 $10^{6} \hat{K}_{25}^{(2)}$ & & --- & --- & 2.4(3.4) & -1.0(1.1) \\
 $10^{6} \left( \hat{K}_{26}^{(2)}+6\hat{K}_{27}^{(2)} \right)$ & & --- & --- & 1.9(6.4) & 10.1(3.1) \\
\hline
\hline
\end{tabular}
}
\caption{$SU(2)$ PQChPT LECs fit at two different chiral scales --- $\Lambda_{\chi} = 1 \, \mathrm{GeV}$ and $\Lambda_{\chi} = 770 \, \mathrm{MeV}$ --- in units of the canonical size at a given order in the chiral expansion. The LECs $\hat{L}_{7}^{(2)}$ and $\hat{L}_{8}^{(2)}$ have no scale dependence. The value of $B$ quoted here is unrenormalized.}
\label{tab:su2_pqchpt_lecs}
\end{table}

\begin{table}[h]
\centering
\resizebox{\columnwidth}{!}{
\begin{tabular}{c||cc|cc}
\hline
\hline
\rule{0cm}{0.4cm} Parameter & NLO ($370 \, \mathrm{MeV}$ cut) & NLO ($450 \, \mathrm{MeV}$ cut) & NNLO ($370 \, \mathrm{MeV}$ cut) & NNLO ($450 \, \mathrm{MeV}$ cut) \\
\hline
 \rule{0cm}{0.4cm}$m^{(K)}$ & 0.4863(27) GeV & 0.4861(43) GeV & 0.4862(24) GeV & 0.4862(25) GeV \\
 $f^{(K)}$ & 0.1501(17) GeV & 0.1535(22) GeV & 0.1490(17) GeV & 0.1488(16) GeV \\
\hline
\rule{0cm}{0.4cm}$10^{3} \lambda_1$ & 3.2(1.0) & 3.64(98) & 3.2(1.0) & 3.3(1.0) \\
 $10^{3} \lambda_2$ & 28.17(65) & 28.45(65) & 28.27(78) & 28.76(74) \\
 $10^{3} \lambda_3$ & -3.9(1.1) & -3.22(98) & -3.8(1.1) & -3.9(1.0) \\
 $10^{3} \lambda_4$ & 5.69(31) & 5.82(32) & 5.70(31) & 5.83(33) \\
\hline
 \rule{0cm}{0.4cm}$c_{f}^{\scriptscriptstyle I}$ & 0.059(47) $\mathrm{GeV}^{2}$ & -0.028(51) $\mathrm{GeV}^{2}$ & 0.081(48) $\mathrm{GeV}^{2}$ & 0.065(45) $\mathrm{GeV}^{2}$ \\
 $c_{f}^{\scriptscriptstyle ID}$ & -0.013(17) $\mathrm{GeV}^{2}$ & -0.058(19) $\mathrm{GeV}^{2}$ & 0.013(15) $\mathrm{GeV}^{2}$ & 0.012(16) $\mathrm{GeV}^{2}$ \\
 $c_{f^{(K)}}^{\scriptscriptstyle I}$ & 0.049(39) $\mathrm{GeV}^{2}$ & -0.035(38) $\mathrm{GeV}^{2}$ & 0.070(41) $\mathrm{GeV}^{2}$ & 0.069(36) $\mathrm{GeV}^{2}$ \\
 $c_{f^{(K)}}^{\scriptscriptstyle ID}$ & -0.005(15) $\mathrm{GeV}^{2}$ & -0.044(14) $\mathrm{GeV}^{2}$ & 0.011(15) $\mathrm{GeV}^{2}$ & 0.019(15) $\mathrm{GeV}^{2}$ \\
\hline
 \rule{0cm}{0.4cm}$c_{m_h,m_{\pi}^{2}}$ & 1.6(2.7) & 0.1(2.2) & 1.4(2.7) & 0.9(2.1) \\
 $c_{m_h,f_{\pi}}$ & 0.14(11) & 0.061(89) & 0.221(97) & 0.257(80) \\
 $c_{m_y,m_{K}^{2}}$ & 3.915(22) GeV & 3.981(34) GeV & 3.895(20) GeV & 3.884(20) GeV \\
 $c_{m_h,m_{K}^{2}}$ & 0.008(52) GeV & 0.046(58) GeV & 0.022(51) GeV & 0.026(56) GeV \\
 $c_{m_y,f_{K}}$ & 0.2926(62) & 0.2983(59) & 0.2906(64) & 0.2987(56) \\
 $c_{m_h,f_{K}}$ & 0.067(50) & 0.073(52) & 0.062(51) & 0.096(48) \\
\hline
 \rule{0cm}{0.4cm}$m^{(\Omega)}$ & 1.6646(47) GeV & 1.6643(91) GeV & 1.6643(37) GeV & 1.6644(36) GeV \\
 $c_{m_l,m_{\Omega}}$ & 3.54(74) & 3.73(67) & 3.68(74) & 3.66(76) \\
 $c_{m_y,m_{\Omega}}$ & 5.650(59) & 5.794(67) & 5.585(55) & 5.550(55) \\
 $c_{m_h,m_{\Omega}}$ & 2.31(62) & 3.19(55) & 1.83(61) & 1.64(63) \\
\hline
\hline
\end{tabular}
}
\caption{Additional fit parameters in physical units and adjusted to the physical strange quark mass. Here $\{m^{(K)},f^{(K)}\}$ and $\{\lambda_{i}\}$ are the LO and NLO LECs of heavy-meson $SU(2)$ PQChPT evaluated at the chiral scale $\Lambda_{\chi} = 1 \, \mathrm{GeV}$. $c_{f}^{I}$ and $c_{f}^{ID}$ are the $a^{2}$ coefficients of $f_{\pi}$ for the Iwasaki and Iwasaki+DSDR gauge actions, respectively, and likewise for $c_{f^{(K)}}^{I}$ and $c_{f^{(K)}}^{ID}$. The
notation $c_{m_{q}, X}$ denotes the coefficient of a term linear in $m_{q}$ for quantity $X$, and $m^{(\Omega)}$ is the constant term in the (linear) $m_{\Omega}$ ansatz. We emphasize that the distinction between ``NLO'' and ``NNLO'' fits, as well as the mass cut, applies only to $m_{\pi}$ and $f_{\pi}$: the kaon and $\Omega$ baryon data and fit forms are the same in all of the fits.}
\label{tab:more_su2_parameters}
\end{table}

\FloatBarrier

\subsection{Histograms}
\label{sec:su2_histograms}

\begin{figure}[h]
\centering
\subfloat[NLO, $m_{\pi}^{\rm cut} = 370 \, \mathrm{MeV}$]{\includegraphics[width=0.44\textwidth]{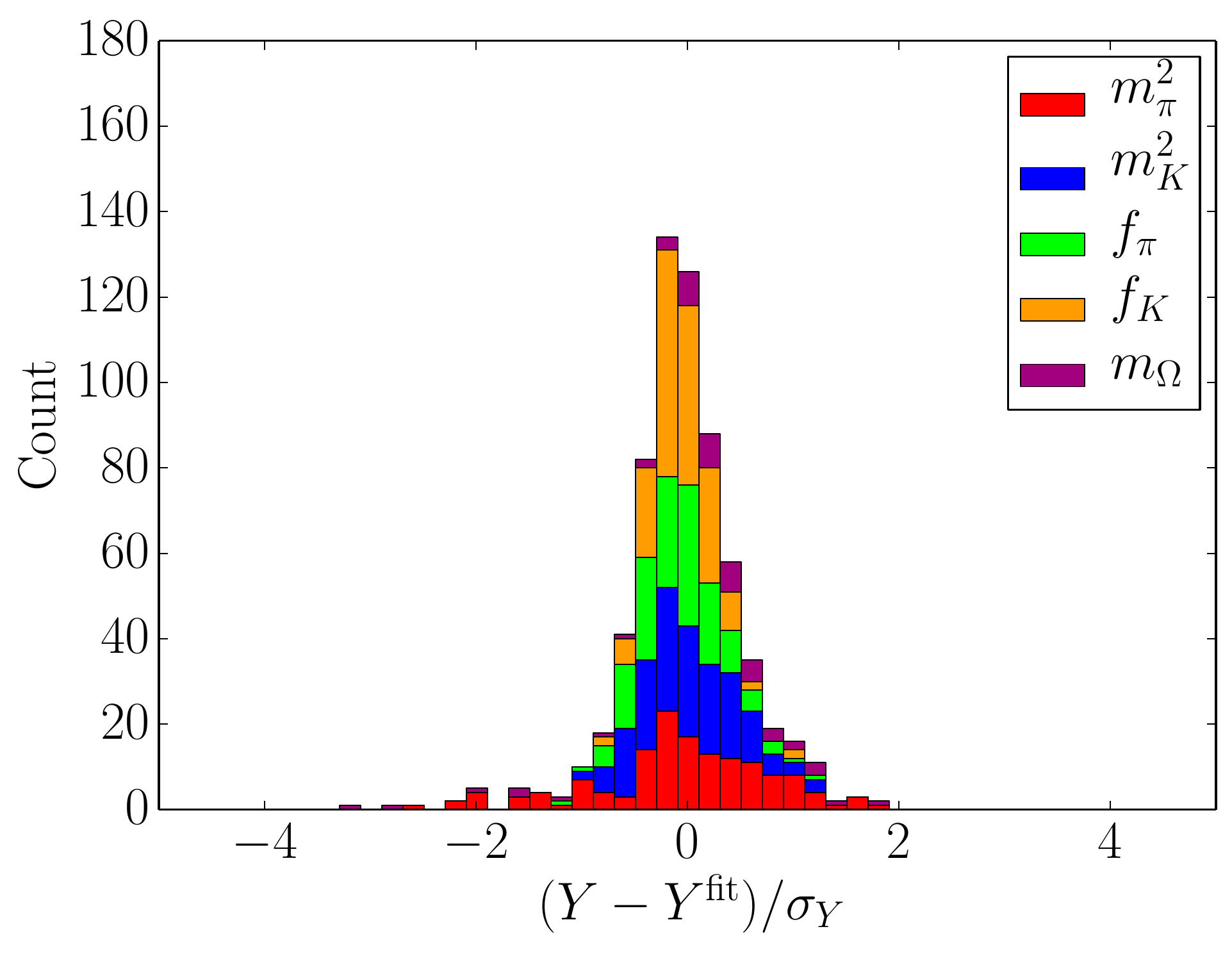}}
\subfloat[NLO, $m_{\pi}^{\rm cut} = 450 \, \mathrm{MeV}$]{\includegraphics[width=0.44\textwidth]{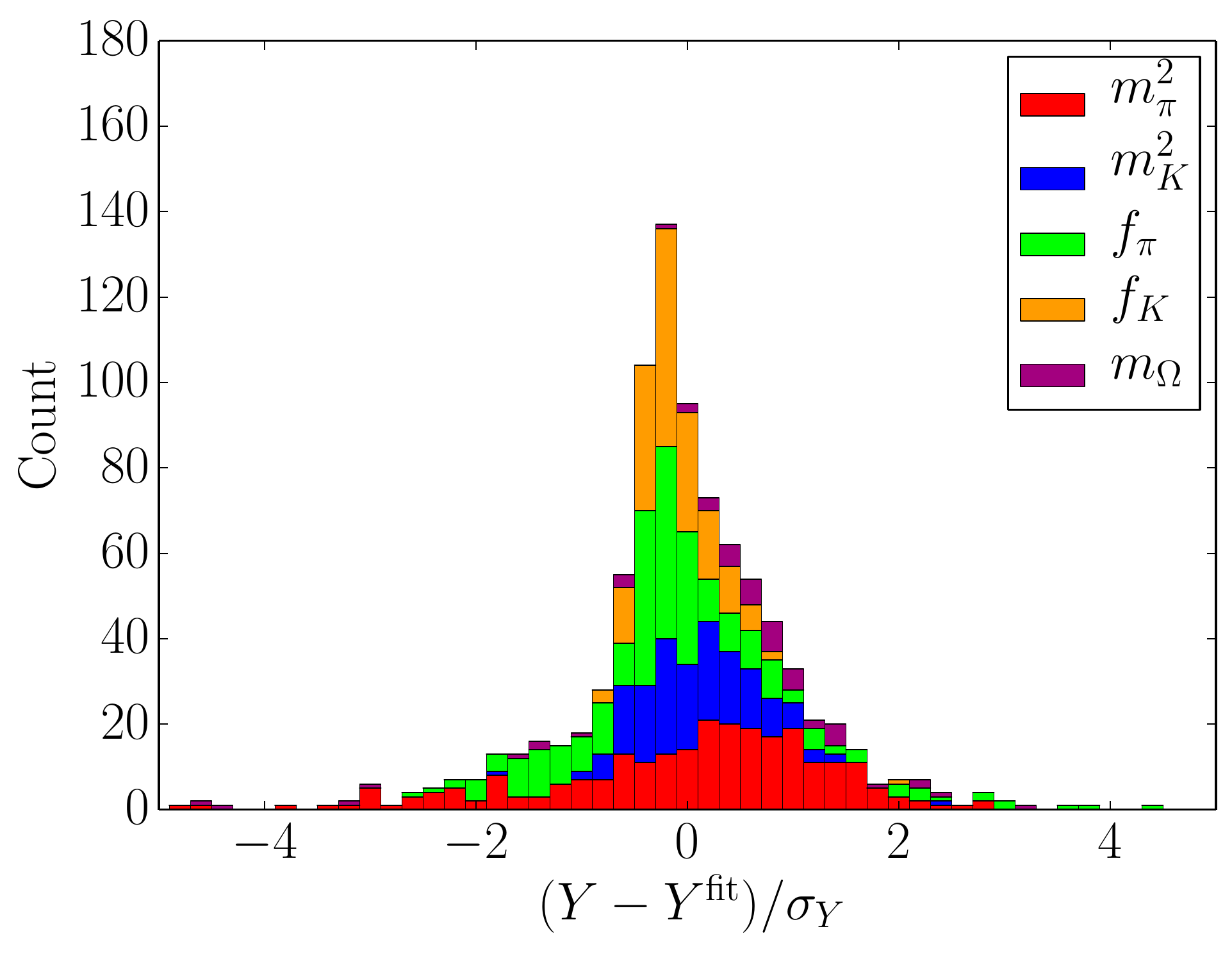}} \\
\subfloat[NNLO, $m_{\pi}^{\rm cut} = 370 \, \mathrm{MeV}$]{\includegraphics[width=0.44\textwidth]{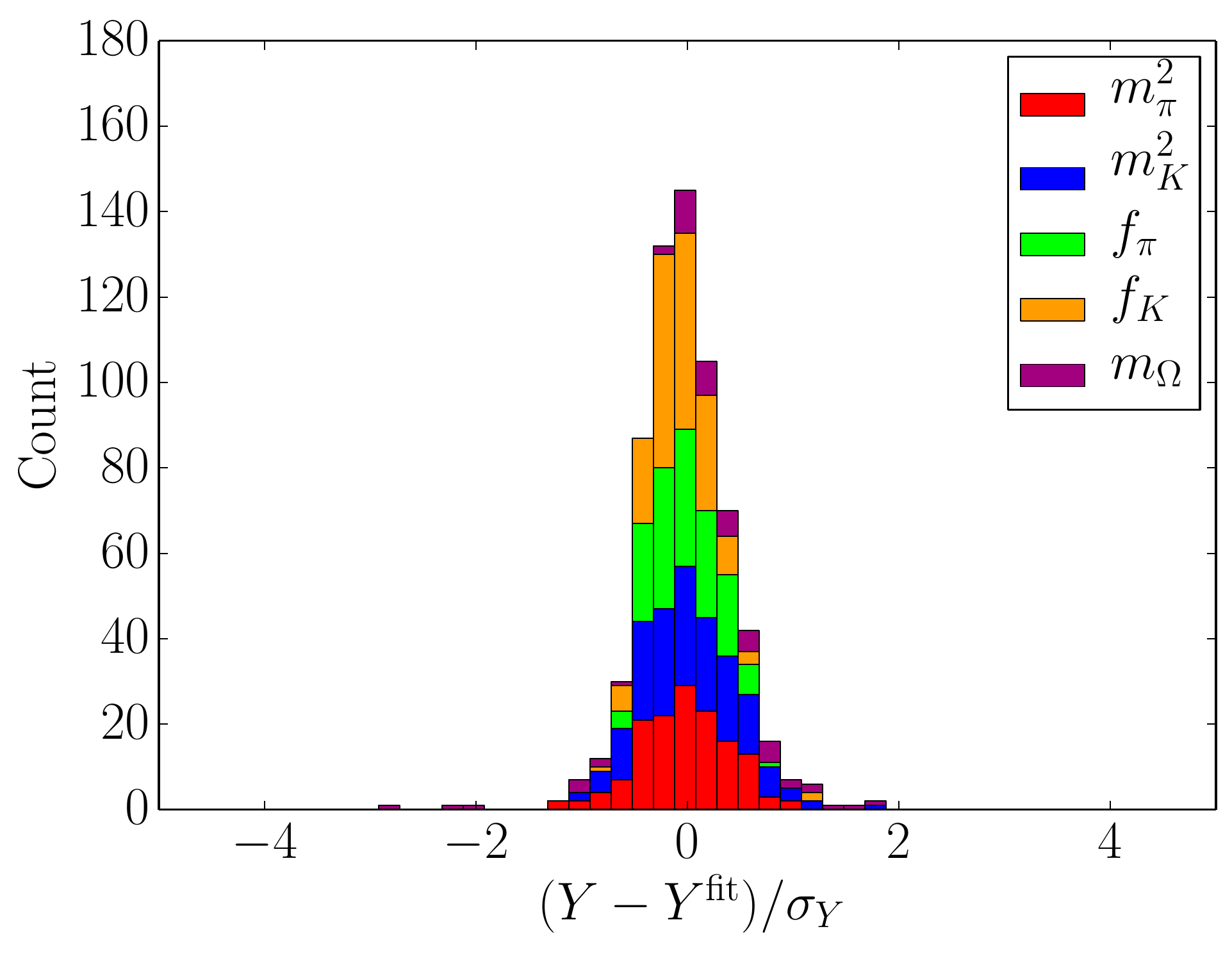}}
\subfloat[NNLO, $m_{\pi}^{\rm cut} = 450 \, \mathrm{MeV}$]{\includegraphics[width=0.44\textwidth]{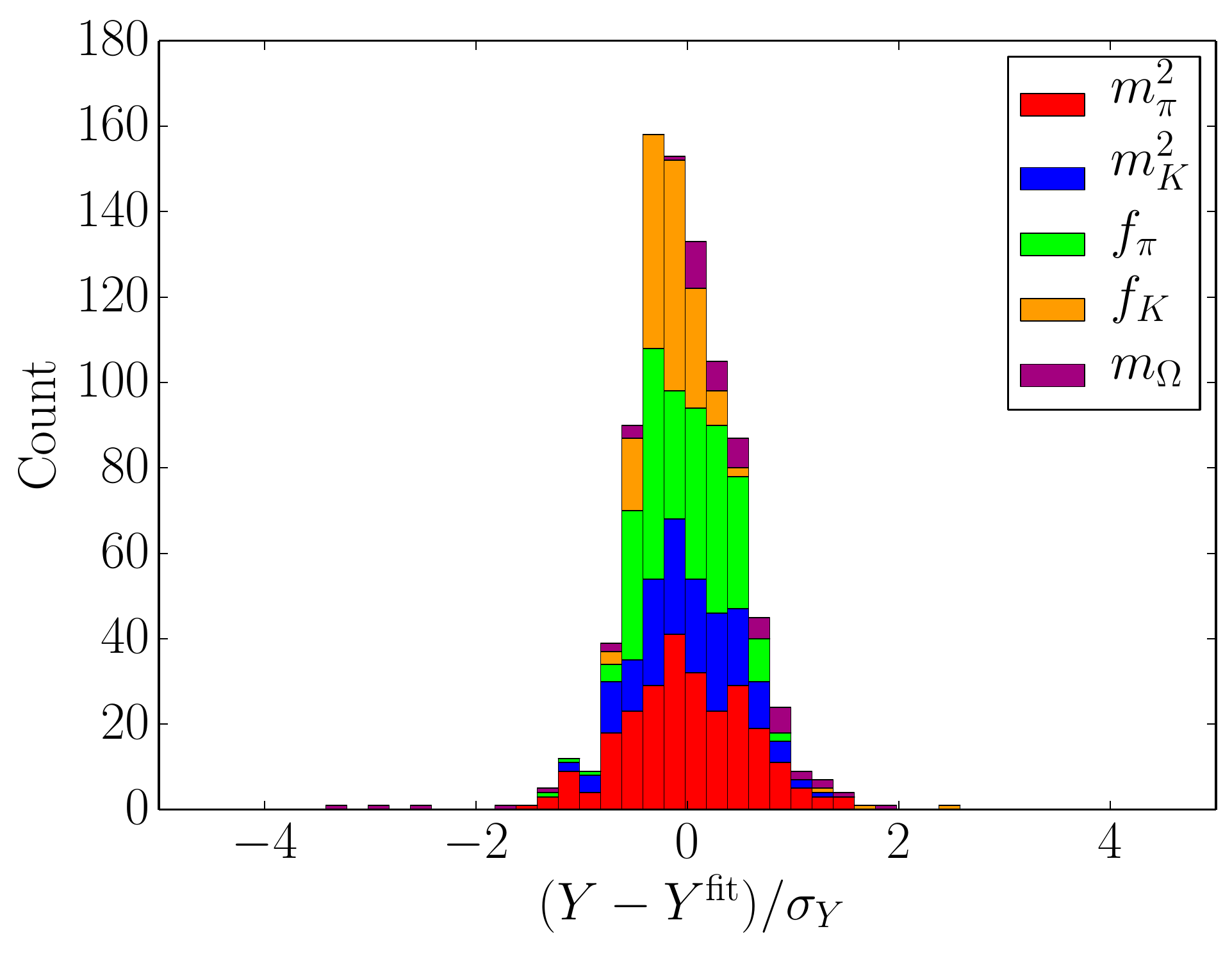}}
\caption{Stacked histograms of the signed deviation of the data from the fit in units of the standard deviation.}
\label{fig:su2_hists}
\end{figure}

In Figure~\ref{fig:su2_hists} we plot stacked histograms of the deviation of each data point $Y_{i}$ from the fit prediction $Y^{\rm fit}_{i}$ in units of the standard deviation of the data $\sigma_{Y_{i}}$: 
\begin{equation}
X_{i} \equiv \frac{Y_{i}-Y^{\rm fit}_{i}}{\sigma_{Y_{i}}}.
\end{equation}
This can be thought of as the signed square root of the contribution to $\chi^{2}$ from each data point, where the sign indicates whether the fit is overestimating (-) or underestimating (+) the data. The distributions of $m_{\pi}^{2}$ and $f_{\pi}$, in particular, give an overall impression of how well partially quenched $SU(2)$ chiral perturbation theory truncated to a given order is able to describe all of our (in general partially quenched) lattice data. We observe excellent agreement between the data and the NLO fit when we use a pion mass cut of 370 MeV, however, when we raise the mass cut to 450 MeV, the NLO fit clearly starts to break down, as evidenced by the larger $\chi^{2}/\mathrm{dof}$ and broader histogram with many $3 \sigma$ and $4 \sigma$ outliers. The NNLO ansatz appears to have no difficulty describing our full data set.

\FloatBarrier

\subsection{Unitary Chiral Extrapolation}
\label{sec:su2_chiral_fit}

\begin{figure}[h]
\centering
\subfloat[NLO, $m_{\pi}^{\rm cut} = 370 \, \mathrm{MeV}$]{\includegraphics[width=0.49\textwidth]{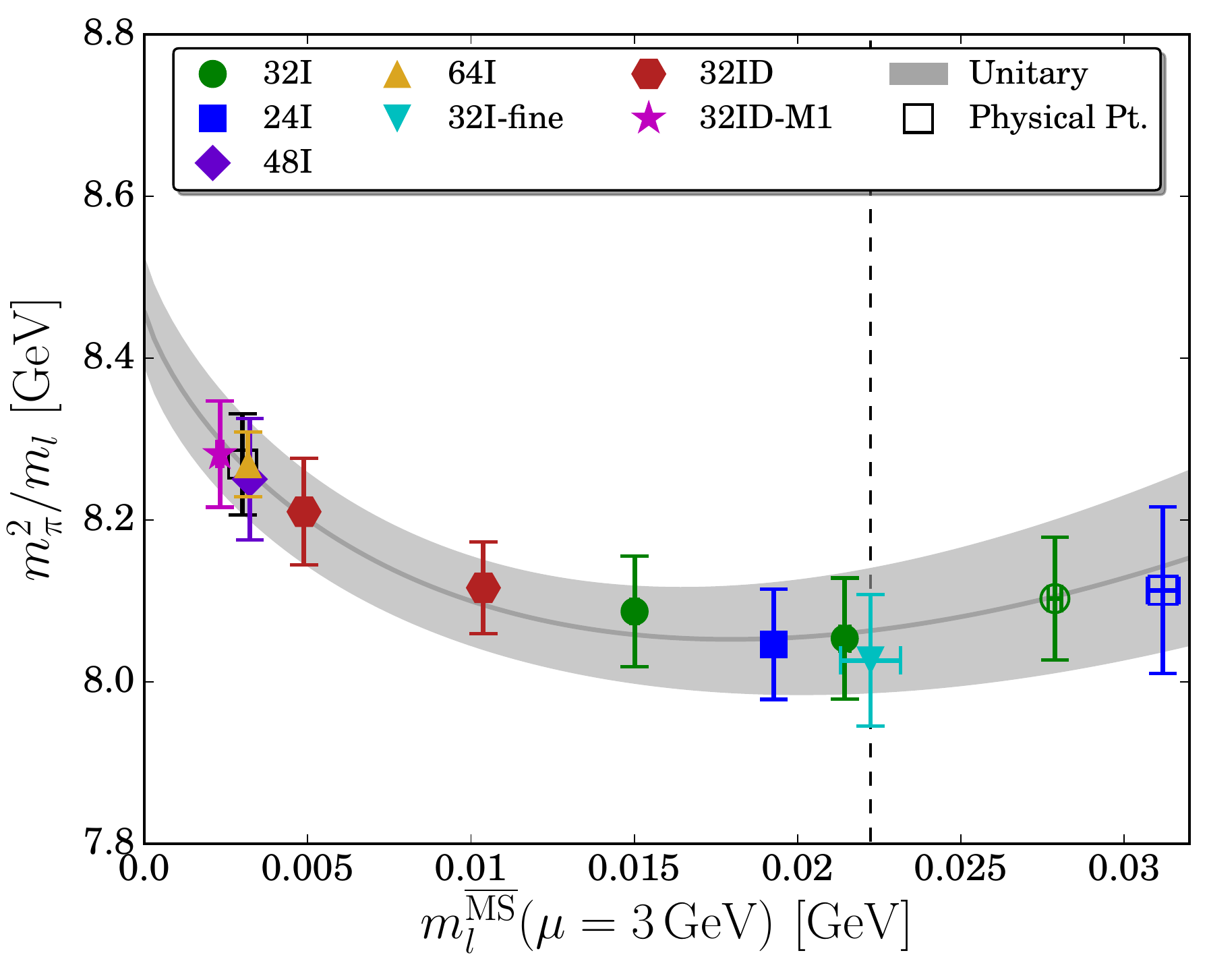}}
\subfloat[NLO, $m_{\pi}^{\rm cut} = 450 \, \mathrm{MeV}$]{\includegraphics[width=0.49\textwidth]{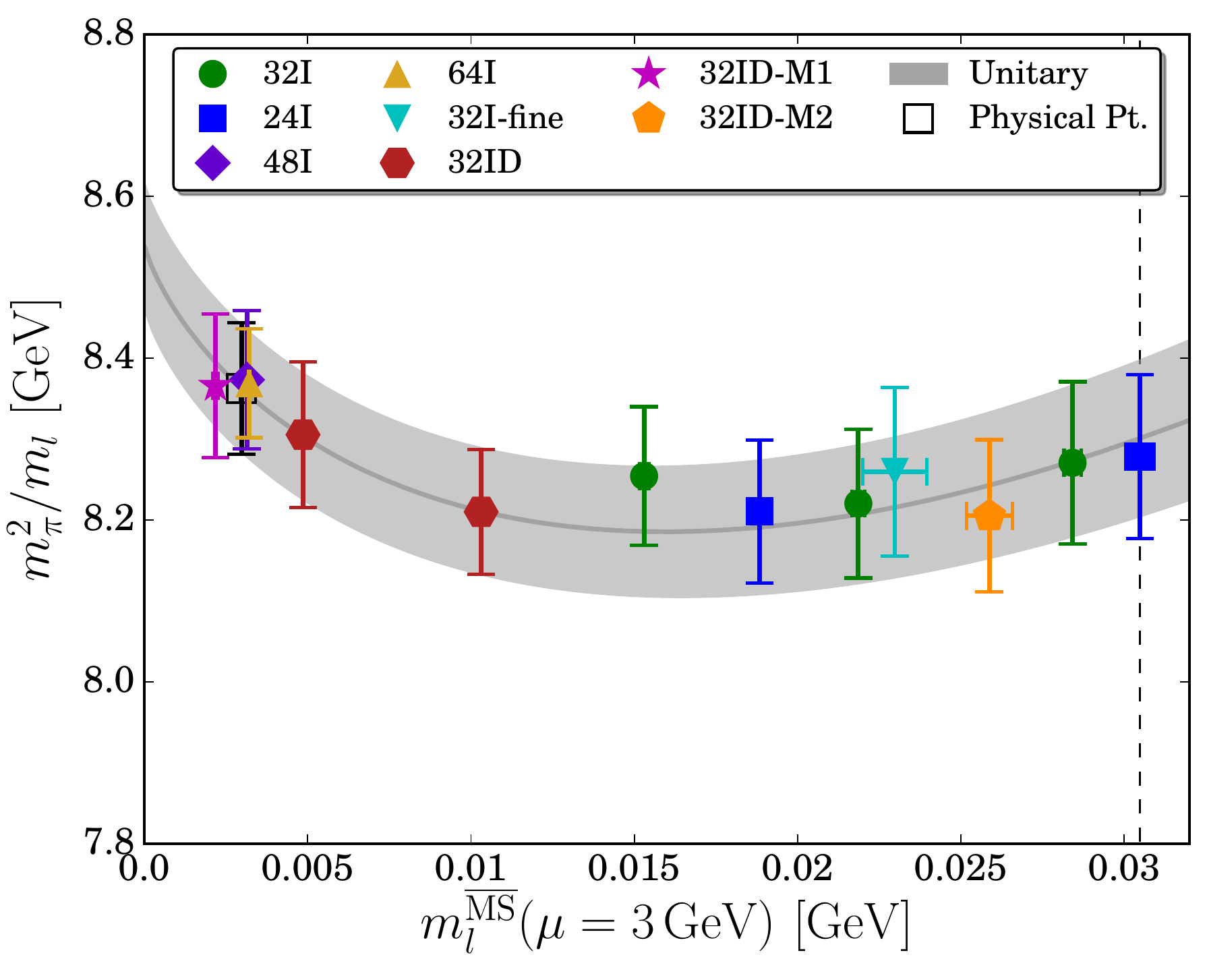}} \\
\subfloat[NNLO, $m_{\pi}^{\rm cut} = 370 \, \mathrm{MeV}$]{\includegraphics[width=0.49\textwidth]{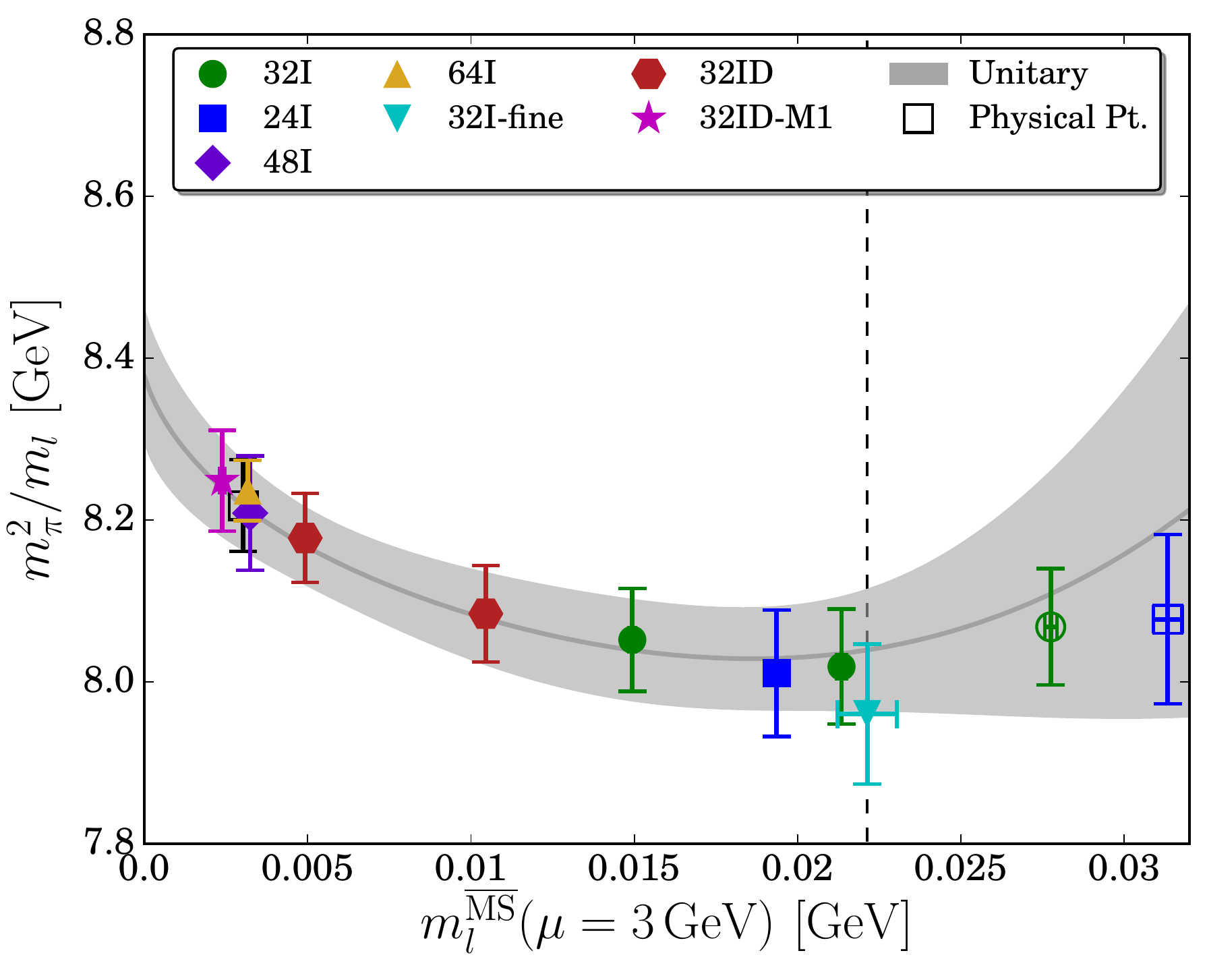}}
\subfloat[NNLO, $m_{\pi}^{\rm cut} = 450 \, \mathrm{MeV}$]{\includegraphics[width=0.49\textwidth]{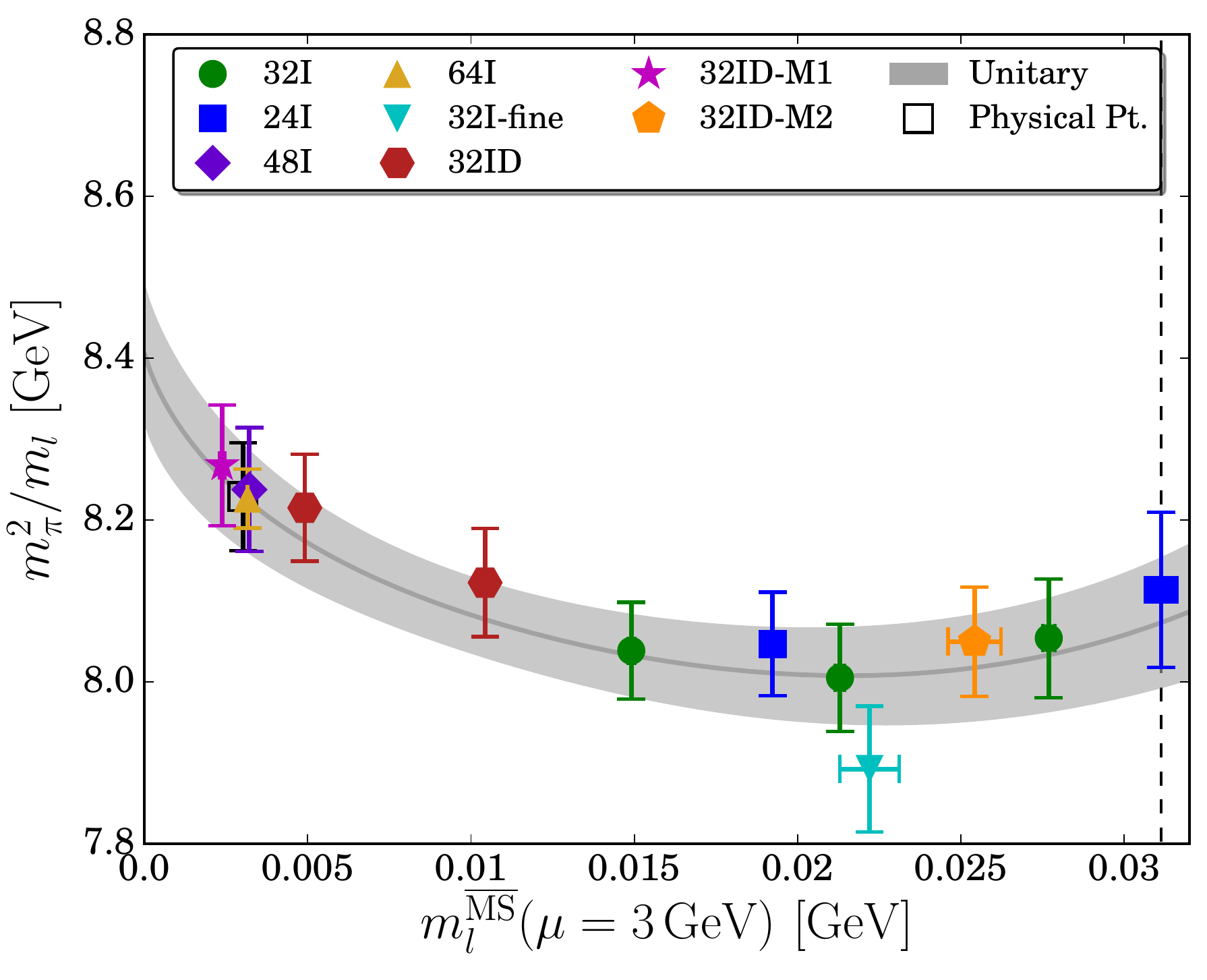}}
\caption{Chiral extrapolation of unitary $m_{\pi}^{2}$ data. The fit has been used to correct each data point from the simulated strange quark mass to the physical strange quark mass, as well as to take the infinite volume limit. Filled symbols correspond to sub-ensembles which were included in the fit, and open symbols correspond to sub-ensembles which were excluded from the fit based on the pion mass cut. The dashed vertical line corresponds to the heaviest unitary point included in the fit. ``Physical point'' is the prediction for the physical pion mass obtained by interpolating the fit to $m_{l}^{\rm phys}$.}
\label{fig:su2_unitary_mpi2}
\end{figure}

\begin{figure}[h]
\centering
\subfloat[NLO, $m_{\pi}^{\rm cut} = 370 \, \mathrm{MeV}$]{\includegraphics[width=0.49\textwidth]{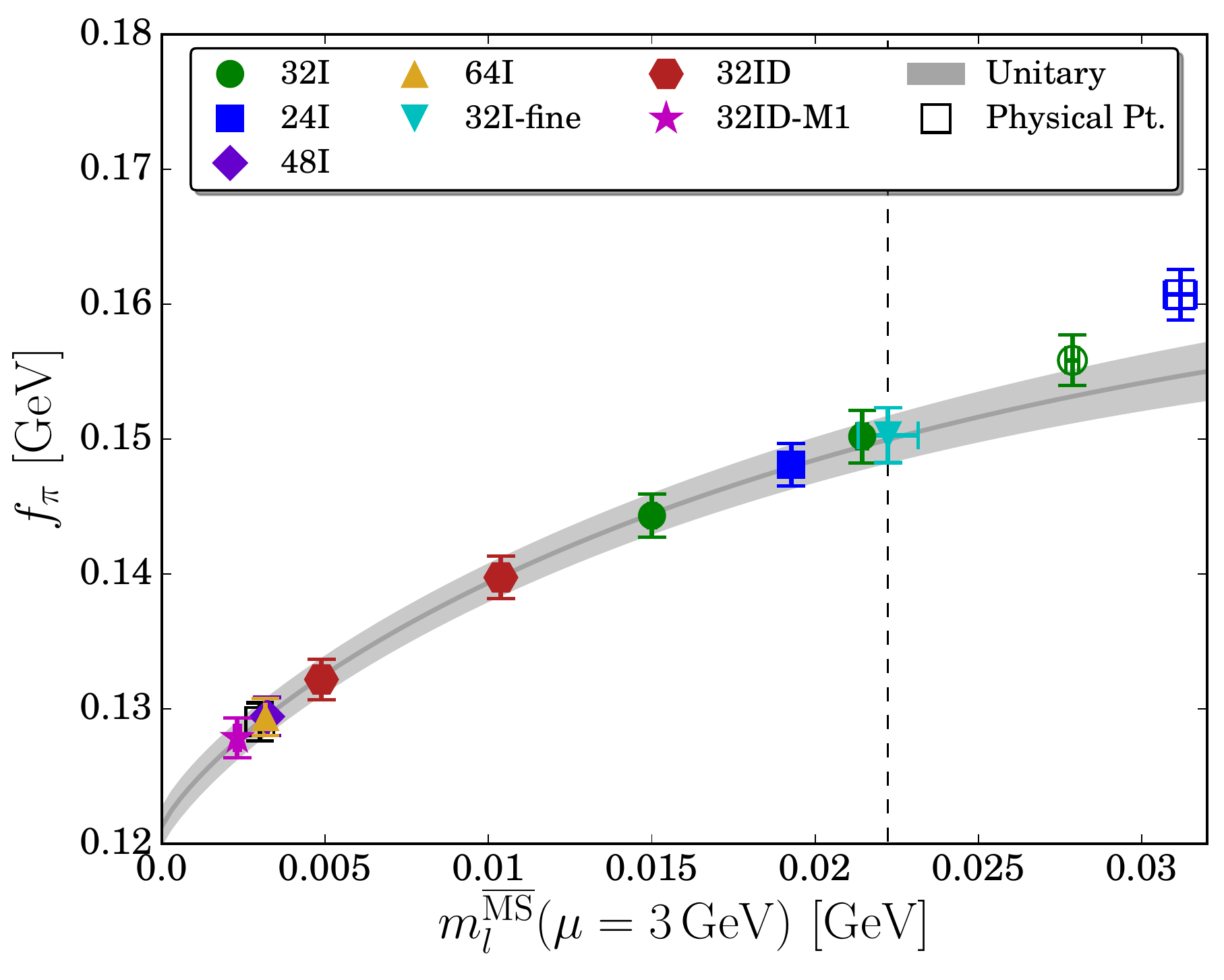}}
\subfloat[NLO, $m_{\pi}^{\rm cut} = 450 \, \mathrm{MeV}$]{\includegraphics[width=0.49\textwidth]{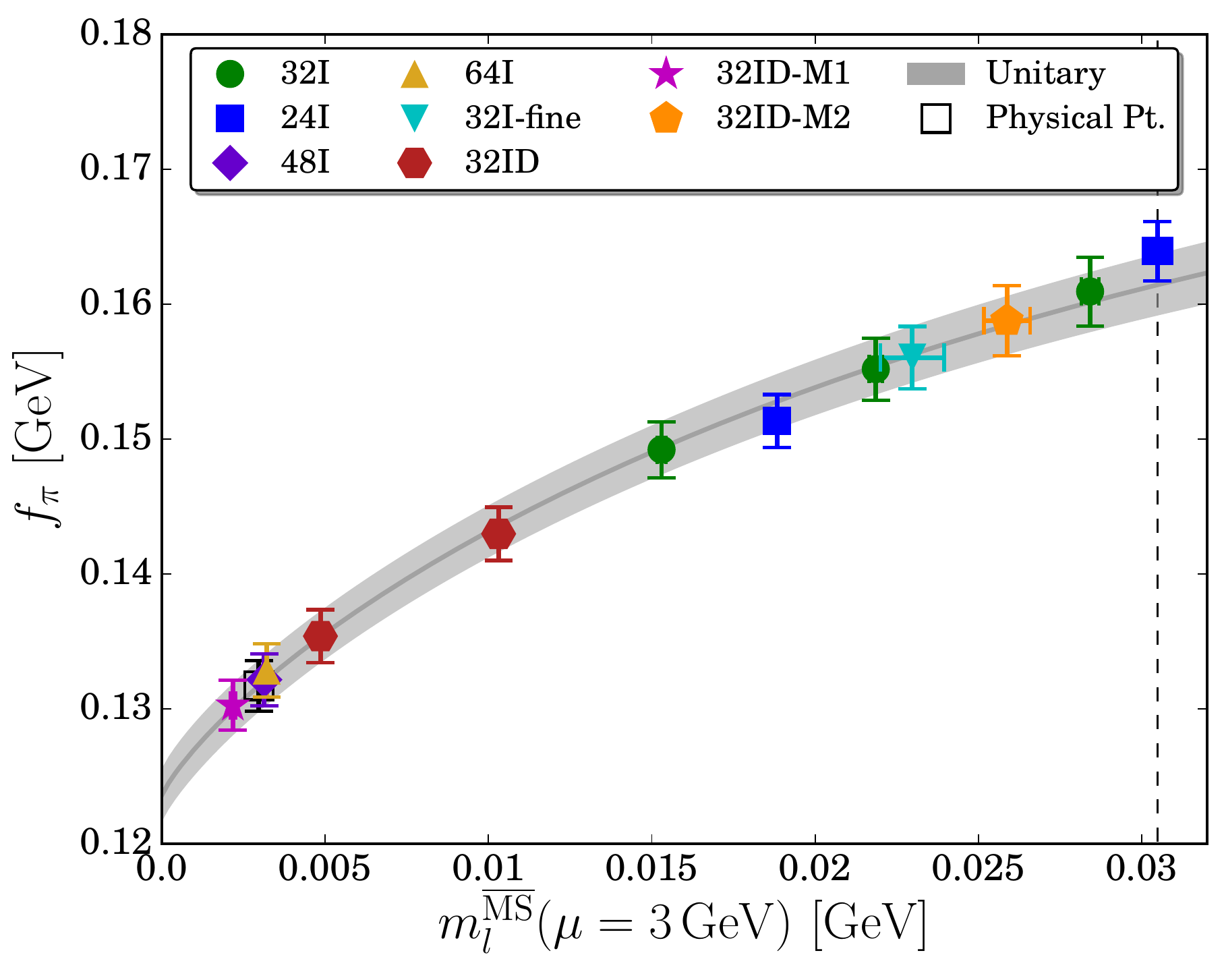}} \\
\subfloat[NNLO, $m_{\pi}^{\rm cut} = 370 \, \mathrm{MeV}$]{\includegraphics[width=0.49\textwidth]{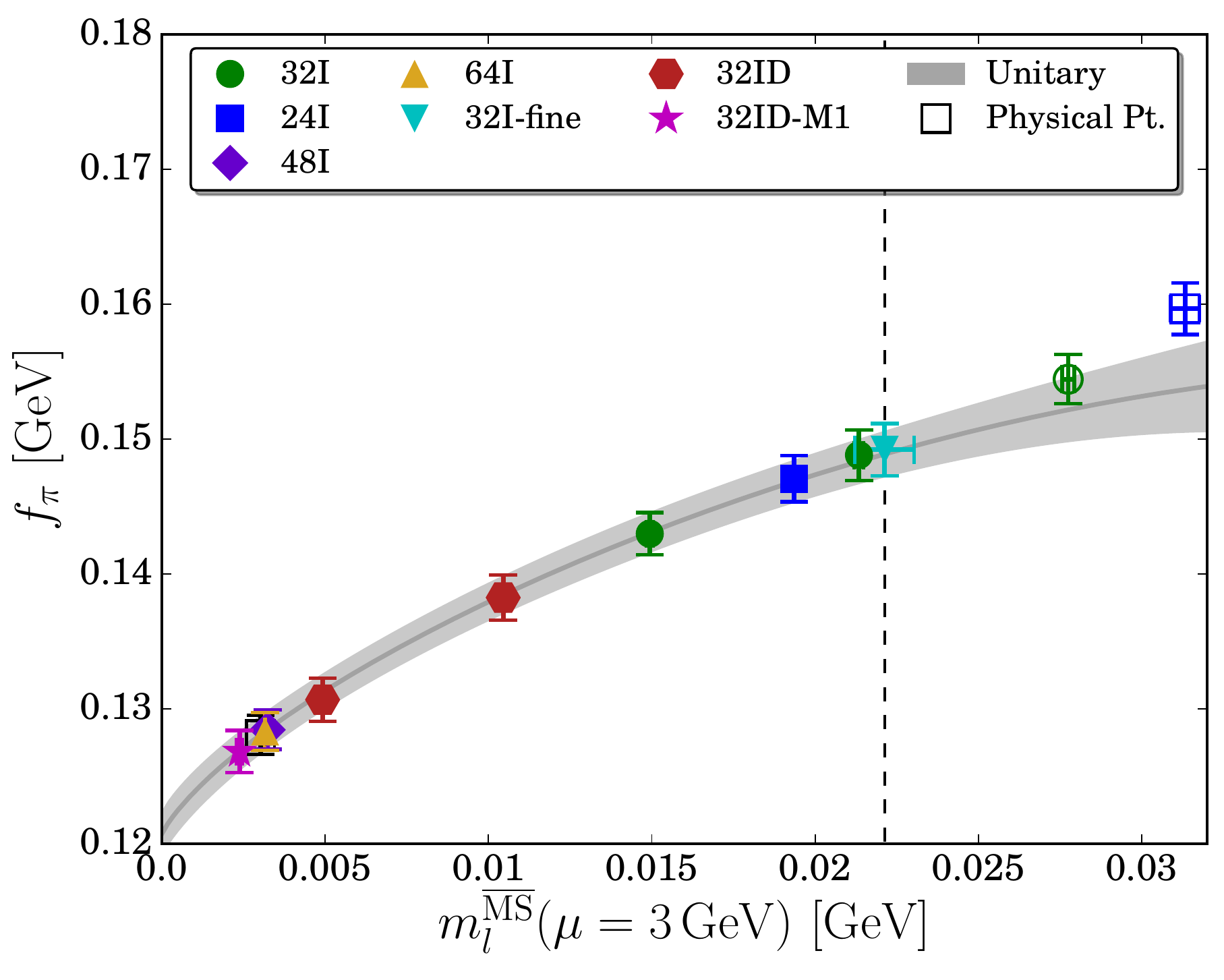}}
\subfloat[NNLO, $m_{\pi}^{\rm cut} = 450 \, \mathrm{MeV}$]{\includegraphics[width=0.49\textwidth]{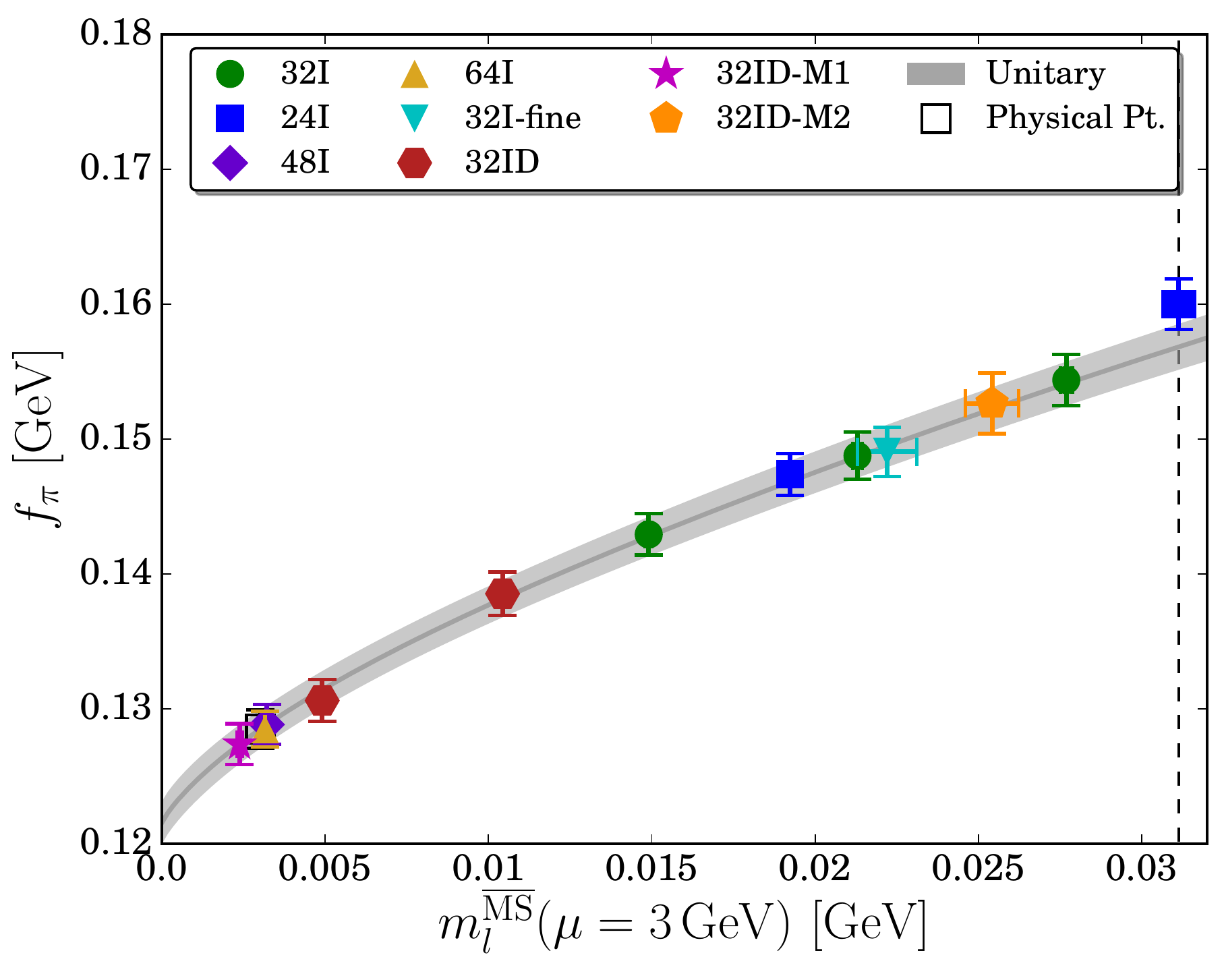}}
\caption{Chiral extrapolation of unitary $f_{\pi}$ data. The fit has been used to correct each data point from the simulated strange quark mass to the physical strange quark mass, as well as to take the infinite volume and continuum limits. Filled symbols correspond to sub-ensembles which were included in the fit, and open symbols correspond to sub-ensembles which were excluded from the fit based on the pion mass cut. The dashed vertical line corresponds to the heaviest unitary point included in the fit. ``Physical point'' is the prediction for the physical pion decay constant obtained by interpolating the fit to $m_{l}^{\rm phys}$.}
\label{fig:su2_unitary_fpi}
\end{figure}

In Figures~\ref{fig:su2_unitary_mpi2} and~\ref{fig:su2_unitary_fpi} we overlay the unitary measurements of $m_{\pi}^{2}/m_{l}$ and $f_{\pi}$ on each ensemble with the ChPT prediction obtained using the LECs from each fit. The fit results have also been used to correct each lattice measurement from the simulated point to the continuum, infinite volume, and physical strange quark mass limit. The light quark mass has been renormalized in the $\overline{\rm MS}$ scheme at 3 GeV using the renormalization coefficient computed in Ref.~\cite{Blum:2014tka}. \\

The influence of the NNLO terms is most clear in the chiral fits to $f_{\pi}$ (Figure~\ref{fig:su2_unitary_fpi}), which, in general, exhibit a more pronounced nonlinearity in the light quark mass than the chiral fits to $m_{\pi}^{2}$. While we observe that both $m_{\pi}^{2}$ and $f_{\pi}$ are consistent between the NLO and NNLO fits with a mass cut of 370 MeV, when the mass cut is raised to 450 MeV the NLO and NNLO ans\"{a}tze accommodate the additional heavy data differently. For the NLO case the entire $m_{\pi}^{2}$ and $f_{\pi}$ curves are systematically shifted upward to higher energy --- as one can see by comparing this fit to the adjacent NLO fit with $m_{\pi}^{\rm cut} = 370 \, \mathrm{MeV}$ in Figures~\ref{fig:su2_unitary_mpi2} and~\ref{fig:su2_unitary_fpi} --- providing further evidence that this heavy data has extended into a regime where NLO PQChPT is no longer reliable. A similar comparison between the NNLO fits suggests that the heavy data influences these fits by smoothing out the curvature of $f_{\pi}$ in the heavy mass regime $m_{l}^{\rm \overline{MS}} \gtrsim 0.025 \, \mathrm{MeV}$. \\ 

\vspace*{\fill}
\vspace*{\fill}

\vspace*{\fill}
\vspace*{\fill}
\FloatBarrier

\subsection{Chiral Expansion}
\label{sec:su2_convergence}

Chiral perturbation theory is an effective field theory with an asymptotic series expansion. For ChPT to have any practical use it must be applied in a regime where the expansion is \textit{well-ordered} in the sense that $|\mathrm{LO}| > |\mathrm{NLO}| > |\mathrm{NNLO}| > \cdots$, since calculations beyond one or two loops are generally intractable, and higher order terms must be neglected. This in turn restricts the range of quark masses for which ChPT is applicable. While the very light masses of the up and down quarks suggest that the $SU(2)$ expansion ought to be well-ordered at the physical point, one expects that there is an upper limit, beyond which the $\mathrm{N}^{3}\mathrm{LO}$ and higher order terms can no longer be discarded if one expects ChPT to describe low-energy QCD with high precision. In this section we use our NNLO fits to probe this scale. \\

\begin{figure}[h]
\centering
\subfloat{\includegraphics[width=0.49\textwidth]{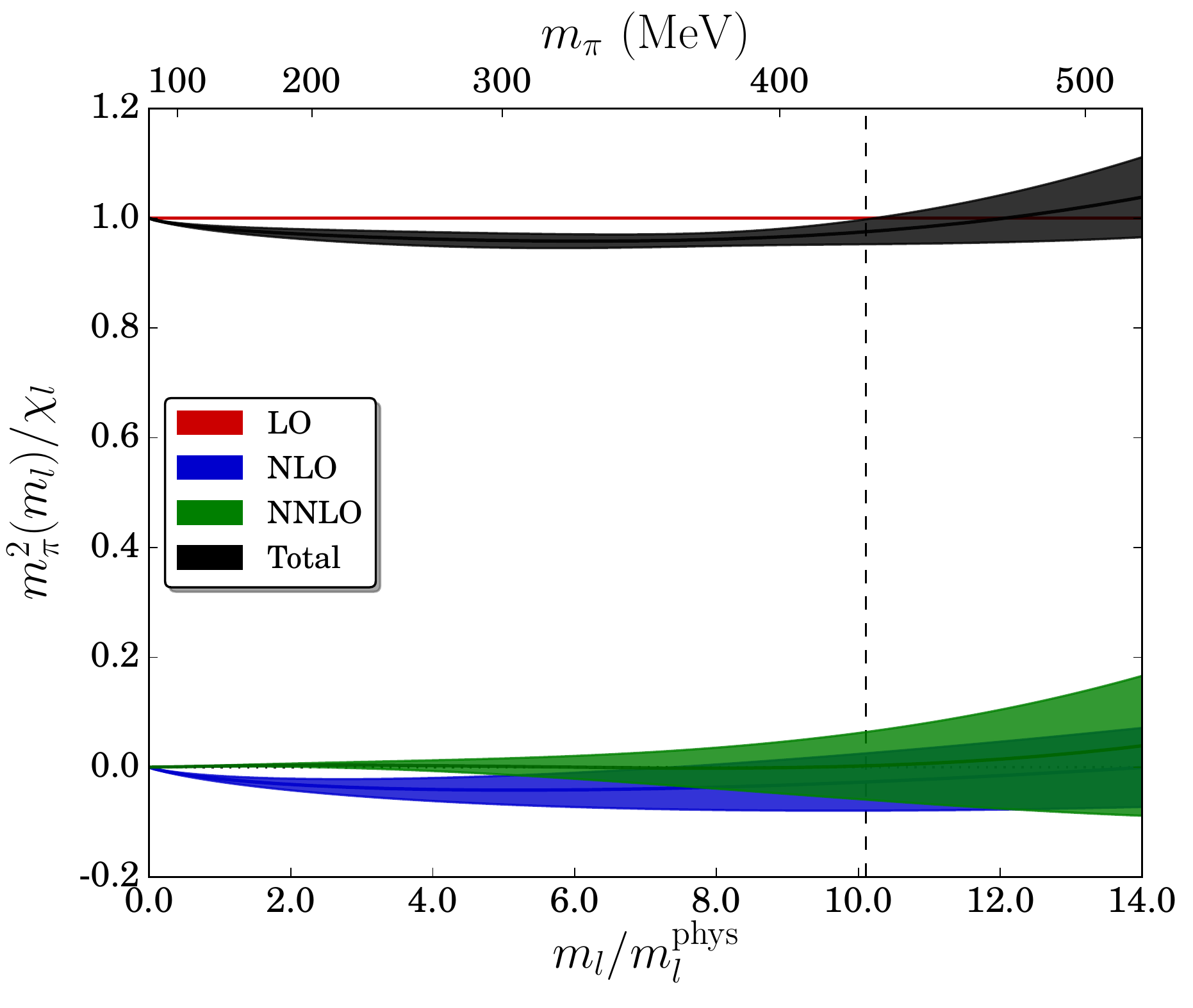}}
\subfloat{\includegraphics[width=0.49\textwidth]{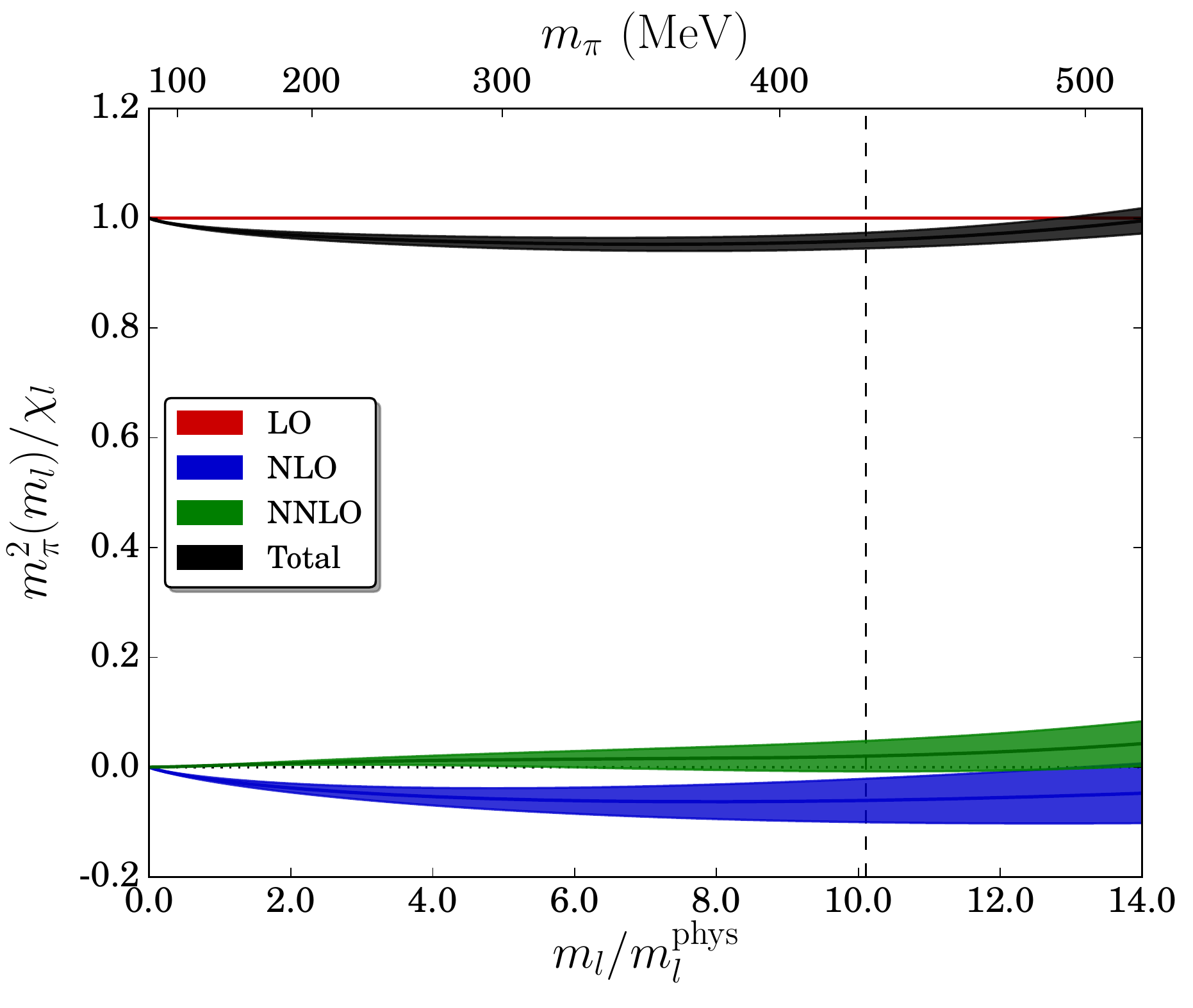}} \\
\subfloat{\includegraphics[width=0.49\textwidth]{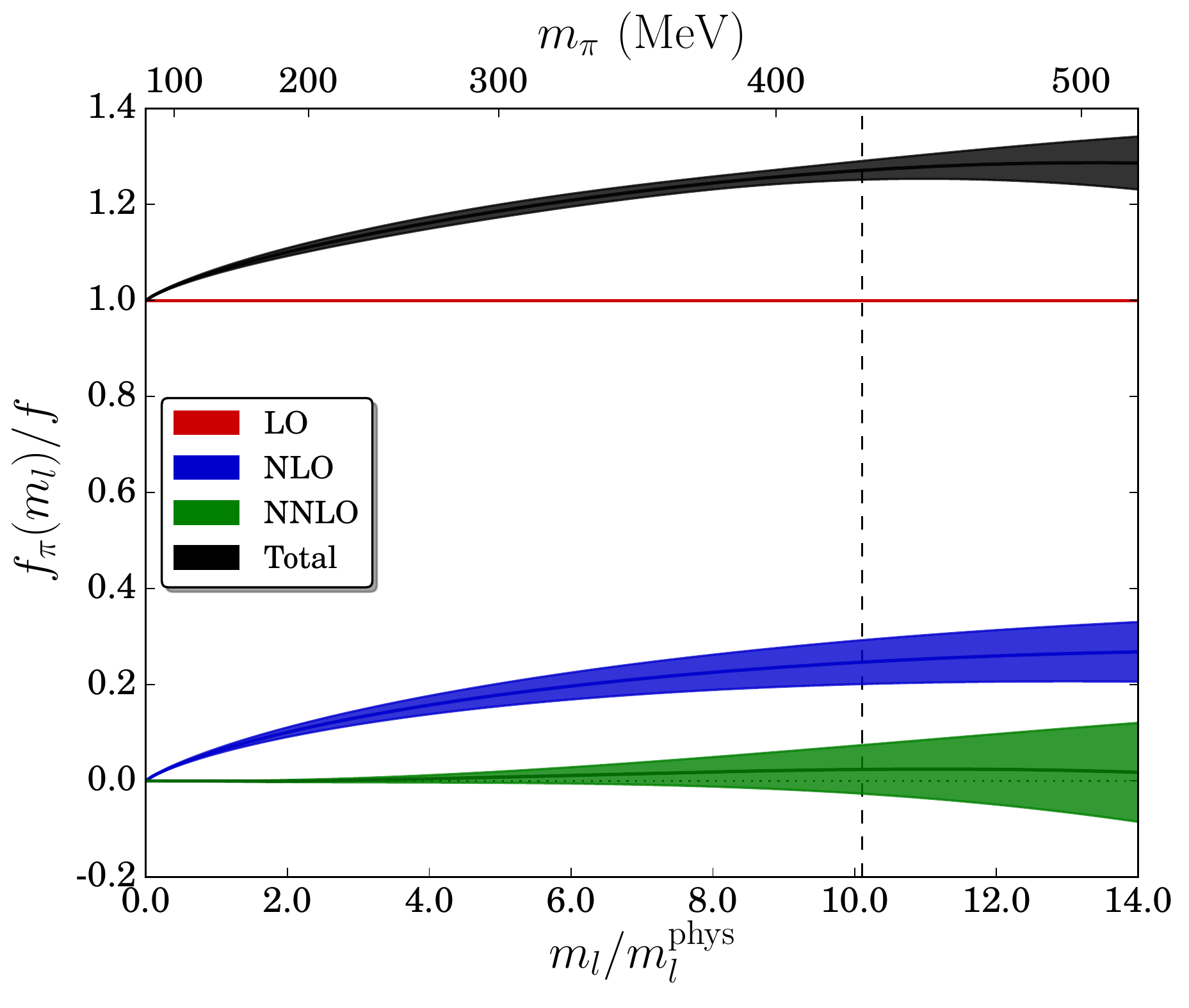}}
\subfloat{\includegraphics[width=0.49\textwidth]{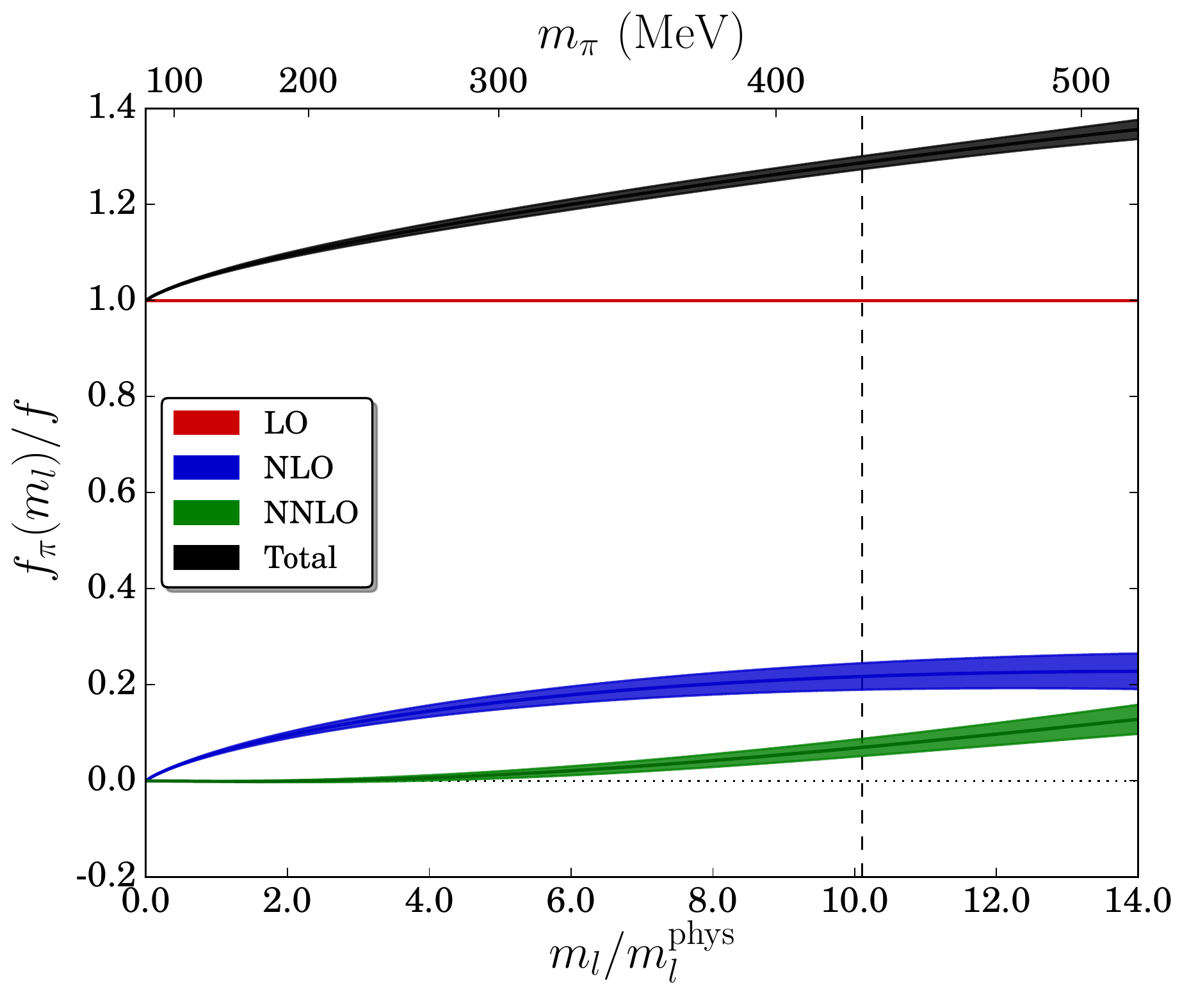}}
\caption{Decomposition of the $SU(2)$ chiral expansion into LO, NLO, and NNLO terms, normalized by LO. The pion mass (top) and pion decay constant (bottom) are plotted as a function of the light quark mass, using the LECs obtained from a fit with a pion mass cut of 370 MeV (left) and 450 MeV (right). The vertical dashed line corresponds to the heaviest unitary point included in the fit, and the horizontal dotted line marks zero.}
\label{fig:su2_chiral_expansion}
\end{figure}

In Figure~\ref{fig:su2_chiral_expansion} we plot the relative sizes of the LO, NLO, and NNLO terms for the pion mass and decay constant as a function of the liqht quark mass, using the LECs from Table~\ref{tab:su2_pqchpt_lecs}. The heaviest unitary ensemble included in the fit is indicated with a dashed vertical line. We observe that the NLO and NNLO terms contribute to $m_{\pi}^{2}$ with opposite sign, but to $f_{\pi}$ with the same sign: this behavior is expected from the lattice data, which suggests that the tree-level prediction $m_{\pi}^{2} \propto m_{l}$ works reasonably well even for heavier-than-physical $m_{l}$, but not for the markedly nonlinear $f_{\pi}$. We also observe that the NNLO terms are generally statistically consistent with zero for the fit with the lighter mass cut, indicating that the ensembles with $m_{\pi} \gtrsim 350 \, \mathrm{MeV}$ are important for constraining the NNLO terms in our fits. This should be viewed as an artifact of our data set rather than a statement about $SU(2)$ chiral perturbation theory; one ought to be able to constrain the LECs to any order with data arbitrarily close to the chiral limit provided one has enough high-precision measurements\footnote{In fact, one could argue that the mass cut should be taken so that only the lightest quark masses are used since systematic deviations between the predictions of ChPT and full QCD vanish in the chiral limit.}. Both mass cuts give consistent results for $m_{l} / m_{l}^{\rm phys} \lesssim 8.0$, where the fits are directly constrained by lattice data. At the physical point we find
\begin{align}
\label{eqn:su2_pion_expansion}
\begin{split}
\frac{m_{\pi}^{2}}{\chi_{l}} &= 1.0000 - 0.0245(41) + 0.0034(10) \\
\frac{f_{\pi}}{f} &= 1.0000 + 0.0586(35) - 0.0011(7)
\end{split}
\end{align}
for the decomposition into $\mathrm{LO} + \mathrm{NLO} + \mathrm{NNLO}$, normalized by LO. The errors on the more restrictive fit quickly grow when we extrapolate to heavier $m_{l}$, so we focus on the $m_{\pi}^{\rm cut} = 450 \, \mathrm{MeV}$ result to test the breakdown of the expansion at heavy quark masses. \\

\begin{figure}[h]
\centering
\subfloat{\includegraphics[width=0.49\textwidth]{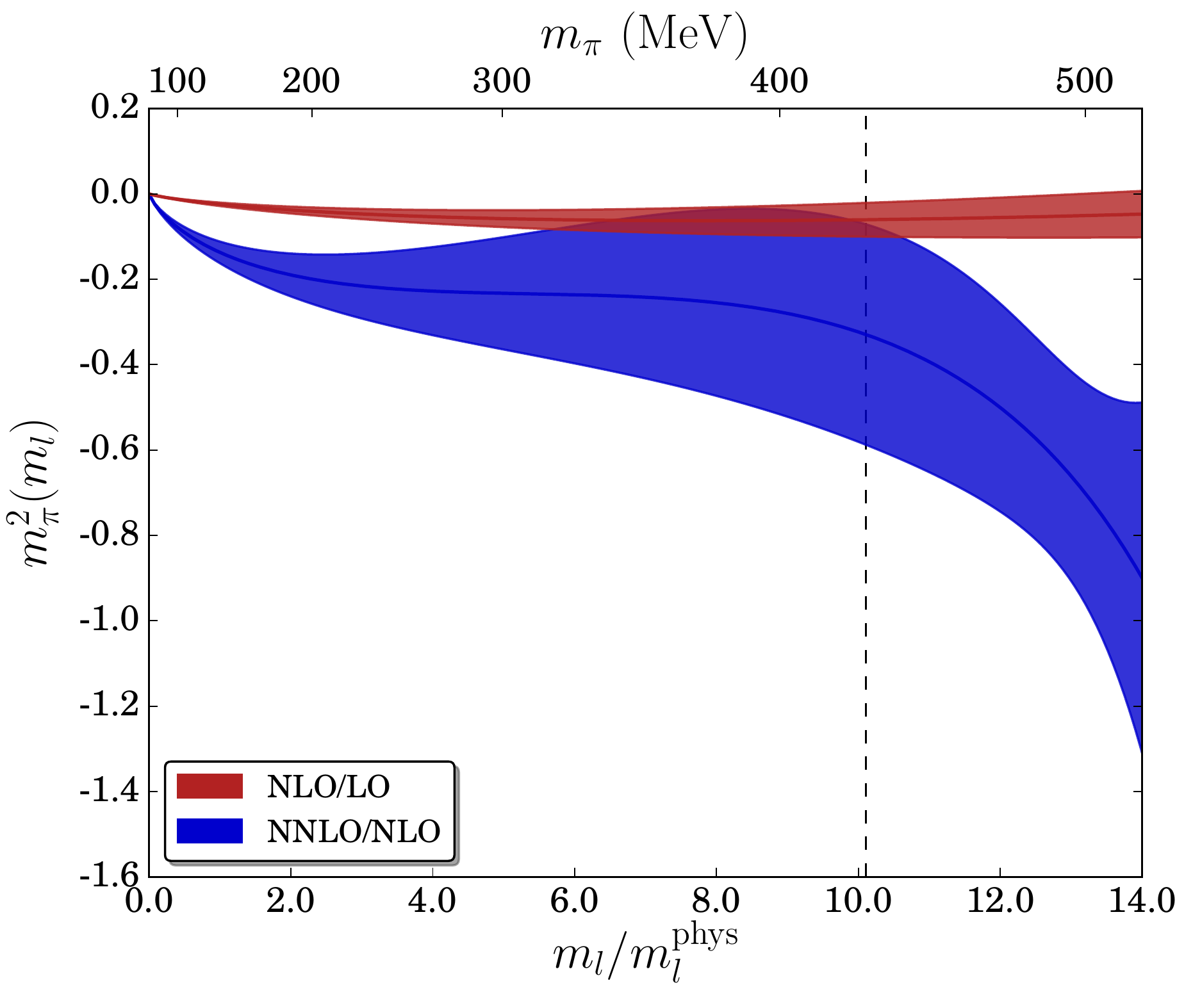}}
\subfloat{\includegraphics[width=0.49\textwidth]{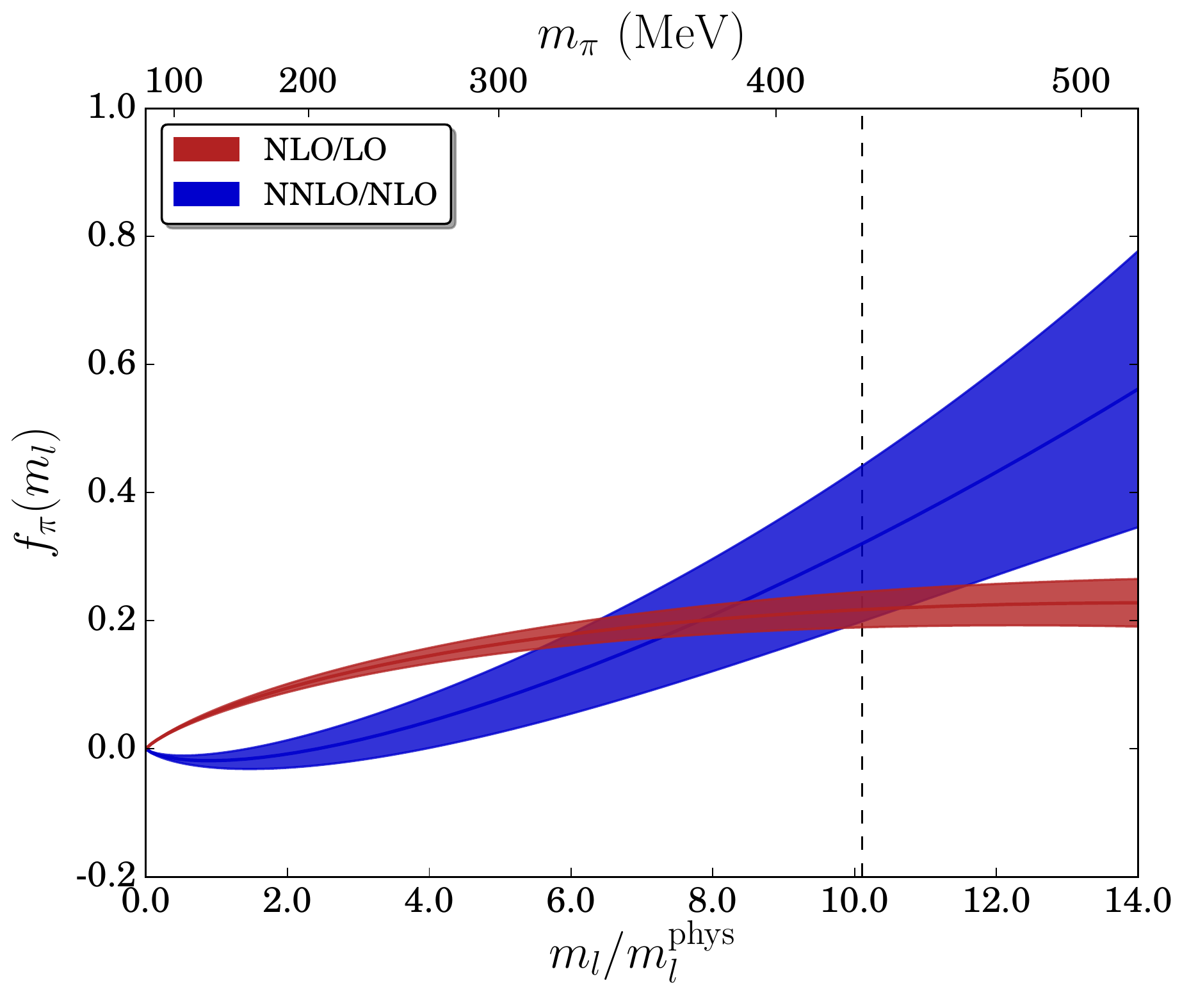}}
\caption{Relative sizes of the LO, NLO, and NNLO terms in the $SU(2)$ chiral expansion for $m_{\pi}^{2}$ (left) and $f_{\pi}$ (right) using the LECs obtained from a fit with a pion mass cut of 450 MeV. The vertical dashed line corresponds to the heaviest unitary point included in the fit.}
\label{fig:su2_breakdown}
\end{figure}

While both the NLO and NNLO terms remain small relative to LO --- at most $\bigO(20\%)$ --- even up to very heavy $m_{\pi} \sim 500 \, \mathrm{MeV}$, the NLO and NNLO terms start to become comparable in size for $m_{\pi} \gtrsim 450 \, \mathrm{MeV}$. In figure~\ref{fig:su2_breakdown} we plot the ratios $\mathrm{NLO}/\mathrm{LO}$ and $\mathrm{NNLO}/\mathrm{NLO}$ as a function of the light quark mass. If we conservatively define ``distress'' in the chiral expansion as $|\mathrm{NNLO}| \simeq 0.5 |\mathrm{NLO}|$ within statistical error, we find that this corresponds to $m_{l} / m_{l}^{\rm phys} \approx 10.9 $ ($m_{\pi} \approx 445 \, \mathrm{MeV}$) for $f_{\pi}$. A more relaxed definition of $|\mathrm{NNLO}| \simeq 0.8 |\mathrm{NLO}|$ corresponds to $m_{l} / m_{l}^{\rm phys} \approx 14.2 $ ($m_{\pi} \approx 520 \, \mathrm{MeV}$). The situation for $m_{\pi}^{2}$ is more subtle: while it is true that we similarly observe an increase in the relative sizes of the NNLO and NLO terms as the light quark mass is increased, they are contributing with opposite sign, and the sum $\mathrm{NLO} + \mathrm{NNLO}$ remains less than $10 \%$ of the LO contribution even at very heavy $m_{\pi} \gtrsim 500 \, \mathrm{MeV}$. We conclude that it is $f_{\pi}$, which exhibits stronger nonlinearity than $m_{\pi}^{2}$, that sets an upper limit on the applicability of NNLO $SU(2)$ ChPT, of roughly $m_{\pi} \sim 450-500 \, \mathrm{MeV}$. We note that the BMW collaboration has performed a similar test by fitting $SU(2)$ ChPT to unitary lattice data computed with $\bigO(a)$-improved Wilson fermions up to $m_{\pi} \sim 500 \, \mathrm{MeV}$, and finds results consistent with our own~\cite{Durr:2013goa}. 

\vspace*{\fill}
\vspace*{\fill}

\FloatBarrier

\subsection{Predictions}
\label{sec:su2_predictions}

\subsubsection{Unquenched LECs}

In Table~\ref{tab:unquenched_su2_lecs} we use the relations listed in Appendix~\ref{appendix:pqchpt_to_chpt} to compute the unquenched $SU(2)$ LECs $\{ l_{i} \}_{i=1}^{7}$ which can be determined from the partially quenched LECs in Table~\ref{tab:su2_pqchpt_lecs}. Traditionally, values for the scale independent LECs $\{ \overline{\ell}_{i} \}_{i=1}^{6}$ are quoted rather than $\{ l_{i} \}_{i=1}^{6}$; we also compute these using relations listed explicitly in Appendix~\ref{appendix:su2_lbar}. There is no analogous $\overline{\ell}_{7}$ since $l_{7}$ is already scale independent. We also compute the renormalized leading order LEC $B$ in the $\overline{\rm MS}$ scheme at $\mu = 2.0 \, \mathrm{GeV}$, and the $\overline{\rm MS}$ renormalized quark condensate
\begin{equation}
\Sigma = - \left. \langle \overline{\psi}_{l} \psi_{l} \rangle \right|_{m_{l} \rightarrow 0} = \frac{B f^{2}}{2}.
\end{equation}
We use the renormalization coefficients $Z_{ml}$ computed in Ref.~\cite{Blum:2014tka} to first renormalize $B$ and $\Sigma$ in the $\SMOM$ and $\SMOMg$ schemes, which are then matched perturbatively to $\overline{\rm MS}$. The difference in central value between the two intermediate schemes is used to assign a systematic error associated with the renormalization procedure. \\

\begin{table}[h]
\centering
\resizebox{\columnwidth}{!}{
\begin{tabular}{cc||cc|cc}
\hline
\hline
\rule{0cm}{0.4cm} LEC & $\Lambda_{\chi}$ & NLO ($370 \, \mathrm{MeV}$ cut) & NLO ($450 \, \mathrm{MeV}$ cut) & NNLO ($370 \, \mathrm{MeV}$ cut) & NNLO ($450 \, \mathrm{MeV}$ cut) \\
\hline
\rule{0cm}{0.4cm}$B^{\overline{\mathrm{MS}}}(\mu = 2 \, \mathrm{GeV})$ & \multirow{3}{*}{---} & 2.804(34)(30) GeV & 2.831(37)(30) GeV & 2.778(40)(30) GeV & 2.787(40)(30) GeV \\
 $f$ & & 121.3(1.5) MeV & 123.6(2.0) MeV & 120.7(1.7) MeV & 121.5(1.6) MeV \\
$\Sigma^{1/3, \, \overline{\mathrm{MS}}}(\mu = 2 \, \mathrm{GeV})$ & & 274.2(2.8)(1.0) MeV & 278.6(3.8)(1.0) MeV & 272.5(3.0)(1.0) MeV & 274.0(2.8)(1.0) MeV \\
\hline
 \rule{0cm}{0.4cm}$10^{3} l_{1}$ & \multirow{5}{*}{1 GeV} & --- & --- & 11.9(9.6) & -7.6(3.9) \\
 $10^{3} l_{2}$ & & --- & --- & -32(17) & 4.3(6.8) \\
 $10^{3} l_{3}$ & & 1.89(30) & 2.08(21) & 2.1(1.0) & 1.46(78) \\
 $10^{3} l_{4}$ & & 0.06(51) & 1.70(34) & -1.0(1.6) & -2.07(94) \\
 $10^{3} l_{7}$ & & --- & --- & 16.6(7.3) & 6.5(3.8) \\
\hline
 \rule{0cm}{0.4cm}$10^{3} l_{1}$ & \multirow{5}{*}{770 MeV} & --- & --- & 13(11) & -7.1(4.0) \\
 $10^{3} l_{2}$ & & --- & --- & -31(19) & 5.4(6.9) \\
 $10^{3} l_{3}$ & & 1.07(30) & 1.25(21) & 1.3(1.0) & 0.63(78) \\
 $10^{3} l_{4}$ & & 3.38(51) & 5.01(34) & 2.3(1.6) & 1.24(95) \\
 $10^{3} l_{7}$ & & --- & --- & 16.6(7.9) & 6.5(3.7) \\
\hline
 \rule{0cm}{0.4cm}$\overline{\ell}_{1}$ & \multirow{4}{*}{---} & --- & --- & 15.3(9.1) & -3.2(3.7) \\
 $\overline{\ell}_{2}$ & & --- & --- & -11.0(7.9) & 6.0(3.2) \\
 $\overline{\ell}_{3}$ & & 2.81(19) & 2.69(13) & 2.66(64) & 3.08(49) \\
 $\overline{\ell}_{4}$ & & 4.015(81) & 4.274(54) & 3.84(25) & 3.68(15) \\
\hline
\hline
\end{tabular}
}
\caption{Unquenched $SU(2)$ LECs computed from partially quenched $SU(2)$ fits. Missing entries are not constrained by the fits at a given order. For $B$ and $\Sigma$ the first error is statistical and the second is a systematic uncertainty in the perturbative matching to $\rm \overline{MS}$.}
\label{tab:unquenched_su2_lecs}
\end{table}

\begin{figure}[h]
\centering
\includegraphics[width=0.8\textwidth]{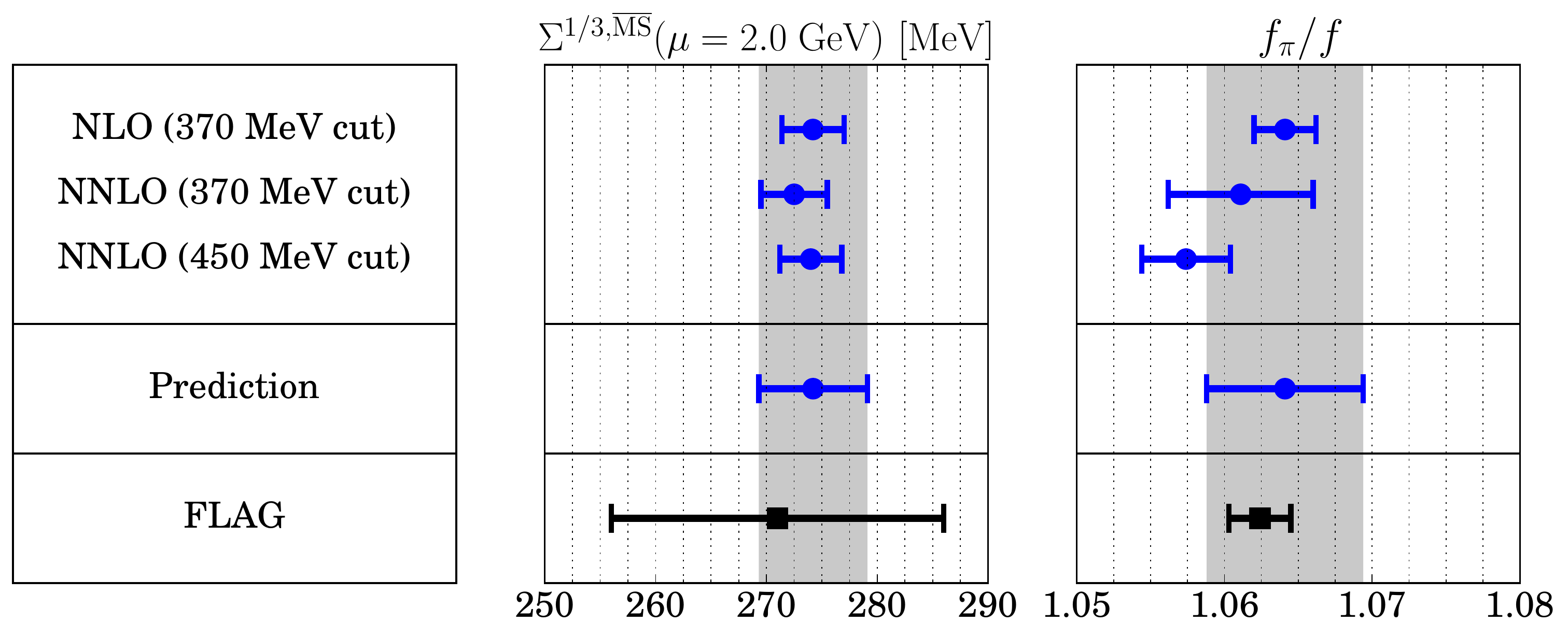}
\caption{Leading order $SU(2)$ ChPT LECs compared to the 2013 FLAG lattice averages.}
\label{fig:su2_LO_LECs}
\end{figure}

\begin{figure}[!ht]
\centering
\includegraphics[width=\textwidth]{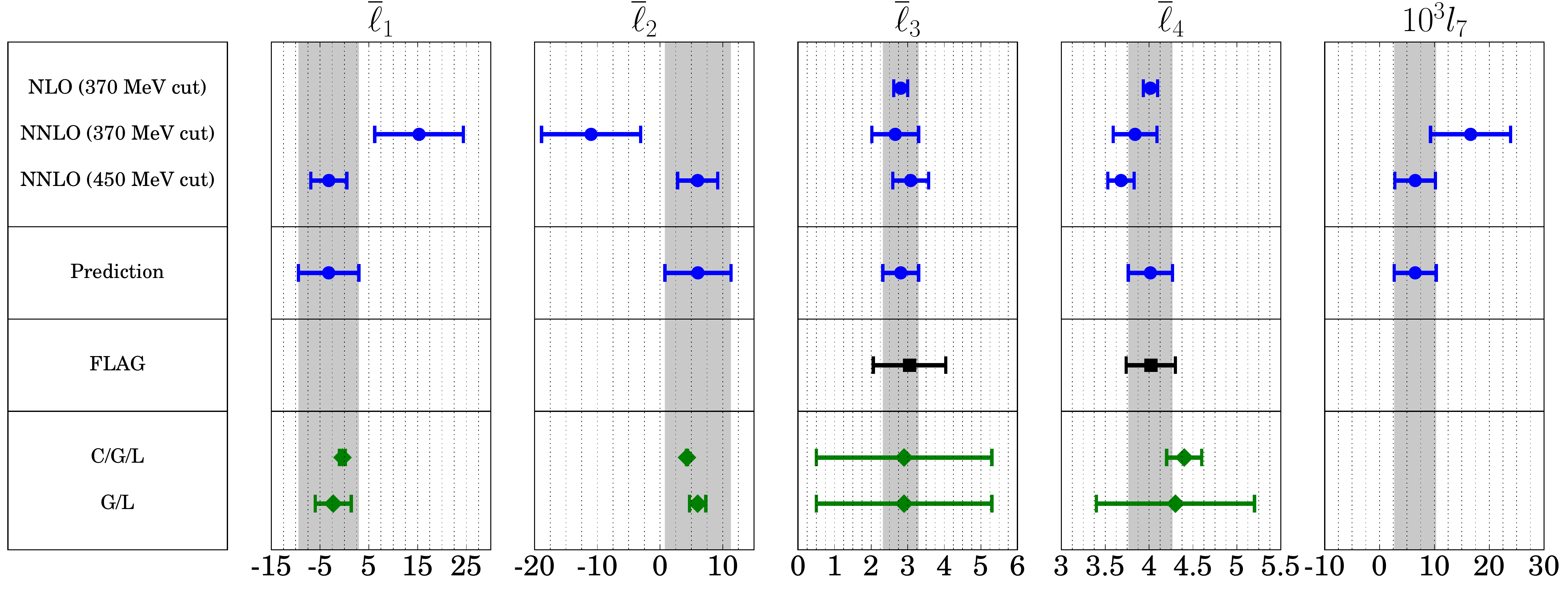}
\caption{Next-to leading order $SU(2)$ ChPT LECs compared to the 2013 FLAG lattice averages and two phenomenological determinations.}
\label{fig:su2_NLO_LECs}
\end{figure}

In Figures~\ref{fig:su2_LO_LECs} and~\ref{fig:su2_NLO_LECs} we compare our preferred determinations of the leading order and next-to leading order unquenched $SU(2)$ LECs (blue circles) to the 2013 $N_{f} = 2 + 1$ FLAG lattice averages~\cite{FLAG2014, Arthur:2012opa, MILC2009A, Bazavov:2010yq, Aoki:2010dy, Borsanyi:2012zv, Beane:2011zm} (black squares) and two phenomenological fits (green diamonds): the first is Gasser and Leutwyler's original determination of the $SU(2)$ LECs in Ref.~\cite{Gasser:1983yg}, and the second is Colangelo et al.'s updated fit of experimental pion scattering and scalar charge radius data to NNLO $SU(2)$ ChPT and the Roy equations~\cite{Colangelo:2001df}. We also include our final prediction for each LEC, including the full statistical and systematic error budget discussed in Section~\ref{sec:error_budget} summed in quadrature (``prediction''). For consistency with FLAG we quote our values for the dimensionless ratio $f_{\pi} / f$ rather than $f$. \\

We generally observe excellent consistency between our fits, and find that our results for the LO LECs, $\overline{\ell}_{3}$, and $\overline{\ell}_{4}$ --- which by now are standard lattice calculations --- compare favorably with the FLAG averages and phenomenological fits. We find that $\overline{\ell}_{3}$ and $\overline{\ell}_{4}$ are determined more precisely by the NLO fits than the NNLO fits, which is not surprising: at two-loop order the NLO LECs can enter into the expressions for the pion mass and decay constant quadratically or as terms which are a product of an LEC and a chiral logarithm, whereas at one-loop order they enter only as simple linear, analytic terms.

\FloatBarrier

From our NNLO fits we are also able to constrain $\overline{\ell}_{1}$, $\overline{\ell}_{2}$, and the scale-independent NLO LEC $l_{7}$. This is, to the authors' knowledge, the first direct prediction for $l_{7}$: Gasser and Leutwyler provide the order of magnitude estimate $l_{7} \sim 5 \times 10^{-3}$~\cite{Gasser:1983yg}, which is consistent with our predictions (\emph{e.g.} $l_{7} = 6.5(3.7) \times 10^{-3}$ from the fit with a 450 MeV cut). While our results for $\overline{\ell}_{1}$ and $\overline{\ell}_{2}$ are consistent with the phenomenological results, these LECs are determined much more precisely by the $\pi \pi$ scattering-based phenomenological fits. In this sense the lattice and phenomenological results are nicely complementary. We have begun to sharpen our predictions for $\overline{\ell}_{1}$ and $\overline{\ell}_{2}$ by including additional observables --- \emph{e.g.} $\pi \pi$ scattering lengths and pion form factors --- which can be computed on the lattice and provide stronger constraints on these LECs \cite{su2_proceedings}. \\

\FloatBarrier

\subsubsection{Other Physical Predictions}

Table~\ref{tab:su2_predictions} summarizes a number of predictions based on our results for the $SU(2)$ LECs from the previous section: $f_{\pi}$, $f_{K}$, and the ratios $f_{K} / f_{\pi}$ and $f_{\pi} / f$ are obtained directly from the global fit by interpolating our lattice results to the physical point. The final three quantities --- the $I=0$ ($a_{0}^{0}$) and $I=2$ ($a_{0}^{2}$) $\pi \pi$ scattering lengths, and the pion mass splitting due to QCD isospin breaking effects --- are one-loop ChPT predictions computed using Appendix~\ref{appendix:one_loop_su2_pred} and the values of the LECs $\{ \overline{\ell}_{i} \}_{i=1}^{4}$ and $l_{7}$ from Table~\ref{tab:unquenched_su2_lecs}.

\begin{table}[!ht]
\centering
\resizebox{\columnwidth}{!}{
\begin{tabular}{c||cc|cc}
\hline
\hline
\rule{0cm}{0.4cm} & NLO ($370 \, \mathrm{MeV}$ cut) & NLO ($450 \, \mathrm{MeV}$ cut) & NNLO ($370 \, \mathrm{MeV}$ cut) & NNLO ($450 \, \mathrm{MeV}$ cut) \\
\hline
 \rule{0cm}{0.4cm}$f_{\pi}$ & 0.1290(14) GeV & 0.1317(19) GeV & 0.1281(14) GeV & 0.1285(14) GeV \\
 $f_{K}$ & 0.1540(15) GeV & 0.1575(18) GeV & 0.1530(15) GeV & 0.1527(14) GeV \\
 $f_{K}/f_{\pi}$ & 1.1937(54) & 1.1962(74) & 1.1944(78) & 1.1884(67) \\
 $f_{\pi}/f$ & 1.0641(21) & 1.0658(21) & 1.0611(49) & 1.0574(30) \\
\hline
 \rule{0cm}{0.4cm}$m_{\pi} a_{0}^{0}$ & --- & --- & 0.170(20) & 0.1987(86) \\
 $m_{\pi} a_{0}^{2}$ & --- & --- & -0.0577(90) & -0.0404(33) \\
\hline
 \rule{0cm}{0.4cm}$[ m_{\pi^{\pm}}^{2} - m_{\pi^{0}}^{2} ]_{\rm QCD} / \Delta m_{du}^{2}$ & --- & --- & 80(35) & 31(17) \\
\hline
\hline
\end{tabular}
}
\caption{Predictions from NLO and NNLO fits and $SU(2)$ ChPT. $\Delta m_{du} \equiv m_{d} - m_{u}$. We emphasize that the distinction between ``NLO'' and ``NNLO'' fits, as well as the mass cut, applies only to $m_{\pi}$ and $f_{\pi}$: the kaon and $\Omega$ baryon data and fit forms are the same in all of these fits.} 
\label{tab:su2_predictions}
\end{table}
\FloatBarrier

\noindent The RBC-UKQCD collaboration has historically observed that, if $f_{\pi}$ and $f_{K}$ are determined from fits to heavy lattice data which is extrapolated down to the physical point, the predictions for $f_{\pi}$ and $f_{K}$ are systematically low compared to the physical values $f_{\pi}^{\rm phys} = 130.7 \, \mathrm{MeV}$ and $f_{K}^{\rm phys} = 156.1 \, \mathrm{MeV}$, which we observe in Table~\ref{tab:su2_predictions} as well. We have also found, however, that either overweighting the contributions to $\chi^{2}$ from the physical pion mass 48I and 64I ensembles\footnote{This procedure was introduced in Ref.~\cite{Blum:2014tka} to make small corrections for quark mass mistunings on the physical point ensembles.} or normalizing the contributions to $\chi^{2}$ from each ensemble by the number of partially quenched measurements performed on that ensemble --- effectively underweighting the heavy pion mass 24I and 32I ensembles --- as we explore in Appendix~\ref{appendix:chisq_normalized}, removes this discrepancy, and results in predictions for $f_{\pi}$ and $f_{K}$ consistent with their physical values. We conclude that two effects are responsible: 1) the large number of partially quenched measurements on the 24I, 32I, and 32ID ensembles causes the heavier data to dominate an unweighted, uncorrelated fit, and 2) chiral fits which are dominated by heavy data can exhibit excessive curvature near the physical point, leading to predictions which are systematically low. \\

The last three predictions in Table~\ref{tab:su2_predictions} allow for an interesting test of chiral perturbation theory. Since our NNLO fits determine the LECs $\overline{\ell}_{1}$, $\overline{\ell}_{2}$, and $l_{7}$ without containing any direct information about $\pi \pi$ scattering or isospin breaking, we can compute these quantities to NLO as predictions from our fits. The $\pi \pi$ scattering lengths computed from our preferred NNLO fit with a 450 MeV cut can be compared to recent experimental results based on measurements of $K_{e4}$ and $K^{\pm} \rightarrow \pi^{\pm} \pi^{0} \pi^{0}$ decays: $m_{\pi} a_{0}^{0} = 0.221(5)$ and $m_{\pi} a_{0}^{2} = -0.043(5)$~\cite{BlochDevaux:2009zzb}. Our prediction for the $\pi^{\pm}-\pi^{0}$ mass splitting is less straightforward to interpret since the majority of this splitting arises from electromagnetic effects which we do not take into account; ChPT predicts that the contribution from electromagnetism enters at $\bigO(\Delta m_{du})$ rather than $\bigO(\Delta m_{du}^{2})$. As far as the authors are aware no direct lattice calculations of the QCD isospin breaking contribution to the $\pi^{\pm}-\pi^{0}$ mass difference have been performed\footnote{Full calculations which include both isospin breaking and electromagnetic effects have been performed, however.}. If we take a reasonable estimate of the up/down mass difference $\Delta m_{du} \equiv m_{d} - m_{u} \sim 2.5 \, \mathrm{MeV}$, we can compare our prediction --- $m_{\pi^{\pm}}^{2} - m_{\pi^{0}}^{2} = 195(112) \, \mathrm{MeV}^{2}$ from the fit with the heavier mass cut --- to the physical mass difference $m_{\pi^{\pm}}^{2} - m_{\pi^{0}}^{2} = 1261 \, \mathrm{MeV}^{2}$~\cite{1674-1137-38-9-090001}, which suggests that $\sim 15(9) \%$ of the total mass splitting arises from QCD isospin breaking effects.  


\section{Error Budget and Final Results for the Unquenched $SU(2)$ LECs}
\label{sec:error_budget}

In this section we discuss the error budget for our determination of the leading and next-to leading order unquenched $SU(2)$ low energy constants, and report our final values including all systematics. In particular, we assign the following error to each LEC in table~\ref{tab:lec_predictions}:
\begin{itemize}
\item \textit{Influence of heavy data as determined by underweighting correlated data in the fits}: While our global fits are uncorrelated, we know that the partially quenched measurements on a given ensemble are highly correlated since they are computed with the same set of field configurations. If we were fitting to a function which exactly represented our data, as opposed to an expansion with some limited precision, our uncorrelated fits would not introduce any systematic bias into our answers. Since this is not the case, changing the weighting of the heavy mass ensembles, which contain highly correlated partially quenched measurements, gives us an estimate of the systematic effects on our results due to the worsening systematic disagreement betweeen PQChPT and QCD at heavier quark masses. We estimate the impact on our fits by taking the difference in central value between the LECs of an unweighted, uncorrelated fit (Section~\ref{sec:su2_fits}) and the LECs of a fit where the contributions to $\chi^{2}$ from ensembles with multiple partially quenched measurements have been systematically underweighted to capture the dominant effects of correlations (Appendix~\ref{appendix:chisq_normalized}). 
\end{itemize}
We also assign additional errors to the LECs which are determined by both NLO and NNLO fits ($B$, $f$, $\Sigma$, $\overline{\ell}_{3}$, and $\overline{\ell}_{4}$):
\begin{itemize}
\item \textit{Influence of heavy data as determined by varying the mass cut}: We also estimate the dependence of the LECs on the choice of mass cut by taking the difference in central value between an NNLO fit with a unitary pion mass cut of 370 MeV and an NNLO fit with a cut of 450 MeV where applicable. For the LECs where we can estimate the influence of the heavy data using both methods we take the larger estimate as the systematic included in our error budget.
\item \textit{Truncation of the (continuum) chiral expansion}: We estimate the influence of truncating $\rm N^{3}LO$ and higher terms by taking the difference in central value between an NLO fit and an NNLO fit, both with a unitary pion mass cut of 370 MeV.
\item \textit{Finite volume effects}: As a conservative bound on the influence of NNLO and higher order FV corrections, as well as neglected cross terms --- \emph{e.g.} $ (\mathrm{NLO} \,\, \mathrm{continuum} \,\, \mathrm{ChPT}) \times (\mathrm{NLO} \,\, \mathrm{FV} \,\, \mathrm{correction})$ --- we compute the difference in central value between an NLO PQChPT fit with NLO FV corrections and an NLO PQChPT fit with no FV corrections, both with a unitary pion mass cut of 370 MeV.
\end{itemize} 
We do not attempt to quantify the latter set of systematics for the LECs which only enter into the $SU(2)$ ChPT expressions for the pion mass and decay constant at two loop order --- $\overline{\ell}_{1}$, $\overline{\ell}_{2}$, and $l_{7}$ --- since these LECs typically have $\mathcal{O}(50 \%)$ or larger statistical errors, and are perhaps more accurately regarded as bounds than high-precision determinations. Likewise, we do not attempt to quantify systematic errors for the partially quenched LECs (Section~\ref{sec:su2_fit_parameters}) or for our predictions of the $\pi \pi$ scattering lengths and isospin breaking effects (Section~\ref{sec:su2_predictions}), but one could, in principle, assign an analogous error budget. \\

\begin{table}[h]
\centering
\begin{tabular}{cc}
\hline
\hline
\rule{0cm}{0.4cm}$\bm{B^{\overline{\rm MS}}(\mu = 2 \, \mathrm{GeV})}$ & $\bm{2.804(34)(40) \, \mathrm{GeV}}$ \\
 $\bm{f}$ & $\bm{121.3(1.5)(2.1) \, \mathrm{MeV}}$ \\
$\bm{\Sigma^{1/3, \, \overline{\mathrm{MS}}}(\mu = 2 \, \mathrm{GeV})}$ & $\bm{274.2(2.8)(4.0) \, \mathrm{MeV}}$ \\
 $\overline{\ell}_{1}$ & $-3.2(3.7)(5.0)$ \\
 $\overline{\ell}_{2}$ & $6.0(3.2)(4.2)$ \\
 $\bm{\overline{\ell}_{3}}$ & $\bm{2.81(19)(45)}$ \\
 $\bm{\overline{\ell}_{4}}$ & $\bm{4.02(8)(24)}$ \\
 $10^{3} l_{7}$ & $6.5(3.8)(0.2)$ \\
\hline
\hline
\end{tabular}
\caption{Final predictions for the unquenched $SU(2)$ LECs including all statistical and systematic errors. The reported errors are the statistical (left) and the total systematic (right) obtained by summing the contributions we discuss in the text in quadrature. Bold entries correspond to LECs which enter into both NLO and NNLO fits, for which we assign the full error budget; for the other entries the mass cut, chiral truncation, and finite volume systematics are assumed to be negligible compared to the statistical error and are not quantified. The central values and statistical errors of $B$, $f$, $\Sigma^{1/3}$, $\overline{\ell}_{3}$, and $\overline{\ell}_{4}$ are from an NLO fit with a 370 MeV cut, while the central values and statistical errors of $\overline{\ell}_{1}$, $\overline{\ell}_{2}$, and $l_{7}$ are from an NNLO fit with a 450 MeV cut.}
\label{tab:lec_predictions}
\end{table}
\FloatBarrier


\section{Conclusions}
\label{sec:conclusion}

\begin{figure}[h]
\centering
\subfloat[NLO, $m_{\pi}^{\rm cut} = 370 \, \mathrm{MeV}$]{\includegraphics[width=0.49\textwidth]{{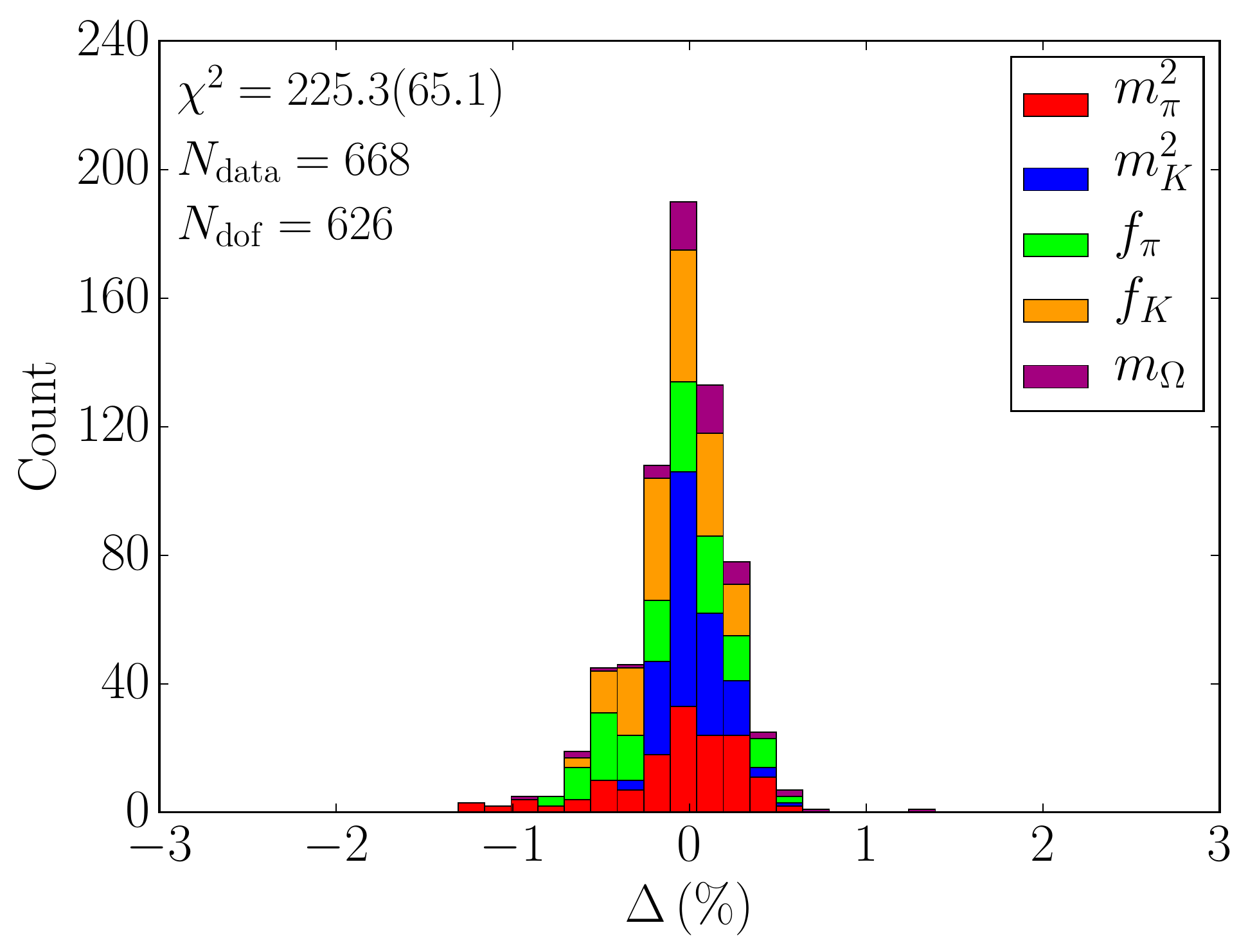}}}
\subfloat[NNLO, $m_{\pi}^{\rm cut} = 450 \, \mathrm{MeV}$]{\includegraphics[width=0.49\textwidth]{{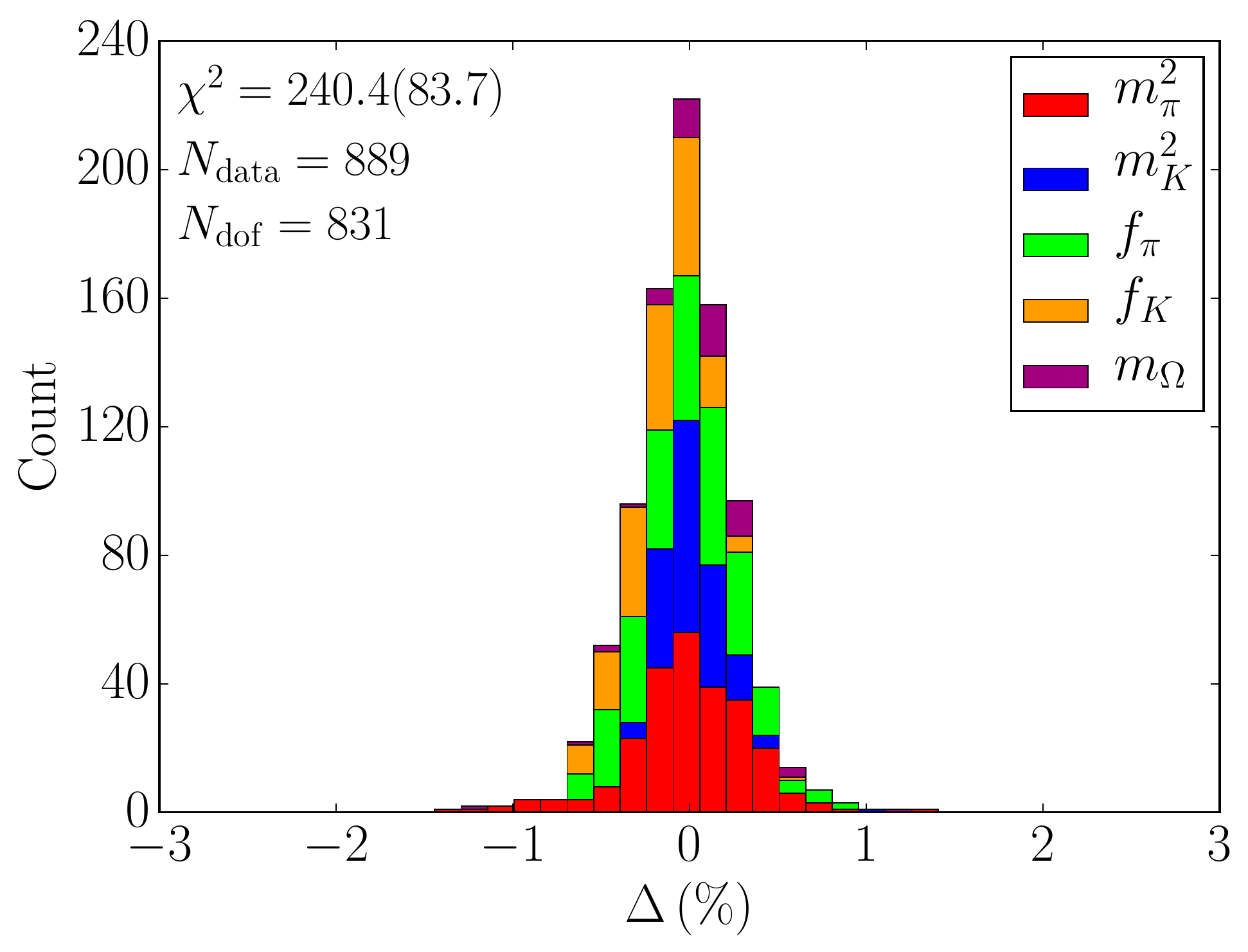}}}
\caption{Percent deviation between fits and data. We plot stacked histograms of the quantity $\Delta \equiv 200 \times (Y - Y^{\rm fit})/(Y + Y^{\rm fit})$.}
\label{fig:pdev_preferred_fits}
\end{figure}

In this work we have performed fits of pseudoscalar masses and decay constants from a series of RBC-UKQCD domain wall fermion ensembles to the corresponding formulae in next-to-next-to leading order $SU(2)$ partially quenched chiral perturbation theory. We reported values for a large set of partially quenched low-energy constants, and used these values to compute the unquenched leading and next-to leading order LECs. We also examined the range of quark masses for which NLO and NNLO ChPT accurately describe our lattice data, and used the newly determined LECs from NNLO fits to make one-loop predictions for isospin breaking effects and $\pi \pi$ scattering lengths, which we compare to other lattice and experimental results. We have observed that $SU(2)$ PQChPT generally describes the included range of partially quenched data with percent-scale accuracy: to emphasize this point we plot in Figure~\ref{fig:pdev_preferred_fits} histograms of the percent deviation between the data and fit
\begin{equation}
\Delta \equiv \frac{(Y - Y^{\rm fit})}{(Y + Y^{\rm fit})/2} \times 100
\end{equation} 
for our preferred fits, NLO PQChPT with a unitary pion mass cut of 370 MeV and NNLO PQChPT with a 450 MeV cut. \\

We have observed that NNLO $SU(2)$ PQChPT can be reliably fit to our data for the pion mass and decay constant without the need for additional terms or constraints to stabilize the fits, and we determine values for 8 linear combinations of NNLO low energy constants. The values we obtained for the unquenched $SU(2)$ LECs were consistent between our NLO and NNLO fits and with other lattice and phenomenological determinations reported in the literature. At the physical light quark mass we found that the chiral expansions for the pion mass and decay constant behave like rapidly convergent series. After probing the breakdown of the chiral expansion at heavy light quark mass we concluded that NLO $SU(2)$ PQChPT is sufficient to describe our lattice data up to $m_{\pi} \sim \mathcal{O}(350 \, \mathrm{MeV})$, beyond which we observe some deviation between the NLO prediction for the pion decay constant and our lattice data. Likewise, we concluded that NNLO $SU(2)$ PQChPT remains consistent with our data up to $m_{\pi} \sim \mathcal{O}(450 \, \mathrm{MeV})$. By 500 MeV, the NNLO corrections to the pion decay constant have grown to the point that they are comparable in size to the NLO corrections, indicating that the chiral expansion truncated to NNLO is unreliable at this scale. Of course, all statements regarding the values of LECs and the behavior of the $SU(2)$ chiral expansion made in this work are subject to the statistical precision, finite volume errors, and cutoff effects inherent in our lattice data. These points will need to be revisited and reassessed in the future as more and increasingly precise data becomes available. \\

We also note that our fits in this work only make use of the pseudoscalar masses and decay constants. Future work will incorporate a calculation of the $I = 2$ $\pi \pi$ scattering length and the pion vector form factor on many of the domain wall fermion ensembles considered here. Including these results in our chiral fits will give first-principles determinations of the scattering length $a_{0}^{2}$, the pion charge radius $\langle r^{2} \rangle_{V}^{\pi}$, and the $SU(2)$ LEC $\overline{\ell}_{6}$, as well as sharpen the predictions for $\overline{\ell}_{1}$ and $\overline{\ell}_{2}$, which are currently determined most precisely by phenomenological fits to experimental data. A forthcoming paper will also explore analogous fits of the pseudoscalar masses and decay constants to $SU(3)$ partially quenched chiral perturbation theory at next-to-next-to leading order.


\section*{Acknowledgements}
\label{sec:acknowledgements}

We thank members of the RBC-UKQCD collaboration for helpful discussions and support of this work. \\

The 32ID-M1 and 32ID-M2 ensembles were generated using the IBM Blue Gene/Q (BG/Q) ``Mira'' system at the Argonne Leadership Class Facility (ALCF), and the ``Vulcan'' BG/Q system at Lawrence Livermore National Laboratory (LLNL). Measurements, on these ensembles as well as older ensembles used in this work, were performed using Vulcan, the ``DiRAC'' BG/Q system in the Advanced Computing Facility at the University of Edinburgh, and the BG/Q computers of the RIKEN-BNL Research Center and Brookhaven National Lab. \\

The software used includes the CPS QCD code (\url{http://qcdoc.phys.columbia.edu/cps.html}) \cite{cps}, supported in part by the USDOE SciDAC program, and the BAGEL (\url{http://www2.ph.ed.ac.uk/~paboyle/bagel/Bagel.html}) assembler kernel generator for high-performance optimized kernels and fermion solvers~\cite{Boyle20092739}. Gauge fixing was performed using the ``GLU" (Gauge Link Utility) codebase (\url{https://github.com/RJhudspith/GLU}). Next-to-next-to leading order partially-quenched $SU(2)$ and $SU(3)$ chiral perturbation theory expressions for the pseudoscalar masses and decay constants were computed using Fortran routines provided by J. Bijnens. Gauge fixing and fitting were performed using the Columbia University CUTH cluster. \\

N.H.C., R.D.M., G.M., and D.J.M.~are supported in part by U.S.~DOE grant \#DE-SC0011941. C.K.~is supported by a RIKEN foreign postdoctoral research (FPR) grant. A.P.~and C.T.S.~were supported in part by UK STFC Grant ST/L000296/1. P.B.~has been in part supported by STFC Grants ST/M006530/1,
ST/L000458/1, ST/K005790/1, and ST/K005804/1. C.J.~is supported by DOE contract \#AC-02-98CH10886(BNL). N.G.~is supported by the Leverhulme Research grant RPG-2014-118. A.J.~acknowledges funding from the European Research Council under the European Community's Seventh Framework Programme (FP7/2007-2013) ERC grant agreement No.~279757.


\bibliography{chpt_paper}

\appendix


\clearpage
\section{ChPT Relations}
\label{appendix:chpt_eqns}

In this appendix we collect various relations used in the analysis in the body of the paper. We do not explicitly reprint the expressions for the pseudoscalar masses and decay constants and the corresponding finite volume corrections used in the chiral fits: instead we refer the reader to the appendices of Ref.~\cite{Allton:2008pn}. The NNLO pseudoscalar masses and decay constants were computed using Fortran routines provided by J. Bijnens.

\subsection{Relations Between PQChPT and ChPT LECs at NLO}
\label{appendix:pqchpt_to_chpt}

The $SU(N_{f})$ ChPT Lagrangian can be recovered from the more general $SU(N_{f})$ PQChPT Lagrangian in the limit of equal sea and valence quark masses. Here we have collected the explicit expressions relating the NLO LECs in this limit from Ref.~\cite{Bijnens:1999hw}. The analogous expressions for the NNLO LECs can be found in the same reference, but we do not use them here. For $N_{f} = 2$, the NLO ChPT LECs $\{ l_{i} \}_{i=1}^{7}$ are related to the NLO PQChPT LECs $\{ \hat{L}_{i}^{(2)} \}_{i=0}^{12}$ by
\begin{equation}
\begin{array}{lll}
l_{1} = -2 \hat{L}_{0}^{(2)} + 4 \hat{L}_{1}^{(2)} + 2 \hat{L}_{3}^{(2)} &
l_{4} = 4 \left( 2 \hat{L}_{4}^{(2)} + \hat{L}_{5}^{(2)} \right) & \\
l_{2} = 4 \left( \hat{L}_{0}^{(2)} + \hat{L}_{2}^{(2)} \right) &
l_{5} = \hat{L}_{10}^{(2)} &
l_{7} = - 8 \left( 2 \hat{L}_{7}^{(2)} + \hat{L}_{8}^{(2)} \right) \\
l_{3} = 4 \left( -2 \hat{L}_{4}^{(2)} - \hat{L}_{5}^{(2)} + 4 \hat{L}_{6}^{(2)} + 2 \hat{L}_{8}^{(2)} \right) &
l_{6} = - 2 \hat{L}_{9}^{(2)} \\
\end{array}
\end{equation}
and the additional constraints $\hat{L}_{11}^{(2)} = - l_{4}/4$ and $\hat{L}_{12}^{(2)} = 0$.

\subsection{Scale Independent $SU(2)$ LECs}
\label{appendix:su2_lbar}

Conventionally, one quotes values of the scale independent $SU(2)$ LECs $\{ \overline{\ell}_{i} \}_{i=1}^{6}$ rather than $\{ l_{i} \}_{i=1}^{6}$. These are obtained by running the $\{ l_{i} \}_{i=1}^{6}$ from the energy scale at which they are defined, $\mu$, to the physical pion mass using
\begin{equation}
\overline{\ell}_{i} = \gamma_{i} l_{i} - \log \left( \frac{m_{\pi}^{2}}{\mu^{2}} \right),
\end{equation}
where the coefficients
\begin{equation}
\begin{array}{cccccc}
\gamma_{1} = 96 \pi^{2},  &
\gamma_{2} = 48 \pi^{2},  &
\gamma_{3} = -64 \pi^{2}, &
\gamma_{4} = 16 \pi^{2},  &
\gamma_{5} = -192 \pi^{2}, &
\gamma_{6} = -96 \pi^{2},
\end{array}
\end{equation}
were computed in Ref.~\cite{Gasser:1983yg}. The remaining LEC $l_{7}$ has no scale dependence.

\subsection{One-Loop $SU(2)$ Predictions}
\label{appendix:one_loop_su2_pred}

While NLO fits to the pion mass and decay constant constrain the unquenched $SU(2)$ LECs $l_{3}$ and $l_{4}$, NNLO fits also constrain $l_{1}$, $l_{2}$, and $l_{7}$, allowing us to make additional one-loop predictions~\cite{Gasser:1983yg}. At NLO $l_{1}$ and $l_{2}$ determine quantities related to $\pi \pi$ scattering. The s-wave scattering lengths $a_{0}^{I}$ in the isospin channels $I=0$ and $I=2$, for example, are given by\footnote{Note: for consistency with the chiral interpolations in our global fits we choose to parametrize the expansions for the scattering lengths in terms of the light quark mass $m_{l}$ rather than the more commonly used ratio $m_{\pi} / f_{\pi}$.}
\begin{align}
\begin{split}
m_{\pi} a_{0}^{0} &= \frac{7 \chi_{l}}{16 \pi f^{2}} \left[ 1 + \frac{16 \chi_{l}}{7 f^{2}} \left( 5 l_{1} + 5 l_{2} + 3 l_{3} \right) + \frac{\chi_{l}}{16 \pi^{2} f^{2}} \left( 5 - 4 \log \left( \frac{\chi_{l}}{\Lambda_{\chi}^{2}} \right) \right) \right] \\
m_{\pi} a_{0}^{2} &= -\frac{\chi_{l}}{8 \pi f^{2}} \left[ 1 - \frac{16 \chi_{l}}{f^{2}} \left( l_{1} + l_{2} \right) - \frac{\chi_{l}}{16 \pi^{2} f^{2}} \left( 1 - 8 \log \left( \frac{\chi_{l}}{\Lambda_{\chi}^{2}} \right) \right) \right]
\end{split}.
\end{align}
The LEC $l_{7}$ controls the size of the pion mass splitting due to the difference between the up and down quark masses,
\begin{equation}
\left[ m_{\pi^{\pm}}^{2} - m_{\pi^{0}}^{2} \right]_{\rm QCD} =  \left( m_{d} - m_{u} \right)^{2} \frac{4 B^{2}}{f^{2}} l_{7}.
\end{equation}
We use the subscript ``QCD'' to emphasize that this is only the contribution to the mass splitting from QCD isospin breaking. The dominant contribution is due to electromagnetic effects, and enters at $\bigO(m_{d}-m_{u})$.


\clearpage
\section{Summary of Lattice Data Included in Chiral Fits}
\label{appendix:lattice_fits}

In this appendix we collect the results for fits of the pseudoscalar masses and decay constants, the $\Omega$ baryon mass, the ratio $R(t)$ (Eqn.~\eqref{eqn:mres}) which determines $m_{\rm res}$ in the chiral limit, and the Wilson flow scales on each ensemble in lattice units. Earlier results for the 24I ensemble can be found in Ref.~\cite{Allton:2008pn}, but differ from the current work in that the number of configurations has been approximately doubled and the spectrum re-analyzed in later works. For the other ensembles, these fits are identical to results we have published in earlier analyses: these can be found in Ref.~\cite{Aoki:2010dy} for the 32I ensembles, Ref.~\cite{Arthur:2012opa} for the 32ID ensembles, and Ref.~\cite{Blum:2014tka} for the 48I, 64I, and 32I-fine ensembles. The 32ID-M1 and 32ID-M2 ensembles have not appeared in any of our earlier global fits.

\subsection{Pseudoscalar Masses, Decay Constants, and $\Omega$ Baryon Mass}

\vspace*{\fill}
\begin{table}[h]
\centering
\begin{tabular}{cc|cc||ccc}
\hline
\hline
$a m_{l}$ & $a m_{h}$ & $a m_{x}$ & $a m_{y}$ & $a m_{xy}$ & $a f_{xy}$ & $a m_{xxx}$ \\
\hline
0.005 & 0.04 & 0.001 & 0.001 & 0.13914(63) & 0.08140(46) & --- \\
0.005 & 0.04 & 0.001 & 0.005 & 0.16693(60) & 0.08316(41) & --- \\
0.005 & 0.04 & 0.001 & 0.01 & 0.19602(59) & 0.08526(40) & --- \\
0.005 & 0.04 & 0.001 & 0.02 & 0.24402(61) & 0.08897(41) & --- \\
0.005 & 0.04 & 0.001 & 0.03 & 0.28430(64) & 0.09222(45) & --- \\
0.005 & 0.04 & 0.001 & 0.04 & 0.31990(69) & 0.09511(49) & --- \\
0.005 & 0.04 & 0.005 & 0.005 & 0.19035(56) & 0.08468(38) & --- \\
0.005 & 0.04 & 0.005 & 0.01 & 0.21609(54) & 0.08666(37) & --- \\
0.005 & 0.04 & 0.005 & 0.02 & 0.26026(53) & 0.09027(37) & --- \\
0.005 & 0.04 & 0.005 & 0.03 & 0.29833(54) & 0.09347(39) & --- \\
0.005 & 0.04 & 0.005 & 0.04 & 0.33245(55) & 0.09632(41) & --- \\
0.005 & 0.04 & 0.01 & 0.01 & 0.23894(51) & 0.08858(35) & --- \\
0.005 & 0.04 & 0.01 & 0.02 & 0.27945(49) & 0.09215(36) & --- \\
0.005 & 0.04 & 0.01 & 0.03 & 0.31524(49) & 0.09533(37) & --- \\
0.005 & 0.04 & 0.01 & 0.04 & 0.34777(50) & 0.09816(39) & --- \\
0.005 & 0.04 & 0.02 & 0.02 & 0.31487(47) & 0.09572(36) & --- \\
0.005 & 0.04 & 0.02 & 0.03 & 0.34722(46) & 0.09890(38) & --- \\
0.005 & 0.04 & 0.02 & 0.04 & 0.37722(46) & 0.10175(40) & --- \\
0.005 & 0.04 & 0.03 & 0.03 & 0.37705(45) & 0.10213(40) & 0.9629(37) \\
0.005 & 0.04 & 0.03 & 0.04 & 0.40512(44) & 0.10502(42) & --- \\
0.005 & 0.04 & 0.04 & 0.04 & 0.43165(42) & 0.10796(43) & 1.0134(31) \\
\hline
\hline
\end{tabular}
\caption{Partially quenched pseudoscalar mass, pseudoscalar decay constant, and $\Omega$ baryon mass measurements on the 24I $a m_{l} = 0.005$ ensemble.}
\end{table}
\vspace*{\fill}

\begin{table}[h]
\centering
\begin{tabular}{cc|cc||ccc}
\hline
\hline
$a m_{l}$ & $a m_{h}$ & $a m_{x}$ & $a m_{y}$ & $a m_{xy}$ & $a f_{xy}$ & $a m_{xxx}$ \\
\hline
0.01 & 0.04 & 0.001 & 0.001 & 0.14342(68) & 0.08531(45) & --- \\
0.01 & 0.04 & 0.001 & 0.005 & 0.17087(63) & 0.08712(41) & --- \\
0.01 & 0.04 & 0.001 & 0.01 & 0.19972(60) & 0.08921(42) & --- \\
0.01 & 0.04 & 0.001 & 0.02 & 0.24751(60) & 0.09288(46) & --- \\
0.01 & 0.04 & 0.001 & 0.03 & 0.28773(63) & 0.09609(53) & --- \\
0.01 & 0.04 & 0.001 & 0.04 & 0.32333(70) & 0.09898(60) & --- \\
0.01 & 0.04 & 0.005 & 0.005 & 0.19399(57) & 0.08841(39) & --- \\
0.01 & 0.04 & 0.005 & 0.01 & 0.21954(53) & 0.09024(39) & --- \\
0.01 & 0.04 & 0.005 & 0.02 & 0.26358(50) & 0.09370(41) & --- \\
0.01 & 0.04 & 0.005 & 0.03 & 0.30164(50) & 0.09684(44) & --- \\
0.01 & 0.04 & 0.005 & 0.04 & 0.33577(53) & 0.09969(48) & --- \\
0.01 & 0.04 & 0.01 & 0.01 & 0.24223(49) & 0.09193(38) & --- \\
0.01 & 0.04 & 0.01 & 0.02 & 0.28264(45) & 0.09529(39) & --- \\
0.01 & 0.04 & 0.01 & 0.03 & 0.31839(45) & 0.09838(41) & --- \\
0.01 & 0.04 & 0.01 & 0.04 & 0.35091(46) & 0.10118(43) & --- \\
0.01 & 0.04 & 0.02 & 0.02 & 0.31795(41) & 0.09859(39) & --- \\
0.01 & 0.04 & 0.02 & 0.03 & 0.35023(40) & 0.10165(39) & --- \\
0.01 & 0.04 & 0.02 & 0.04 & 0.38018(40) & 0.10443(40) & --- \\
0.01 & 0.04 & 0.03 & 0.03 & 0.37997(39) & 0.10471(39) & 0.9785(44) \\
0.01 & 0.04 & 0.03 & 0.04 & 0.40797(38) & 0.10751(40) & --- \\
0.01 & 0.04 & 0.04 & 0.04 & 0.43443(38) & 0.11035(40) & 1.0276(36) \\
\hline
\hline
\end{tabular}
\caption{Partially quenched pseudoscalar mass, pseudoscalar decay constant, and $\Omega$ baryon mass measurements on the 24I $a m_{l} = 0.01$ ensemble.}
\end{table}

\begin{table}[h]
\centering
\begin{tabular}{cc|cc||ccc}
\hline
\hline
$a m_{l}$ & $a m_{h}$ & $a m_{x}$ & $a m_{y}$ & $a m_{xy}$ & $a f_{xy}$ & $a m_{xxx}$ \\
\hline
0.004 & 0.03 & 0.002 & 0.002 & 0.09757(38) & 0.05983(30) & --- \\
0.004 & 0.03 & 0.002 & 0.004 & 0.11330(37) & 0.06090(29) & --- \\
0.004 & 0.03 & 0.002 & 0.006 & 0.12707(37) & 0.06192(29) & --- \\
0.004 & 0.03 & 0.002 & 0.008 & 0.13945(37) & 0.06286(30) & --- \\
0.004 & 0.03 & 0.002 & 0.025 & 0.21797(44) & 0.06905(34) & --- \\
0.004 & 0.03 & 0.002 & 0.03 & 0.23631(47) & 0.07048(35) & --- \\
0.004 & 0.03 & 0.004 & 0.004 & 0.12694(35) & 0.06181(29) & --- \\
0.004 & 0.03 & 0.004 & 0.006 & 0.13926(34) & 0.06274(29) & --- \\
0.004 & 0.03 & 0.004 & 0.008 & 0.15058(34) & 0.06363(29) & --- \\
0.004 & 0.03 & 0.004 & 0.025 & 0.22518(37) & 0.06969(32) & --- \\
0.004 & 0.03 & 0.004 & 0.03 & 0.24301(39) & 0.07112(33) & --- \\
0.004 & 0.03 & 0.006 & 0.006 & 0.15051(33) & 0.06363(29) & --- \\
0.004 & 0.03 & 0.006 & 0.008 & 0.16100(33) & 0.06449(30) & --- \\
0.004 & 0.03 & 0.006 & 0.025 & 0.23227(33) & 0.07050(32) & --- \\
0.004 & 0.03 & 0.006 & 0.03 & 0.24963(35) & 0.07193(33) & --- \\
0.004 & 0.03 & 0.008 & 0.008 & 0.17081(32) & 0.06534(30) & --- \\
0.004 & 0.03 & 0.008 & 0.025 & 0.23920(32) & 0.07132(31) & --- \\
0.004 & 0.03 & 0.008 & 0.03 & 0.25614(32) & 0.07276(32) & --- \\
0.004 & 0.03 & 0.025 & 0.025 & 0.29296(27) & 0.07750(32) & 0.7332(23) \\
0.004 & 0.03 & 0.025 & 0.03 & 0.30733(27) & 0.07902(32) & --- \\
0.004 & 0.03 & 0.03 & 0.03 & 0.32118(27) & 0.08058(32) & 0.7597(21) \\
\hline
\hline
\end{tabular}
\caption{Partially quenched pseudoscalar mass, pseudoscalar decay constant, and $\Omega$ baryon mass measurements on the 32I $a m_{l} = 0.004$ ensemble.}
\end{table}

\begin{table}[h]
\centering
\begin{tabular}{cc|cc||ccc}
\hline
\hline
$a m_{l}$ & $a m_{h}$ & $a m_{x}$ & $a m_{y}$ & $a m_{xy}$ & $a f_{xy}$ & $a m_{xxx}$ \\
\hline
0.006 & 0.03 & 0.002 & 0.002 & 0.09888(38) & 0.06070(33) & --- \\
0.006 & 0.03 & 0.002 & 0.004 & 0.11439(32) & 0.06179(32) & --- \\
0.006 & 0.03 & 0.002 & 0.006 & 0.12802(30) & 0.06282(32) & --- \\
0.006 & 0.03 & 0.002 & 0.008 & 0.14031(29) & 0.06377(32) & --- \\
0.006 & 0.03 & 0.002 & 0.025 & 0.21843(31) & 0.06987(35) & --- \\
0.006 & 0.03 & 0.002 & 0.03 & 0.23673(34) & 0.07129(36) & --- \\
0.006 & 0.03 & 0.004 & 0.004 & 0.12782(28) & 0.06263(31) & --- \\
0.006 & 0.03 & 0.004 & 0.006 & 0.14003(27) & 0.06354(31) & --- \\
0.006 & 0.03 & 0.004 & 0.008 & 0.15127(26) & 0.06442(31) & --- \\
0.006 & 0.03 & 0.004 & 0.025 & 0.22559(27) & 0.07038(32) & --- \\
0.006 & 0.03 & 0.004 & 0.03 & 0.24338(28) & 0.07178(33) & --- \\
0.006 & 0.03 & 0.006 & 0.006 & 0.15118(26) & 0.06439(30) & --- \\
0.006 & 0.03 & 0.006 & 0.008 & 0.16160(25) & 0.06523(30) & --- \\
0.006 & 0.03 & 0.006 & 0.025 & 0.23266(25) & 0.07113(31) & --- \\
0.006 & 0.03 & 0.006 & 0.03 & 0.24999(26) & 0.07254(32) & --- \\
0.006 & 0.03 & 0.008 & 0.008 & 0.17136(25) & 0.06605(30) & --- \\
0.006 & 0.03 & 0.008 & 0.025 & 0.23961(25) & 0.07192(31) & --- \\
0.006 & 0.03 & 0.008 & 0.03 & 0.25652(25) & 0.07334(31) & --- \\
0.006 & 0.03 & 0.025 & 0.025 & 0.29338(23) & 0.07793(30) & 0.7392(22) \\
0.006 & 0.03 & 0.025 & 0.03 & 0.30775(23) & 0.07941(31) & --- \\
0.006 & 0.03 & 0.03 & 0.03 & 0.32161(22) & 0.08092(31) & 0.7655(20) \\
\hline
\hline
\end{tabular}
\caption{Partially quenched pseudoscalar mass, pseudoscalar decay constant, and $\Omega$ baryon mass measurements on the 32I $a m_{l} = 0.006$ ensemble.}
\end{table}

\begin{table}[h]
\centering
\begin{tabular}{cc|cc||ccc}
\hline
\hline
$a m_{l}$ & $a m_{h}$ & $a m_{x}$ & $a m_{y}$ & $a m_{xy}$ & $a f_{xy}$ & $a m_{xxx}$ \\
\hline
0.008 & 0.03 & 0.002 & 0.002 & 0.10008(46) & 0.06211(40) & --- \\
0.008 & 0.03 & 0.002 & 0.004 & 0.11564(44) & 0.06310(38) & --- \\
0.008 & 0.03 & 0.002 & 0.006 & 0.12933(43) & 0.06408(36) & --- \\
0.008 & 0.03 & 0.002 & 0.008 & 0.14167(44) & 0.06501(36) & --- \\
0.008 & 0.03 & 0.002 & 0.025 & 0.22029(54) & 0.07127(37) & --- \\
0.008 & 0.03 & 0.002 & 0.03 & 0.23875(58) & 0.07276(39) & --- \\
0.008 & 0.03 & 0.004 & 0.004 & 0.12910(41) & 0.06382(35) & --- \\
0.008 & 0.03 & 0.004 & 0.006 & 0.14134(40) & 0.06467(34) & --- \\
0.008 & 0.03 & 0.004 & 0.008 & 0.15261(40) & 0.06551(33) & --- \\
0.008 & 0.03 & 0.004 & 0.025 & 0.22728(45) & 0.07151(33) & --- \\
0.008 & 0.03 & 0.004 & 0.03 & 0.24519(48) & 0.07296(34) & --- \\
0.008 & 0.03 & 0.006 & 0.006 & 0.15250(39) & 0.06545(33) & --- \\
0.008 & 0.03 & 0.006 & 0.008 & 0.16293(38) & 0.06625(32) & --- \\
0.008 & 0.03 & 0.006 & 0.025 & 0.23419(41) & 0.07212(32) & --- \\
0.008 & 0.03 & 0.006 & 0.03 & 0.25160(42) & 0.07354(33) & --- \\
0.008 & 0.03 & 0.008 & 0.008 & 0.17268(37) & 0.06702(31) & --- \\
0.008 & 0.03 & 0.008 & 0.025 & 0.24099(38) & 0.07280(31) & --- \\
0.008 & 0.03 & 0.008 & 0.03 & 0.25795(39) & 0.07422(32) & --- \\
0.008 & 0.03 & 0.025 & 0.025 & 0.29429(32) & 0.07847(31) & 0.7399(30) \\
0.008 & 0.03 & 0.025 & 0.03 & 0.30862(32) & 0.07993(31) & --- \\
0.008 & 0.03 & 0.03 & 0.03 & 0.32243(31) & 0.08140(31) & 0.7664(27) \\
\hline
\hline
\end{tabular}
\caption{Partially quenched pseudoscalar mass, pseudoscalar decay constant, and $\Omega$ baryon mass measurements on the 32I $a m_{l} = 0.008$ ensemble.}
\end{table}

\begin{table}[h]
\centering
\begin{tabular}{cc|cc||ccc}
\hline
\hline
$a m_{l}$ & $a m_{h}$ & $a m_{x}$ & $a m_{y}$ & $a m_{xy}$ & $a f_{xy}$ & $a m_{xxx}$ \\
\hline
0.001 & 0.046 & 0.0001 & 0.0001 & 0.10423(23) & 0.0938(12) & --- \\
0.001 & 0.046 & 0.0001 & 0.001 & 0.11512(22) & 0.0944(12) & --- \\
0.001 & 0.046 & 0.0001 & 0.0042 & 0.14718(22) & 0.0964(12) & --- \\
0.001 & 0.046 & 0.0001 & 0.008 & 0.17755(24) & 0.0984(12) & --- \\
0.001 & 0.046 & 0.0001 & 0.035 & 0.31783(45) & 0.1090(13) & --- \\
0.001 & 0.046 & 0.0001 & 0.045 & 0.35642(56) & 0.1121(14) & --- \\
0.001 & 0.046 & 0.0001 & 0.055 & 0.39150(67) & 0.1149(14) & --- \\
0.001 & 0.046 & 0.001 & 0.001 & 0.12497(22) & 0.0950(12) & --- \\
0.001 & 0.046 & 0.001 & 0.0042 & 0.15485(21) & 0.0969(12) & --- \\
0.001 & 0.046 & 0.001 & 0.008 & 0.18385(22) & 0.0988(12) & --- \\
0.001 & 0.046 & 0.001 & 0.035 & 0.32120(39) & 0.1092(13) & --- \\
0.001 & 0.046 & 0.001 & 0.045 & 0.35939(47) & 0.1123(14) & --- \\
0.001 & 0.046 & 0.001 & 0.055 & 0.39418(56) & 0.1151(14) & --- \\
0.001 & 0.046 & 0.0042 & 0.0042 & 0.17949(21) & 0.0986(12) & --- \\
0.001 & 0.046 & 0.0042 & 0.008 & 0.20483(21) & 0.1005(12) & --- \\
0.001 & 0.046 & 0.0042 & 0.035 & 0.33342(30) & 0.1107(13) & --- \\
0.001 & 0.046 & 0.0042 & 0.045 & 0.37030(34) & 0.1137(14) & --- \\
0.001 & 0.046 & 0.0042 & 0.055 & 0.40411(38) & 0.1164(14) & --- \\
0.001 & 0.046 & 0.008 & 0.008 & 0.22725(21) & 0.1024(12) & --- \\
0.001 & 0.046 & 0.008 & 0.035 & 0.34760(26) & 0.1126(14) & --- \\
0.001 & 0.046 & 0.008 & 0.045 & 0.38315(28) & 0.1156(14) & --- \\
0.001 & 0.046 & 0.008 & 0.055 & 0.41594(30) & 0.1183(14) & --- \\
0.001 & 0.046 & 0.035 & 0.035 & 0.43684(21) & 0.1231(15) & 1.1608(42) \\
0.001 & 0.046 & 0.035 & 0.045 & 0.46618(22) & 0.1262(15) & --- \\
0.001 & 0.046 & 0.035 & 0.055 & 0.49409(22) & 0.1291(16) & --- \\
0.001 & 0.046 & 0.045 & 0.045 & 0.49404(21) & 0.1294(16) & 1.2130(37) \\
0.001 & 0.046 & 0.045 & 0.055 & 0.52070(21) & 0.1324(16) & --- \\
0.001 & 0.046 & 0.055 & 0.055 & 0.54632(21) & 0.1354(16) & 1.2641(34) \\
\hline
\hline
\end{tabular}
\caption{Partially quenched pseudoscalar mass, pseudoscalar decay constant, and $\Omega$ baryon mass measurements on the 32ID $a m_{l} = 0.001$ ensemble.}
\end{table}

\begin{table}[h]
\centering
\begin{tabular}{cc|cc||ccc}
\hline
\hline
$a m_{l}$ & $a m_{h}$ & $a m_{x}$ & $a m_{y}$ & $a m_{xy}$ & $a f_{xy}$ & $a m_{xxx}$ \\
\hline
0.0042 & 0.046 & 0.0001 & 0.0001 & 0.10581(27) & 0.0973(12) & --- \\
0.0042 & 0.046 & 0.0001 & 0.001 & 0.11668(25) & 0.0977(12) & --- \\
0.0042 & 0.046 & 0.0001 & 0.0042 & 0.14870(26) & 0.0994(12) & --- \\
0.0042 & 0.046 & 0.0001 & 0.008 & 0.17913(27) & 0.1013(13) & --- \\
0.0042 & 0.046 & 0.0001 & 0.035 & 0.31972(52) & 0.1118(14) & --- \\
0.0042 & 0.046 & 0.0001 & 0.045 & 0.35808(62) & 0.1147(15) & --- \\
0.0042 & 0.046 & 0.0001 & 0.055 & 0.39279(71) & 0.1173(15) & --- \\
0.0042 & 0.046 & 0.001 & 0.001 & 0.12654(24) & 0.0981(12) & --- \\
0.0042 & 0.046 & 0.001 & 0.0042 & 0.15638(24) & 0.0997(12) & --- \\
0.0042 & 0.046 & 0.001 & 0.008 & 0.18544(26) & 0.1015(12) & --- \\
0.0042 & 0.046 & 0.001 & 0.035 & 0.32302(44) & 0.1118(14) & --- \\
0.0042 & 0.046 & 0.001 & 0.045 & 0.36102(52) & 0.1148(14) & --- \\
0.0042 & 0.046 & 0.001 & 0.055 & 0.39549(59) & 0.1173(15) & --- \\
0.0042 & 0.046 & 0.0042 & 0.0042 & 0.18099(25) & 0.1011(12) & --- \\
0.0042 & 0.046 & 0.0042 & 0.008 & 0.20634(26) & 0.1028(13) & --- \\
0.0042 & 0.046 & 0.0042 & 0.035 & 0.33502(32) & 0.1129(14) & --- \\
0.0042 & 0.046 & 0.0042 & 0.045 & 0.37182(35) & 0.1158(14) & --- \\
0.0042 & 0.046 & 0.0042 & 0.055 & 0.40549(40) & 0.1184(15) & --- \\
0.0042 & 0.046 & 0.008 & 0.008 & 0.22872(26) & 0.1044(13) & --- \\
0.0042 & 0.046 & 0.008 & 0.035 & 0.34906(27) & 0.1144(14) & --- \\
0.0042 & 0.046 & 0.008 & 0.045 & 0.38459(28) & 0.1174(14) & --- \\
0.0042 & 0.046 & 0.008 & 0.055 & 0.41736(31) & 0.1200(15) & --- \\
0.0042 & 0.046 & 0.035 & 0.035 & 0.43813(22) & 0.1243(15) & 1.1695(48) \\
0.0042 & 0.046 & 0.035 & 0.045 & 0.46748(21) & 0.1274(15) & --- \\
0.0042 & 0.046 & 0.035 & 0.055 & 0.49540(21) & 0.1302(16) & --- \\
0.0042 & 0.046 & 0.045 & 0.045 & 0.49534(21) & 0.1305(16) & 1.2220(41) \\
0.0042 & 0.046 & 0.045 & 0.055 & 0.52200(20) & 0.1334(16) & --- \\
0.0042 & 0.046 & 0.055 & 0.055 & 0.54759(19) & 0.1363(16) & 1.2735(36) \\
\hline
\hline
\end{tabular}
\caption{Partially quenched pseudoscalar mass, pseudoscalar decay constant, and $\Omega$ baryon mass measurements on the 32ID $a m_{l} = 0.0042$ ensemble.}
\end{table}

\begin{table}[h]
\centering
\begin{tabular}{c||cc|cc|cc|c}
\hline
\hline
Ensemble & $a m_{l}$ & $a m_{h}$ & $a m_{ll}$ & $a m_{lh}$ & $a f_{ll}$ & $a f_{lh}$ & $a m_{hhh}$ \\
\hline
\rule{0cm}{0.4cm} 32I-fine & 0.0047 & 0.0186 & 0.1179(13) & 0.1772(12) & 0.04846(32) & 0.05358(22) & 0.5522(29)\\
\hline
\rule{0cm}{0.4cm} 48I & 0.00078 & 0.0362 & 0.08049(13) & 0.28853(14) & 0.075799(84) & 0.090396(86) & 0.97018(96)\\
\hline
\rule{0cm}{0.4cm} 64I & 0.000678 & 0.02661 & 0.05903(13) & 0.21531(17) & 0.055505(95) & 0.066534(99) & 0.71811(73)\\
\hline
\rule{0cm}{0.4cm} 32ID-M1 & 0.00022 & 0.0596 & 0.11812(46) & 0.42313(49) & 0.12489(23) & 0.14673(33) & 1.5290(31)\\
\hline
\rule{0cm}{0.4cm} 32ID-M2 & 0.00478 & 0.03297 & 0.19487(64) & 0.30792(64) & 0.07771(22) & 0.08716(21) & 0.9148(34)\\
\hline
\hline
\end{tabular}
\caption{Unitary pseudoscalar mass, pseudoscalar decay constant, and $\Omega$ baryon mass measurements.}
\end{table}
\FloatBarrier

\subsection{$R$ (Equation~\eqref{eqn:mres})}

\begin{table}[h]
\centering
\begin{tabular}{c||cc|c}
\hline
\hline
Ensemble & $a m_{l}$ & $a m_{h}$ & $a R$ \\
\hline
\rule{0cm}{0.4cm} \multirow{2}{*}{24I} & 0.005 & 0.04 & 0.003154(15) \\
 & 0.01 & 0.04 & 0.003187(24) \\
\hline
\rule{0cm}{0.4cm} \multirow{3}{*}{32I} & 0.004 & 0.03 & 0.0006697(34) \\
 & 0.006 & 0.03 & 0.0006589(30) \\
 & 0.008 & 0.03 & 0.0006676(34) \\
\hline
\rule{0cm}{0.4cm} \multirow{2}{*}{32ID} & 0.001 & 0.046 & 0.0018510(43) \\
 & 0.0042 & 0.046 & 0.0018735(48) \\
\hline
\rule{0cm}{0.4cm} 32I-fine & 0.0047 & 0.0186 & 0.0006300(59) \\
\hline
\rule{0cm}{0.4cm} 48I & 0.00078 & 0.0362 & 0.0006102(40) \\
\hline
\rule{0cm}{0.4cm} 64I & 0.000678 & 0.02661 & 0.0003116(23) \\
\hline
\rule{0cm}{0.4cm} 32ID-M1 & 0.00022 & 0.0596 & 0.002170(16) \\
\hline
\rule{0cm}{0.4cm} 32ID-M2 & 0.00478 & 0.03297 & 0.0044660(46) \\
\hline
\hline
\end{tabular}
\caption{Summary of measurements of $R$ (Equation \eqref{eqn:mres}) at the simulated quark masses on each ensemble. This quantity is equal to $m_{\rm res}$ in the chiral limit.}
\end{table}
\FloatBarrier

\subsection{Wilson Flow Scales}

\begin{table}[h]
\centering
\begin{tabular}{c||cc|cc}
\hline
\hline
Ensemble & $a m_{l}$ & $a m_{h}$ & ${t_{0}}^{1/2}$ & $w_{0}$ \\
\hline
\rule{0cm}{0.4cm} \multirow{2}{*}{24I} & 0.005 & 0.04 & 1.31625(57) & 1.4911(15) \\
 & 0.01 & 0.04 & 1.30501(65) & 1.4653(14) \\
\hline
\rule{0cm}{0.4cm} \multirow{3}{*}{32I} & 0.004 & 0.03 & 1.7422(11) & 2.0124(26) \\
 & 0.006 & 0.03 & 1.73622(86) & 1.9963(19) \\
 & 0.008 & 0.03 & 1.7286(11) & 1.9793(24) \\
\hline
\rule{0cm}{0.4cm} \multirow{2}{*}{32ID} & 0.001 & 0.046 & 1.02682(25) & 1.21778(72) \\
 & 0.0042 & 0.046 & 1.02245(27) & 1.20420(73) \\
\hline
\rule{0cm}{0.4cm} 32I-fine & 0.0047 & 0.0186 & 2.2860(63) & 2.664(16) \\
\hline
\rule{0cm}{0.4cm} 48I & 0.00078 & 0.0362 & 1.29659(39) & 1.5013(10) \\
\hline
\rule{0cm}{0.4cm} 64I & 0.000678 & 0.02661 & 1.74448(98) & 2.0502(26) \\
\hline
\rule{0cm}{0.4cm} 32ID-M1 & 0.00022 & 0.0596 & 0.78719(16) & 0.88865(78) \\
\hline
\rule{0cm}{0.4cm} 32ID-M2 & 0.00478 & 0.03297 & 1.4841(16) & 1.7151(33) \\
\hline
\hline
\end{tabular}
\caption{Summary of Wilson flow measurements.}
\end{table}
\FloatBarrier


\clearpage
\section{Analysis of the 32ID-M1 and 32ID-M2 Ensembles}
\label{appendix:new_ensembles}

Here we present details of an analysis of the 32ID-M1 and 32ID-M2 ensembles. These lattices were originally generated for scale setting in the context of QCD thermodynamics calculations, and have not appeared in any of our previous chiral fits.

\subsection{Evolution}

The M\"{o}bius domain wall action~\cite{Brower:2012vk} introduces two new scaling parameters, $b$ and $c$, into the kernel of the domain wall action. If $b-c=1$, the kernel is identical to the Shamir kernel of conventional domain wall fermions up to a scaling coefficient $\alpha = b+c$. In Ref.~\cite{Blum:2014tka} we show that a M\"{o}bius DWF simulation with $b-c=1$, a fifth-dimensional extent of $L_{s}$, and a scaling coefficient $\alpha$ is directly equivalent to a simulation with Shamir DWF and fifth-dimensional extent $\alpha L_{s}$ up to small terms that vanish in the $L_{s} \rightarrow \infty$ limit. For the same cost we can therefore use M\"{o}bius DWF to simulate with substantially reduced explicit chiral symmetry breaking simply by increasing $\alpha$, without deviating from the scaling trajectory of our conventional Shamir ensembles. \\

In Table~\ref{tab:ensemble_params} we summarize the M\"{o}bius scale $\alpha = b + c$, the average plaquette and quark condensates, and evolution parameters for the 32ID-M1 and 32ID-M2 ensembles. Both ensembles were generated using an exact hybrid Monte Carlo algorithm with five intermediate Hasenbusch masses --- (0.008, 0.04, 0.12, 0.30, 0.60) --- for the two, degenerate flavors of light quarks, and a rational approximation for the strange quark determinant. Integration of the gauge and fermion fields was performed using a three-level nested force gradient integrator (FGI QPQPQ): the top level corresponds to updates of the fermion force, the middle level corresponds to DSDR updates, and the bottom level corresponds to gauge field updates, with equal numbers of updates of each level per HMC trajectory. Details regarding the implementation of the DSDR term can be found in Ref.~\cite{Arthur:2012opa}. 

\begin{table}[h]
\centering
\begin{tabular}{c||c|c}
\hline
\hline
\rule{0cm}{0.4cm} & 32ID-M1 & 32ID-M2 \\
\hline
\rule{0cm}{0.4cm} $\alpha$ & 4.0 & 4.0 \\
\hline
\rule{0cm}{0.4cm}Steps per HMC traj. & 18 & 10 \\
$\Delta \tau$ & 0.056 & 0.1 \\
Metropolis acceptance & $89\%$ & $68\%$ \\
\hline
\rule{0cm}{0.4cm} $\langle \mathrm{Plaquette} \rangle$ & 0.4681561(65) & 0.5671088(24) \\
$\langle \overline{\psi}_{l} \psi_{l} \rangle$ & 0.0019387(73) & 0.0010403(9) \\
$\langle \overline{\psi}_{l} \gamma_{5} \psi_{l} \rangle$ & -0.000008(13) & -0.000007(2) \\
\hline
\hline
\end{tabular}
\caption{The M\"{o}bius scale ($\alpha = b+c$), integration parameters, and the measured ensemble averages of the plaquette and quark condensates on the 32ID-M1 and 32ID-M2 ensembles. Here $\Delta \tau$ is the MD time step.}
\label{tab:ensemble_params}
\end{table}

\begin{figure}[h]
\centering
\subfloat{\includegraphics[width=0.32\textwidth]{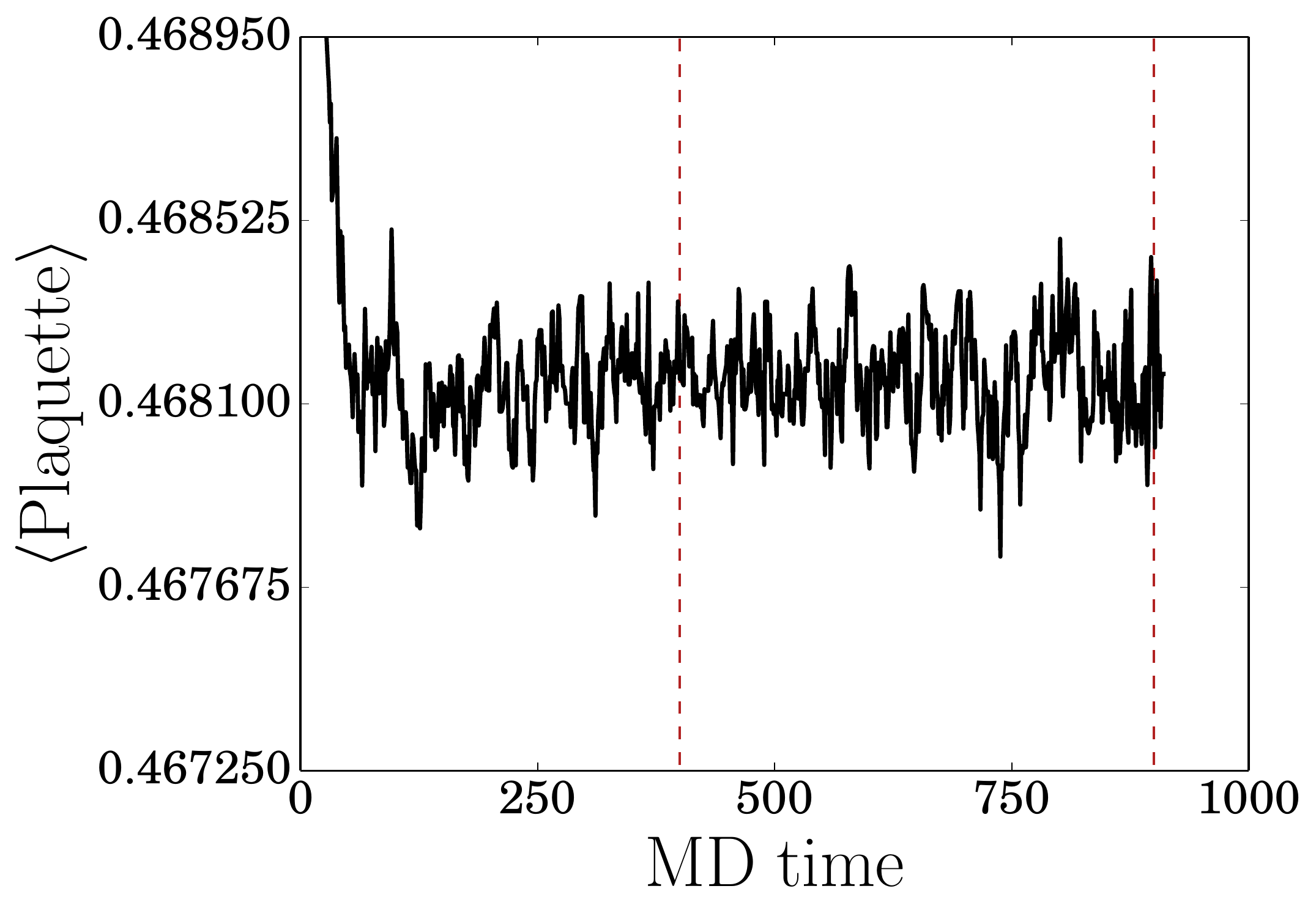}}
\subfloat{\includegraphics[width=0.29\textwidth]{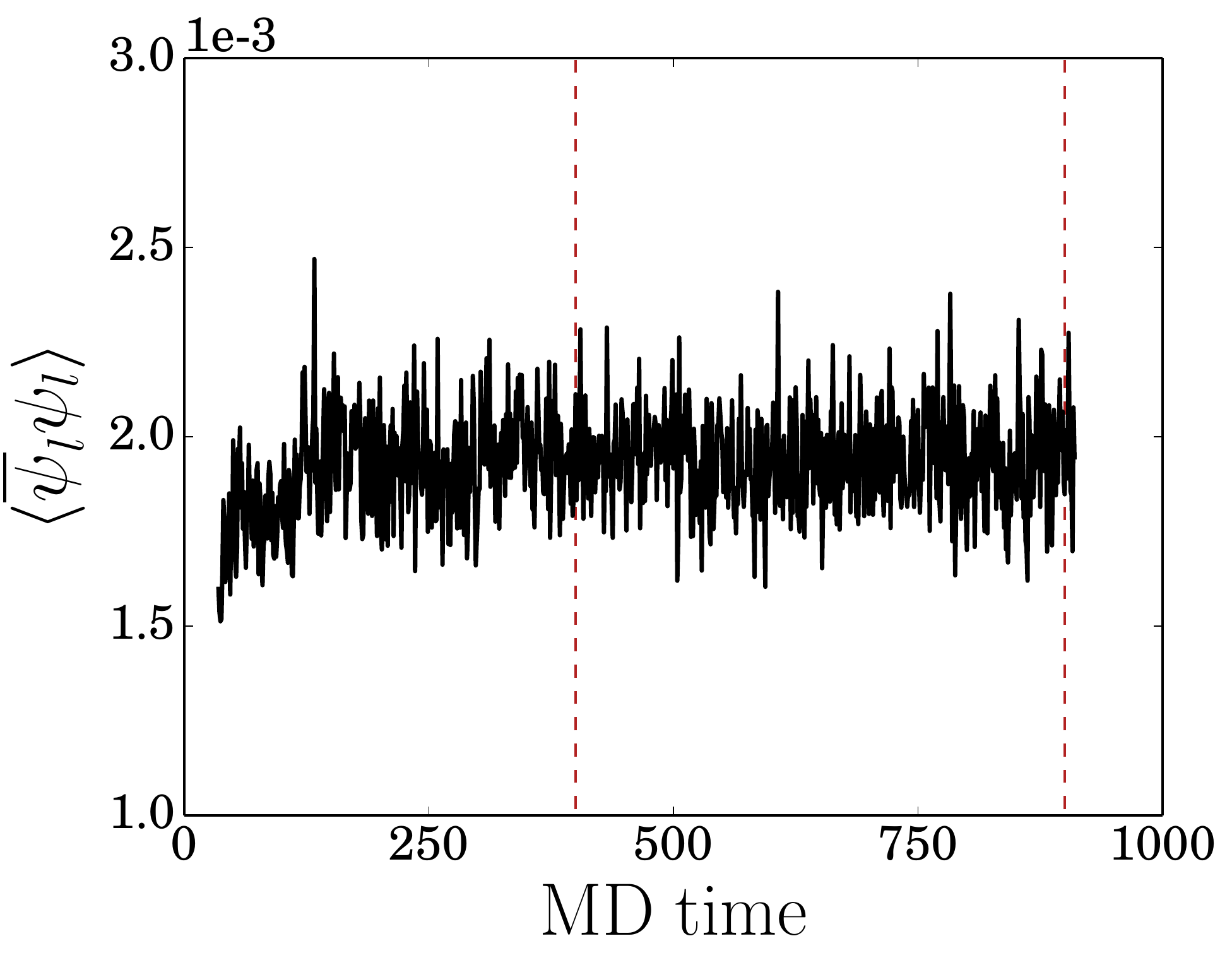}}
\subfloat{\includegraphics[width=0.3\textwidth]{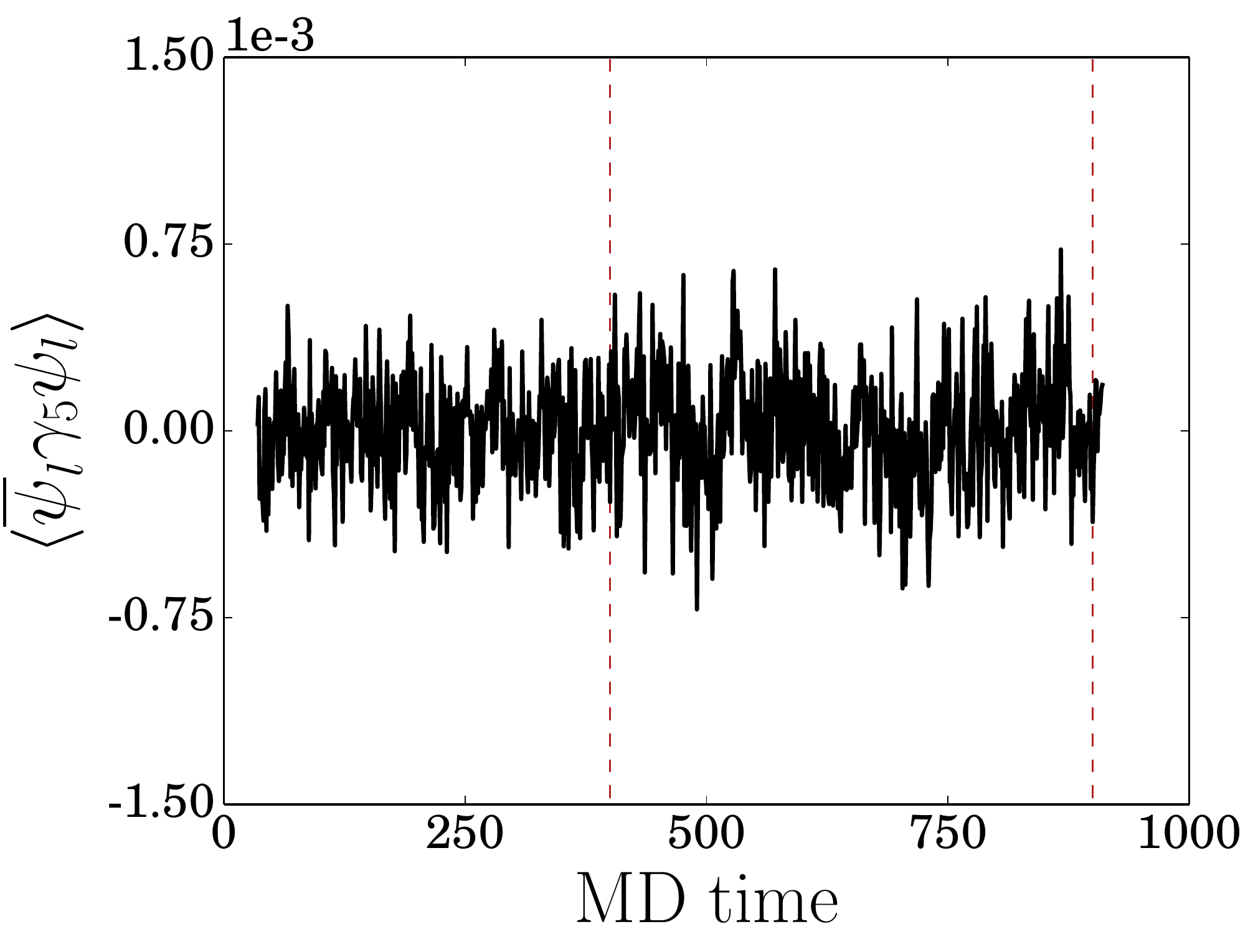}} \\
\subfloat{\includegraphics[width=0.31\textwidth]{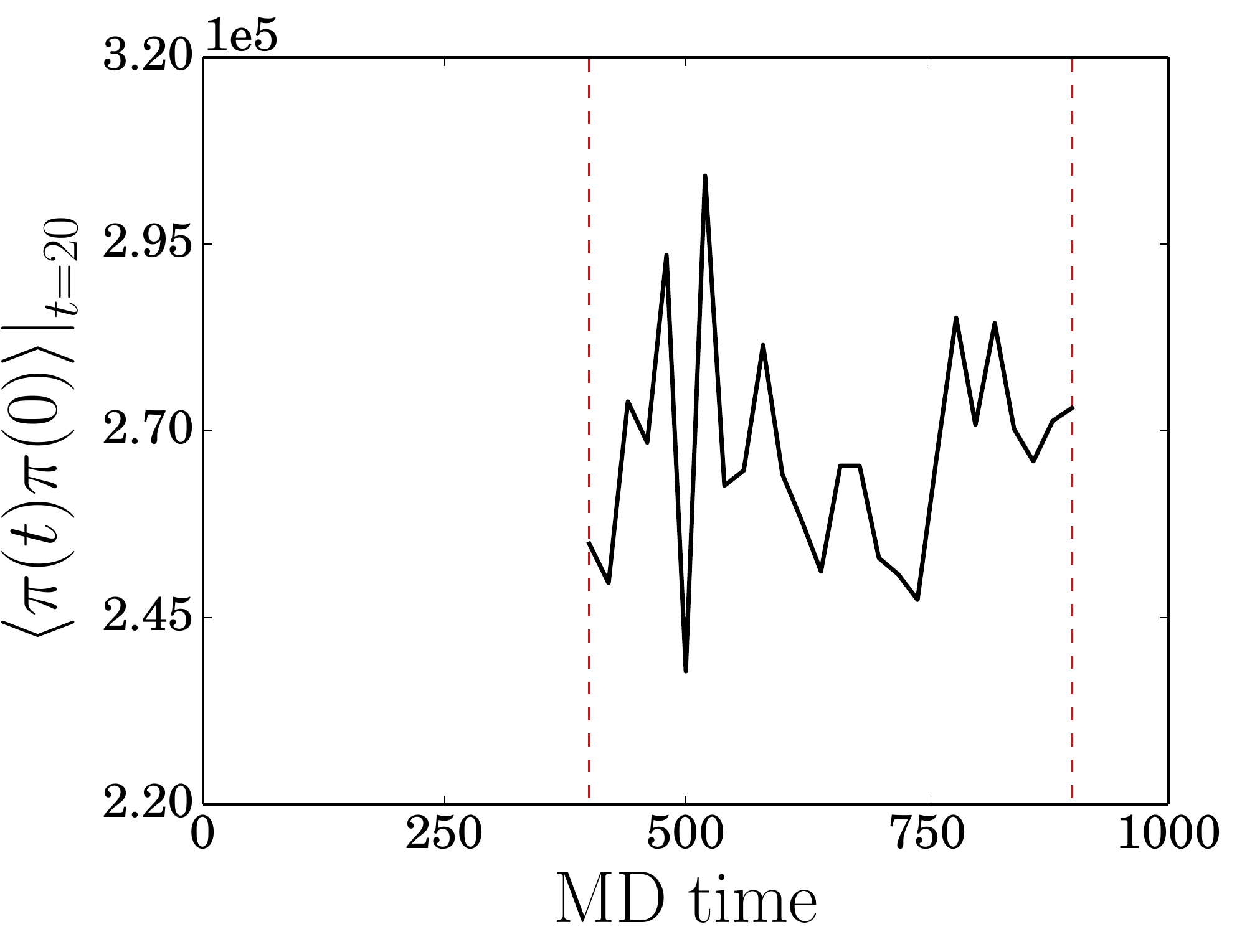}}
\subfloat{\includegraphics[width=0.32\textwidth]{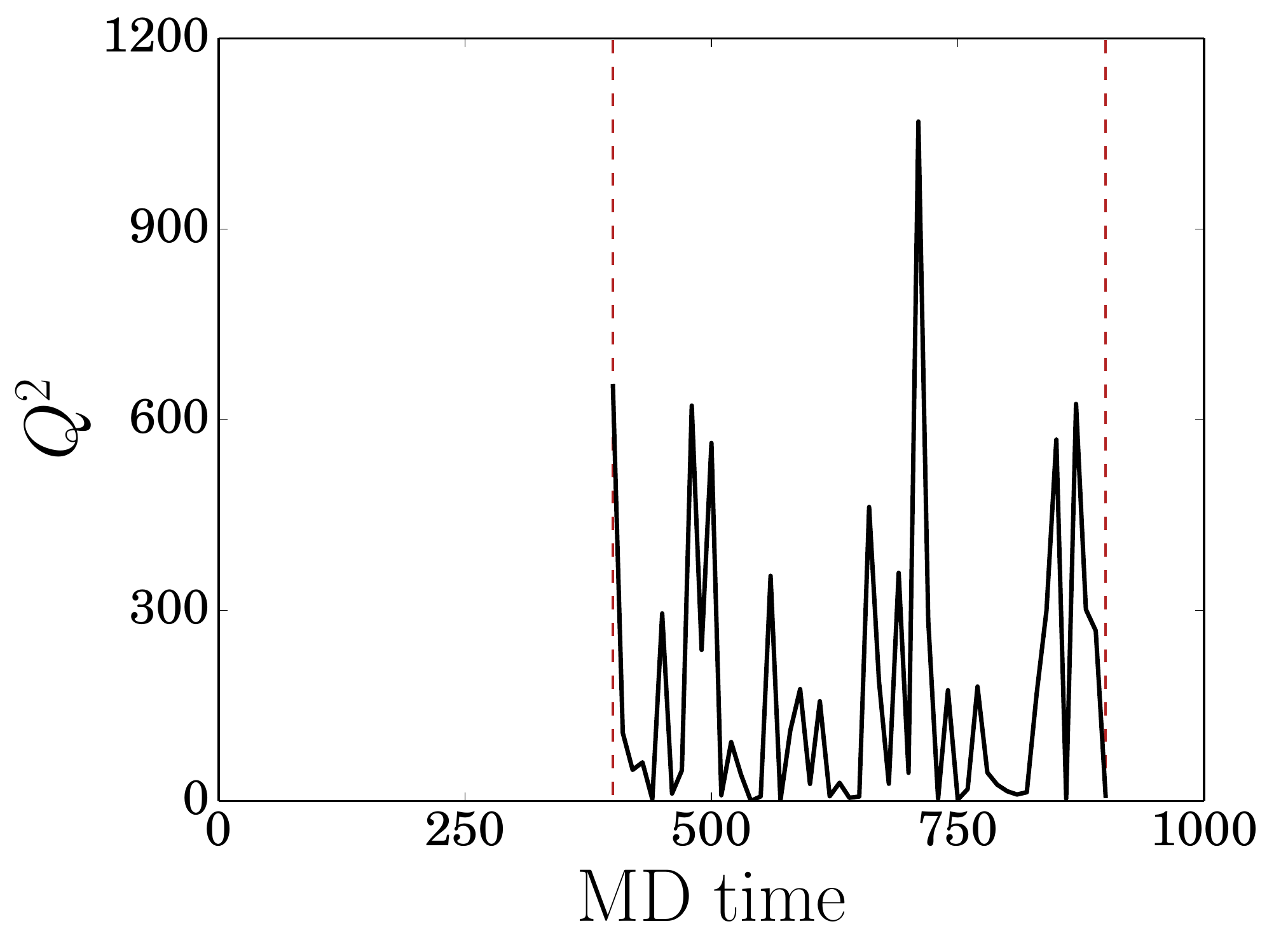}} \\
\subfloat{\includegraphics[width=0.32\textwidth]{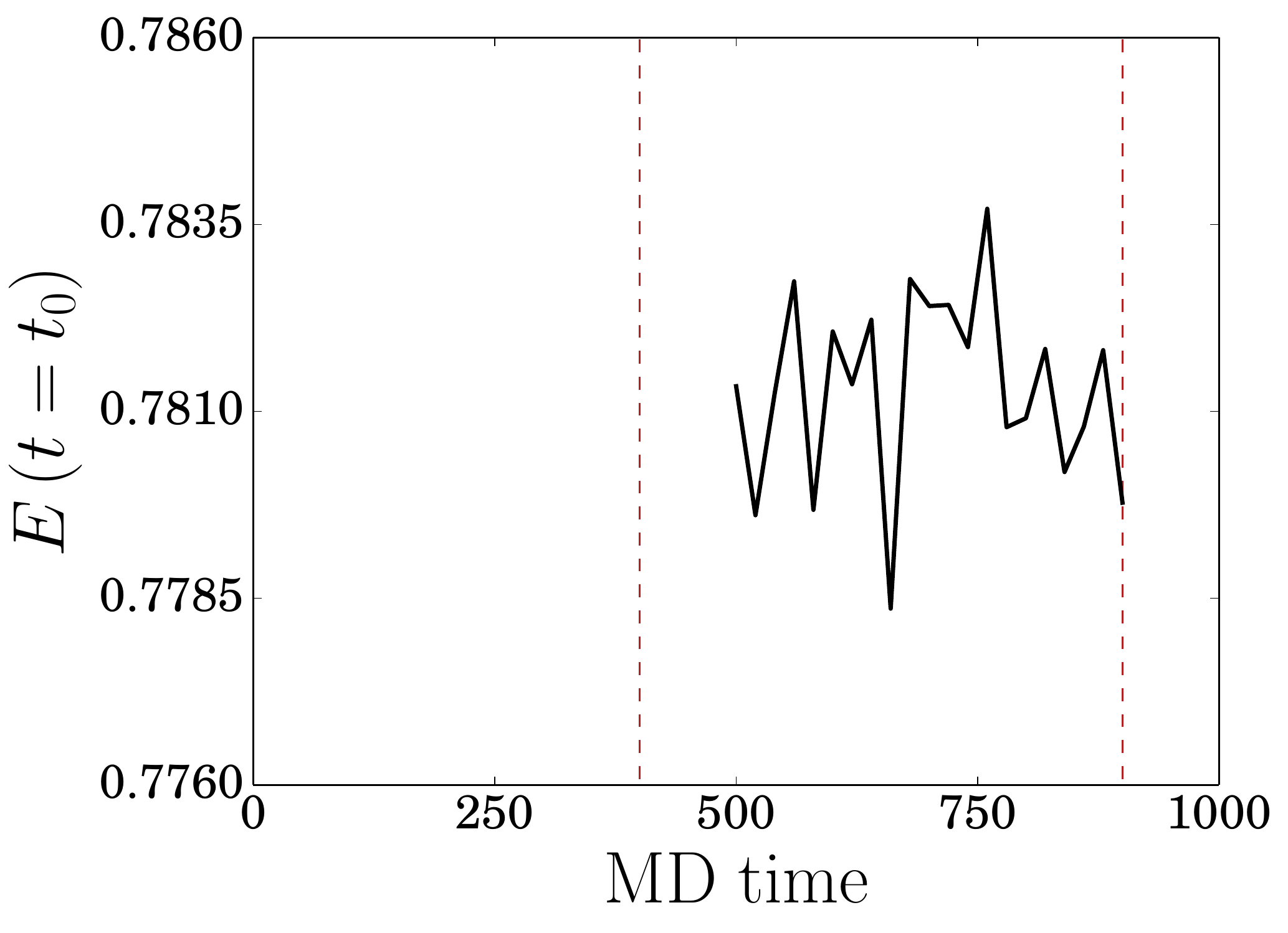}}
\subfloat{\includegraphics[width=0.32\textwidth]{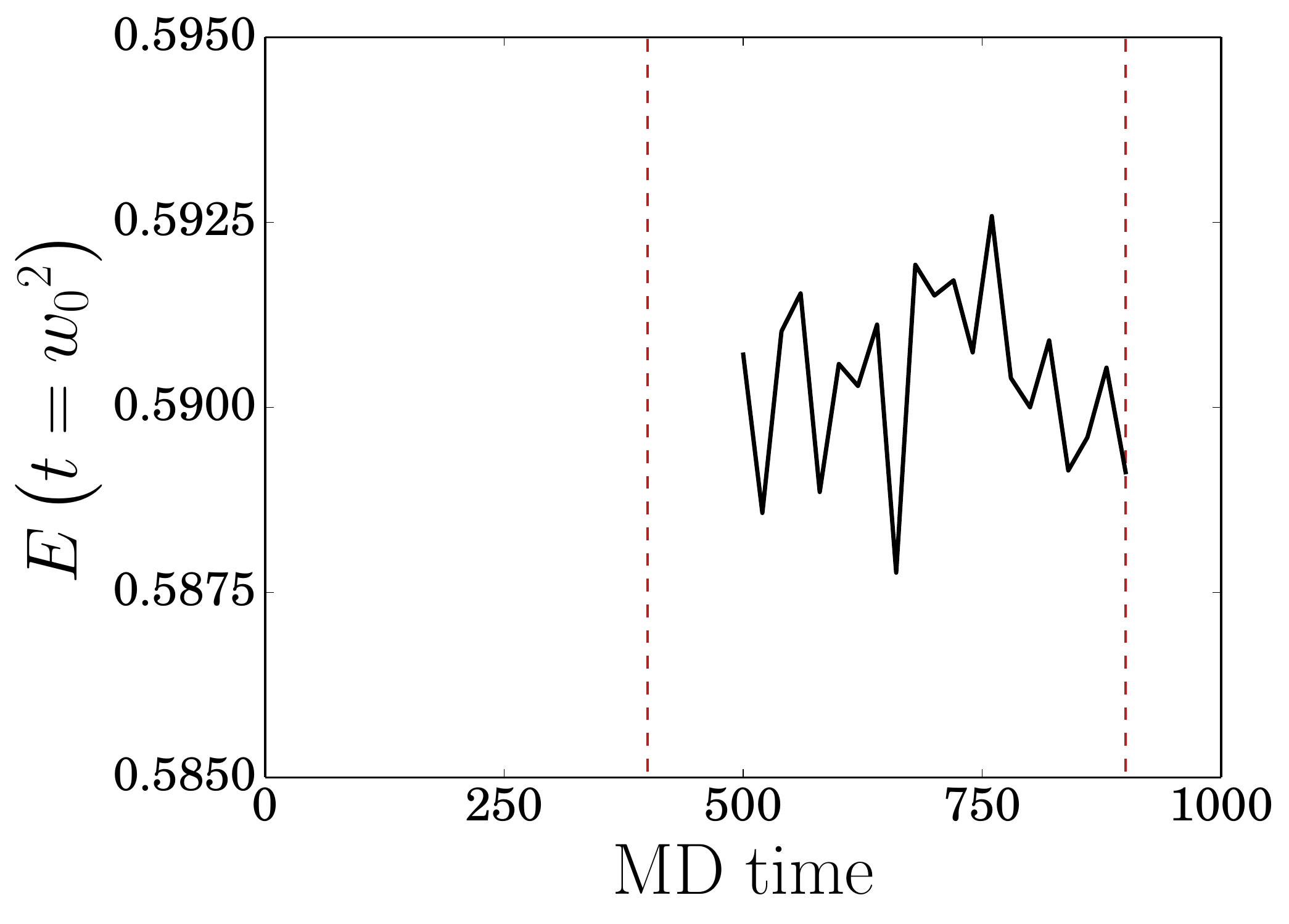}}
\caption{Molecular dynamics evolution of the plaquette, chiral and pseudoscalar condensates, pion propagator at $t/a = 20$, square of the topological charge, and clover discretized action density computed at the Wilson flow times $t_{0}$ and $w_{0}^{2}$ as a function of MD time on the 32ID-M1 ensemble. The first three quantities were computed every MD time step as part of the evolution. The topological charge and Wilson flow scales were computed every 10 and 20 MD time steps, respectively, after the ensemble was thermalized. The dashed vertical lines mark the range of MD times used to perform calculations of the spectrum.}
\label{fig:evol_32ID_M1}
\end{figure}

\begin{figure}[h]
\centering
\subfloat{\includegraphics[width=0.32\textwidth]{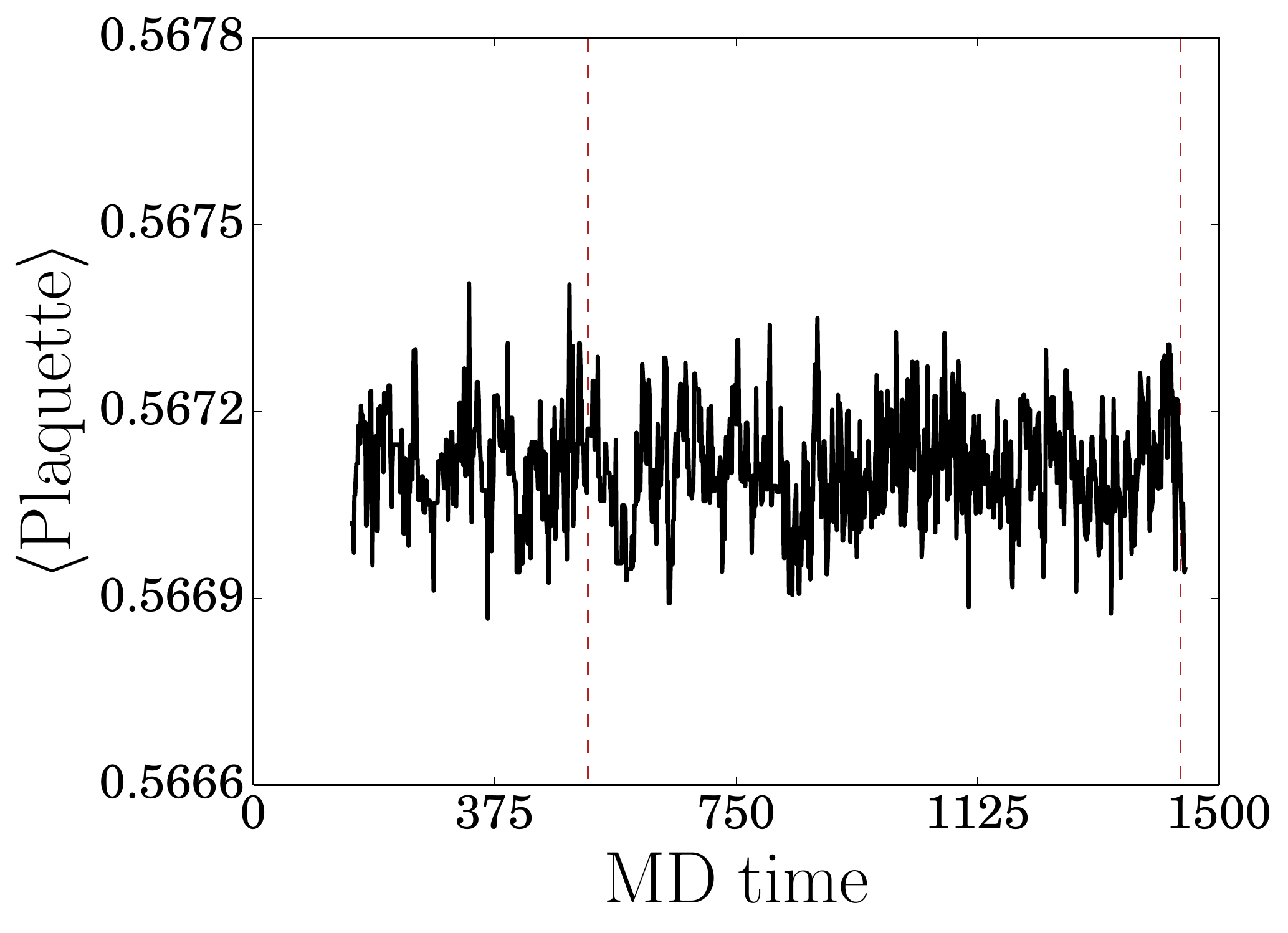}} 
\subfloat{\includegraphics[width=0.31\textwidth]{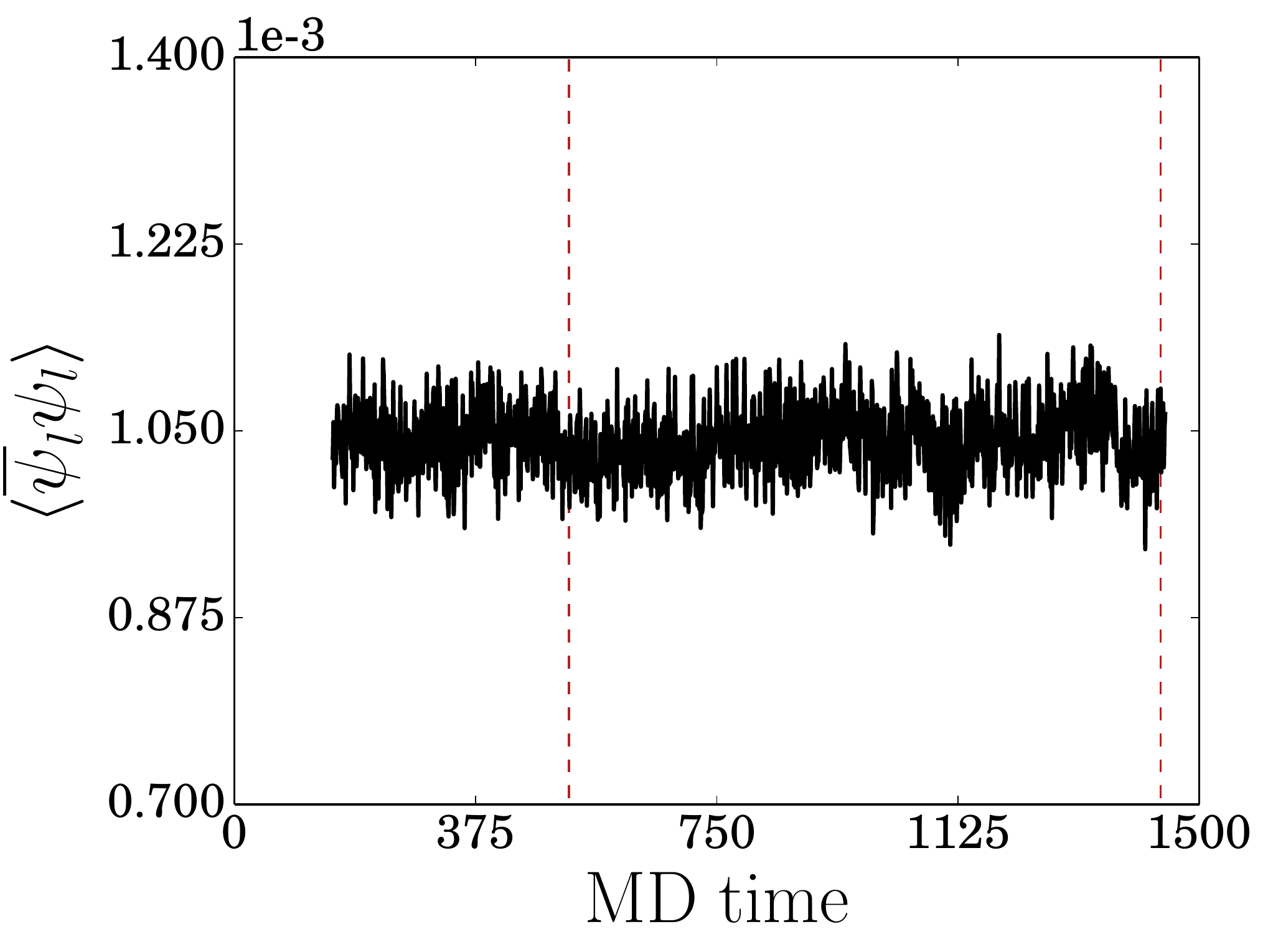}}
\subfloat{\includegraphics[width=0.3\textwidth]{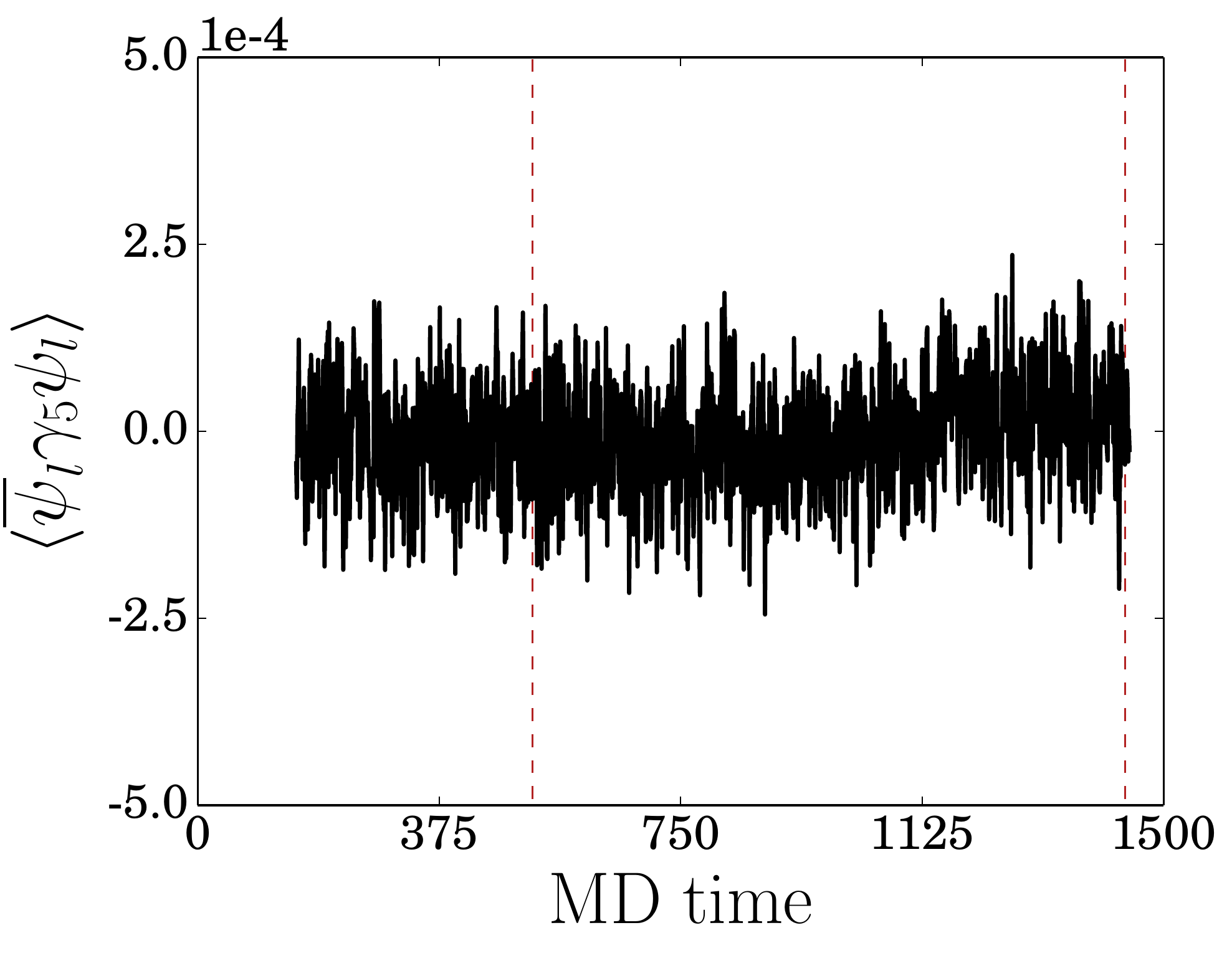}} \\
\subfloat{\includegraphics[width=0.32\textwidth]{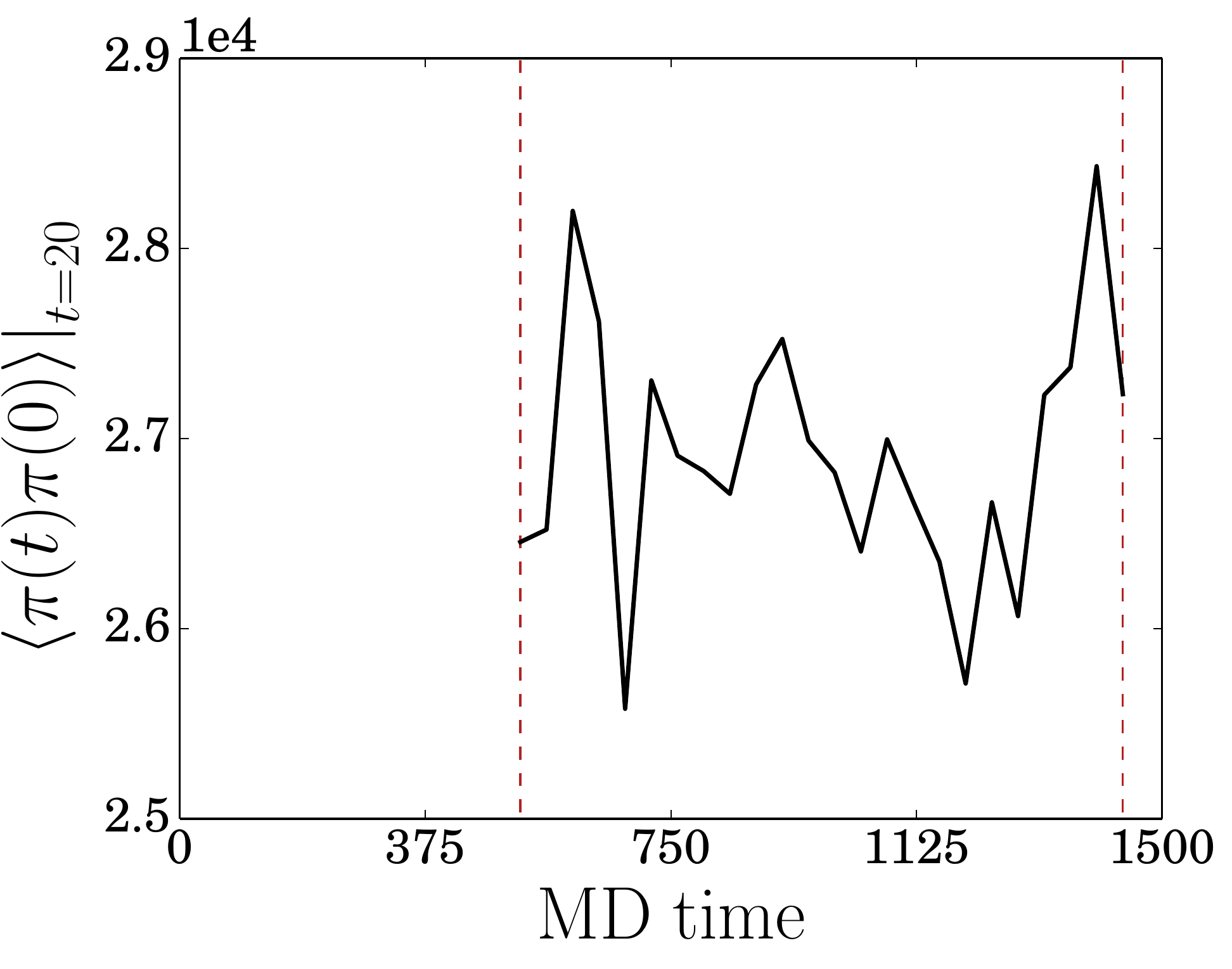}}
\subfloat{\includegraphics[width=0.32\textwidth]{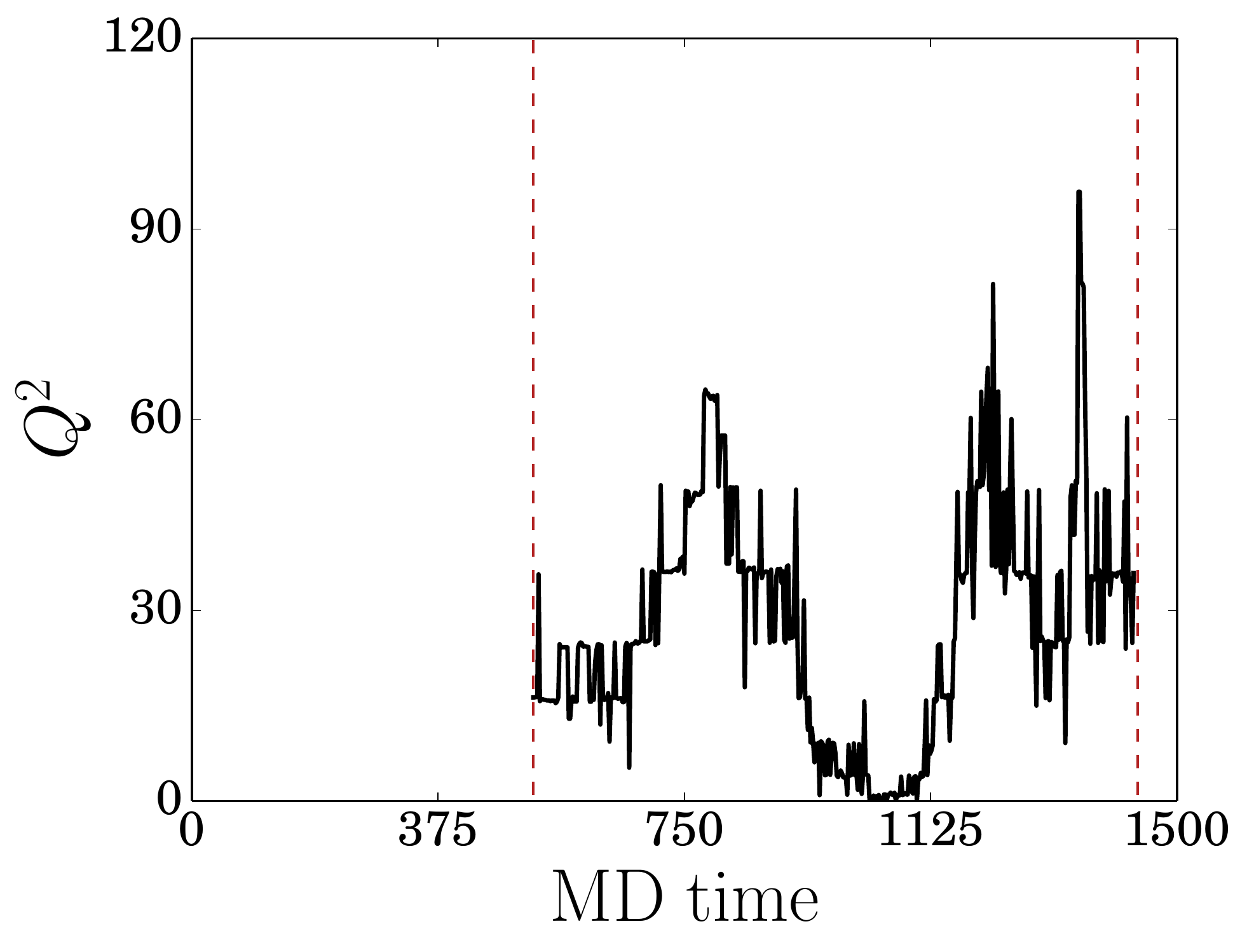}} \\
\subfloat{\includegraphics[width=0.32\textwidth]{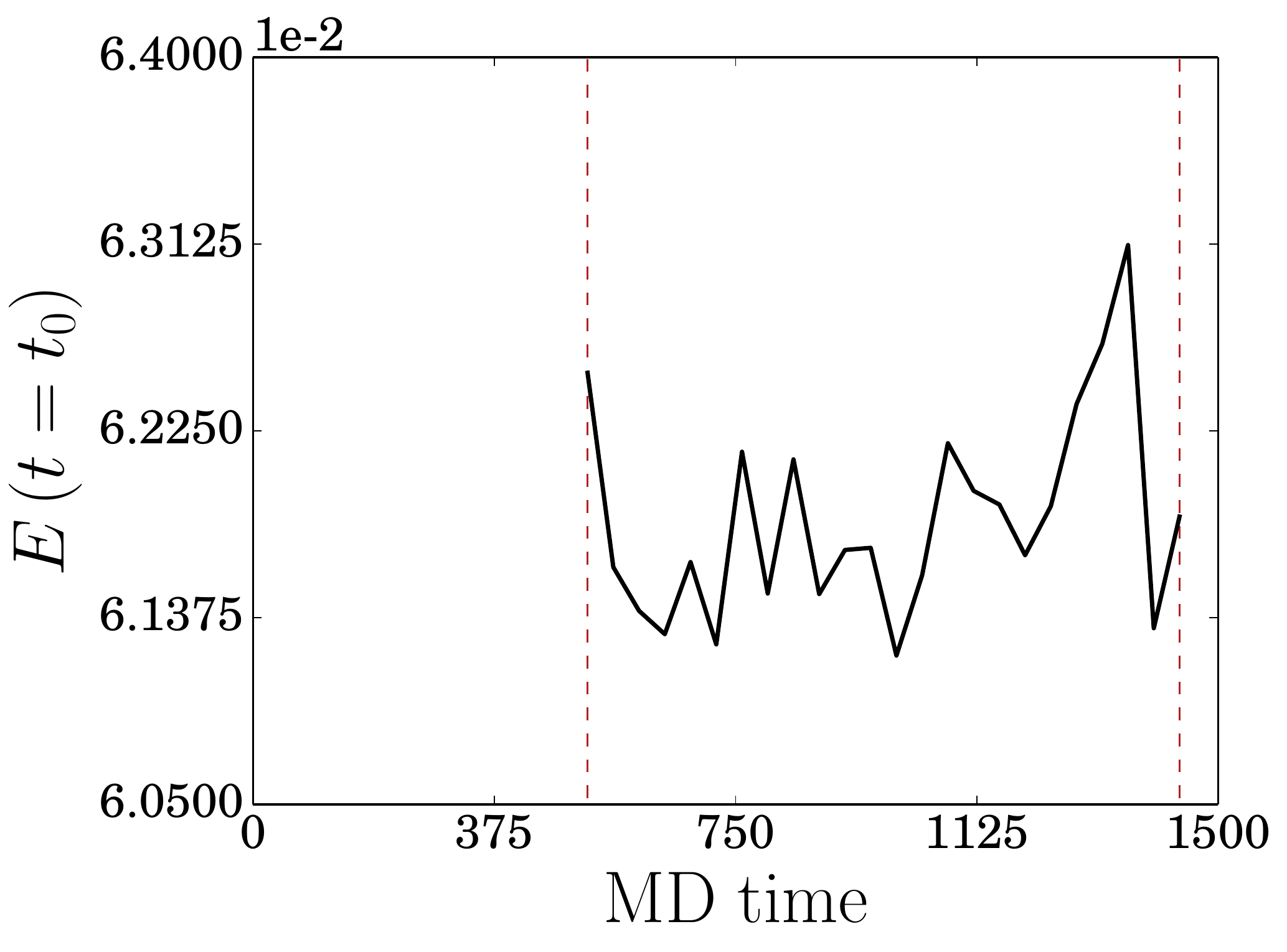}}
\subfloat{\includegraphics[width=0.32\textwidth]{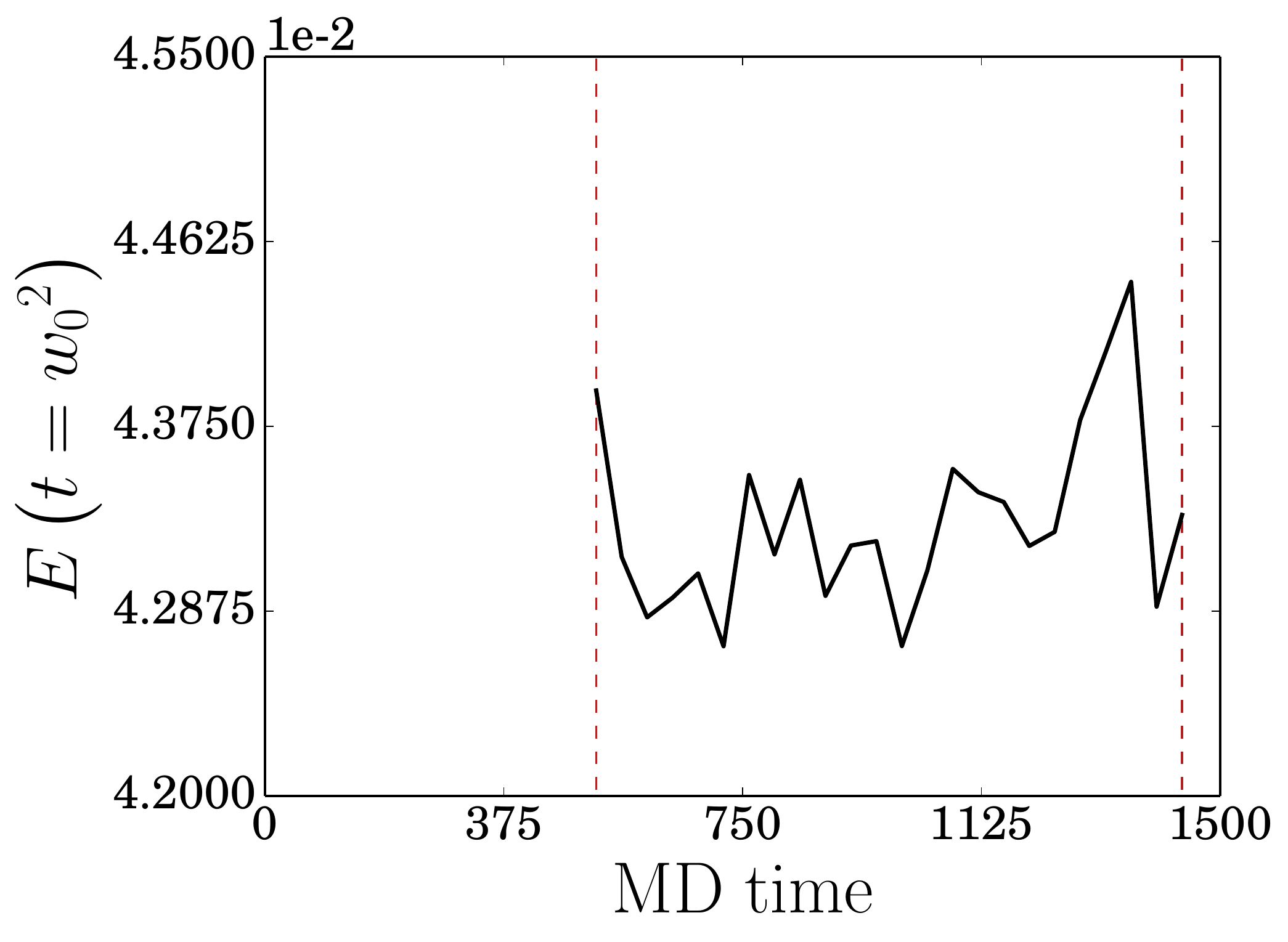}}
\caption{Molecular dynamics evolution of the plaquette, chiral and pseudoscalar condensates, pion propagator at $t/a = 20$, square of the topological charge, and clover discretized action density computed at the Wilson flow times $t_{0}$ and $w_{0}^{2}$ as a function of MD time on the 32ID-M2 ensemble. The first three quantities were computed every MD time step as part of the evolution. The topological charge and Wilson flow scales were computed every 2 and 40 MD time steps, respectively, after the ensemble was thermalized. The dashed vertical lines mark the range of MD times used to perform calculations of the spectrum.}
\label{fig:evol_32ID_M2}
\end{figure}

In Figures~\ref{fig:evol_32ID_M1} and~\ref{fig:evol_32ID_M2} we plot the evolution of the average plaquette, light quark chiral condensate $\langle \overline{\psi}_{l} \psi_{l} \rangle$, light quark pseudoscalar condensate $\langle \overline{\psi}_{l} \gamma_{5} \psi_{l} \rangle$, pion propagator evaluated at the fixed time slice $t/a = 20$, square of the topological charge $Q^{2}$, and the clover discretized Yang-Mills action density $E = \tr ( F_{\mu \nu} F_{\mu \nu} )$ evaluated at the Wilson flow times $t = t_{0}$ and $t = w_{0}^{2}$, as a function of the molecular dynamics simulation time (MD time). Following~\cite{Schaefer201193} and our most recent analysis~\cite{Blum:2014tka} we consider the square of the topological charge rather than the topological charge itself, since this is a parity even observable and our HMC algorithm is parity invariant. We measured the topological charge by cooling the gauge fields with 60 rounds of APE smearing using a smearing coefficient of 0.45, and then measured the topological charge density using the five-loop-improved discretization introduced in Ref.~\cite{deForcrand:1997sq}. \\

\begin{figure}[h]
\centering
\subfloat[32ID-M1]{\includegraphics[width=0.49\textwidth]{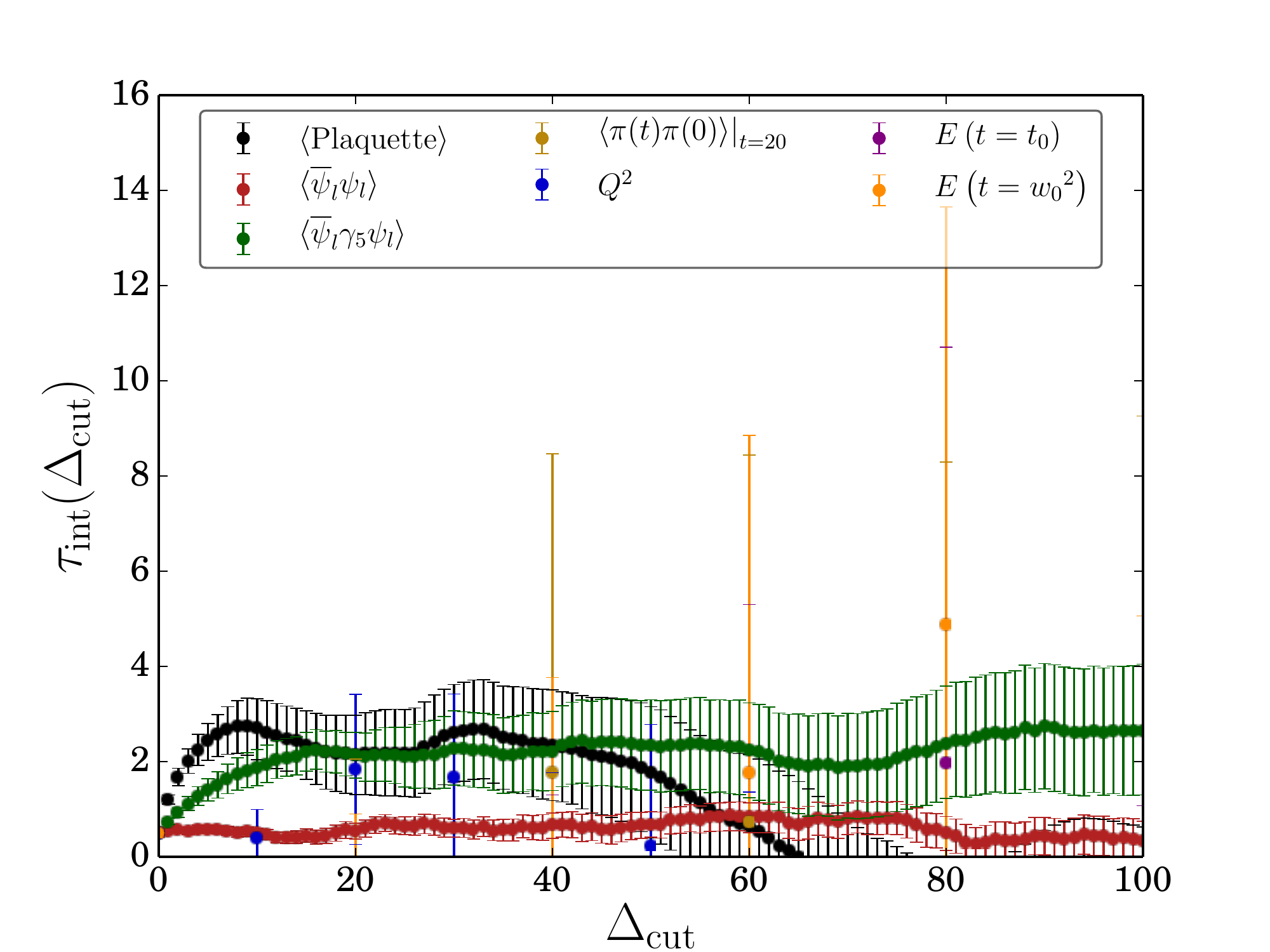}}
\subfloat[32ID-M2]{\includegraphics[width=0.49\textwidth]{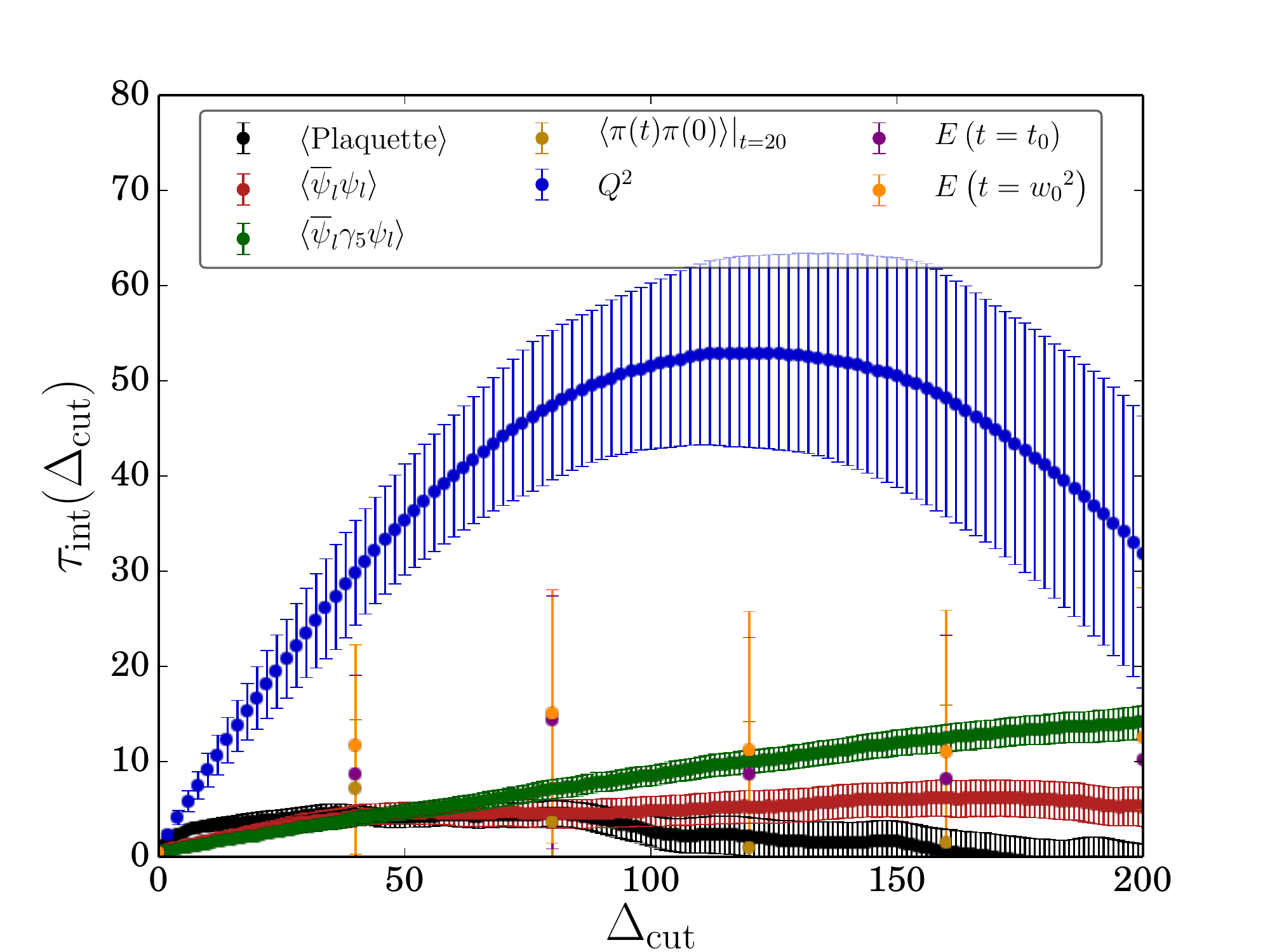}}
\caption{Integrated autocorrelation times for the observables plotted in Figures~\ref{fig:evol_32ID_M1} and~\ref{fig:evol_32ID_M2}.}
\label{fig:autocorrelation_time}
\end{figure}

In Figure~\ref{fig:autocorrelation_time} we plot the integrated autocorrelation times obtained from each of these observables. The integrated autocorrelation time for an observable $Y(t)$ with mean $\overline{Y}$ and variance $\sigma_{Y}^{2}$ is defined to be
\begin{equation}
\tau_{\rm int}(\Delta_{\rm cut}) = \frac{1}{2} + \sum\limits_{\Delta=1}^{\Delta_{\rm cut}} C(\Delta),
\end{equation}
where
\begin{equation}
C(\Delta) = \left< \frac{\left( Y(t) - \overline{Y} \right) \left( Y(t + \Delta) - \overline{Y} \right)}{\sigma_{Y}^{2}} \right>_{t}
\end{equation}
is the autocorrelation at lag $\Delta$, and $\Delta_{\rm cut}$ is a cutoff on the maximum lag. The quantity $2 \tau_{\rm int}$ estimates the number of MD time units separating statistically uncorrelated measurements of $Y$. The error on the integrated autocorrelation time is estimated by bootstrap resampling the set of measurements of $(Y(t) - \overline{Y}) (Y(t+\Delta)-\overline{Y})$ with fixed $\Delta$, binned over 20 (40) MD time units on the 32ID-M1 (32ID-M2) ensembles. This choice of binning corresponds to the separation between measurements of the spectrum, and was chosen based on increasing the bin size until the error bars in Figure~\ref{fig:autocorrelation_time} were observed to stabilize and stop growing. More detail regarding this procedure can be found in Ref.~\cite{Arthur:2012opa}. \\

We conclude from the autocorrelation analysis that our separation of 20 (40) MD time units between measurements of the spectrum on the 32ID-M1 (32ID-M2) ensemble is sufficient to ensure that the measurements are uncorrelated, and so we do not perform any further binning. While one should worry about the long autocorrelation time associated with the topological charge on the 32ID-M2 ensemble, we note that our ChPT fits depend only on the measured values of masses and decay constants, and the long range observables in Figure~\ref{fig:autocorrelation_time} --- the pion propagator and quark condensates, for example --- suggest an autocorrelation time well within our measurement separation. One should additionally worry that this significant autocorrelation time associated with $Q^{2}$ and the poor sampling of topological sectors evidenced by Figure~\ref{fig:evol_32ID_M2} suggests statistical errors on the 32ID-M2 ensemble may be underestimated. We choose to still include this ensemble in some of our fits\footnote{Because of the heavy pion mass $m_{\pi} \sim 400~\mathrm{MeV}$ this ensemble is excluded completely from the fits with a 370 MeV mass cut.} for a number of reasons: in particular, it allows us to overconstrain the linear $a^{2}$-scaling terms associated with the DSDR gauge action since it provides an additional DSDR ensemble with a third, independent lattice spacing. In addition, we observe that our results for the LECs of $SU(2)$ PQChPT are completely consistent when we consider the same fit performed with and without the 32ID-M2 ensemble, suggesting that the influence of any undesirable effects of undersampling on our conclusions regarding ChPT are negligible.

\subsection{Spectrum}

We measure and fit the spectrum with the same analysis package previously used to analyze the 48I, 64I, and 32I-fine ensembles in Ref.~\cite{Blum:2014tka}. This analysis package uses the all-mode averaging (AMA) technique introduced by Blum, Izubuchi, and Shintani~\cite{Blum:2012uh}. Five \textit{exact} light quark propagators were computed per trajectory using a deflated mixed-precision conjugate gradient solver~\cite{Stathopoulos:2007zi} with 1000 low-mode deflation vectors and a tight stopping precision $r = 10^{-8}$, while \textit{sloppy} light quark propagators with a reduced stopping precision $r = 10^{-4}$ were computed for all time slices. The cheaper strange quark propagators were computed to the tight residual $r = 10^{-8}$ on all time slices using the ordinary conjugate gradient algorithm with no deflation. AMA correlation functions were then computed by time-translational averaging of the sloppy propagators, using the available exact propagators to correct for bias. In all cases we use Coulomb gauge-fixed wall sources (W), and either local (L) or wall sinks. \\

We have computed the low-energy QCD spectrum for 21 configurations separated by 20 MD time units each on the 32ID-M1 ensemble, and 24 configurations separated by 40 MD time units each on the 32ID-M2 ensemble. These measurements include the residual mass ($m_{\rm res}$), light-light and heavy-light pseudoscalar masses ($m_{ll}$, $m_{lh}$) and decay constants ($f_{ll}$, $f_{lh}$), the axial and vector current renormalization coefficients ($Z_{A}$, $Z_{V}$), the $\Omega$ baryon mass ($m_{hhh}$), and the Wilson flow scales ($t_{0}^{1/2}$, $w_{0}$). Since the analysis package has been discussed in detail in our previous work we paraphrase the fits which were preformed below, and refer the reader to~\cite{Blum:2014tka} for additional detail. In the following we use the notation ``$\simeq$'' to denote equality up to excited state contamination for a suitably chosen plateau range. These fits are performed by minimizing an uncorrelated $\chi^{2}$ (Eqn.~\ref{eqn:uncorr_chi2}) where the correlation functions and fit forms are listed explicitly below. 

\begin{enumerate}
\item The ratio
\begin{equation}
\label{eqn:mres}
R(t) = \frac{\langle 0 | \Sigma_{\vec{x}} j_{5 q}^{a}(\vec{x},t) | \pi \rangle}{\langle 0 | \Sigma_{\vec{x}} j_{5}^{a}(\vec{x},t) | \pi \rangle},
\end{equation}
where $j^{a}_{5 q}$ is the pseudoscalar density evaluated at the midpoint of the fifth dimension, and $j^{a}_{5}$ is the physical pseudoscalar density constructed from the surface fields. The residual mass is obtained by averaging over a range of values of $t$ and extrapolating $R$ to the chiral limit.
\item The light-light and heavy-light pseudoscalar masses from
\begin{equation}
\langle 0 | \mathscr{O}_{1}^{s_{1}}(t) \mathscr{O}_{2}^{s_{2}}(0) | 0 \rangle \simeq \frac{\langle 0 | \mathscr{O}_{1}^{s_{1}} | X \rangle \langle X | \mathscr{O}_{2}^{s_{2}} | 0 \rangle}{2 m_{X} V} \left( e^{-m_{X} t} \pm e^{-m_{X} (T - t)} \right).
\end{equation}
Here $\mathscr{O}_{i}^{s_{i}}$ denotes the interpolating operator and smearing, and $X$ denotes the state to which the interpolating operator couples. We perform simultaneous fits to the $\langle PP^{LW} \rangle$, $\langle PP^{WW} \rangle$, and $\langle AP^{LW} \rangle$ correlators for both the light-light and heavy-light pseudoscalar states. The sign is +(-) for the PP(AP) correlator.
\item The ratio $Z_{A} / Z_{\mathscr A}$ --- where $Z_{A}$ ($Z_{\mathscr A}$) is a renormalization coefficient relating the local four-dimensional (non-local five-dimensional) axial current to the Symanzik-improved axial current --- from
\begin{equation}
\label{eqn:ZA}
\frac{1}{2} \left[ \frac{C_{\mathscr{A}}(t-1) + C_{\mathscr{A}}(t)}{2 \, C_{A}(t-\frac{1}{2})} + \frac{2 \, C_{\mathscr{A}}(t)}{C_{A}(t+\frac{1}{2}) + C_{A}(t-\frac{1}{2})} \right] \simeq \frac{Z_{A}}{Z_{\mathscr A}},
\end{equation}
where $C_{\mathscr{A}}(t) \equiv \langle 0 | \sum_{\vec{x}} \partial_{\mu} \mathscr{A}_{\mu}^{a}(\vec{x},t) | \pi \rangle$ and $C_{A}(t - \frac{1}{2}) \equiv \langle 0 | \sum_{\vec{x}} \partial_{\mu} A_{\mu}^{a}(\vec{x},t) | \pi \rangle$. This is the procedure we introduced in~\cite{Blum:2014tka} to extract $Z_{A}$ on our M\"{o}bius domain wall fermion ensembles; in our earlier analyses with plain domain wall fermions we extracted $Z_{A}$ directly from matrix elements of the four-dimensional and five-dimensional axial currents.
\item The renormalization coefficient $Z_{V}$ relating the local four-dimensional vector current to the Symanzik-improved vector current from
\begin{equation}
\label{eqn:ZV}
\frac{\langle \pi(\Delta t) | \pi(0) \rangle}{\langle \pi(\Delta t) | V_{0}(t) | \pi(0) \rangle} \simeq Z_{V}.
\end{equation}
Here $V_{0}$ is the temporal component of the light quark electromagnetic current $V_{\mu} = \overline{q}_{l} \gamma_{\mu} q_{l}$. While Eqn.~\eqref{eqn:ZV} is technically equal to the ratio $Z_{V} / Z_{\mathscr V}$, where $Z_{\mathscr V}$ relates the non-local five-dimensional vector current to the Symanzik current, the five-dimensional current is exactly conserved on the lattice, implying $Z_{\mathscr V} = 1$.
\item The renormalized light-light and heavy-light pseudoscalar decay constants
\begin{equation}
\label{eqn:decay}
f_{X} = Z_{V} \sqrt{ \frac{2}{m_{X} V} \frac{\left(\mathscr{N}_{AP}^{LW} \right)^{2}}{\mathscr{N}_{PP}^{WW}} },
\end{equation}
where we have defined
\begin{equation}
\mathscr{N}_{\mathscr{O}_{1} \mathscr{O}_{2}}^{s_{1} s_{2}} \equiv \frac{\langle 0 | \mathscr{O}_{1}^{s_{1}} | X \rangle \langle X | \mathscr{O}_{2}^{s_{2}} | 0 \rangle}{2 m_{X} V}.
\end{equation}
We choose to renormalize the decay constants by $Z_{V}$ rather than $Z_{A}$, which differ by small terms of $\bigO(m_{\rm res}^{2})$ since the five-dimensional axial current differs from unity by terms of $\bigO(m_{\rm res})$, introducing $\bigO(m_{\rm res})$ errors into the determination of $Z_{A}$ via Eqn.~\eqref{eqn:ZA}. This point is discussed in further detail in Ref.~\cite{Aoki:2010dy}.
\item The $\Omega$ baryon mass from the two-point correlation function
\begin{equation}
\mathscr{C}_{\Omega \Omega}^{s_{1} s_{2}}(t) = \sum\limits_{i=1}^{3} \sum\limits_{\vec{x}} \langle 0 | \mathscr{O}_{\Omega}^{s_{1}}(\vec{x},t)_{i} \overline{\mathscr{O}}_{\Omega}^{s_{2}}(0)_{i} | 0 \rangle
\end{equation}
with the interpolating operator $\mathscr{O}_{\Omega}(x)_{i} = \epsilon_{a b c} \left( s_{a}^{\top}(x) C \gamma_{i} s_{b}(x) \right) s_{c}(x)$. This correlator was computed for both a Coulomb gauge-fixed wall source and a $Z_{3}$ box source ($Z_{3}B$), and, in both cases, a local sink. The correlators were then projected onto the positive parity component
\begin{equation}
\mathscr{P}_{+} \mathscr{C}_{\Omega \Omega}^{s_{1} s_{2}} = \frac{1}{4} \tr \left[ \frac{1}{2} \left( 1 + \gamma_{4} \right) \mathscr{C}_{\Omega \Omega}^{s_{1} s_{2}} \right]
\end{equation}
and simultaneously fit to a double exponential ansatz with common mass terms
\begin{equation}
\label{eqn:omega_fit}
\begin{dcases}
\mathscr{C}_{\Omega \Omega}^{LW}(t) = \mathscr{N}_{\Omega \Omega}^{LW} e^{-m_{hhh} t} + {\mathscr{N}_{\Omega \Omega}^{LW}}' e^{-m_{hhh}' t} \\
\mathscr{C}_{\Omega \Omega}^{L Z_{3} B}(t) = \mathscr{N}_{\Omega \Omega}^{L Z_{3} B} e^{-m_{hhh} t} + {\mathscr{N}_{\Omega \Omega}^{L Z_{3} B}}' e^{-m_{hhh}' t}
\end{dcases},
\end{equation}
where $m_{h h h}$ is the $\Omega$ baryon mass and $m_{h h h}'$ is the mass of the first excited state in the positive parity channel.
\item The Wilson flow scales, $t_{0}^{1/2}$ and $w_{0}$, defined by
\begin{equation}
\left. t^{2} \langle E(t) \rangle \right|_{t=t_{0}} = 0.3
\end{equation}
and
\begin{equation}
\left. t \frac{d}{d t} \left( t^{2} \langle E(t) \rangle \right) \right|_{t=w_{0}^{2}} = 0.3
\end{equation}
respectively, where $E = \frac{1}{2} \tr ( F_{\mu \nu} F_{\mu \nu} )$ is the clover discretized Yang-Mills action density.
\end{enumerate}

\begin{figure}[h]
\centering
\subfloat{\includegraphics[width=0.49\textwidth]{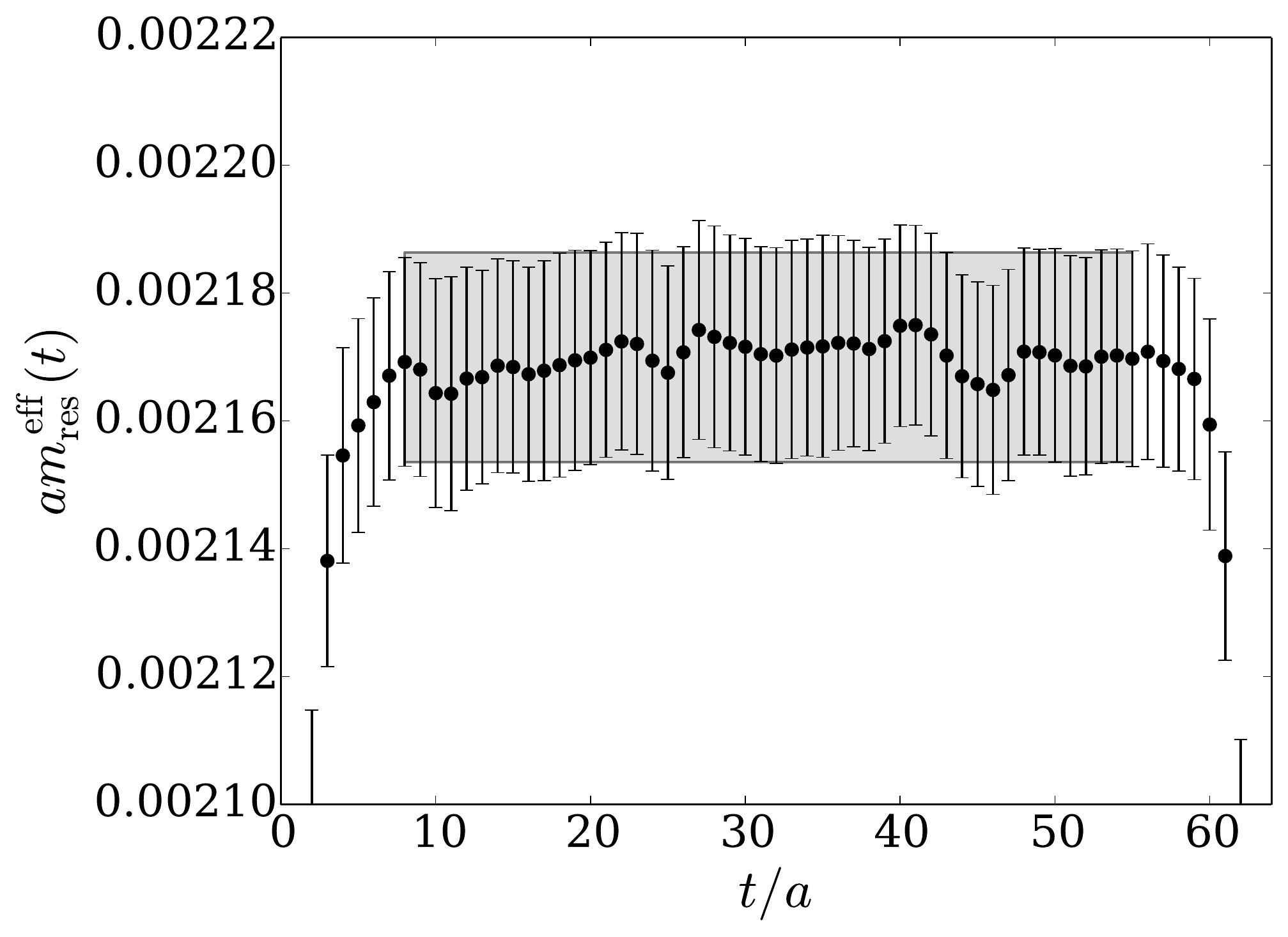}}
\subfloat{\includegraphics[width=0.49\textwidth]{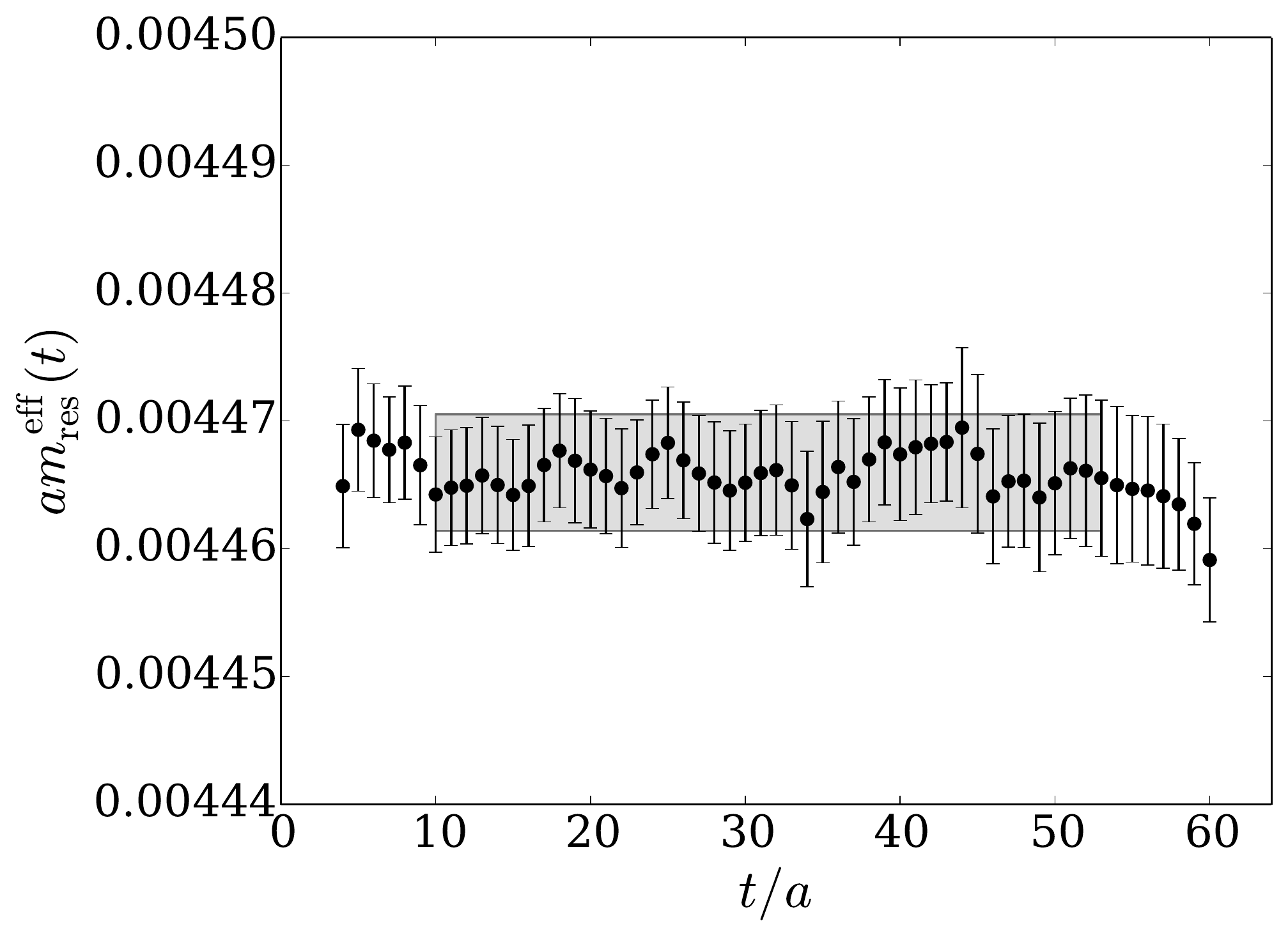}}
\caption{The residual mass, from Eqn.~\eqref{eqn:mres}, on the 32ID-M1 (left) and 32ID-M2 (right) ensembles.}
\label{fig:mres_fits}
\end{figure}

\begin{figure}[h]
\centering
\subfloat{\includegraphics[width=0.49\textwidth]{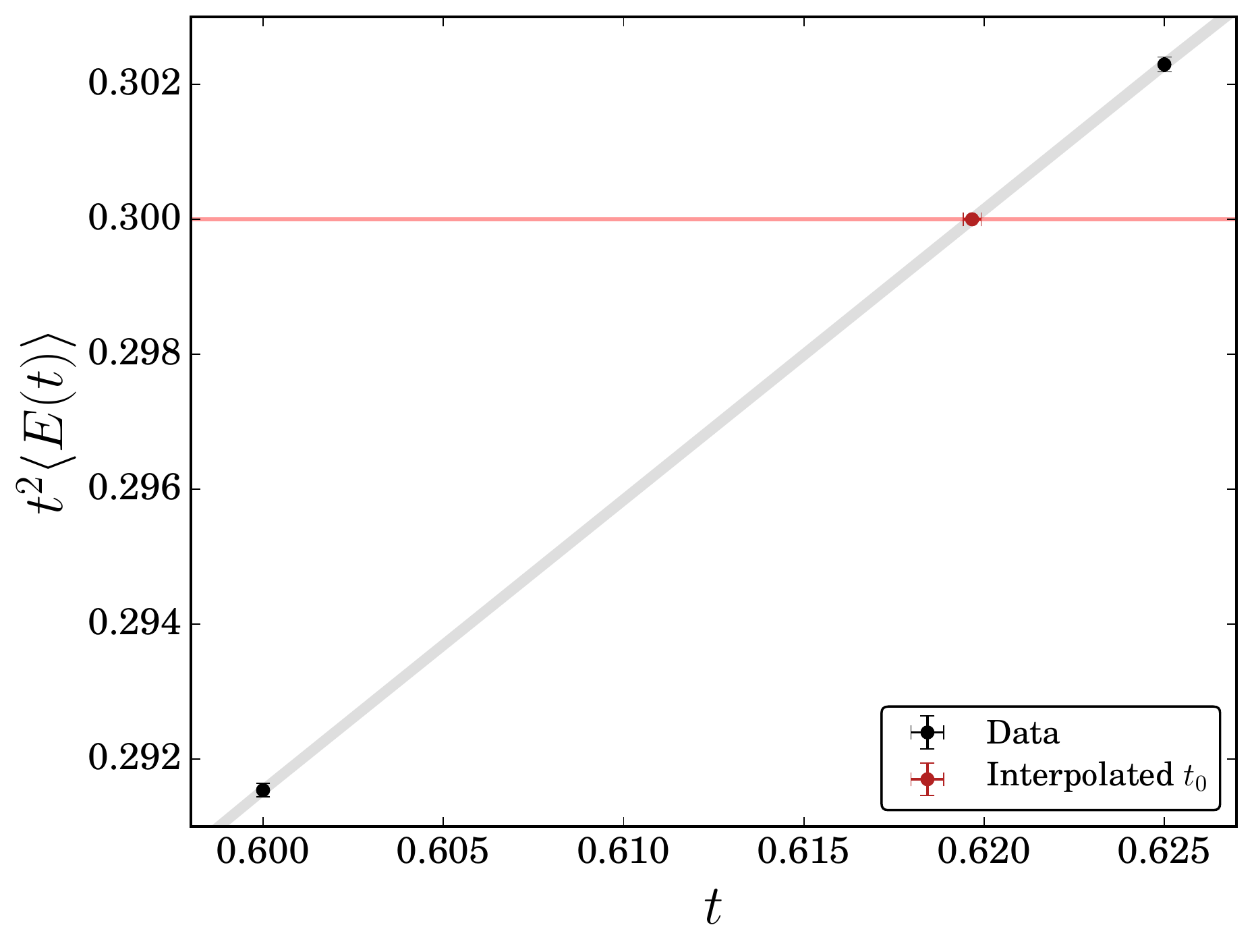}}
\subfloat{\includegraphics[width=0.49\textwidth]{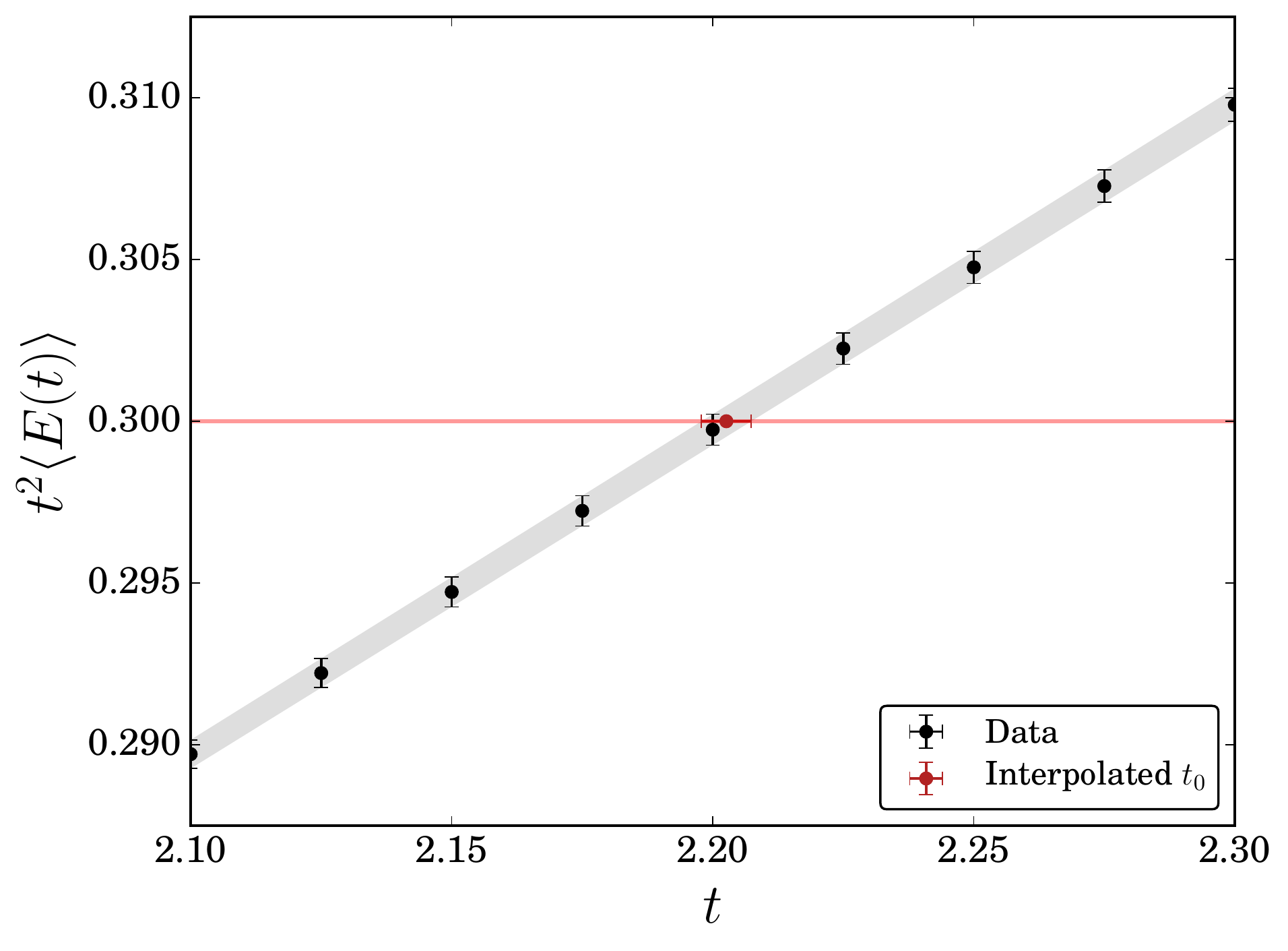}} \\
\subfloat{\includegraphics[width=0.49\textwidth]{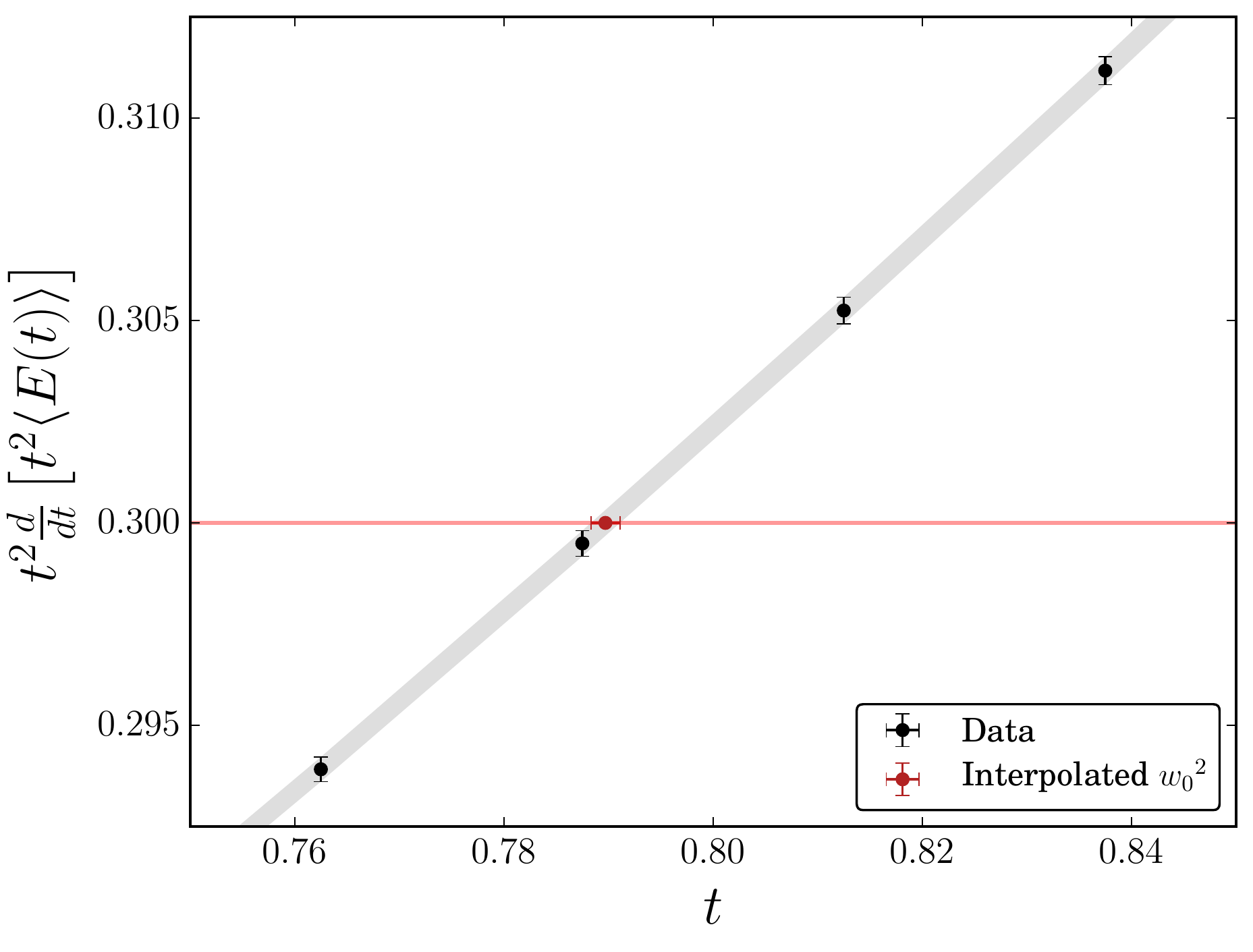}}
\subfloat{\includegraphics[width=0.49\textwidth]{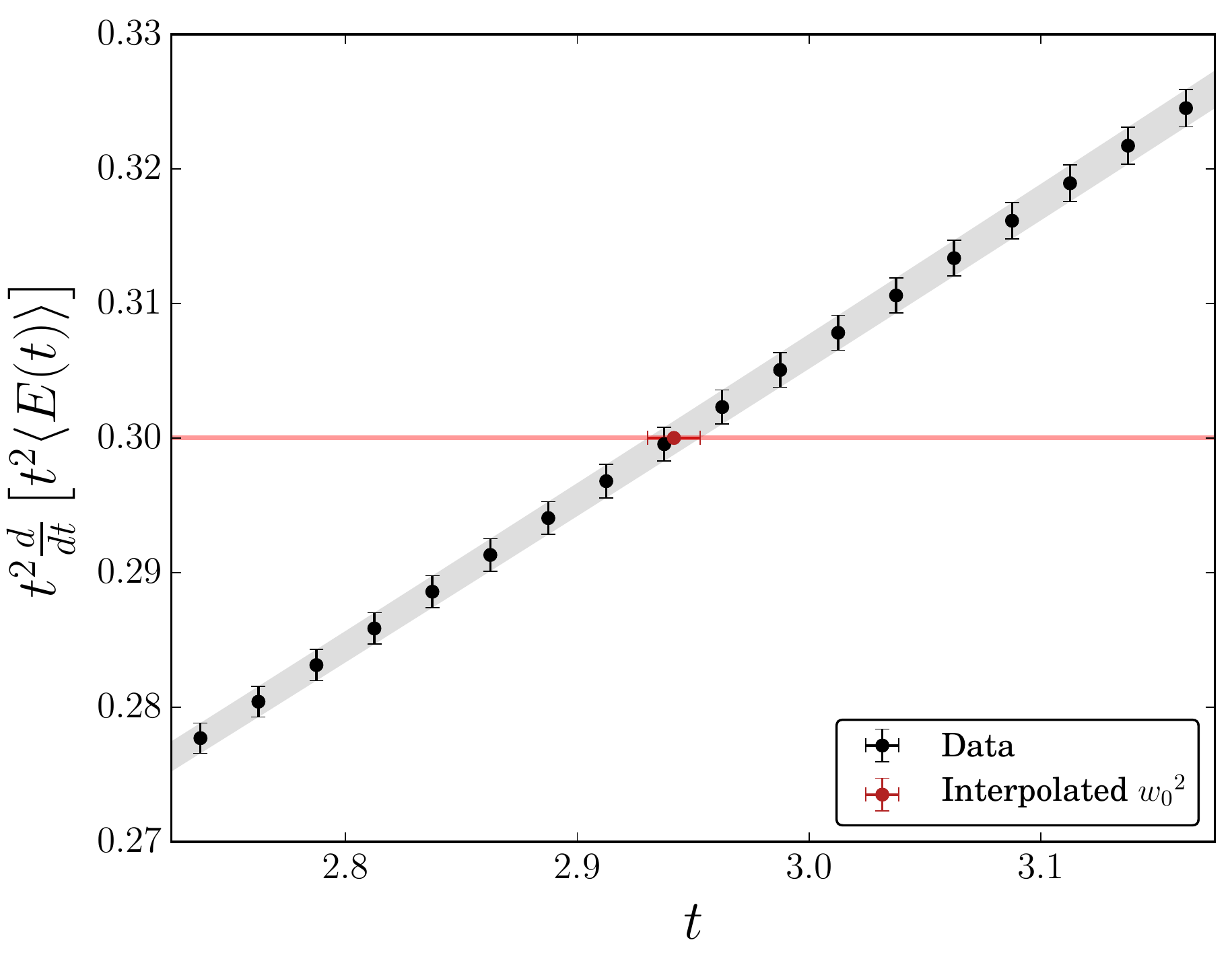}}
\caption{The Wilson flow scales $t_{0}^{1/2}$ (top) and $w_{0}$ (bottom) on the 32ID-M1 (left) and 32ID-M2 (right) ensembles.}
\label{fig:wflow_fits}
\end{figure}

The fit results are summarized in Table~\ref{tab:fit_results}. The corresponding effective mass plots are shown in Figures~\ref{fig:mres_fits}-\ref{fig:wflow_fits}.

\begin{table}[h]
\centering
\begin{tabular}{c||c|c}
\hline
\hline
\rule{0cm}{0.4cm} & 32ID-M1 & 32ID-M2 \\
\hline
\rule{0cm}{0.4cm} $ a m_{ll} $ & 0.11812(46) & 0.19487(64) \\
$ a m_{lh} $ & 0.42313(49) & 0.30792(64) \\
$ a f_{ll} $ & 0.12489(23) & 0.07771(22) \\
$ a f_{lh} $ & 0.14673(33) & 0.087164(21) \\
$ Z_{A} $ & 0.73195(39) & 0.70087(14) \\
$ Z_{V} $ & 0.72482(52) & 0.70593(92) \\
$a m_{hhh} $ & 1.5290(31) & 0.9148(34) \\
$a m_{hhh}' $ & 1.917(39) & 1.215(36) \\
$a R $ & 0.002170(16) & 0.0044660(46) \\
\hline
\rule{0cm}{0.4cm} $t_{0}^{1/2} / a$ & 0.78719(16) & 1.4841(16) \\
$w_{0} / a$ & 0.88865(78) & 1.7151(33) \\
\hline
\rule{0cm}{0.4cm} $ m_{ll} / m_{hhh} $ & 0.07725(34) & 0.21303(90) \\
$ m_{lh} / m_{hhh} $ & 0.27673(65) & 0.3366(12) \\
$ f_{ll} / m_{hhh} $ & 0.08248(14) & 0.08496(41) \\
$ f_{lh} / m_{hhh} $ & 0.09690(26) & 0.09529(40) \\
\hline
\hline
\end{tabular}
\caption{Summary of fit results in lattice units. Here $R$ is defined by Equation~\eqref{eqn:mres}, which becomes $m_{\rm res}$ when extrapolated to the chiral limit.}
\label{tab:fit_results}
\end{table}

\begin{figure}[h]
\centering
\subfloat{\includegraphics[width=0.49\textwidth]{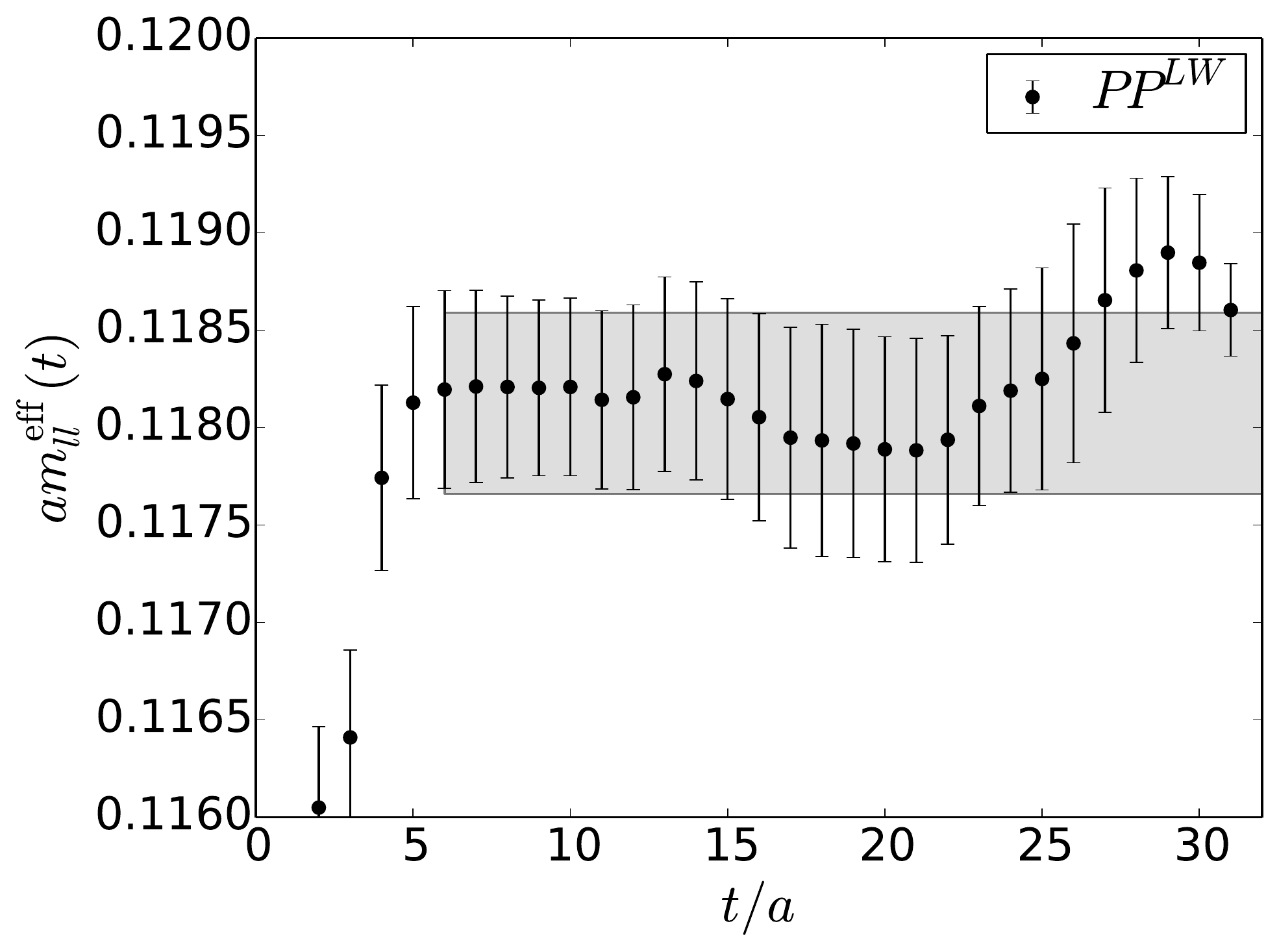}}
\subfloat{\includegraphics[width=0.49\textwidth]{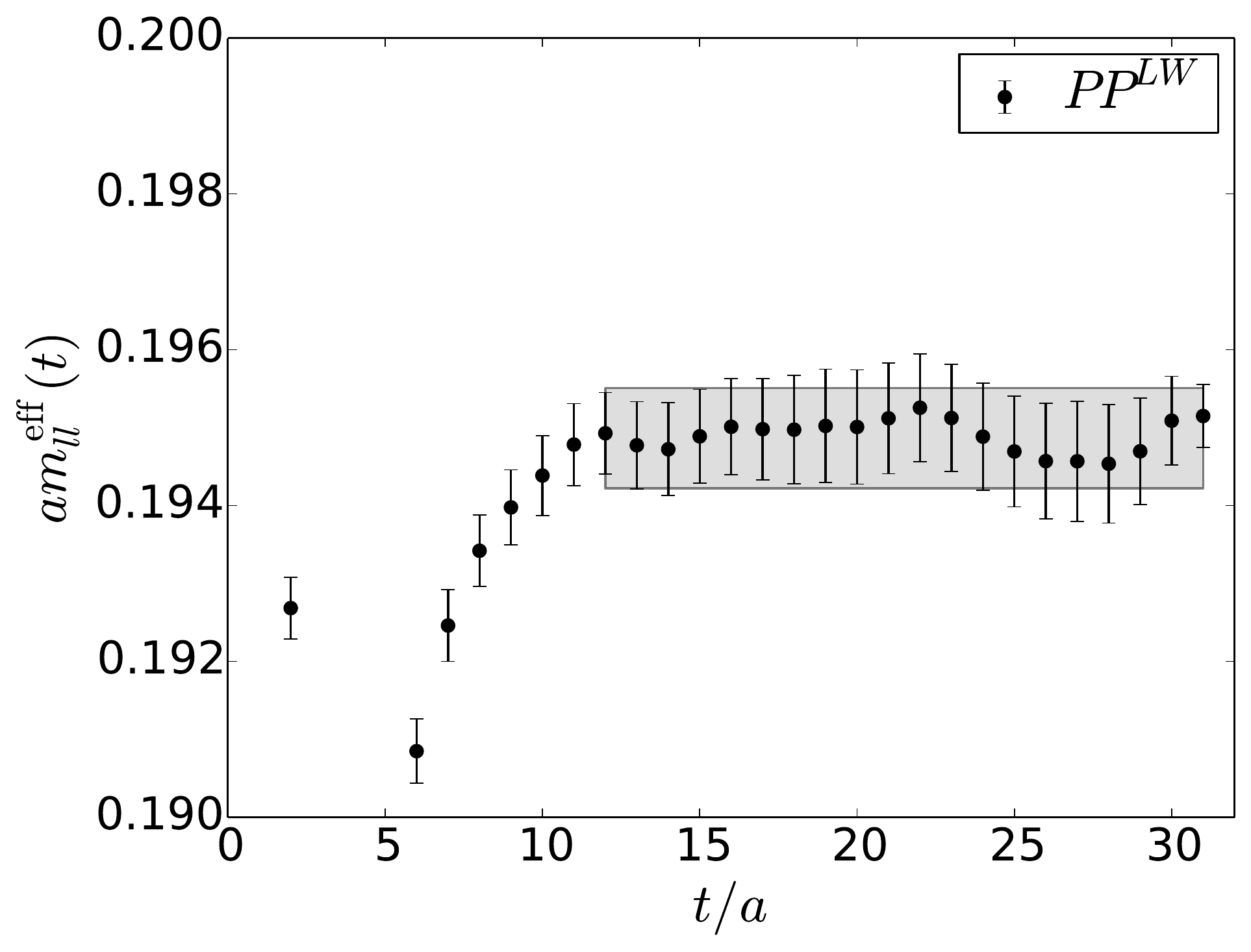}} \\
\subfloat{\includegraphics[width=0.49\textwidth]{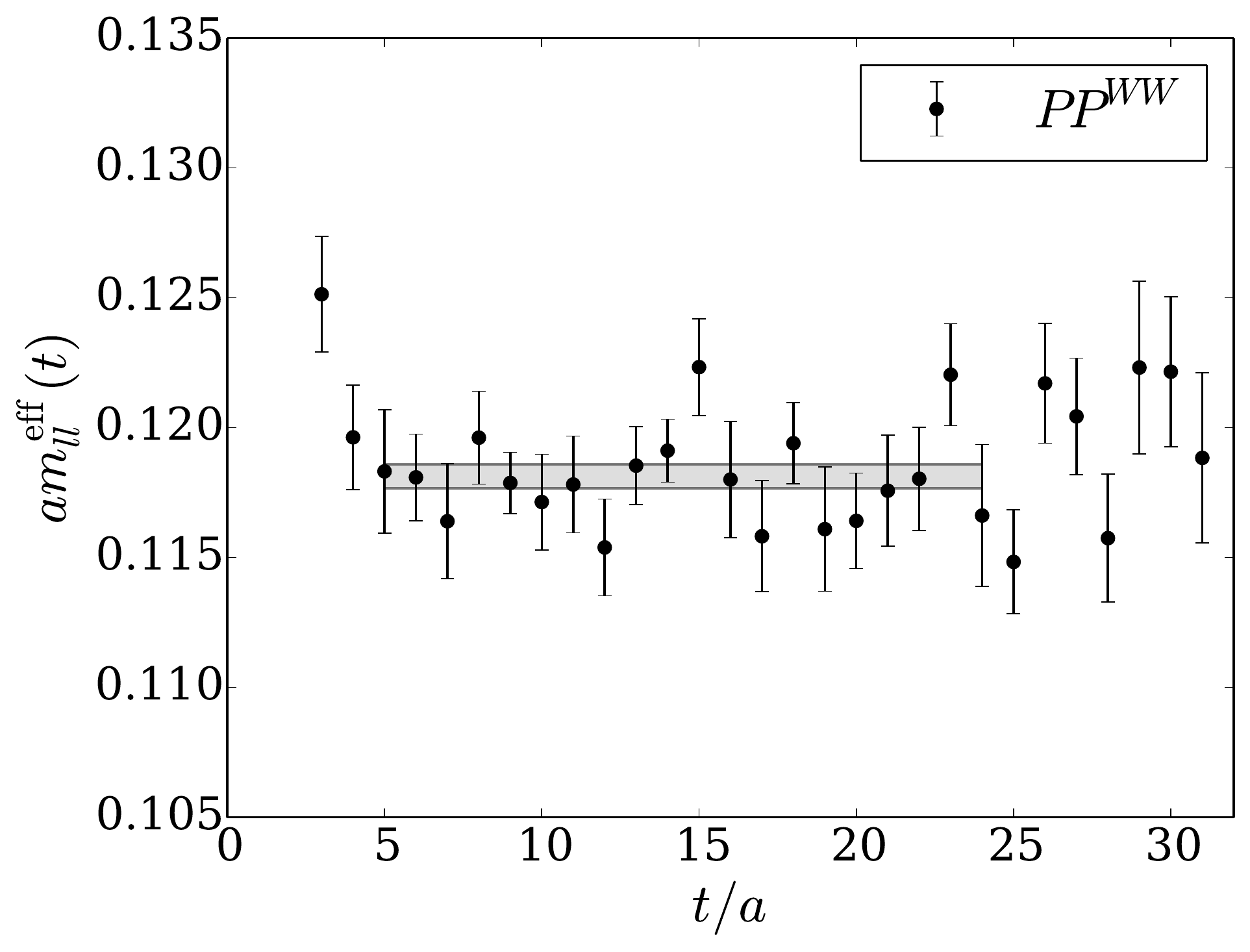}}
\subfloat{\includegraphics[width=0.49\textwidth]{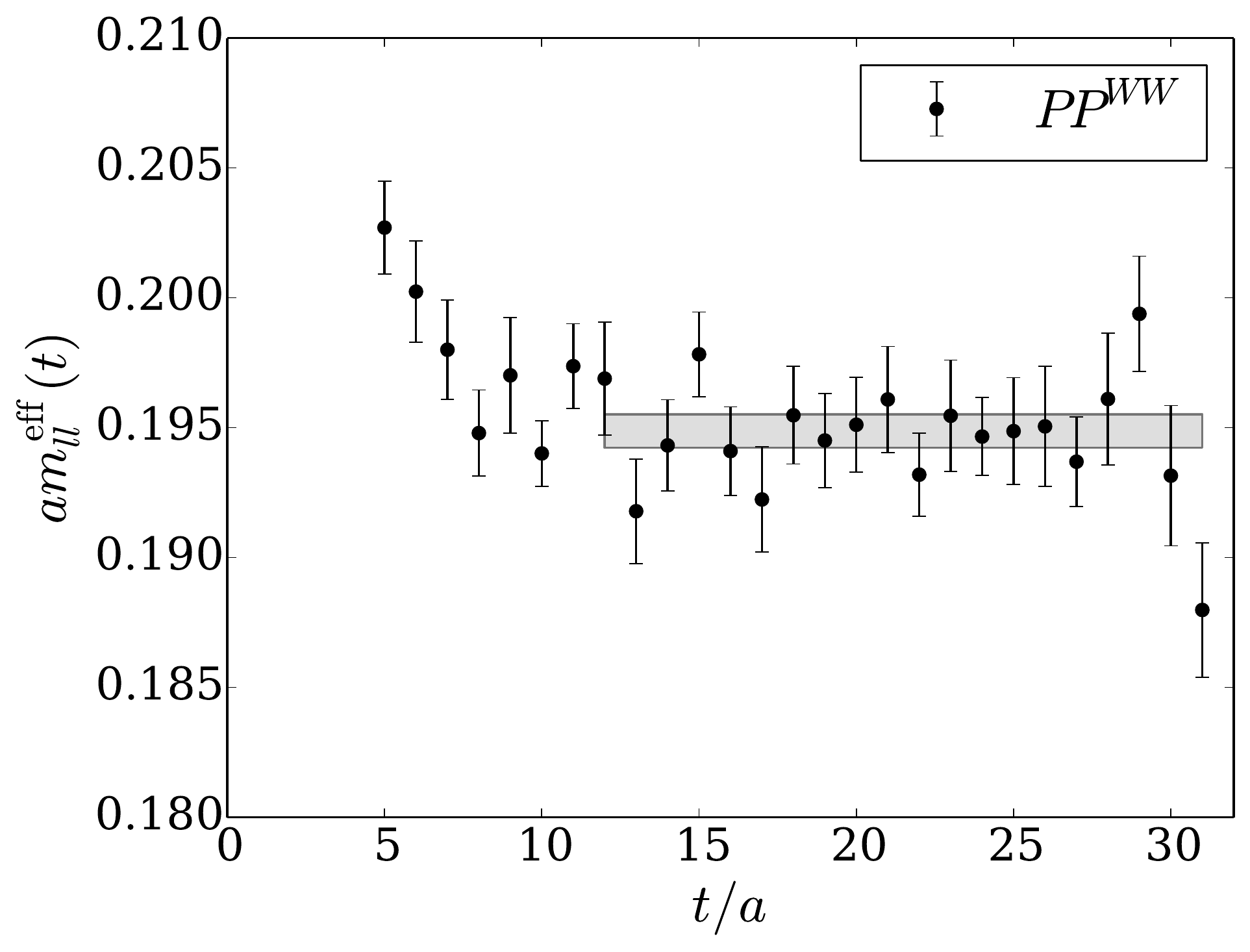}} \\
\subfloat{\includegraphics[width=0.49\textwidth]{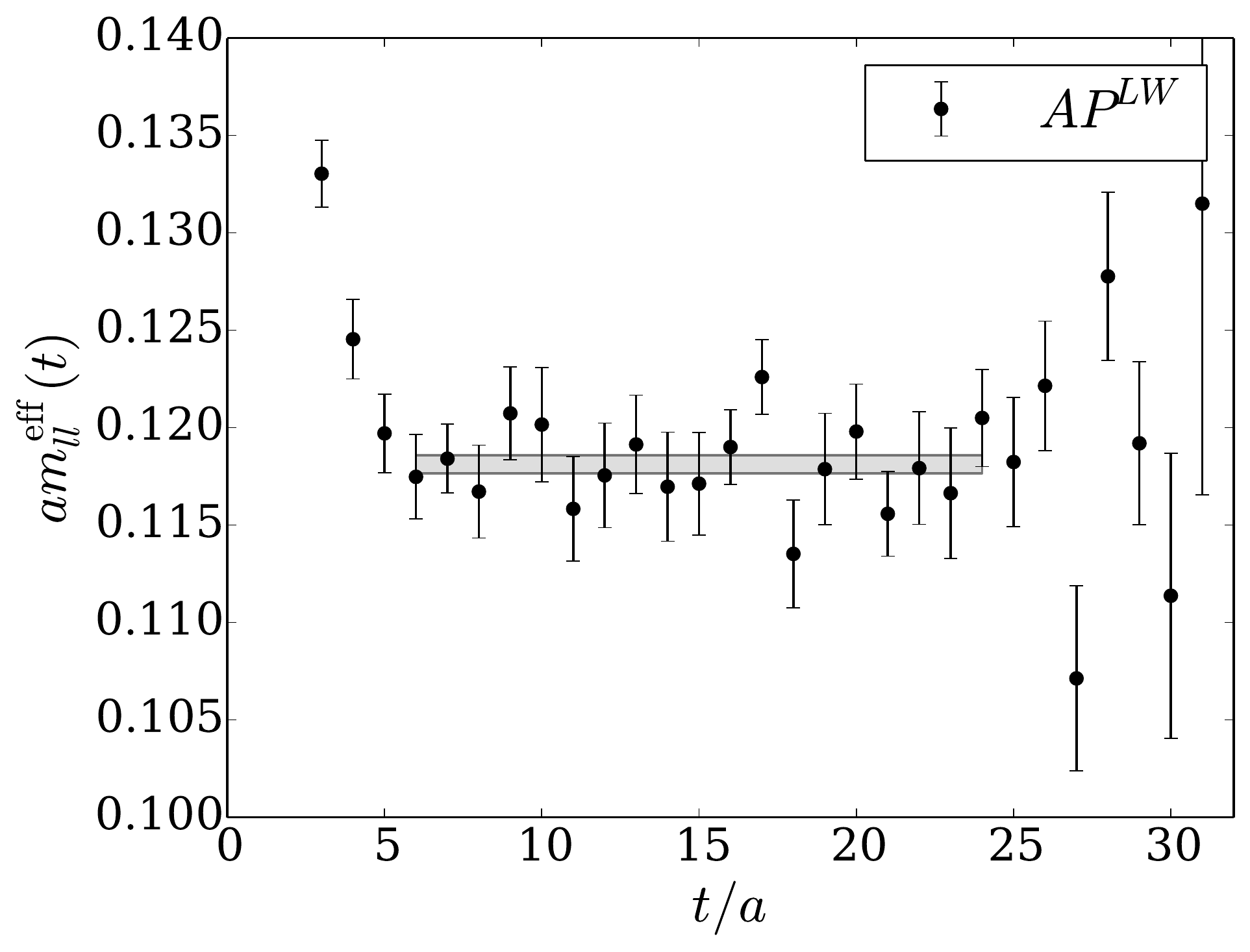}}
\subfloat{\includegraphics[width=0.49\textwidth]{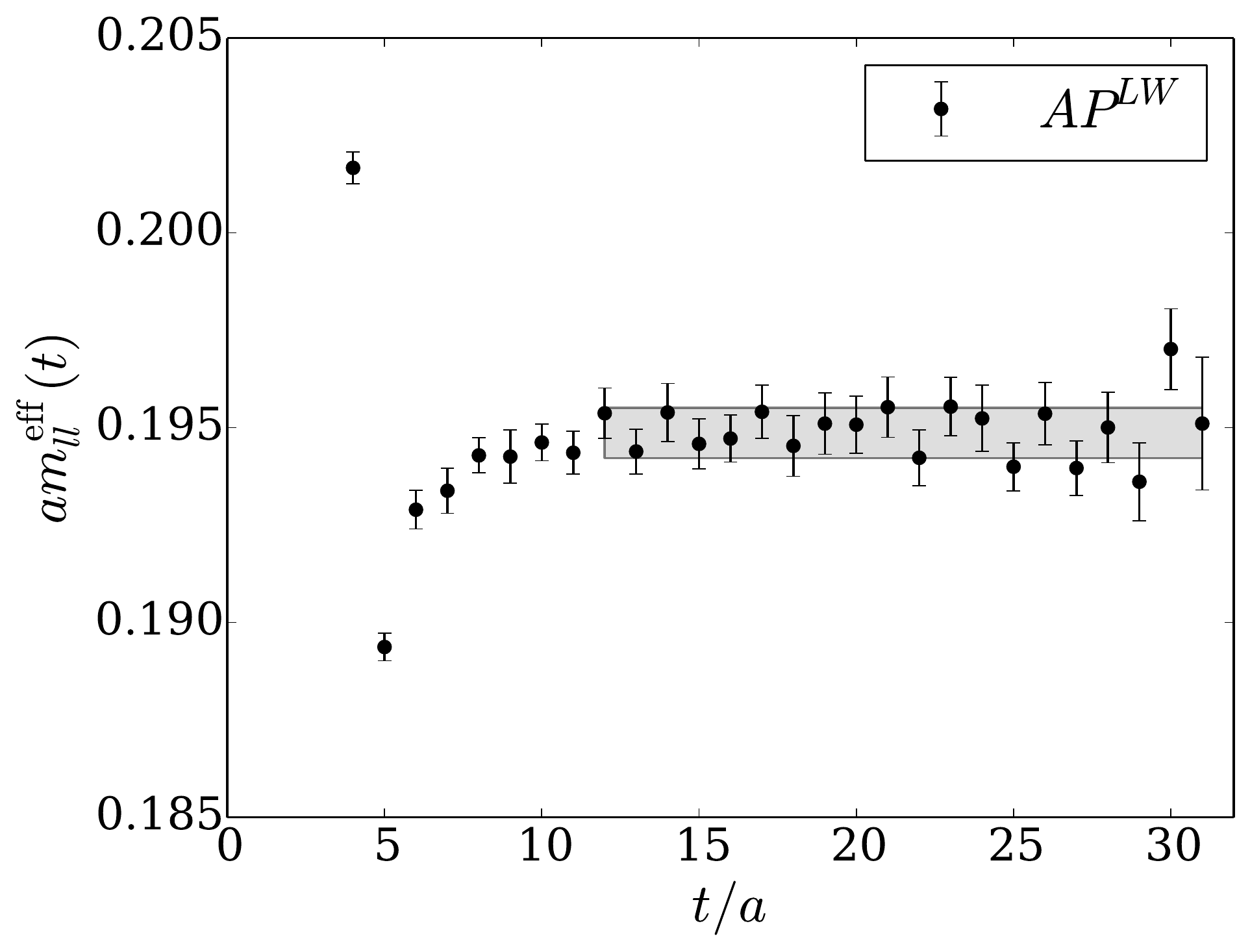}}
\caption{Light-light pseudoscalar mass on the 32ID-M1 (left) and 32ID-M2 (right) ensembles. We simultaneously fit a common mass $m_{ll}$ to the three correlators $\langle PP^{LW} \rangle$, $\langle PP^{WW} \rangle$, and $\langle AP^{LW} \rangle$ on each ensemble.}
\end{figure}

\begin{figure}[h]
\centering
\subfloat{\includegraphics[width=0.49\textwidth]{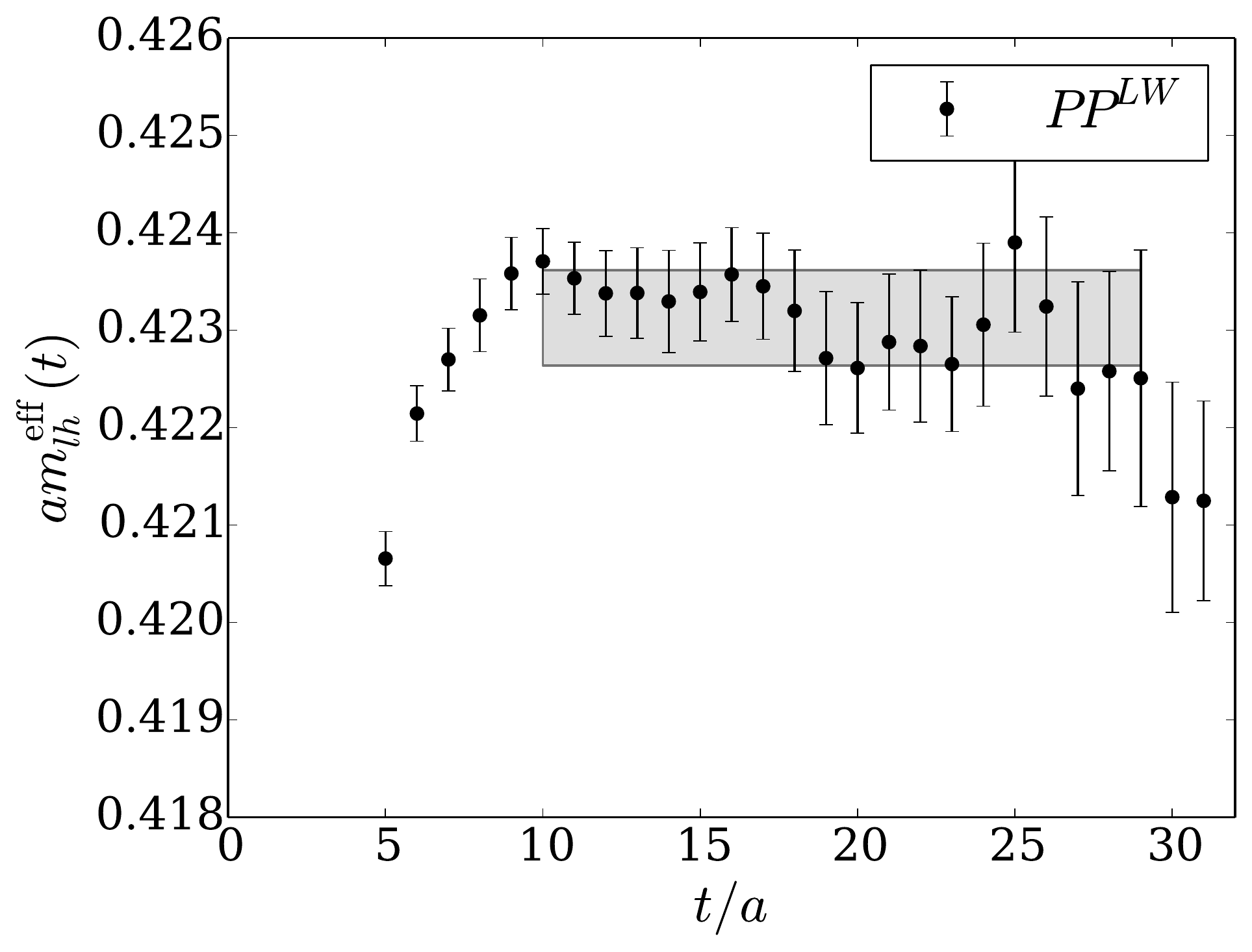}}
\subfloat{\includegraphics[width=0.49\textwidth]{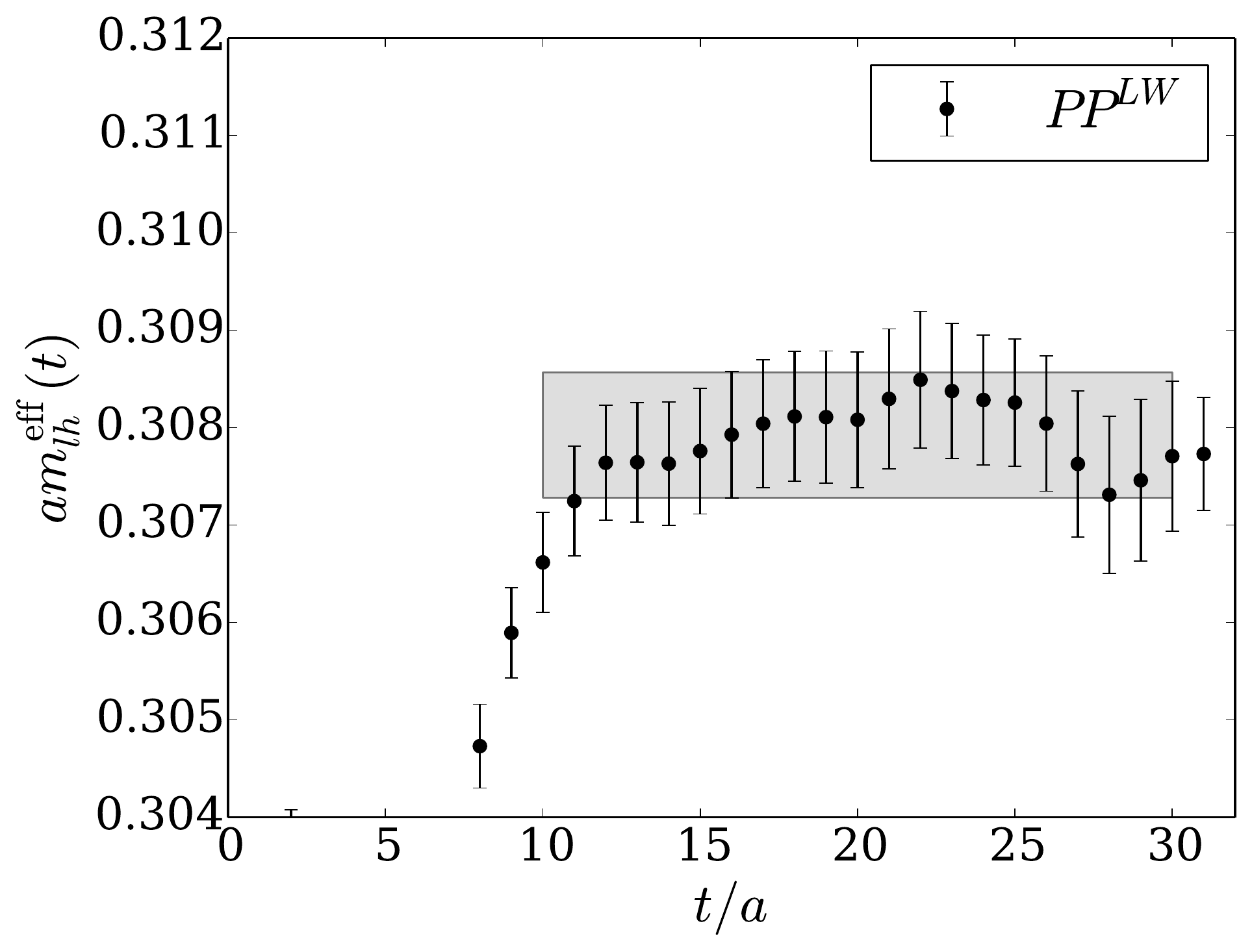}} \\
\subfloat{\includegraphics[width=0.49\textwidth]{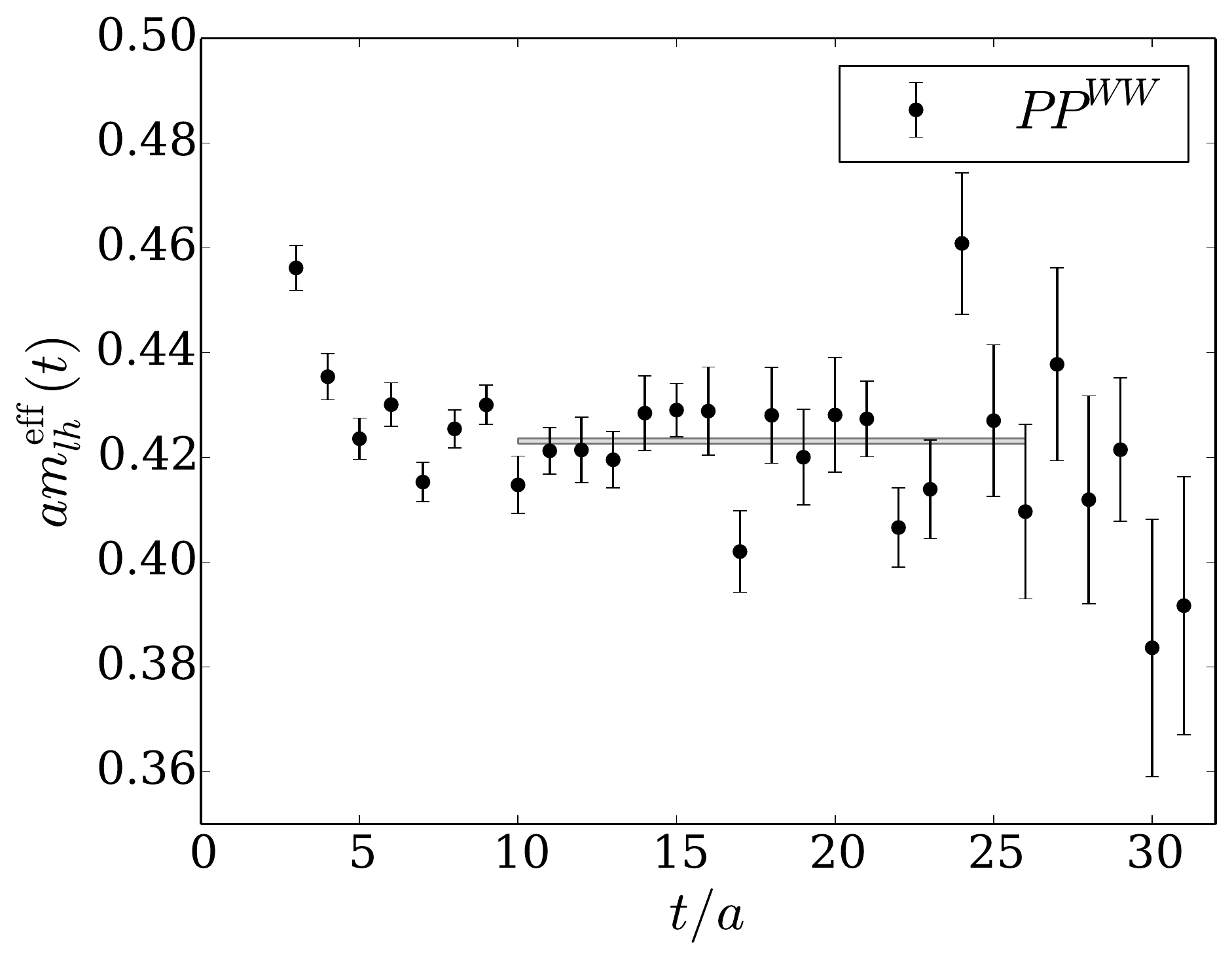}}
\subfloat{\includegraphics[width=0.49\textwidth]{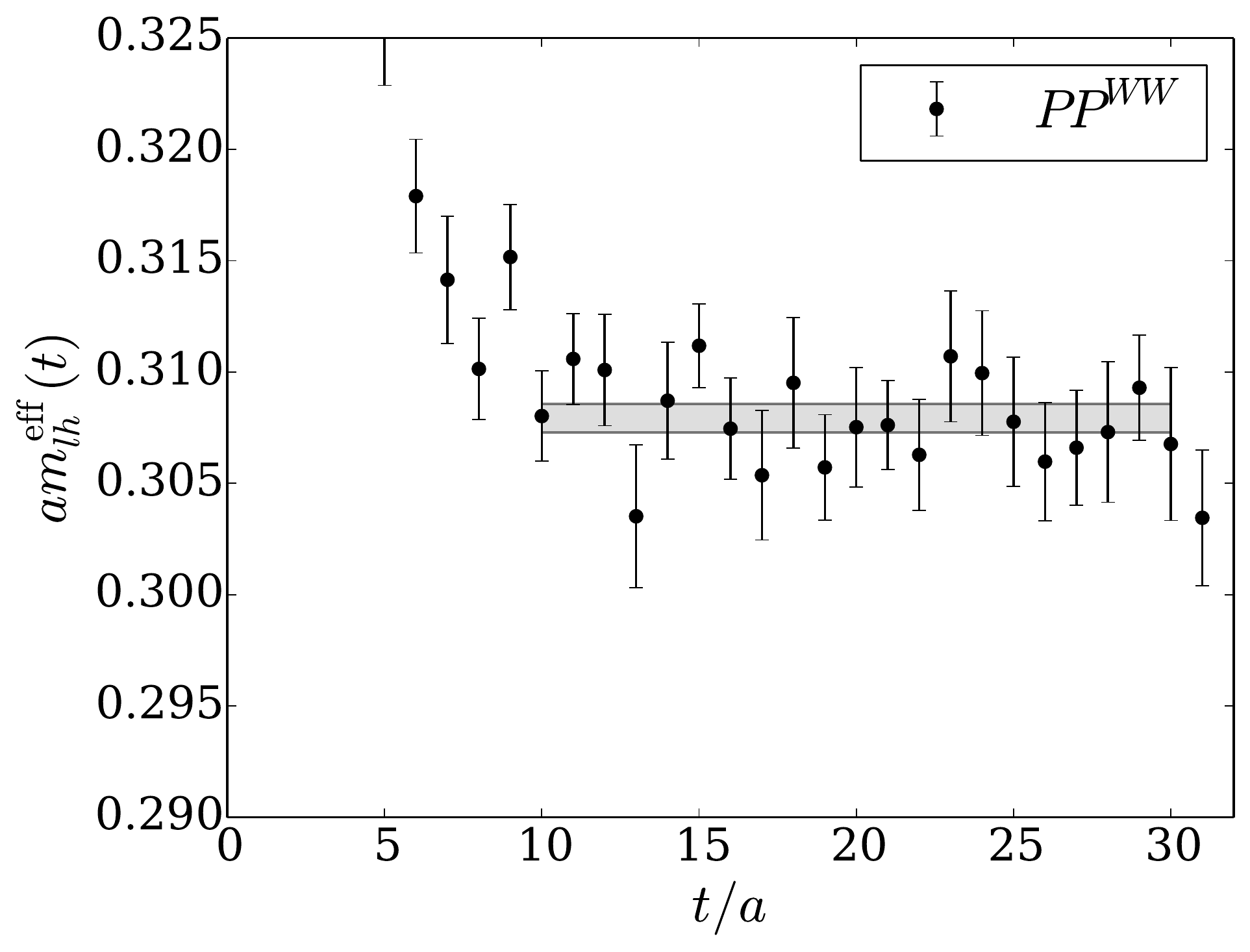}} \\
\subfloat{\includegraphics[width=0.49\textwidth]{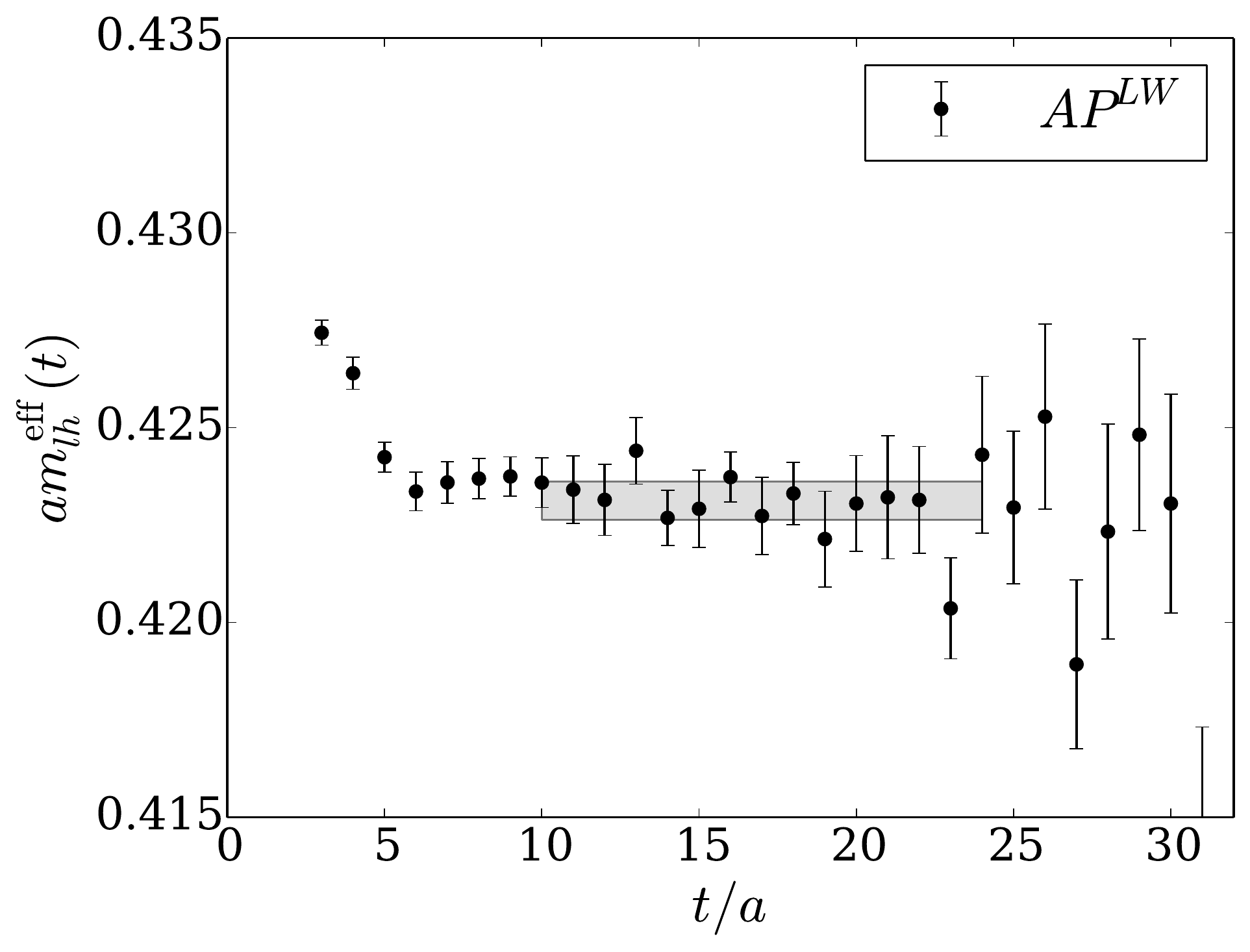}}
\subfloat{\includegraphics[width=0.49\textwidth]{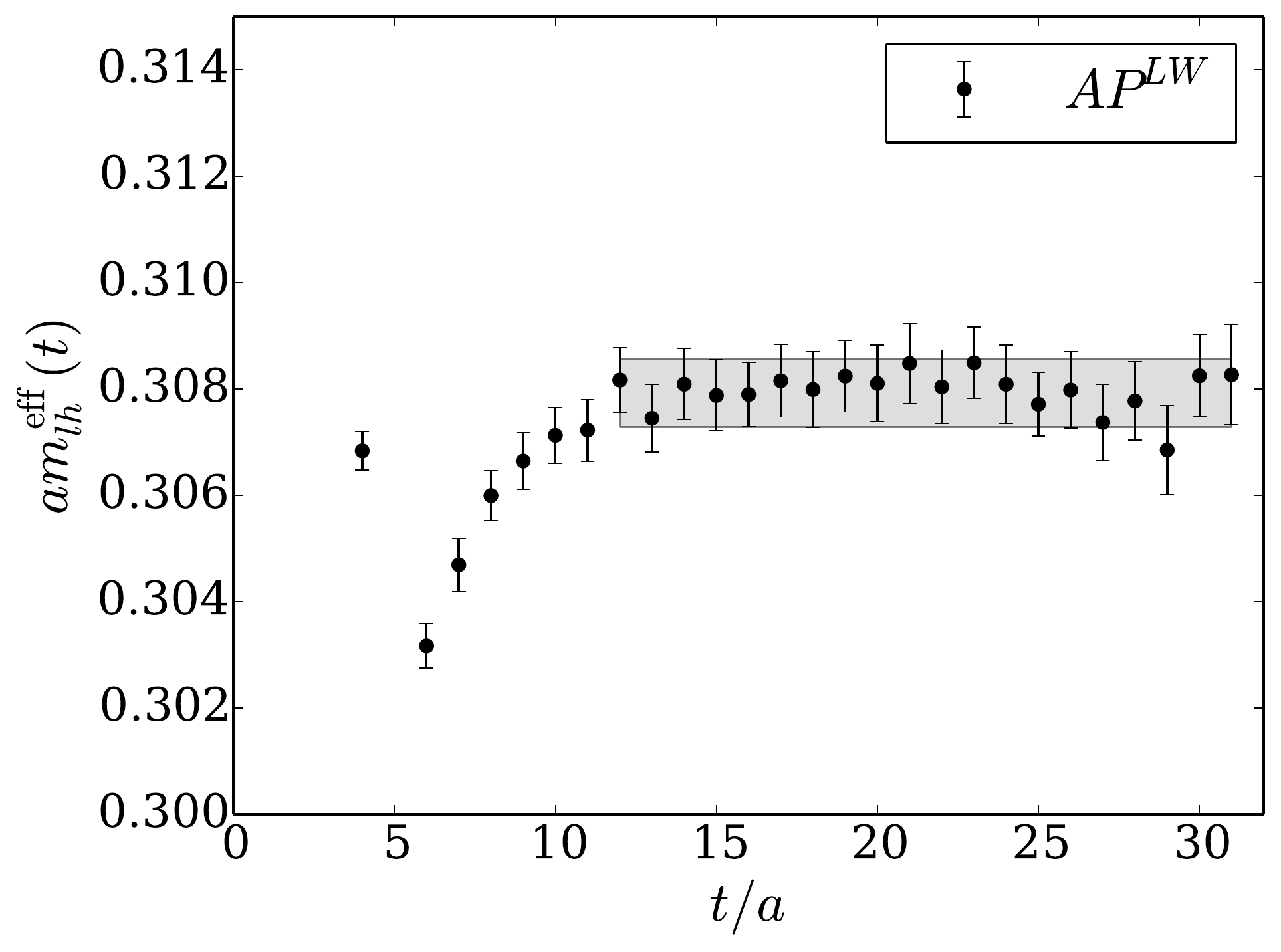}}
\caption{Heavy-light pseudoscalar mass on the 32ID-M1 (left) and 32ID-M2 (right) ensembles. We simultaneously fit a common mass $m_{lh}$ to the three correlators $\langle PP^{LW} \rangle$, $\langle PP^{WW} \rangle$, and $\langle AP^{LW} \rangle$ on each ensemble.}
\end{figure}

\begin{figure}[h]
\centering
\subfloat{\includegraphics[width=0.49\textwidth]{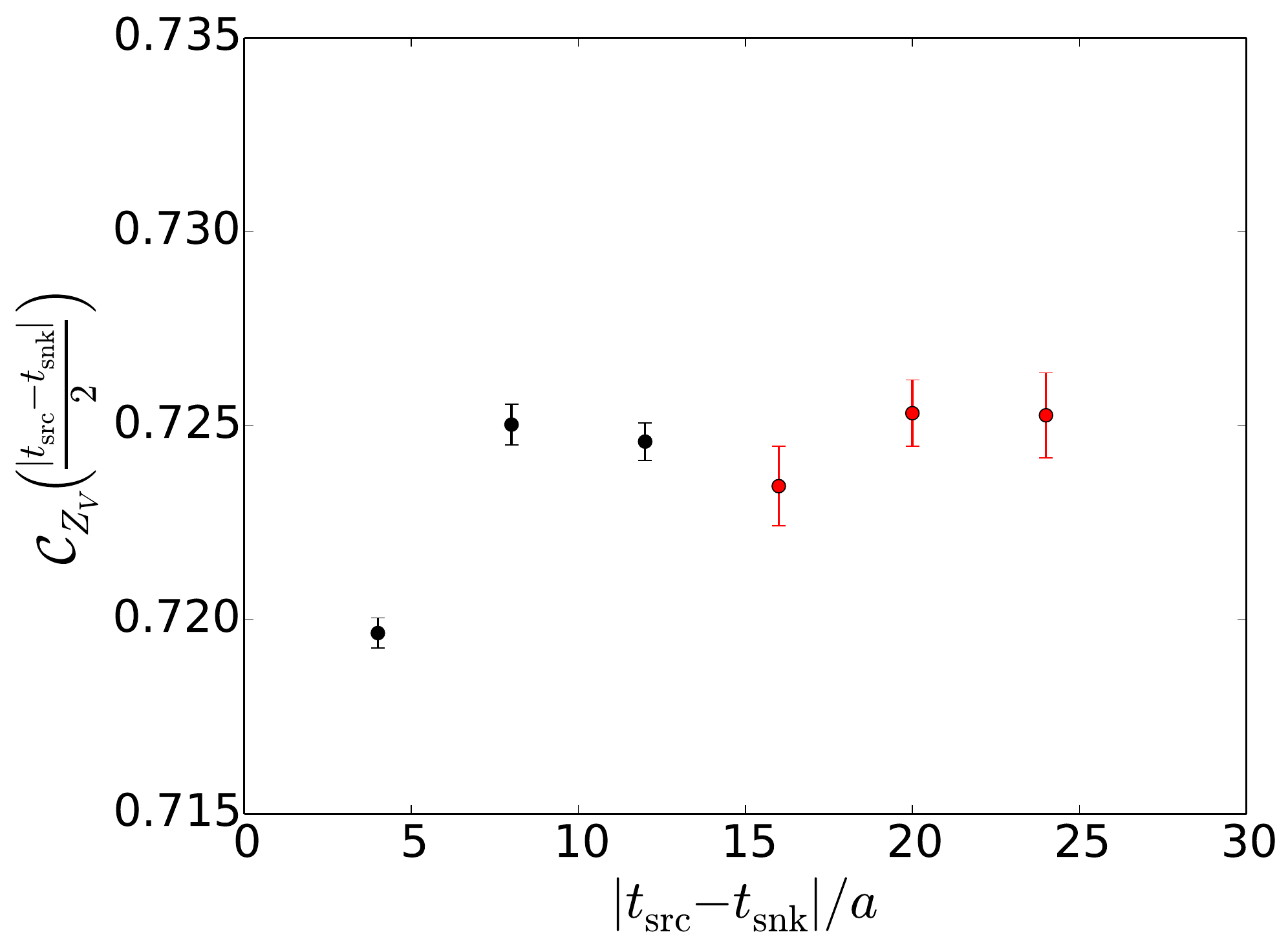}}
\subfloat{\includegraphics[width=0.49\textwidth]{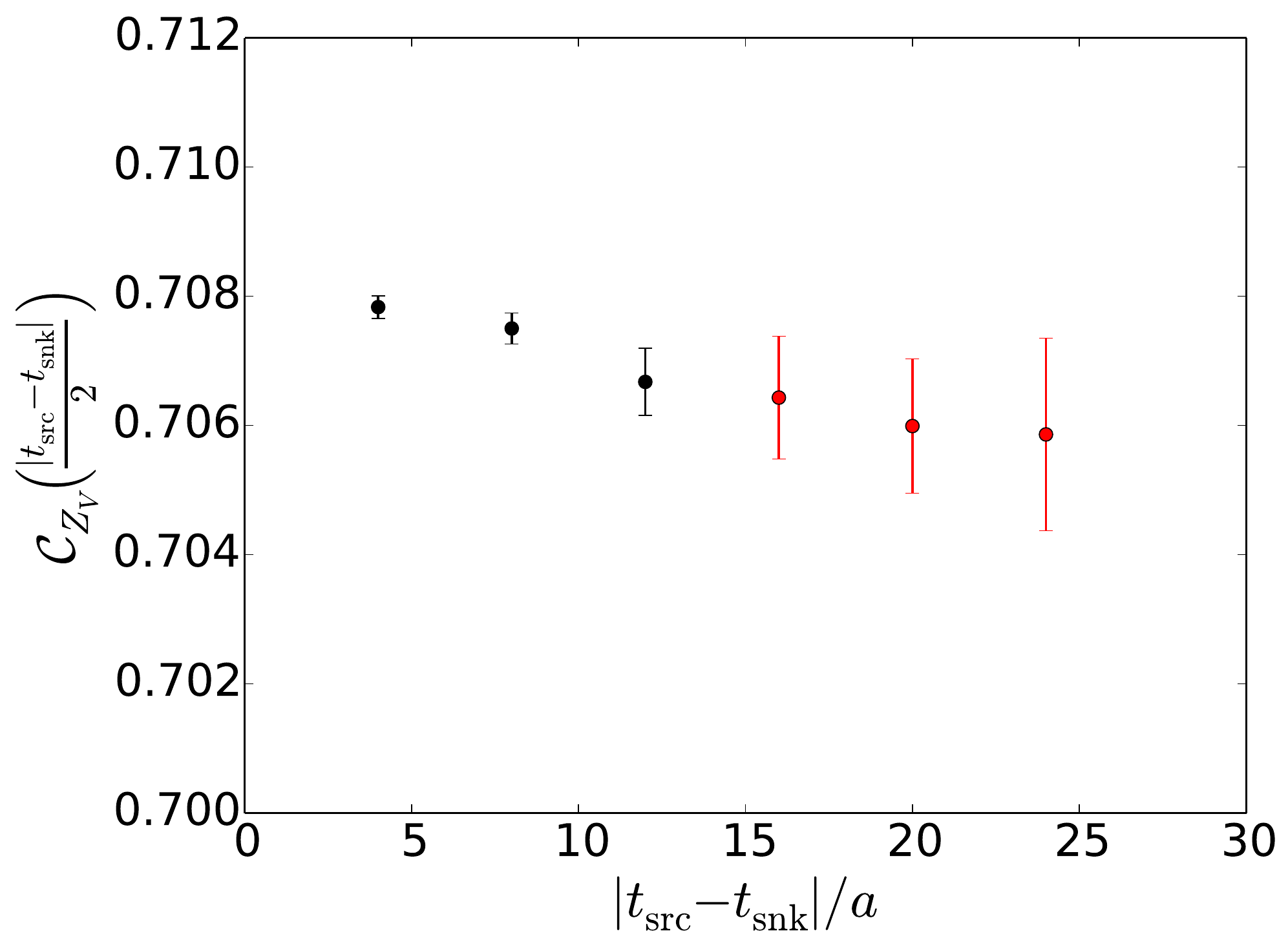}} \\
\subfloat{\includegraphics[width=0.49\textwidth]{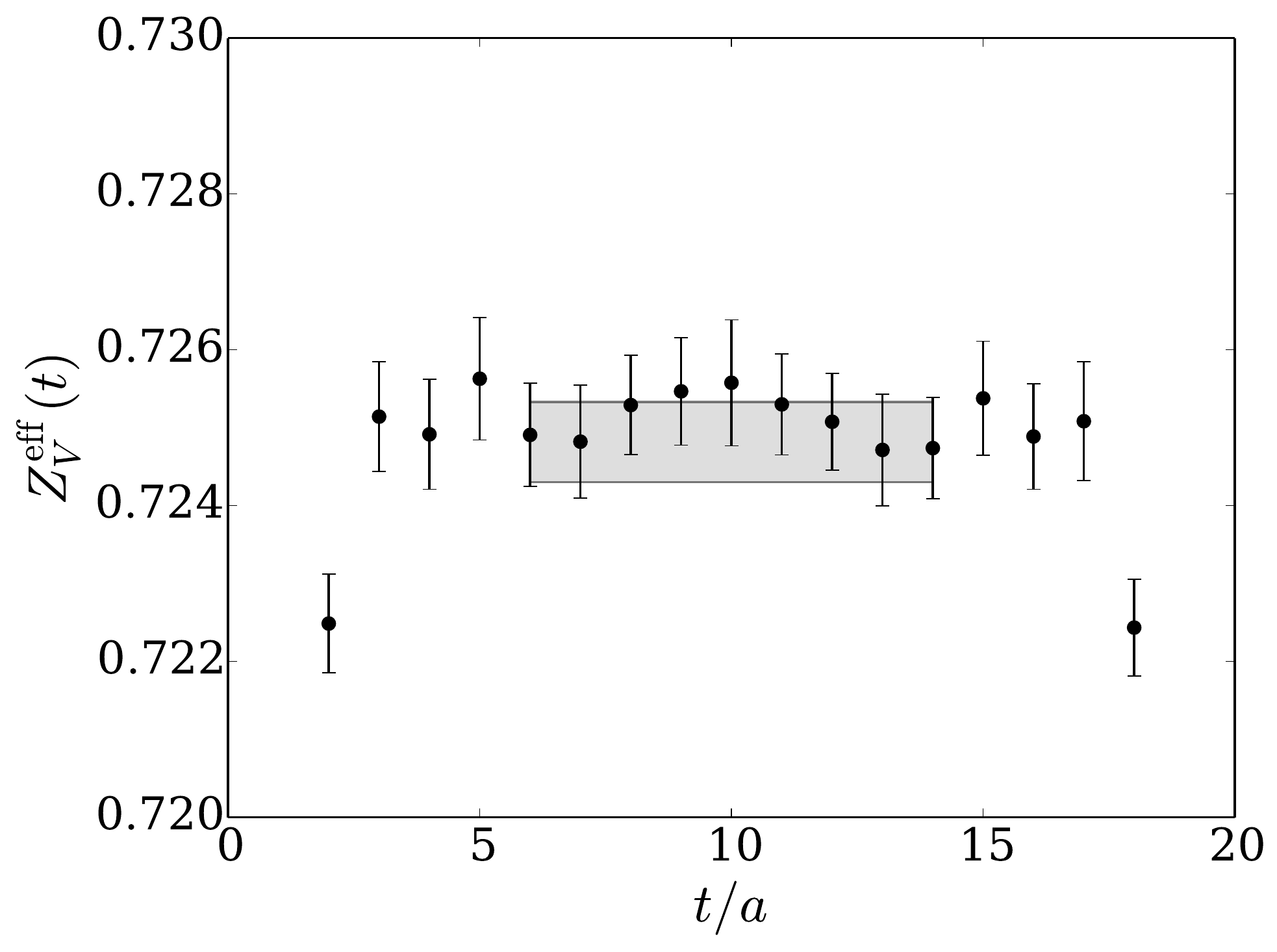}}
\subfloat{\includegraphics[width=0.49\textwidth]{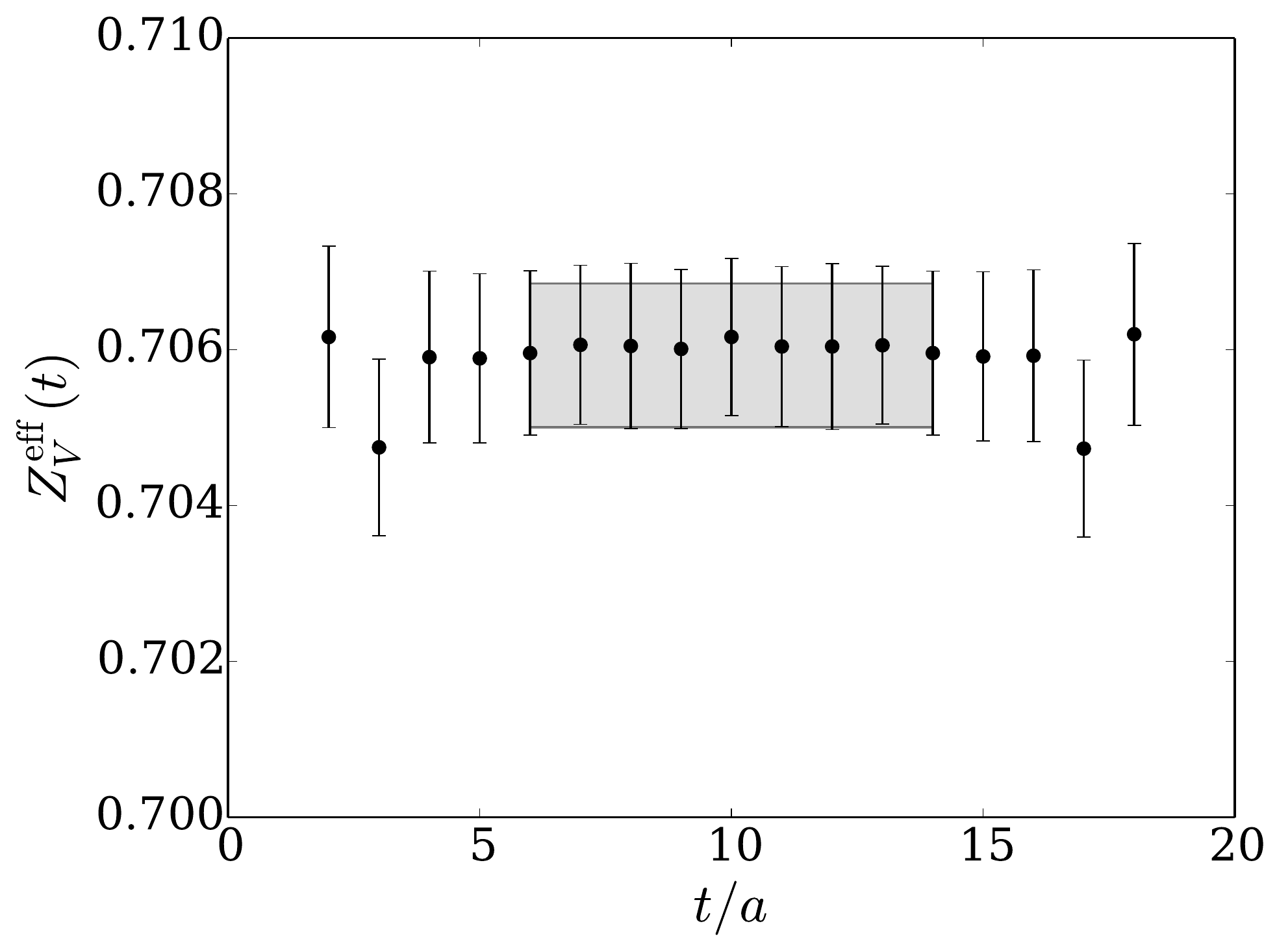}} \\
\caption{The vector current renormalization coefficient on the 32ID-M1 (left) and 32ID-M2 (right) ensembles. In the upper plot we show the dependence of the ratio~\eqref{eqn:ZV} on the source-sink separation: the point plotted for each separation is evaluated at the midpoint $t = | t_{\rm src} - t_{\rm snk} | / 2 a$. Points which were included in the fit are marked in red. In the lower plot we show an example of the fit to $Z_{V}$ overlaying the ratio~\eqref{eqn:ZV} for one of the source-sink separations included in the fit.}
\end{figure}

\begin{figure}[h]
\centering
\subfloat{\includegraphics[width=0.49\textwidth]{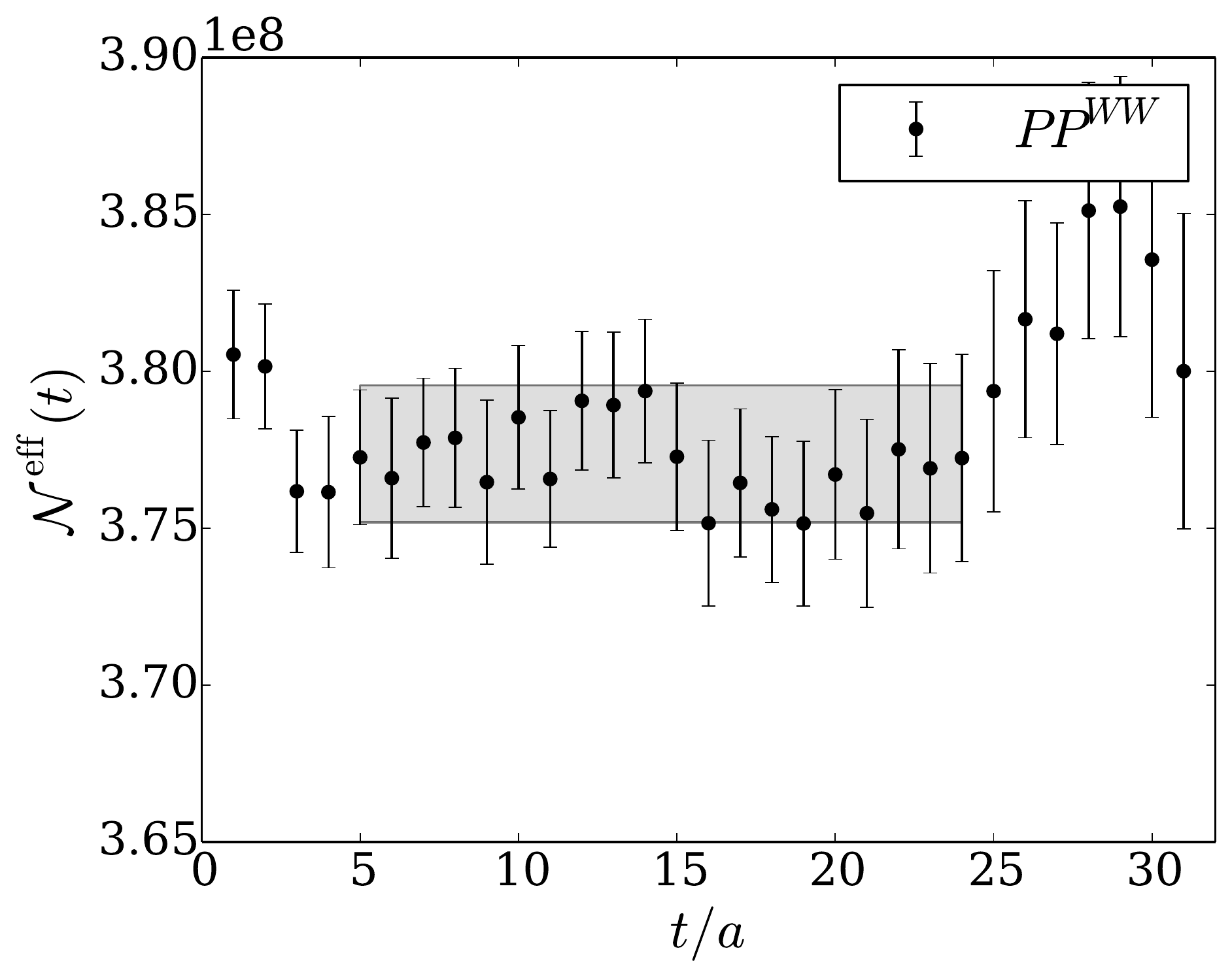}}
\subfloat{\includegraphics[width=0.49\textwidth]{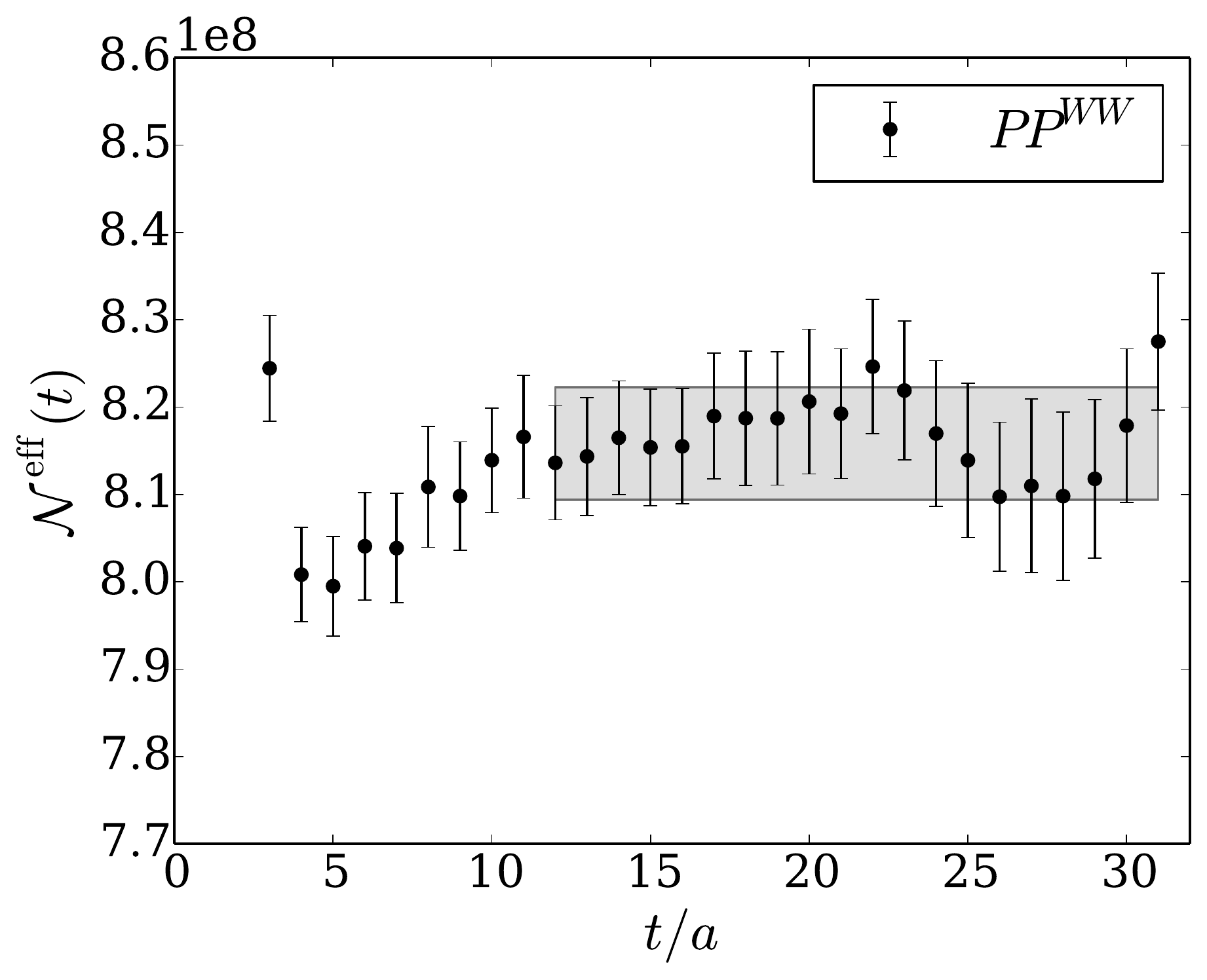}} \\
\subfloat{\includegraphics[width=0.49\textwidth]{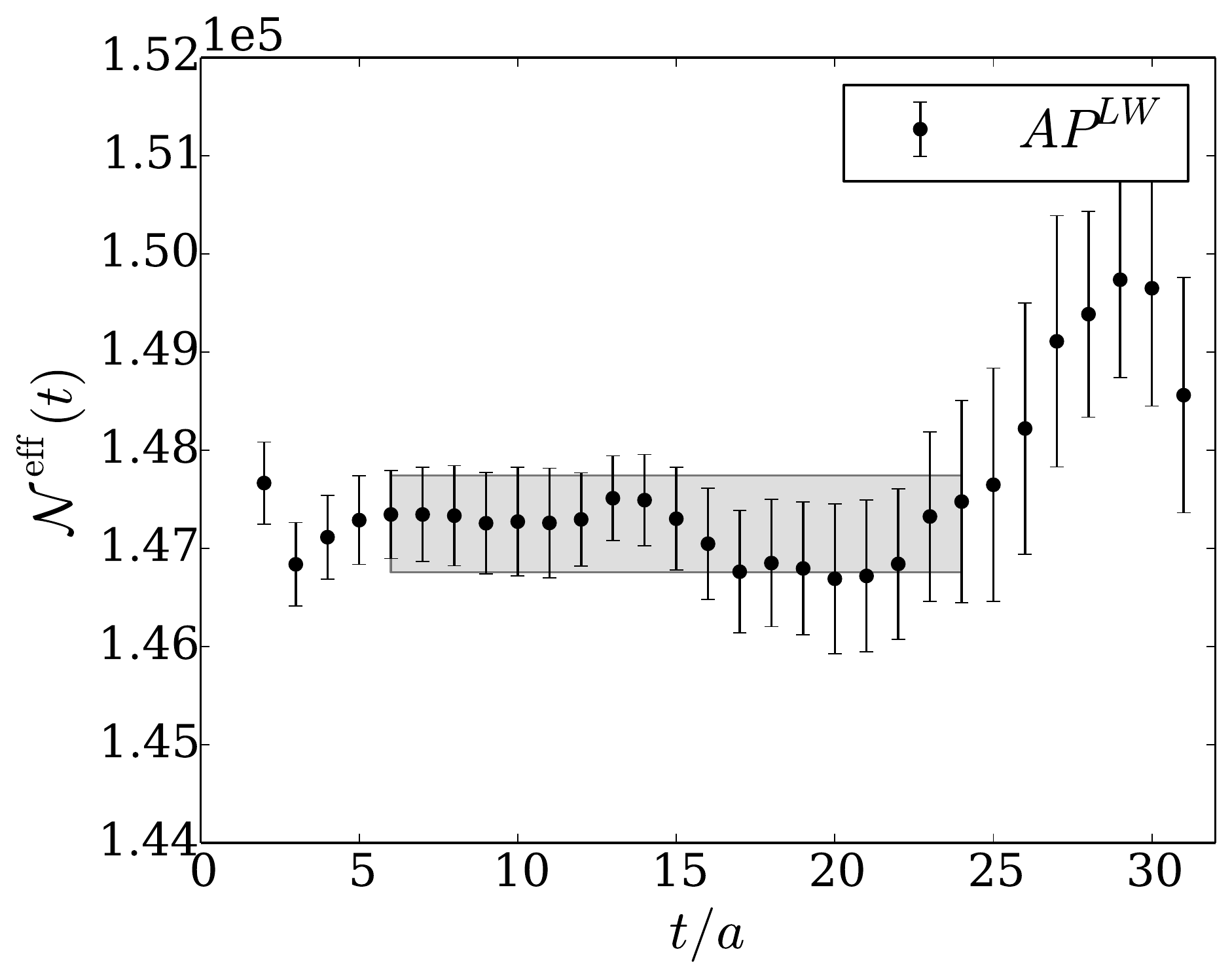}}
\subfloat{\includegraphics[width=0.49\textwidth]{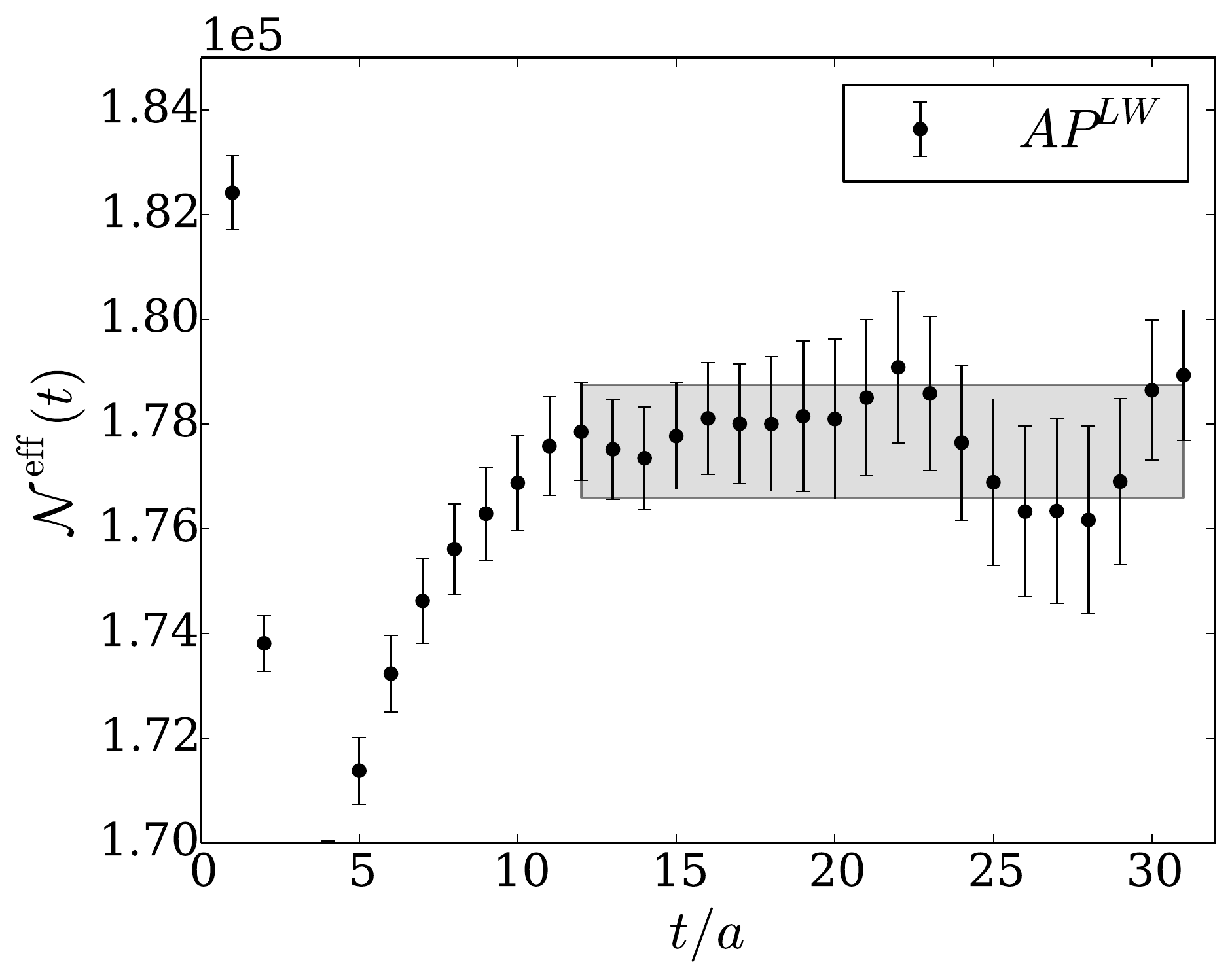}}
\caption{Light-light effective amplitudes $\mathscr{N}^{\rm eff}_{\mathscr{O}_{1} \mathscr{O}_{2}}(t) \equiv \langle \mathscr{O}_{1}(t) \mathscr{O}_{2}(0) \rangle / ( e^{-m_{\rm eff} t} \pm e^{-m_{\rm eff}(T-t)} )$ on the 32ID-M1 (left) and 32ID-M2 (right) ensembles. The sign is +(-) for the PP(AP) correlator. These are related to the light-light pseudoscalar decay constant according to Eqn.~\eqref{eqn:decay}.}
\end{figure}

\begin{figure}[h]
\centering
\subfloat{\includegraphics[width=0.49\textwidth]{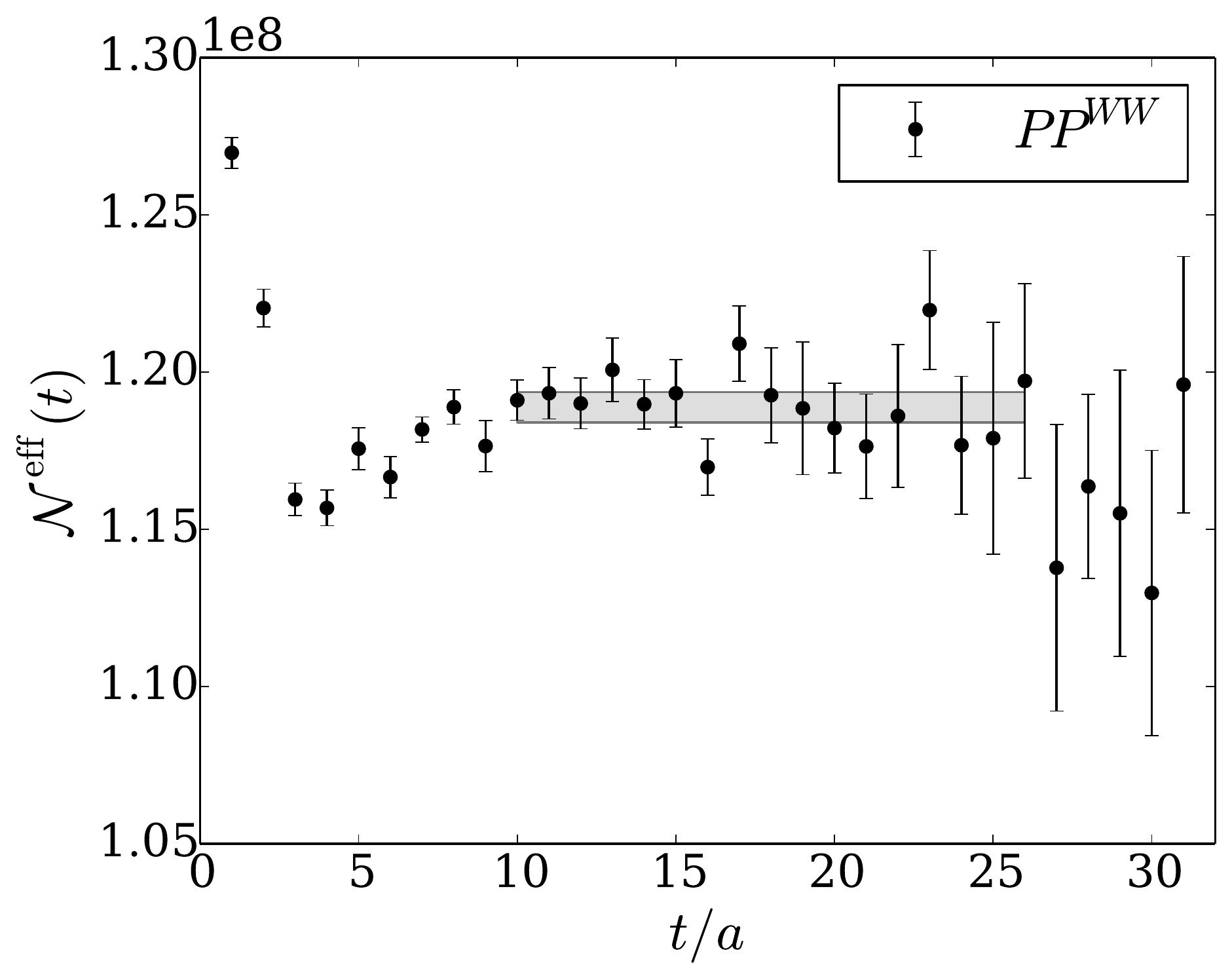}}
\subfloat{\includegraphics[width=0.49\textwidth]{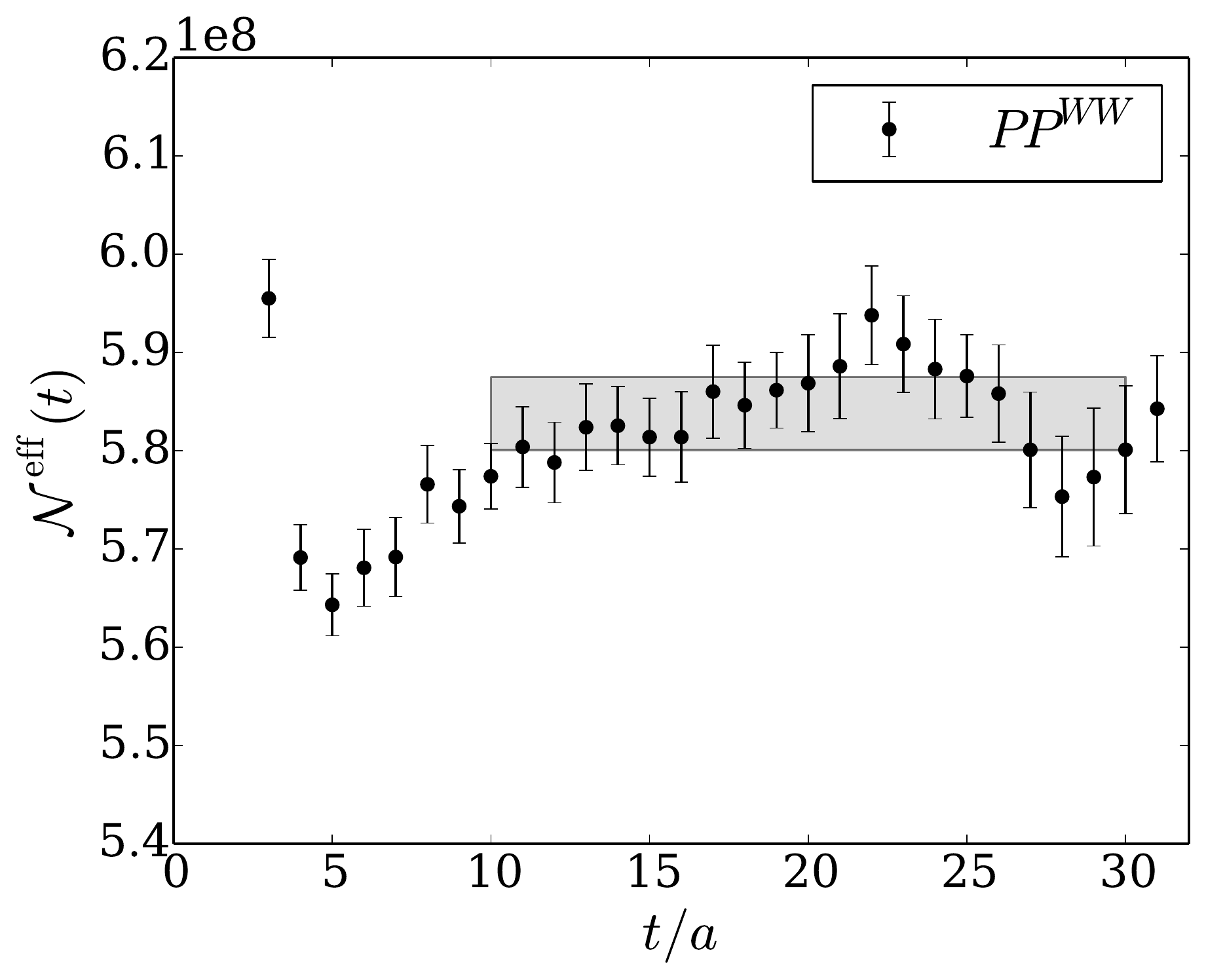}} \\
\subfloat{\includegraphics[width=0.49\textwidth]{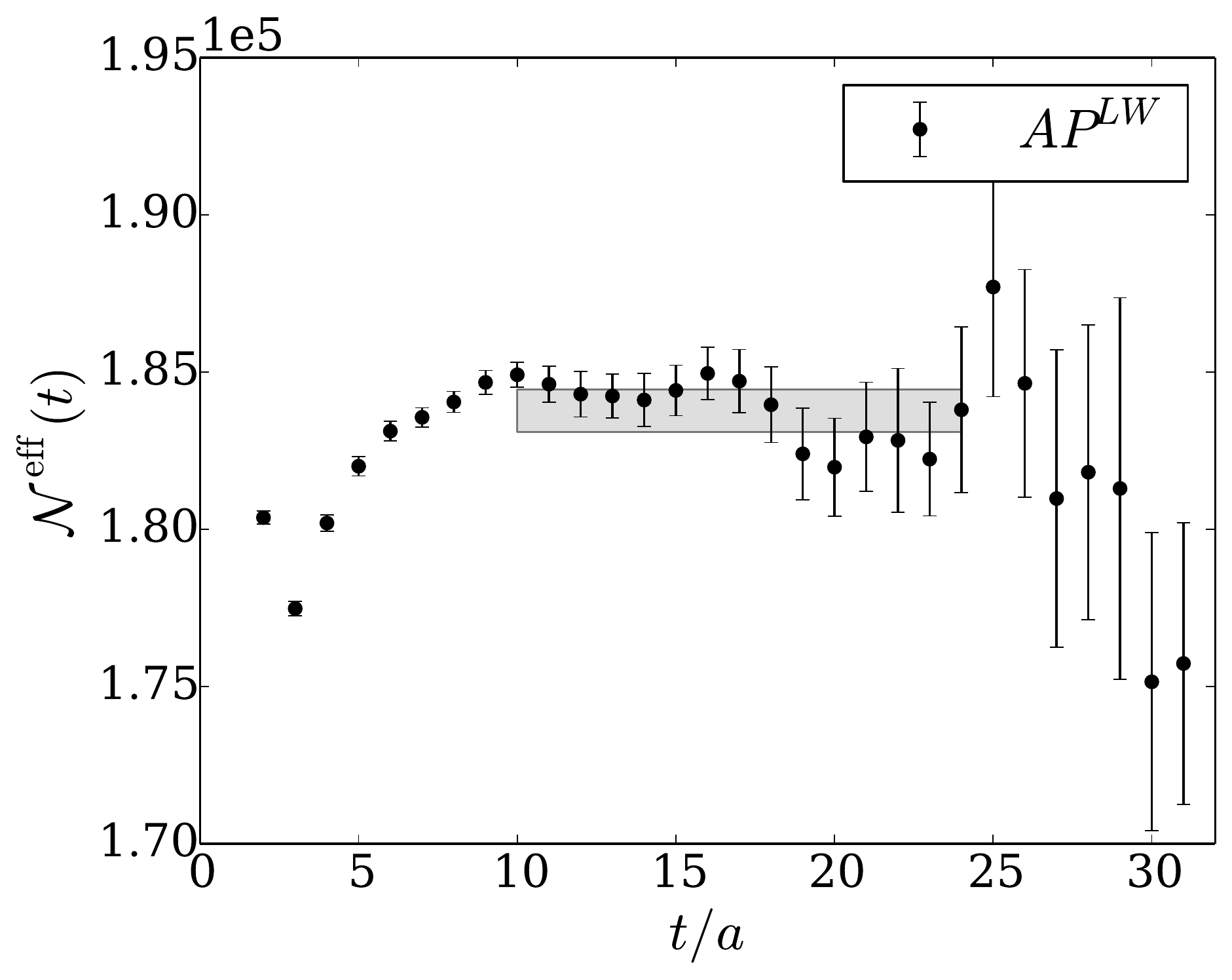}}
\subfloat{\includegraphics[width=0.49\textwidth]{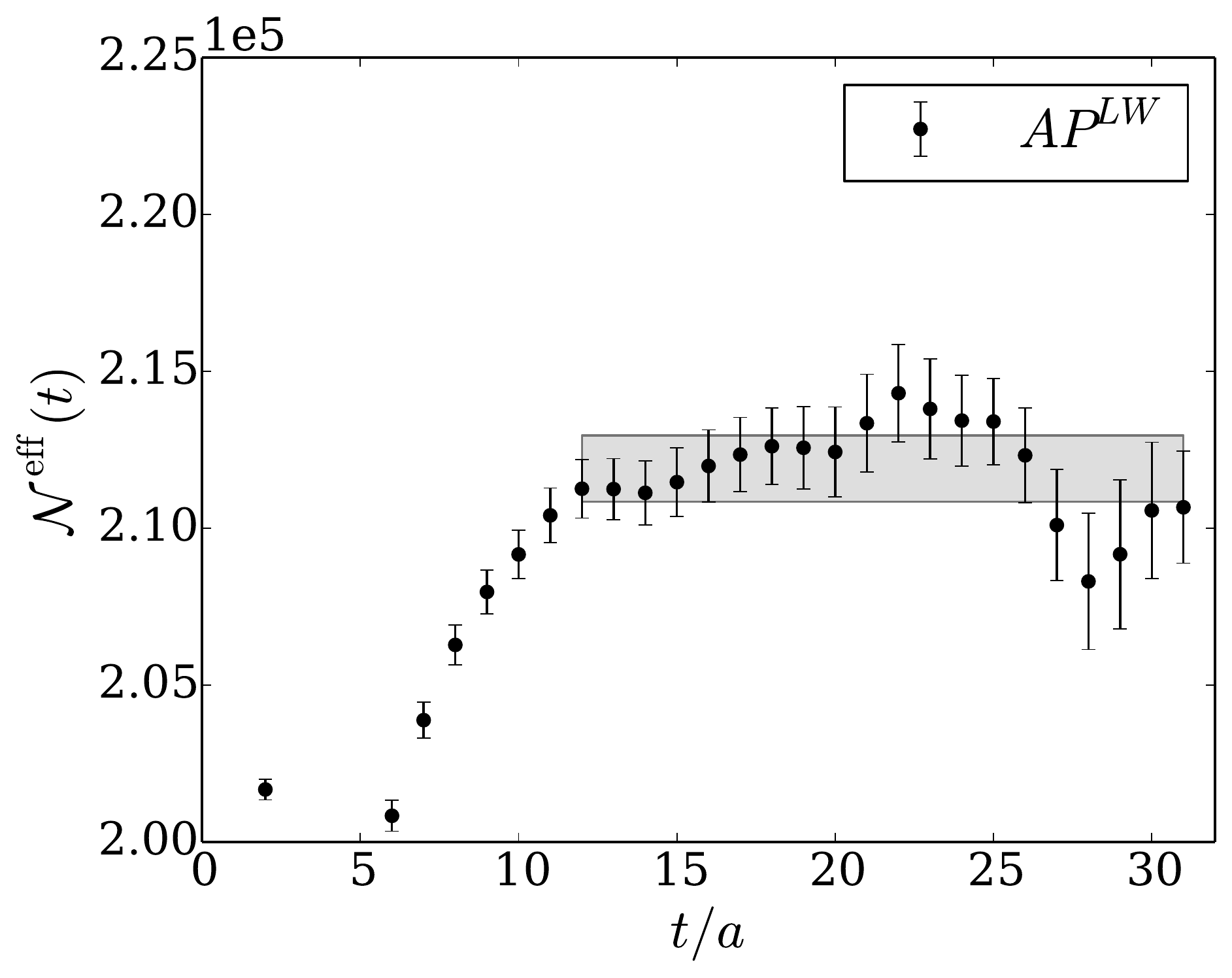}}
\caption{Heavy-light effective amplitudes on the 32ID-M1 (left) and 32ID-M2 (right) ensembles.}
\end{figure}

\clearpage
\vspace*{\fill}
\begin{figure}[h]
\centering
\subfloat{\includegraphics[width=0.49\textwidth]{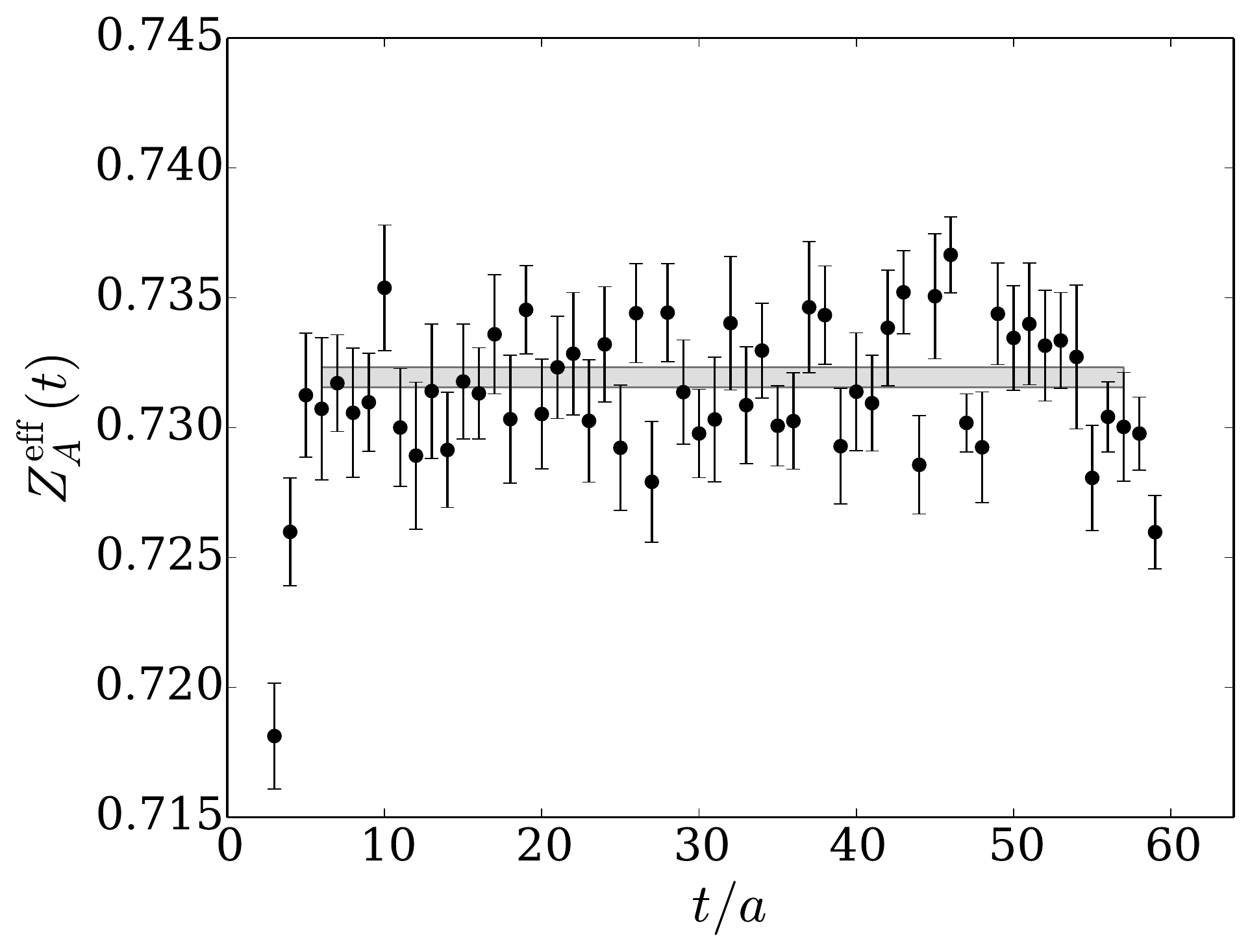}}
\subfloat{\includegraphics[width=0.49\textwidth]{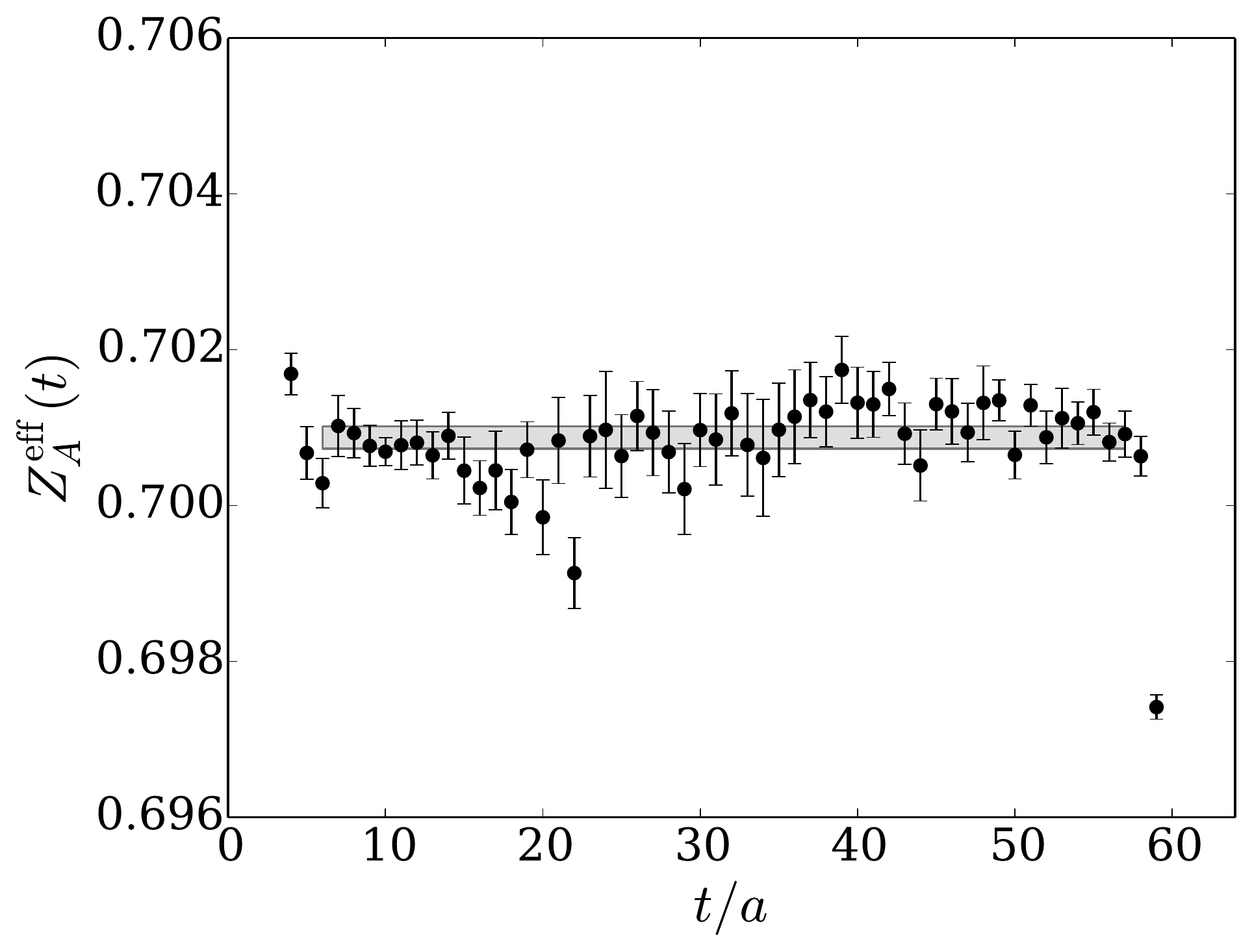}}
\caption{The axial current renormalization coefficient, from Eqn.~\eqref{eqn:ZA}, on the 32ID-M1 (left) and 32ID-M2 (right) ensembles.}
\end{figure}
\vspace{2.5cm}
\begin{figure}[h]
\centering
\subfloat{\includegraphics[width=0.49\textwidth]{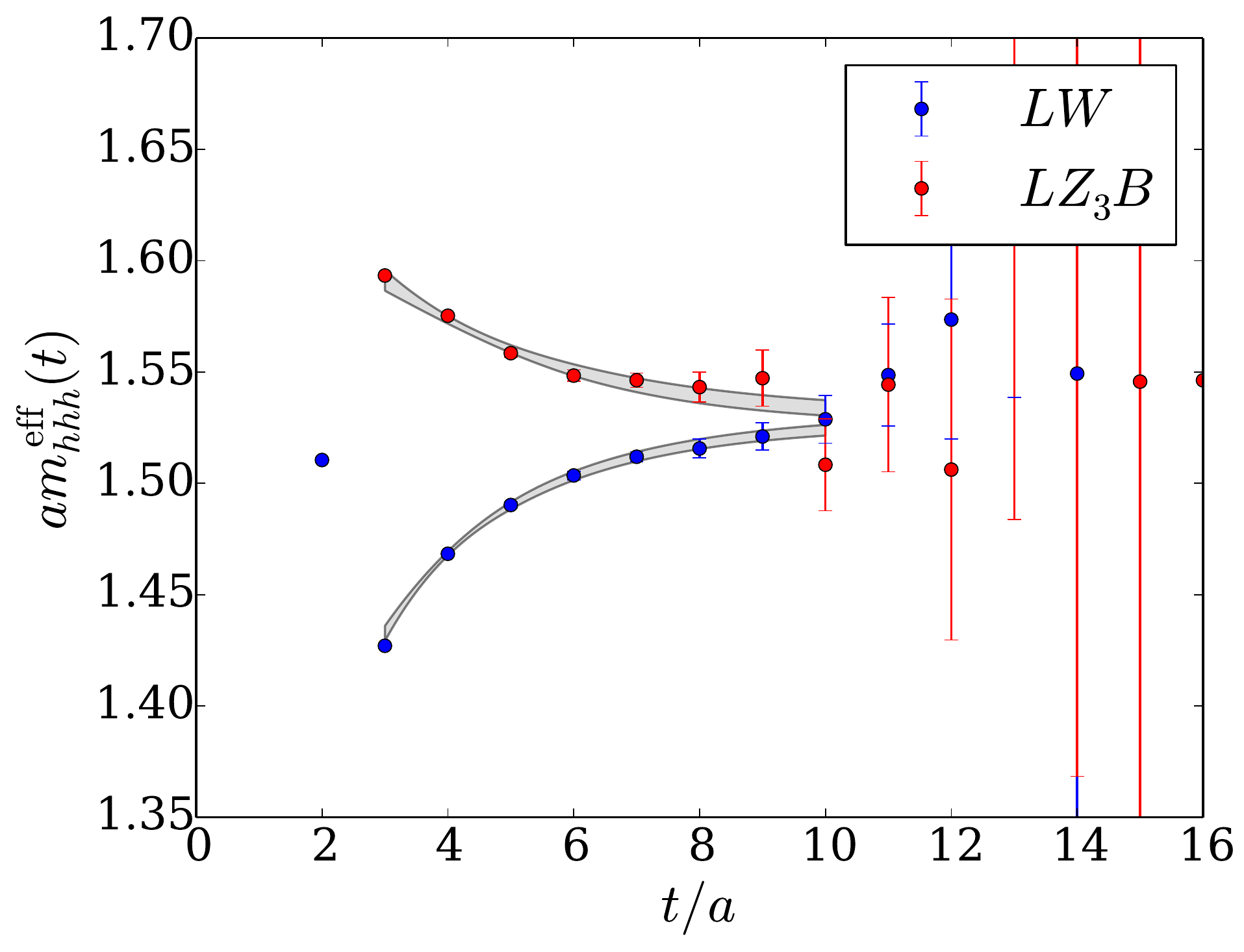}}
\subfloat{\includegraphics[width=0.49\textwidth]{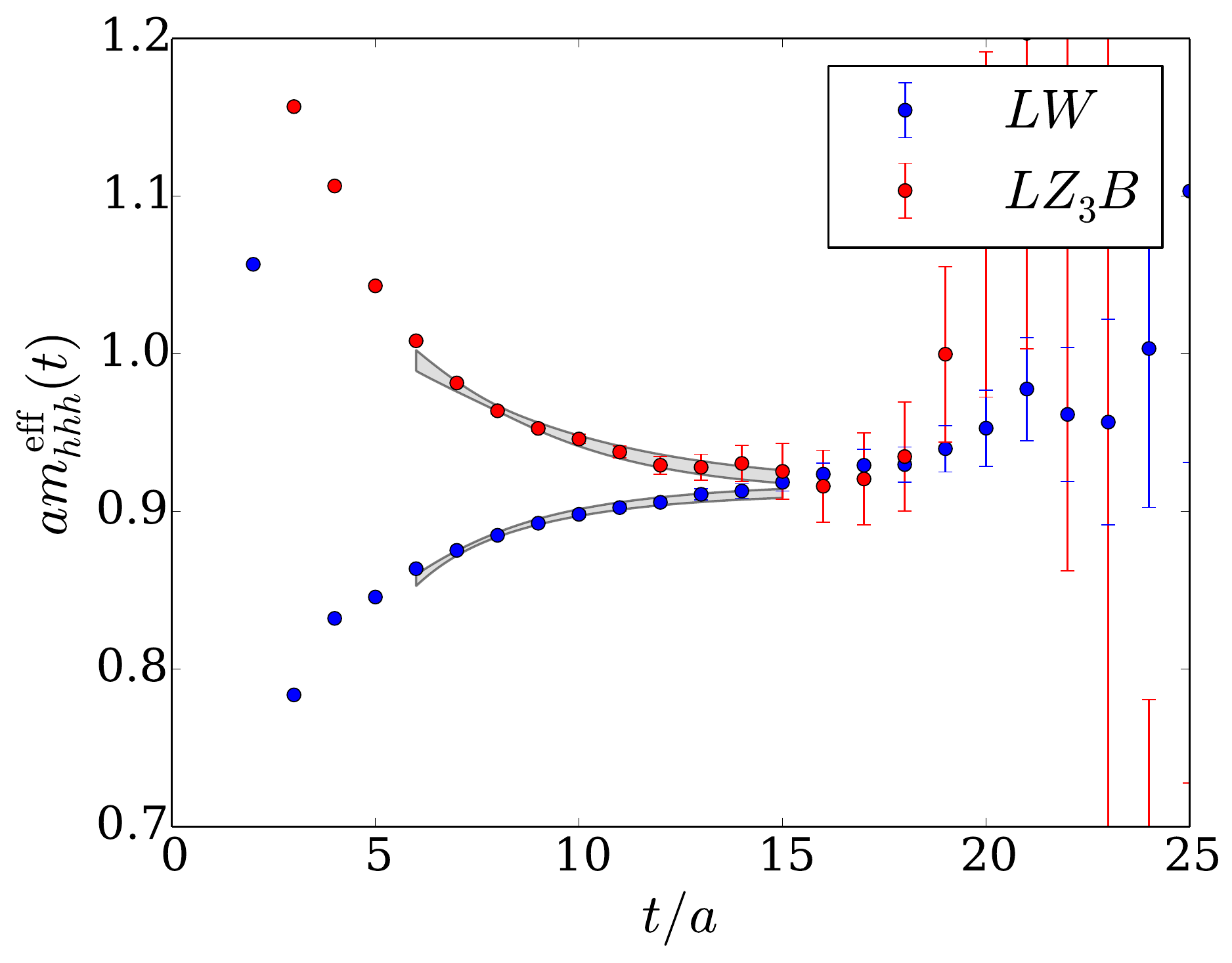}}
\caption{The $\Omega$ baryon mass on the 32ID-M1 (left) and 32ID-M2 (right) ensembles. The wall source and $Z_{3}$ box source correlators are simultaneously fit to double exponential ans\"{a}tze with common mass terms (Eqn.~\eqref{eqn:omega_fit}). Here we overlay the data with the effective mass curves obtained from the fit.}
\end{figure}
\vspace*{\fill}

\FloatBarrier


\clearpage
\section{Fits with Weighted $\chi^{2}$}
\label{appendix:chisq_normalized}

Correlations among data in a non-linear least squares fit are included by minimizing the \textit{correlated} $\chi^{2}$
\begin{equation}
\label{eqn:corr_chi2}
\chi^{2} = \sum_{ij} \left( \frac{y^{i} - f(\vec{\beta})^{i}}{\sigma^{i}} \right) \left( \rho^{-1} \right)^{i j} \left( \frac{y^{j} - f(\vec{\beta})^{j}}{\sigma^{j}} \right)
\end{equation}
over the the space of model parameters $\vec{\beta}$, where
\begin{equation}
\label{eqn:rho}
\rho^{i j} = \frac{\langle \left( y^{i} - \mu^{i} \right) \left( y^{j} - \mu^{j} \right) \rangle}{\sigma^{i} \sigma^{j}}
\end{equation}
is the correlation matrix. In the limit that the data is completely uncorrelated $\rho^{ij} = \delta^{ij}$, and we recover the familiar \textit{uncorrelated} $\chi^{2}$
\begin{equation}
\label{eqn:uncorr_chi2}
\chi^{2} = \sum_{i} \left( \frac{y^{i} - f(\vec{\beta})^{i}}{\sigma^{i}} \right)^{2}.
\end{equation}
In practice, correlations between data points computed on the same ensemble are often so strong that $\rho^{i j} \approx 1 \, \forall i,j$ is nearly singular, and minimization of the correlated $\chi^{2}$ defined by Eqn.~\eqref{eqn:corr_chi2} is numerically unstable. This pathology can be tamed by ignoring the correlations and instead minimizing the uncorrelated $\chi^{2}$, or by removing modes with small eigenvalues from $\rho^{i j}$ until the minimization algorithm becomes stable. In either case one loses a rigorous interpretation of $\chi^{2}$ as a statistical measure of the goodness-of-fit. \\

\begin{figure}[!ht]
\centering
\subfloat{\includegraphics[width=0.464\textwidth]{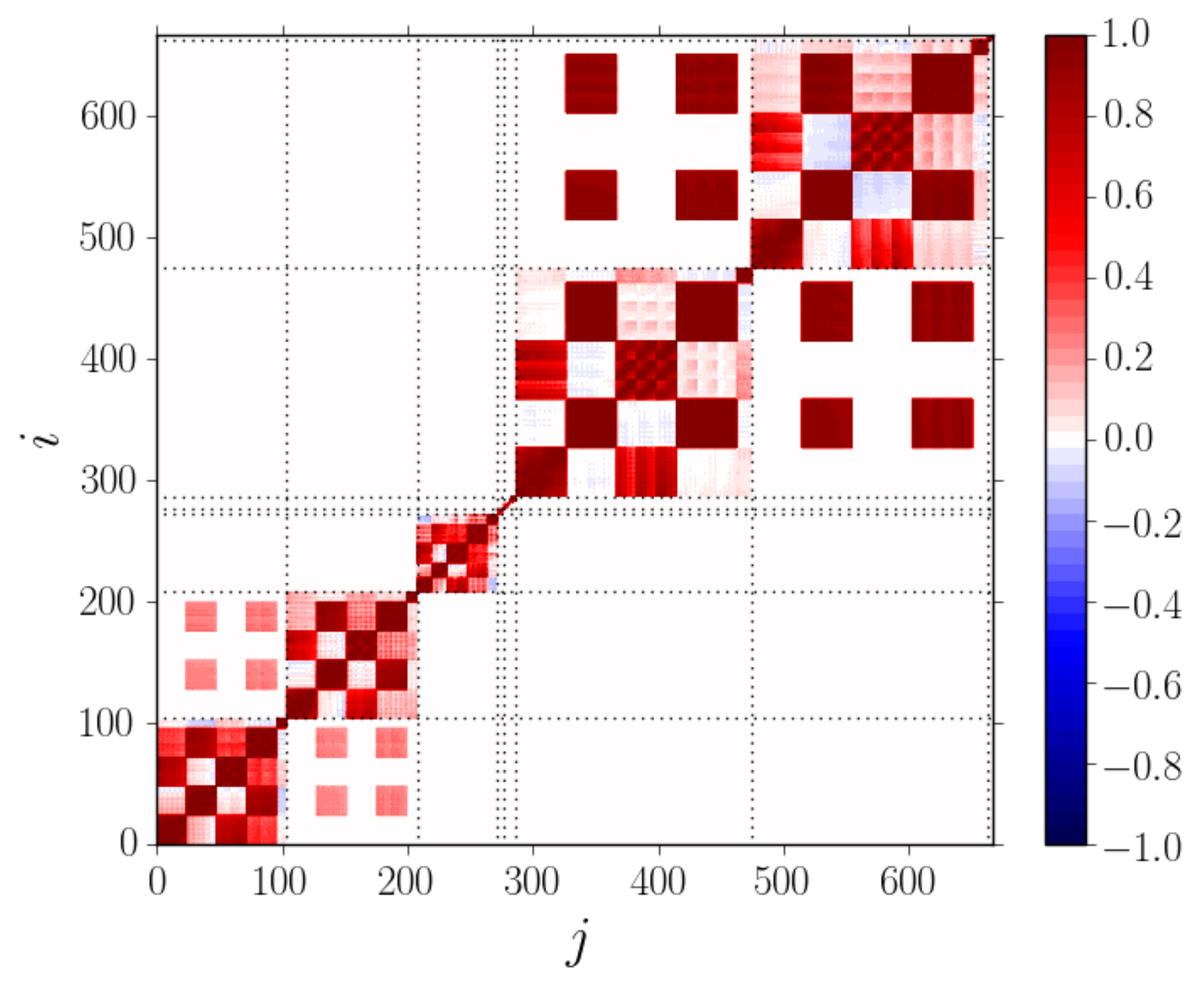}}
\subfloat{\includegraphics[width=0.49\textwidth]{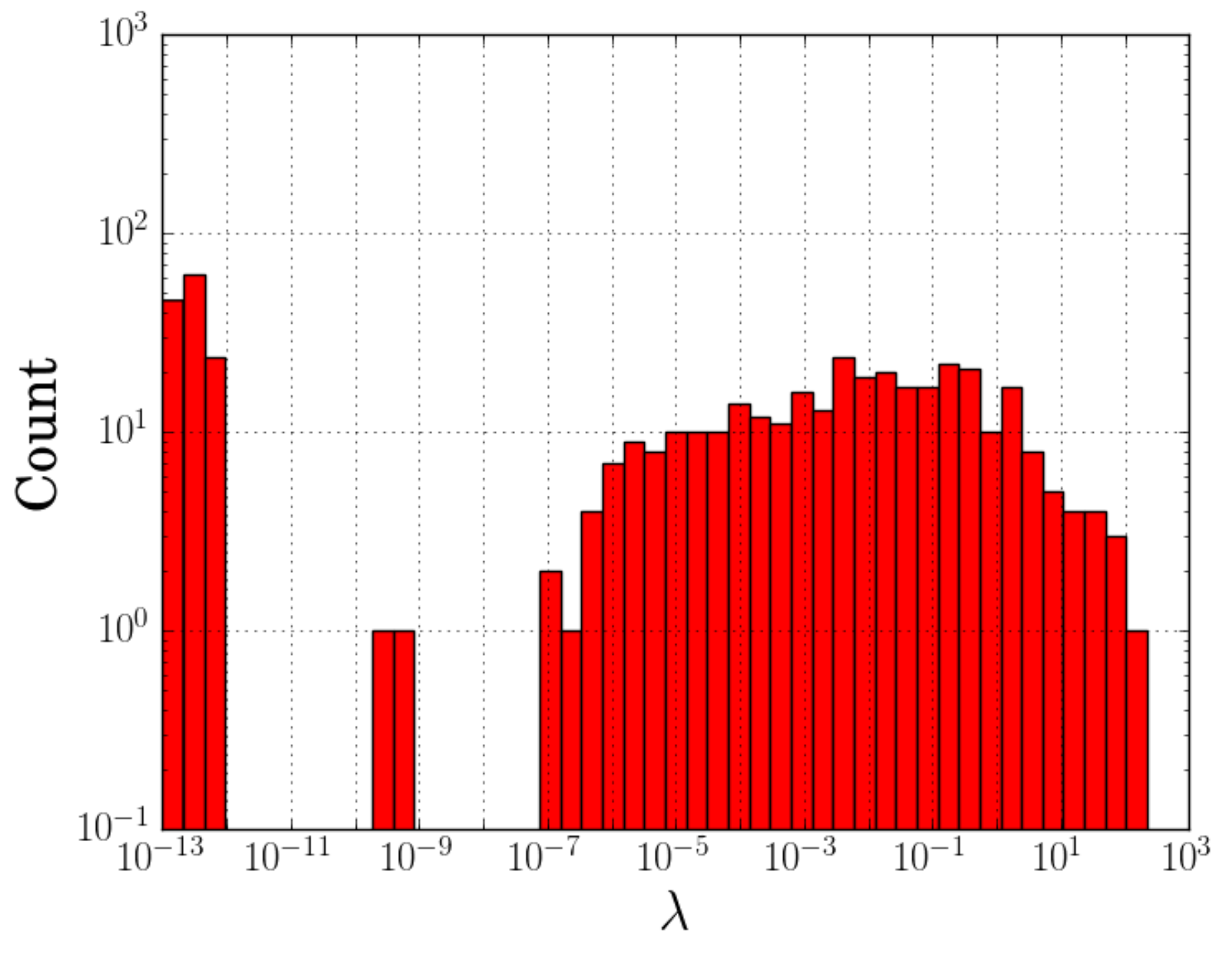}}
\caption{Left: the correlation matrix $\rho^{ij}$ corresponding to fits with a 370 MeV cut. The dashed lines mark the division into sub-blocks by ensemble. From left to right these are: 32I ($m_{l} = 0.004$), 32I ($m_{l} = 0.006$), 24I ($m_{l} = 0.005$), 48I, 64I, 32I-fine, 32ID ($m_{l} = 0.001$), 32ID ($m_{l} = 0.0042$), and 32ID-M1. Right: the eigenvalue spectrum of $\rho^{ij}$.} 
\label{fig:full_corr_matrix}
\end{figure}

\begin{figure}[!ht]
\centering
\subfloat[32I, $m_{l} = 0.004$]{\includegraphics[width=0.464\textwidth]{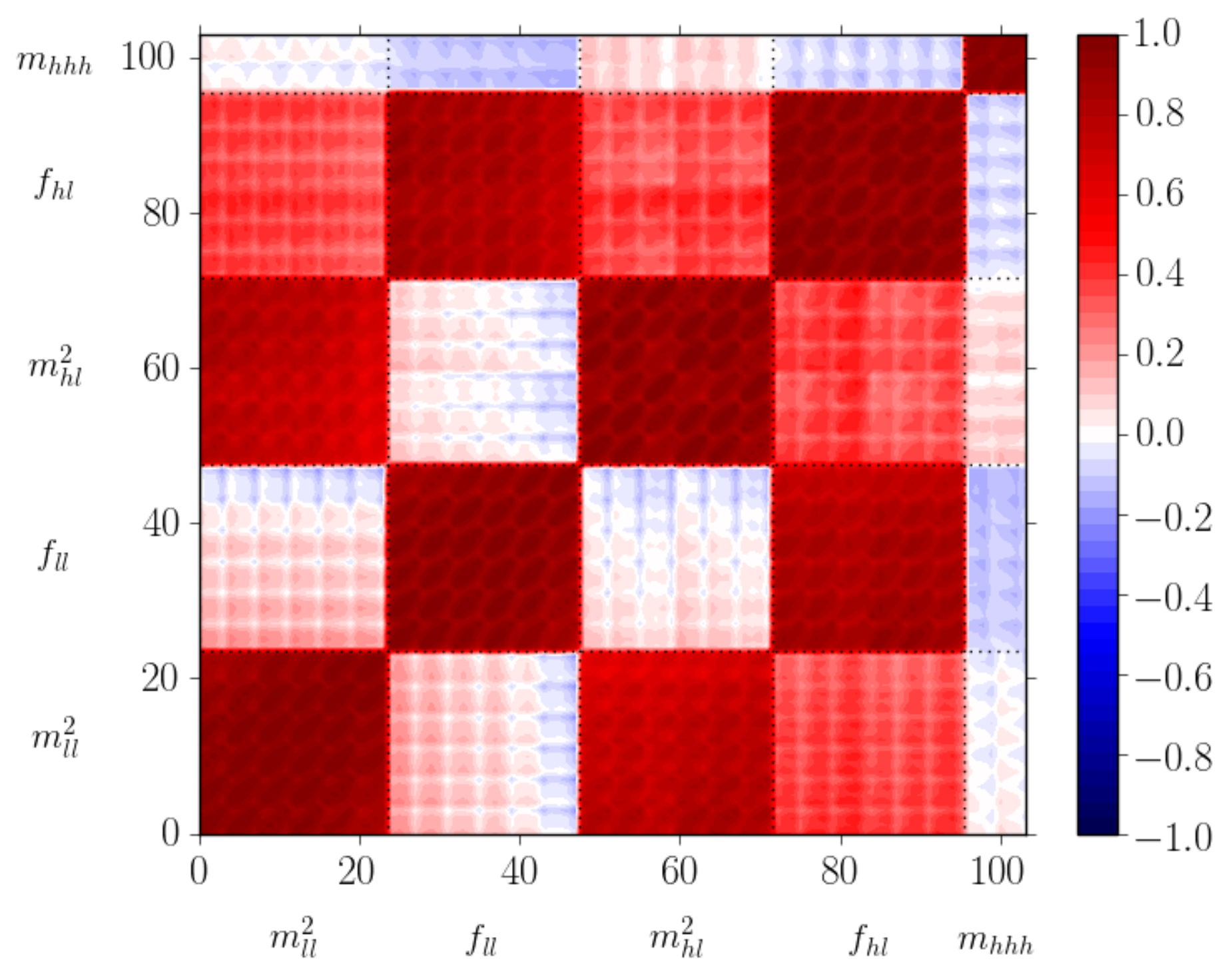}}
\subfloat[32I, $m_{l} = 0.006$]{\includegraphics[width=0.464\textwidth]{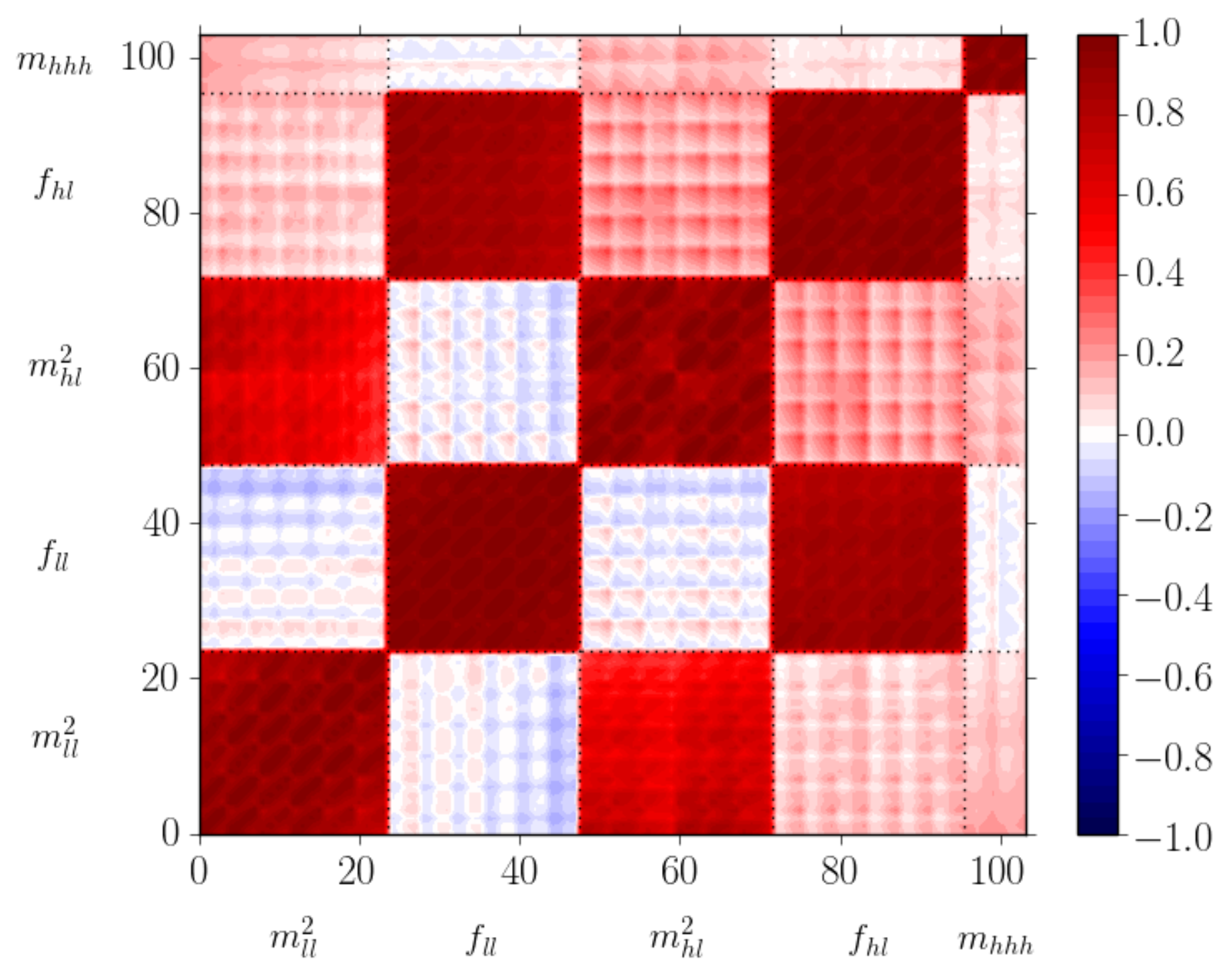}} \\
\subfloat[32I, $m_{l} = 0.004~\times~m_{l} = 0.006$]{\includegraphics[width=0.464\textwidth]{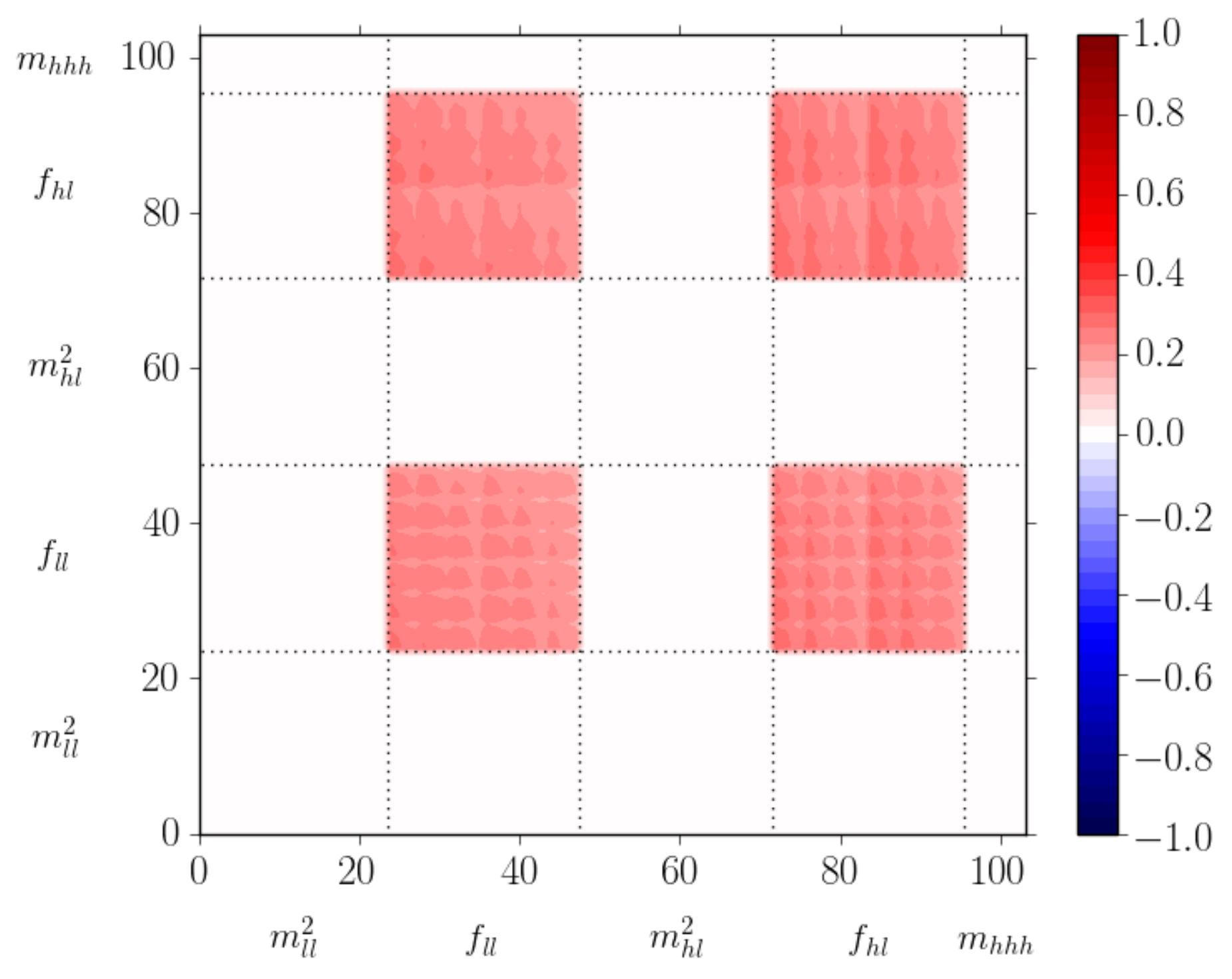}}
\caption{Sub-blocks of the correlation matrix corresponding to the 32I ensembles. Panel (c) shows the cross-correlations between the $m_{l} = 0.004$ and $m_{l} = 0.006$ ensembles induced by the use of $Z_{V}$ extrapolated to the chiral limit to normalize the decay constants.} 
\label{fig:32I_sub_corr_matrix}
\end{figure}

In Figure~\ref{fig:full_corr_matrix} we plot the correlation matrix $\rho^{ij}$ and its eigenvalue spectrum computed from the data included in fits with a 370 MeV cut. We find, as expected, that the correlation matrix is extremely singular due to strong correlations associated with partial quenching and reweighting: the eigenvalues span 15 orders of magnitude, and the condition number is $\mathrm{cond}(\rho^{ij}) = 1.85 \times 10^{17}$. In Figure~\ref{fig:32I_sub_corr_matrix} we further plot the sub-blocks of $\rho^{ij}$ corresponding to the 32I ensembles as an example of the cross-correlations present in our data, for example, between the light-light and heavy-light pseudoscalar masses. We conclude that we are unable to accurately invert the correlation matrix, much less attempt fully correlated fits as defined by Equation~\ref{eqn:corr_chi2}. \\

The fits discussed in Section~\ref{sec:su2_fits} were performed by minimizing the uncorrelated $\chi^{2}$ (Eqn.~\eqref{eqn:uncorr_chi2}). We expect, however, that our data is highly correlated, in particular between measurements of partially quenched observables on the same ensemble but with different choices of the valence quark masses, and between different reweightings in $m_{h}$ of the same observable. These particular classes of correlations are especially troublesome since our partially quenched measurements and $m_{h}$ reweightings were performed on the relatively heavy pion mass ensembles (24I, 32I, and 32ID) --- a naive uncorrelated fit will tend to give too much weight to this data, which is far from the chiral limit where ChPT is exact. In this appendix we repeat these fits, normalizing the contributions to $\chi^{2}$ by the number of nondegenerate pseudoscalar mass measurements ($N_{e}$) associated with a given ensemble ($e$):
\begin{equation}
\label{eqn:normalized_chi2}
\chi^{2}_{e} = \frac{1}{N_{e}} \sum\limits_{i} \left( \frac{y^{i}_{e} - f^{i}_{e}}{\sigma^{i}_{e}} \right)^{2},
\end{equation}
where $\chi^{2} = \sum_{e} \chi^{2}_{e}$ and the $N_{e}$ are listed in Table~\ref{tab:Ne}. This can be loosely regarded as the limit of extreme correlation, in which all of the partially quenched measurements on a given ensemble are effectively weighted as a single point by inflating their statistical errors $\sigma_{e}^{i} \rightarrow \sqrt{N_{e}} \sigma_{e}^{i}$. We use the difference in central values between these two schemes to assign a systematic error associated with our inability to fully resolve the true correlation matrix to our fits. \\

\begin{table}[h]
\centering
\begin{tabular}{c||c|c|c|c|c|c|c|c}
\hline
\hline
\rule{0cm}{0.4cm}Mass Cut & 24I & 32I & 32ID & 32I-fine & 48I & 64I & 32ID-M1 & 32ID-M2 \\
\hline
\rule{0cm}{0.4cm}370 MeV & 12 & 48 & 80 & 1 & 1 & 1 & 1 & --- \\
450 MeV & 48 & 120 & 80 & 1 & 1 & 1 & 1 & 1 \\
\hline
\hline
\end{tabular}
\caption{The value of $N_{e}$ --- the number of non-degenerate quark mass combinations $(m_{x},m_{y},m_{l},m_{h})$ used for pseudoscalar  measurements entering into the fits --- for each ensemble and mass cut. There are four values of $m_{h}$ for each fixed $(m_{x},m_{y},m_{l}$) obtained by reweighting in the heavy sea quark determinant.}
\label{tab:Ne}
\end{table}

This scheme for weighting $\chi^{2}$ can be further understood by analyzing the correlation matrix in the limit that the off-diagonal terms are completely dominated by the correlations between partially quenched measurements on the same ensemble. To clarify this discussion, we write the full correlation matrix (Eqn.~\eqref{eqn:rho}) as $\rho_{(e,i,a);(e',i',a')}$, where $e$ indexes the ensemble, $i$ indexes the valence quark mass combination, and $a$ indexes the observable. If the data is both highly correlated and dominated by the correlations between the partially quenched data the correlation matrix will have a block structure 
\begin{equation}
\rho_{(e,i,a),(e',i',a')} = (\rho_{PQ})_{(e,a)}^{i'j'} \delta_{ee'} \delta_{aa'}
\end{equation} 
where $(\rho_{PQ})_{(e,a)}^{i'j'}$ is the $N_{(e,a)} \times N_{(e,a)}$ sub-matrix describing the correlations between partially quenched measurements of observable $a$ on ensemble $e$; in the left panel of Figure \ref{fig:full_corr_matrix}, for example, these are the extremely correlated blocks lying along the main diagonal. We expect these correlations to be sufficiently strong that the blocks $(\rho_{PQ})_{(e,a)}^{i'j'}$ will be nearly singular, which we can write in general as
\begin{equation}
\label{eqn:rho_pq_e}
(\rho_{\rm PQ})_{(e,a)}^{i' j'} = \left( \begin{array}{cccc} 1 & 1 - \epsilon^{12}_{(e,a)} & \cdots & 1 - \epsilon^{1 N_{(e,a)}}_{(e,a)} \\ 1 - \epsilon^{21}_{(e,a)} & 1 & \cdots & 1 - \epsilon^{2  N_{(e,a)}}_{(e,a)} \\ \vdots & \vdots & \ddots & \vdots \\ 1 - \epsilon^{N_{(e,a)} 1}_{(e,a)} & 1 - \epsilon^{N_{(e,a)} 2}_{(e,a)} & \cdots & 1 \end{array} \right)
\end{equation}
where the $\epsilon_{(e,a)}^{i'j'} \ll 1$ measure the small deviations from unity of the off-diagonal entries. To simplify the analysis we set $\epsilon^{i'j'}_{(e,a)} = \epsilon$ everywhere and work to leading order in $\epsilon$. In this limit each of the $(\rho_{\rm PQ})^{i'j'}_{(e,a)}$ has a single eigenvector $(1, 1, \ldots, 1)$ with eigenvalue $N_{(e,a)} - (N_{(e,a)} - 1) \epsilon$, representing the mode where all $N_{(e,a)}$ data points are completely correlated. The remaining $N_{(e,a)} - 1$ eigenvectors are degenerate with eigenvalue $\epsilon$ and span the subspace of correlations in the data orthogonal to the completely correlated mode; their poor statistical resolution can be understood as a source of the numerical instabilities observed in fully correlated fits. Since $(\rho_{\rm PQ}^{ij})_{(e,a)}$ is a real, symmetric matrix it can be diagonalized by an orthogonal transformation:
\begin{equation}
(\rho_{\rm PQ}^{ij})_{ee} = \left( \begin{array}{c|ccc} 1 & & \multirow{4}{*}{$Q$} & \\ 1 & & & \\ \vdots & & & \\ 1 & & & \end{array} \right) \left( \begin{array}{cccc} N_{(e,a)} - (N_{(e,a)}-1) \epsilon & 0 & \cdots & 0 \\ 0 & \epsilon & \cdots & 0 \\ \vdots & \vdots & \ddots & \vdots \\ 0 & 0 & \cdots & \epsilon \end{array} \right) \left( \begin{array}{cccc} 1 & 1 & \cdots & 1 \\ \hline & & & \\ \multicolumn{4}{c}{Q^{\top}} \\ & & & \end{array} \right).
\end{equation}
Here $Q$ is an orthogonal matrix whose columns correspond to an appropriate choice of the $N_{(e,a)}-1$ degenerate eigenvectors with eigenvalue $\epsilon$. Eqn.~\eqref{eqn:normalized_chi2} follows from the fully correlated Eqn.~\eqref{eqn:corr_chi2} if we define new blocks $(\tilde{\rho}_{\rm PQ})^{i'j'}_{(e,a)}$ by making the replacements $N_{(e,a)} - (N_{(e,a)} - 1) \epsilon \approx N_{(e,a)}$ for the largest eigenvalue, and $\epsilon \rightarrow N_{(e,a)}$ for the remaining $N_{(e,a)} - 1$ eigenvalues:
\begin{equation}
(\tilde{\rho}_{\rm PQ})^{i'j'}_{(e,a)} = \left( \begin{array}{c|ccc} 1 & & \multirow{4}{*}{$Q$} & \\ 1 & & & \\ \vdots & & & \\ 1 & & & \end{array} \right) \left( \begin{array}{cccc} N_{(e,a)} & 0 & \cdots & 0 \\ 0 & N_{(e,a)} & \cdots & 0 \\ \vdots & \vdots & \ddots & \vdots \\ 0 & 0 & \cdots & N_{(e,a)} \end{array} \right) \left( \begin{array}{cccc} 1 & 1 & \cdots & 1 \\ \hline & & & \\ \multicolumn{4}{c}{Q^{\top}} \\ & & & \end{array} \right) = N_{(e,a)} \delta^{i' j'},
\end{equation}
and substitute $(\rho_{\rm PQ})^{i'j'}_{(e,a)} \rightarrow (\tilde{\rho}_{\rm PQ})^{i'j'}_{(e,a)}$. Effectively, in Eqn.~\eqref{eqn:normalized_chi2} we are treating the modes associated with the largest eigenvalue of each of the $(\rho_{\rm PQ})^{i'j'}_{(e,a)}$ exactly up to terms of $\mathcal{O}(\epsilon)$, and underweighting the subdominant modes by a factor $\sim \epsilon / N_{(e,a)}$. We find in practice that this stabilizes the fits while still capturing some of the important effects of correlations in the data. More generally, one expects that the $\epsilon_{(e,a)}^{i' j'}$ in Eqn.~\eqref{eqn:rho_pq_e} are not all equal --- breaking the degeneracy between the $N_{(e,a)} - 1$ smallest eigenvalues of $(\rho_{\rm PQ})^{i'j'}_{(e,a)}$ --- and that some of the off-diagonal entries in the full correlation matrix, representing other kinds of correlations, are non-zero; these effects are $\mathcal{O}(\epsilon)$ and do not change the argument presented here. \\

In the remainder of the appendix we summarize the results of fits performed by minimizing the normalized $\chi^{2}$ (Eqn.~\eqref{eqn:normalized_chi2}).

\clearpage
\subsection{Fit Parameters}
\vspace*{\fill}
\begin{table}[h]
\centering
\resizebox{\columnwidth}{!}{
\begin{tabular}{cc||cc|cc}
\hline
\hline
\rule{0cm}{0.4cm} & & NLO ($370 \, \mathrm{MeV}$ cut) & NLO ($450 \, \mathrm{MeV}$ cut) & NNLO ($370 \, \mathrm{MeV}$ cut) & NNLO ($450 \, \mathrm{MeV}$ cut) \\
\hline 
\rule{0cm}{0.4cm} & $\chi^2/$dof & 0.011(5)  & 0.049(13)  & 0.008(4)  & 0.007(4)  \\
\hline
 \rule{0cm}{0.4cm}\multirow{3}{*}{24I} & $a m_{l}^{\rm phys}$ & -0.001767(79)  & -0.001774(81)  & -0.001765(79)  & -0.001765(79)  \\
 & $a m_{h}^{\rm phys}$ & 0.03236(32)  & 0.03206(29)  & 0.03237(32)  & 0.03238(30)  \\
 & $a^{-1}$ & 1.777(13) GeV & 1.797(12) GeV & 1.777(13) GeV & 1.777(12) GeV \\
\hline
 \rule{0cm}{0.4cm}\multirow{3}{*}{32I} & $a m_{l}^{\rm phys}$ & 0.000263(14)  & 0.000254(13)  & 0.000265(14)  & 0.000266(13)  \\
 & $a m_{h}^{\rm phys}$ & 0.02485(24)  & 0.02469(18)  & 0.02491(23)  & 0.02496(21)  \\
 & $a^{-1}$ & 2.371(16) GeV & 2.398(14) GeV & 2.369(16) GeV & 2.365(15) GeV \\
\hline
 \rule{0cm}{0.4cm}\multirow{3}{*}{32ID} & $a m_{l}^{\rm phys}$ & -0.000131(27)  & -0.000156(25)  & -0.000121(25)  & -0.000120(26)  \\
 & $a m_{h}^{\rm phys}$ & 0.04547(86)  & 0.04496(75)  & 0.04557(80)  & 0.04544(82)  \\
 & $a^{-1}$ & 1.389(13) GeV & 1.400(12) GeV & 1.387(12) GeV & 1.389(12) GeV \\
\hline
 \rule{0cm}{0.4cm}\multirow{3}{*}{32I-fine} & $a m_{l}^{\rm phys}$ & 0.000077(30)  & 0.000060(30)  & 0.000073(31)  & 0.000082(33)  \\
 & $a m_{h}^{\rm phys}$ & 0.01884(60)  & 0.01830(58)  & 0.01881(59)  & 0.01907(65)  \\
 & $a^{-1}$ & 3.110(43) GeV & 3.172(42) GeV & 3.114(43) GeV & 3.094(44) GeV \\
\hline
 \rule{0cm}{0.4cm}\multirow{3}{*}{48I} & $a m_{l}^{\rm phys}$ & 0.0006959(86)  & 0.0007012(75)  & 0.0006983(84)  & 0.0007001(81)  \\
 & $a m_{h}^{\rm phys}$ & 0.03574(18)  & 0.03588(14)  & 0.03575(17)  & 0.03580(16)  \\
 & $a^{-1}$ & 1.731(4) GeV & 1.728(3) GeV & 1.730(4) GeV & 1.729(4) GeV \\
\hline
 \rule{0cm}{0.4cm}\multirow{3}{*}{64I} & $a m_{l}^{\rm phys}$ & 0.0006175(78)  & 0.0006219(64)  & 0.0006192(74)  & 0.0006198(67)  \\
 & $a m_{h}^{\rm phys}$ & 0.02530(17)  & 0.02552(13)  & 0.02535(17)  & 0.02539(14)  \\
 & $a^{-1}$ & 2.362(7) GeV & 2.354(5) GeV & 2.360(7) GeV & 2.358(6) GeV \\
\hline
 \rule{0cm}{0.4cm}\multirow{3}{*}{32ID-M1} & $a m_{l}^{\rm phys}$ & 0.000825(68)  & 0.000731(47)  & 0.000808(65)  & 0.000797(51)  \\
 & $a m_{h}^{\rm phys}$ & 0.0791(16)  & 0.0753(10)  & 0.0784(16)  & 0.0778(12)  \\
 & $a^{-1}$ & 1.020(10) GeV & 1.039(7) GeV & 1.024(10) GeV & 1.029(7) GeV \\
\hline
 \rule{0cm}{0.4cm}\multirow{3}{*}{32ID-M2} & $a m_{l}^{\rm phys}$ & --- & -0.003417(20)  & ---  & -0.003413(23)  \\
 & $a m_{h}^{\rm phys}$ & ---  & 0.02435(48)  & ---  & 0.02422(55)  \\
 & $a^{-1}$ & --- & 2.048(19) GeV & --- & 2.030(22) GeV \\
\hline
\hline
\end{tabular}
}
\caption{The (uncorrelated) $\chi^{2}$/dof, unrenormalized physical quark masses in bare lattice units (without $m_{\mathrm{res}}$ included), and the values of the inverse lattice spacing $a^{-1}$ in physical units, obtained from fits to $SU(2)$ PQChPT with the stated pion mass cuts.}
\end{table}
\vspace*{\fill}

\begin{table}[p]
\centering
\resizebox{\columnwidth}{!}{
\begin{tabular}{cc||cc|cc}
\hline
\hline
\rule{0cm}{0.4cm} & & NLO ($370 \, \mathrm{MeV}$ cut) & NLO ($450 \, \mathrm{MeV}$ cut) & NNLO ($370 \, \mathrm{MeV}$ cut) & NNLO ($450 \, \mathrm{MeV}$ cut) \\
\hline
 \rule{0cm}{0.4cm}\multirow{3}{*}{24I} & $Z_{l}$ & 0.9710(53) & 0.9698(46) & 0.9702(51) & 0.9691(50) \\
 & $Z_{h}$ & 0.9618(39) & 0.9642(32) & 0.9626(38) & 0.9626(37) \\
 & $R_{a}$ & 0.7495(39) & 0.7493(36) & 0.7501(38) & 0.7515(39) \\
 \hline
 \rule{0cm}{0.4cm}\multirow{3}{*}{32I} & $Z_{l}$ & $\equiv 1.0$ & $\equiv 1.0$ & $\equiv 1.0$ & $\equiv 1.0$ \\
 & $Z_{h}$ & $\equiv 1.0$ & $\equiv 1.0$ & $\equiv 1.0$ & $\equiv 1.0$ \\
 & $R_{a}$ & $\equiv 1.0$ & $\equiv 1.0$ & $\equiv 1.0$ & $\equiv 1.0$ \\
 \hline
 \rule{0cm}{0.4cm}\multirow{3}{*}{32ID} & $Z_{l}$ & 0.9225(90) & 0.9310(87) & 0.9189(83) & 0.9170(86) \\
 & $Z_{h}$ & 0.9209(85) & 0.9279(75) & 0.9210(82) & 0.9228(84) \\
 & $R_{a}$ & 0.5857(60) & 0.5838(55) & 0.5855(57) & 0.5872(60) \\
 \hline
 \rule{0cm}{0.4cm}\multirow{3}{*}{32I-fine} & $Z_{l}$ & 0.998(30) & 1.003(31) & 1.003(31) & 0.997(33) \\
 & $Z_{h}$ & 0.999(19) & 1.012(20) & 1.001(19) & 0.994(21) \\
 & $R_{a}$ & 1.311(16) & 1.323(16) & 1.315(16) & 1.308(17) \\
 \hline
 \rule{0cm}{0.4cm}\multirow{3}{*}{48I} & $Z_{l}$ & 0.9710(53) & 0.9698(46) & 0.9702(51) & 0.9691(50) \\
 & $Z_{h}$ & 0.9618(39) & 0.9642(32) & 0.9626(38) & 0.9626(37) \\
 & $R_{a}$ & 0.7299(51) & 0.7205(43) & 0.7304(50) & 0.7311(48) \\
 \hline
 \rule{0cm}{0.4cm}\multirow{3}{*}{64I} & $Z_{l}$ & $\equiv 1.0$ & $\equiv 1.0$ & $\equiv 1.0$ & $\equiv 1.0$ \\
 & $Z_{h}$ & $\equiv 1.0$ & $\equiv 1.0$ & $\equiv 1.0$ & $\equiv 1.0$ \\
 & $R_{a}$ & 0.9963(60) & 0.9816(52) & 0.9963(58) & 0.9968(57) \\
 \hline
 \rule{0cm}{0.4cm}\multirow{3}{*}{32ID-M1} & $Z_{l}$ & 0.719(12) & 0.7291(86) & 0.720(11) & 0.7192(84) \\
 & $Z_{h}$ & 0.7303(100) & 0.7552(71) & 0.7345(98) & 0.7368(78) \\
 & $R_{a}$ & 0.4301(57) & 0.4332(41) & 0.4323(57) & 0.4351(44) \\
 \hline
 \rule{0cm}{0.4cm}\multirow{3}{*}{32ID-M2} & $Z_{l}$ & --- & 1.023(11) & --- & 1.027(12) \\
 & $Z_{h}$ & --- & 1.0300(84) & --- & 1.0405(93) \\
 & $R_{a}$ & --- & 0.8541(59) & --- & 0.8585(64) \\
\hline
\hline
\end{tabular}
}
\caption{Ratios of lattice spacings ($R_{a}$) and light and heavy quark masses ($Z_{l}$, $Z_{h}$) between each ensemble and the reference 32I ensemble.}
\end{table}

\begin{table}[p]
\centering
\resizebox{\columnwidth}{!}{
\begin{tabular}{cc||cc|cc}
\hline
\hline
\rule{0cm}{0.4cm} LEC & $\Lambda_{\chi}$ & NLO ($370 \, \mathrm{MeV}$ cut) & NLO ($450 \, \mathrm{MeV}$ cut) & NNLO ($370 \, \mathrm{MeV}$ cut) & NNLO ($450 \, \mathrm{MeV}$ cut) \\
\hline
 \rule{0cm}{0.4cm}$B$ & \multirow{2}{*}{---} & 4.246(22) GeV & 4.234(18) GeV & 4.235(26) GeV & 4.238(22) GeV \\
 $f$ & & 0.12298(93) GeV & 0.12153(77) GeV & 0.1226(13) GeV & 0.1229(11) GeV \\
\hline
 \rule{0cm}{0.4cm}$10^{3} \hat{L}_{0}^{(2)}$ & \multirow{10}{*}{1 GeV} & --- & --- & -4.5(4.8) & -0.2(2.0) \\
 $10^{3} \hat{L}_{1}^{(2)}$ & & --- & --- & 0.7(1.2) & -0.30(57) \\
 $10^{3} \hat{L}_{2}^{(2)}$ & & --- & --- & -4.4(3.3) & -0.9(1.3) \\
 $10^{3} \hat{L}_{3}^{(2)}$ & & --- & --- & 1.4(2.5) & -0.8(1.2) \\
 $10^{3} \hat{L}_{4}^{(2)}$ & & -0.193(77) & 0.024(55) & -0.36(36) & -0.48(20) \\
 $10^{3} \hat{L}_{5}^{(2)}$ & & 0.479(82) & 0.448(48) & 0.94(49) & 0.69(29) \\
 $10^{3} \hat{L}_{6}^{(2)}$ & & -0.165(48) & -0.004(35) & -0.25(17) & -0.345(99) \\
 $10^{3} \hat{L}_{7}^{(2)}$ & & --- & --- & -1.60(80) & -0.78(36) \\
 $10^{3} \hat{L}_{8}^{(2)}$ & & 0.604(41) & 0.532(24) & 0.81(22) & 0.73(14) \\
\hline
 \rule{0cm}{0.4cm}$10^{3} \hat{L}_{0}^{(2)}$ & \multirow{10}{*}{770 MeV} & --- & --- & -4.5(5.1) & -0.1(2.0) \\
 $10^{3} \hat{L}_{1}^{(2)}$ & & --- & --- & 0.8(1.2) & -0.20(58) \\
 $10^{3} \hat{L}_{2}^{(2)}$ & & --- & --- & -4.2(3.4) & -0.7(1.4) \\
 $10^{3} \hat{L}_{3}^{(2)}$ & & --- & --- & 1.6(2.7) & -0.6(1.3) \\
 $10^{3} \hat{L}_{4}^{(2)}$ & & 0.014(77) & 0.231(55) & -0.15(36) & -0.27(20) \\
 $10^{3} \hat{L}_{5}^{(2)}$ & & 0.893(82) & 0.862(48) & 1.35(48) & 1.11(29) \\
 $10^{3} \hat{L}_{6}^{(2)}$ & & -0.010(48) & 0.151(35) & -0.09(17) & -0.189(99) \\
 $10^{3} \hat{L}_{7}^{(2)}$ & & --- & --- & -1.61(84) & -0.78(36) \\
 $10^{3} \hat{L}_{8}^{(2)}$ & & 0.604(41) & 0.532(24) & 0.81(22) & 0.73(14) \\
\hline
 \rule{0cm}{0.4cm}$10^{6} \left( \hat{K}_{17}^{(2)}-\hat{K}_{39}^{(2)} \right)$ & \multirow{10}{*}{1 GeV} & --- & --- & -10.1(2.7) & -8.2(1.3) \\
 $10^{6} \left( \hat{K}_{18}^{(2)}+6\hat{K}_{27}^{(2)}-\hat{K}_{40}^{(2)} \right)$ & & --- & --- & 20(13) & 18.5(5.2) \\
 $10^{6} \hat{K}_{19}^{(2)}$ & & --- & --- & 6(25) & -2.9(8.0) \\
 $10^{6} \hat{K}_{20}^{(2)}$ & & --- & --- & -15(16) & -3.9(4.4) \\
 $10^{6} \left( \hat{K}_{21}^{(2)}+2K_{22}^{(2)} \right)$ & & --- & --- & -5.3(7.7) & 3.4(3.9) \\
 $10^{6} \hat{K}_{23}^{(2)}$ & & --- & --- & -10.3(5.7) & -2.6(2.3) \\
 $10^{6} \hat{K}_{25}^{(2)}$ & & --- & --- & 3.7(7.1) & -0.0(2.8) \\
 $10^{6} \left( \hat{K}_{26}^{(2)}+6\hat{K}_{27}^{(2)} \right)$ & & --- & --- & 6.3(7.7) & 10.7(3.3) \\
\hline
 \rule{0cm}{0.4cm}$10^{6} \left( \hat{K}_{17}^{(2)}-\hat{K}_{39}^{(2)} \right)$ & \multirow{10}{*}{770 MeV} & --- & --- & -8.3(2.1) & -6.1(1.0) \\
 $10^{6} \left( \hat{K}_{18}^{(2)}+6\hat{K}_{27}^{(2)}-\hat{K}_{40}^{(2)} \right)$ & & --- & --- & 12(10) & 13.7(3.9) \\
 $10^{6} \hat{K}_{19}^{(2)}$ & & --- & --- & -5(17) & -6.4(4.9) \\
 $10^{6} \hat{K}_{20}^{(2)}$ & & --- & --- & -7(11) & -0.1(2.7) \\
 $10^{6} \left( \hat{K}_{21}^{(2)}+2K_{22}^{(2)} \right)$ & & --- & --- & -5.6(8.0) & 4.2(3.5) \\
 $10^{6} \hat{K}_{23}^{(2)}$ & & --- & --- & -5.6(4.8) & -0.0(19) \\
 $10^{6} \hat{K}_{25}^{(2)}$ & & --- & --- & -1.0(5.2) & -2.3(1.9) \\
 $10^{6} \left( \hat{K}_{26}^{(2)}+6\hat{K}_{27}^{(2)} \right)$ & & --- & --- & 4.2(7.0) & 9.5(2.9) \\
\hline
\hline
\end{tabular}
}
\caption{$SU(2)$ PQChPT LECs fit at two different chiral scales --- $\Lambda_{\chi} = 1 \, \mathrm{GeV}$ and $\Lambda_{\chi} = 770 \, \mathrm{MeV}$ --- in units of the canonical size at a given order in the chiral expansion. The LECs $\hat{L}_{7}^{(2)}$ and $\hat{L}_{8}^{(2)}$ have no scale dependence. The value of $B$ quoted here is unrenormalized.}
\end{table}
 
\begin{table}[p]
\centering
\resizebox{\columnwidth}{!}{
\begin{tabular}{c||cc|cc}
\hline
\hline
\rule{0cm}{0.4cm} Parameter & NLO ($370 \, \mathrm{MeV}$ cut) & NLO ($450 \, \mathrm{MeV}$ cut) & NNLO ($370 \, \mathrm{MeV}$ cut) & NNLO ($450 \, \mathrm{MeV}$ cut) \\
\hline
 \rule{0cm}{0.4cm} $m^{(K)}$ & 0.4863(21) GeV & 0.4857(17) GeV & 0.4863(21) GeV & 0.4863(18) GeV \\
 $f^{(K)}$ & 0.15201(94) GeV & 0.15108(81) GeV & 0.15187(92) GeV & 0.15121(86) GeV \\
\hline
 \rule{0cm}{0.4cm}$10^{3} \lambda_1$ & 3.1(1.0) & 4.56(80) & 3.06(99) & 3.2(1.0) \\
 $10^{3} \lambda_2$ & 28.62(45) & 28.36(42) & 28.57(65) & 28.87(57) \\
 $10^{3} \lambda_3$ & -4.01(98) & -2.33(77) & -3.91(97) & -4.04(87) \\
 $10^{3} \lambda_4$ & 5.74(38) & 6.18(48) & 5.74(39) & 5.93(47) \\
\hline
 \rule{0cm}{0.4cm}$c_{f}^{\scriptscriptstyle I}$ & 0.007(22) $\mathrm{GeV}^{2}$ & 0.022(19) $\mathrm{GeV}^{2}$ & 0.018(25) $\mathrm{GeV}^{2}$ & 0.021(24) $\mathrm{GeV}^{2}$ \\
 $c_{f}^{\scriptscriptstyle ID}$ & -0.012(13) $\mathrm{GeV}^{2}$ & 0.016(10) $\mathrm{GeV}^{2}$ & -0.000(16) $\mathrm{GeV}^{2}$ & 0.008(14) $\mathrm{GeV}^{2}$ \\
 $c_{f^{(K)}}^{\scriptscriptstyle I}$ & 0.004(17) $\mathrm{GeV}^{2}$ & 0.009(16) $\mathrm{GeV}^{2}$ & 0.006(17) $\mathrm{GeV}^{2}$ & 0.017(16) $\mathrm{GeV}^{2}$ \\
 $c_{f^{(K)}}^{\scriptscriptstyle ID}$ & -0.003(11) $\mathrm{GeV}^{2}$ & 0.0131(82) $\mathrm{GeV}^{2}$ & -0.001(11) $\mathrm{GeV}^{2}$ & 0.0175(78) $\mathrm{GeV}^{2}$ \\
\hline
 \rule{0cm}{0.4cm}$c_{m_h,m_{\pi}^{2}}$ & 3.5(3.9) & 0.1(3.4) & 3.5(3.3) & 0.2(2.9) \\
 $c_{m_h,f_{\pi}}$ & 0.09(12) & 0.116(96) & 0.14(12) & 0.184(92) \\
 $c_{m_y,m_{K}^{2}}$ & 3.939(18) GeV & 3.953(15) GeV & 3.934(18) GeV & 3.930(16) GeV \\
 $c_{m_h,m_{K}^{2}}$ & 0.040(67) GeV & 0.167(76) GeV & 0.048(66) GeV & 0.017(63) GeV \\
 $c_{m_y,f_{K}}$ & 0.2903(88) & 0.2944(86) & 0.2879(84) & 0.3228(93) \\
 $c_{m_h,f_{K}}$ & 0.067(50) & 0.042(44) & 0.050(57) & 0.108(44) \\
\hline
 \rule{0cm}{0.4cm}$m^{(\Omega)}$ & 1.6645(36) GeV & 1.6614(25) GeV & 1.6643(34) GeV & 1.6651(29) GeV \\
 $c_{m_l,m_{\Omega}}$ & 3.63(64) & 5.05(57) & 3.73(65) & 3.33(63) \\
 $c_{m_y,m_{\Omega}}$ & 5.678(81) & 5.39(12) & 5.633(78) & 5.537(74) \\
 $c_{m_h,m_{\Omega}}$ & 1.99(48) & 1.23(41) & 1.80(52) & 1.52(40) \\
\hline
\hline
\end{tabular}
}
\caption{Additional fit parameters in physical units and adjusted to the physical strange quark mass. Here $\{m^{(K)},f^{(K)}\}$ and $\{\lambda_{i}\}$ are the LO and NLO LECs of heavy-meson $SU(2)$ PQChPT evaluated at the chiral scale $\Lambda_{\chi} = 1 \, \mathrm{GeV}$. $c_{f}^{I}$ and $c_{f}^{ID}$ are the $a^{2}$ coefficients of $f_{\pi}$ for the Iwasaki and Iwasaki+DSDR gauge actions, respectively, and likewise for $c_{f^{(K)}}^{I}$ and $c_{f^{(K)}}^{ID}$. The notation $c_{m_{q}, X}$ denotes the coefficient of a term linear in $m_{q}$ for quantity $X$, and $m^{(\Omega)}$ is the constant term in the (linear) $m_{\Omega}$ ansatz.}
\label{tab:more_su2_parameters_normalized}
\end{table}
\FloatBarrier

\subsection{Predictions}

\vspace*{\fill}
\begin{table}[h]
\centering
\resizebox{\columnwidth}{!}{
\begin{tabular}{cc||cc|cc}
\hline
\hline
\rule{0cm}{0.4cm}LEC & $\Lambda_{\chi}$ & NLO ($370 \, \mathrm{MeV}$ cut) & NLO ($450 \, \mathrm{MeV}$ cut) & NNLO ($370 \, \mathrm{MeV}$ cut) & NNLO ($450 \, \mathrm{MeV}$ cut) \\
\hline
\rule{0cm}{0.4cm}$B^{\overline{\mathrm{MS}}}(\mu = 2 \, \mathrm{GeV})$ & \multirow{3}{*}{---} & 2.815(33)(30) GeV & 2.808(31)(30) GeV & 2.808(36)(30) GeV & 2.811(35)(30) GeV \\
 $f$ & & 123.0(9) MeV & 121.5(8) MeV & 122.6(1.3) MeV & 122.9(1.1) MeV \\
$\Sigma^{1/3, \, \overline{\mathrm{MS}}}(\mu = 2 \, \mathrm{GeV})$ & & 277.2(1.8)(1.0) MeV & 274.7(1.5)(1.0) MeV & 276.3(2.1)(1.0) MeV & 276.9(1.9)(1.0) MeV \\
\hline
 \rule{0cm}{0.4cm}$10^{3} l_{1}$ & \multirow{5}{*}{1 GeV} & --- & --- & 15(19) & -2.4(7.6) \\
 $10^{3} l_{2}$ & & --- & --- & -35(32) & -5(13) \\
 $10^{3} l_{3}$ & & 1.82(26) & 2.22(20) & 1.62(79) & 1.36(56) \\
 $10^{3} l_{4}$ & & 0.37(52) & 1.98(36) & 0.8(1.5) & -1.05(99) \\
 $10^{3} l_{7}$ & & --- & --- & 19(12) & 6.7(5.4) \\
\hline
 \rule{0cm}{0.4cm}$10^{3} l_{1}$ & \multirow{5}{*}{770 MeV} & --- & --- & 16(19) & -1.8(7.7) \\
 $10^{3} l_{2}$ & & --- & --- & -35(33) & -3(13) \\
 $10^{3} l_{3}$ & & 0.99(26) & 1.39(20) & 0.78(79) & 0.54(56) \\
 $10^{3} l_{4}$ & & 3.68(52) & 5.29(36) & 4.2(1.6) & 2.26(99) \\
 $10^{3} l_{7}$ & & --- & --- & 19(13) & 6.7(5.4) \\
\hline
 \rule{0cm}{0.4cm}$\overline{\ell}_{1}$ & \multirow{4}{*}{---} & --- & --- & 18(18) & 1.8(7.2) \\
 $\overline{\ell}_{2}$ & & --- & --- & -13(15) & 1.9(6.2) \\
 $\overline{\ell}_{3}$ & & 2.86(16) & 2.61(12) & 2.98(50) & 3.14(35) \\
 $\overline{\ell}_{4}$ & & 4.064(82) & 4.318(57) & 4.14(24) & 3.84(16) \\
\hline
\hline
\end{tabular}
}
\caption{Unquenched $SU(2)$ LECs computed from partially quenched $SU(2)$ fits. Missing entries are not constrained by the fits at a given order. For $B$ and $\Sigma$ the first error is statistical and the second is a systematic uncertainty in the perturbative matching to $\rm \overline{MS}$.}
\label{tab:unquenched_su2_lecs_normalized}
\end{table}
\vspace{2.5cm}
\begin{table}[h]
\centering
\resizebox{\columnwidth}{!}{
\begin{tabular}{c||cc|cc}
\hline
\hline
\rule{0cm}{0.4cm}& NLO ($370 \, \mathrm{MeV}$ cut) & NLO ($450 \, \mathrm{MeV}$ cut) & NNLO ($370 \, \mathrm{MeV}$ cut) & NNLO ($450 \, \mathrm{MeV}$ cut) \\
\hline
 \rule{0cm}{0.4cm}$f_{\pi}$ & 0.13074(84) GeV & 0.12986(71) GeV & 0.13032(94) GeV & 0.13011(89) GeV \\
 $f_{K}$ & 0.15587(79) GeV & 0.15542(70) GeV & 0.15577(78) GeV & 0.15508(71) GeV \\
 $f_{K}/f_{\pi}$ & 1.1922(41) & 1.1968(39) & 1.1953(59) & 1.1919(56) \\
 $f_{\pi}/f$ & 1.0631(18) & 1.0686(13) & 1.0631(44) & 1.0583(29) \\
\hline
\rule{0cm}{0.4cm}$m_{\pi} a_{0}^{0}$ & --- & --- & 0.153(33) & 0.185(14) \\
 $m_{\pi} a_{0}^{2}$ & --- & --- & -0.057(13) & -0.0431(53) \\
\hline
 \rule{0cm}{0.5cm}$[ m_{\pi^{\pm}}^{2} - m_{\pi^{0}}^{2} ]_{\rm QCD} / \Delta m_{du}^{2}$ & --- & --- & 91(57) & 32(26) \\
\hline
\hline
\end{tabular}
}
\caption{Predictions from NLO and NNLO fits and $SU(2)$ ChPT. $\Delta m_{d u} \equiv m_{d} - m_{u}$.}
\label{tab:su2_predictions_normalized}
\end{table}
\vspace*{\fill}

\end{document}